\pdfoutput=1
\documentclass[11pt]{article}
\usepackage[margin=1.3in]{geometry}
\usepackage{booktabs}
\usepackage[utf8]{inputenc}
\usepackage[UKenglish]{babel}

\usepackage{amsmath}
\usepackage{amsthm}
\usepackage{amsfonts}
\usepackage{amssymb}

\usepackage{algorithm}
\usepackage[noend]{algpseudocode}
\usepackage{float}

\usepackage{enumitem}
\usepackage{multicol}
\usepackage[hypertexnames=false]{hyperref}
\hypersetup{pdftitle={Fast Evaluation of Polynomials with Rational Preprocessing},pdfauthor={Thomas D. Ahle and Jakob B. T. Knudsen}}
\newcommand{\coderepo}{\url{https://github.com/thomasahle/fast-polynomials}}
\makeatletter
\providecommand*{\theHALG@line}{\arabic{algorithm}.\arabic{ALG@line}}
\makeatother
\usepackage{cleveref}
\newtheorem{theorem}{Theorem}
\newtheorem{proposition}[theorem]{Proposition}
\newtheorem{lemma}{Lemma}
\newtheorem{corollary}{Corollary}

\theoremstyle{definition}
\newtheorem{definition}{Definition}
\newtheorem{example}{Example}
\theoremstyle{remark}
\newtheorem{remark}{Remark}
\usepackage{autonum}
\usepackage{graphicx}
\usepackage{tikz}
\usetikzlibrary{arrows.meta,positioning,calc}
\definecolor{fpBlue}{RGB}{53,105,171}
\definecolor{fpOrange}{RGB}{222,132,48}
\definecolor{fpGreen}{RGB}{70,139,96}
\definecolor{fpRed}{RGB}{186,62,55}
\definecolor{fpPurple}{RGB}{119,91,160}
\definecolor{fpGray}{RGB}{108,112,118}

\tikzset{
  fp strip/.style={draw=black!65, line width=.45pt, minimum height=5.5mm,
    inner sep=1.4pt, font=\scriptsize, align=center},
  fp active/.style={fp strip, fill=fpBlue!15, draw=fpBlue!85!black},
  fp second/.style={fp strip, fill=fpOrange!18, draw=fpOrange!85!black},
  fp recursive/.style={fp strip, fill=fpGreen!15, draw=fpGreen!80!black},
  fp known/.style={fp strip, fill=black!7, draw=fpGray, dashed},
  fp unread/.style={fp strip, fill=white, draw=black!45, densely dotted},
  fp seam/.style={fp strip, fill=fpRed!18, draw=fpRed!90!black,
    line width=.7pt},
  fp pivot/.style={fp strip, fill=fpPurple!17, draw=fpPurple!90!black,
    line width=.7pt},
  fp gate/.style={circle, draw=black!70, fill=white, minimum size=6mm,
    inner sep=0pt, font=\small},
  fp flow/.style={-{Latex[length=2.2mm]}, line width=.65pt, draw=black!75},
  fp decode/.style={-{Latex[length=2.4mm]}, line width=1pt, draw=fpBlue!85!black},
  fp dependence/.style={-{Latex[length=1.8mm]}, densely dashed,
    line width=.5pt, draw=fpGray},
  fp label/.style={font=\scriptsize, align=center, inner sep=1pt},
  fp panel title/.style={font=\small\bfseries, anchor=west},
  fp brace label/.style={font=\scriptsize, align=center, text=black!75}
}

\allowdisplaybreaks

\usepackage{mathtools}
\DeclarePairedDelimiterX{\infdivx}[2]{(}{)}{#1 \,\delimsize\|\, #2}

\newcommand{\R}{\mathbb{R}}

\newcommand{\F}{\mathbb{F}}
\newcommand{\C}{\mathbb{C}}
\newcommand{\Q}{\mathbb{Q}}

\newcommand{\abs}[1]{\left\lvert #1\right\rvert}

\newcommand{\floor}[1]{\left\lfloor {#1} \right\rfloor}

\newcommand{\idx}[1]{\{0,\dotsc,#1\}}
\newcommand{\rng}[1]{\{0,\dotsc,#1-1\}}

\newcommand{\coeff}[2]{[x^{#2}]\,#1}
\newcommand{\polyfrom}{\mathrel{\xleftarrow{\mathrm{poly}}}}

\title{Fast Evaluation of Polynomials with Rational Preprocessing}
\author{Thomas D. Ahle\thanks{Normal Computing. Email: \texttt{thomas@ahle.dk}.} \and Jakob B. T. Knudsen\thanks{Alipes ApS.}}

\begin{document}

\maketitle

\begin{abstract}
   Horner's rule evaluates a monic degree-$n$ polynomial using $n-1$ multiplications.
   We show that with rational preprocessing of the coefficients, any such polynomial
   can be evaluated using only $\lfloor n/2 \rfloor + 1$ multiplications over
   fields of characteristic zero or of characteristic $p>n$.
   This resolves the multiplication side of a conjecture of Rabin and Winograd (Comm. Pure Appl. Math 1972), who achieved
   $n/2 + 2\lceil\log_2 n\rceil$ multiplications and conjectured the
   logarithmic overhead was necessary.

   We show that this multiplication count can't be beaten in general,
   proving that three multiplications do not suffice for degree~$6$ when the
   rational preprocessing is everywhere defined.
   This separates rational from algebraic preprocessing: with complex
   preprocessing three multiplications do suffice for monic sextics
   (Motzkin, Belaga, Pan).

   In characteristic~2, for every $n>1$ and every finite field of size at least $2n$, we prove that
   an $n$-multiplication chain cannot parametrise all value vectors at $2n$
   distinct evaluation points, even with arbitrary preprocessing.
   We give $\lfloor n/2 \rfloor + 1$ multiplication schedules over
   characteristic~2, each with an explicit inverse, for every odd degree
   $n\le 21$ and conjecture that this is possible for all $n$.

   We also give an injective polynomial construction for universal hashing that uses
   $N$ multiplications to hash $2N$ values with a single random key.
   This improves the best previous construction by Daniel J. Bernstein (cryp.to).

   Experiments on ARM and x86 processors show up to $2\times$ speedups for
   $k$-wise independent hashing and up to $4.6\times$ speedups for
   the prime-field evaluation kernel of Shamir secret sharing~\cite{shamir1979share}.
\end{abstract}

\section{Introduction}

Polynomial evaluation sits at the bottom of a surprising number of algorithms:
interpolation and approximation,
root-finding and optimization,
coding theory \& cryptography,
simulation, and data analysis.
We therefore revisit the classic task of evaluating
\[
   P = a_0 + a_1 x + a_2 x^2 + \dots + a_{n-1} x^{n-1} + x^{n},
\]
where the coefficients $a_i$ are known to the programmer.
Following Motzkin and the classical literature we take $P$ to be \emph{monic}: a
leading coefficient costs exactly one extra multiplication in every scheme
displayed here, so all counts in this paper are for monic polynomials and
increase by one for general ones.
Horner's method evaluates $P$ using just $n-1$ multiplications:
$P = a_0 + x(a_1 + x(a_2 + \dots + x(a_{n-1} + x)\dots))$ and has been known for centuries.\footnote{Although named after William George Horner, this method is much older, as it has been attributed to Joseph-Louis Lagrange by Horner himself, and can be traced back many hundreds of years to Chinese and Persian mathematicians; see, e.g., Knuth~\cite{knuth1962evaluation}.}
Thus it is surprising to many that this is not optimal.

The trick is to allow some preprocessing of the coefficients $a_0, \dots, a_{n-1}$.
For example, the polynomial
\[
P = 9 + 7 x + 5 x^2 + 3 x^3 + x^4
\] can be evaluated by first computing
$y = x (x + 1)$ and then $P = (x + y - 1)(y + 4) + 13$ using just two multiplications rather than three with Horner.

In the 50s Motzkin~\cite{motzkin1955evaluation} showed that, after preprocessing the coefficients, any monic polynomial of degree $n$ can be evaluated using just $\lceil n/2\rceil$ multiplications (Belaga~\cite{belaga1958some}, see scheme~(0.5) in Pan's survey~\cite{pan1966methods}; Knuth's formulation of the method~\cite{knuth1997seminumerical} uses $\lfloor n/2\rfloor+2$ multiplications for a general polynomial).
The downside is that the method requires solving a polynomial system of equations over the complex numbers, which reduces precision and breaks over finite fields.
Many other methods have been proposed, but rational preprocessing was known only in special cases---Pan's sextic scheme~\cite{pan1961schemes} and individual polynomials for which the preprocessing systems of Motzkin and Belaga~\cite{belaga1958some} happen to have rational solutions, such as the quartic above---until Winograd, and Rabin and Winograd~\cite{rabin1972number}, gave a general way to preprocess the coefficients using \emph{rational operations} (\Cref{sec:related}).
Unfortunately, their method requires $n/2 + 2\lceil\log n\rceil$ multiplications, and they suspected that this logarithmic overhead might be necessary.

\begin{quote}
   We feel that [the Theorems] may be, in a certain sense, the best results possible for rational preconditioning.
   In Theorem 15, we are within $O(\log n)$ of the in bound $n/2$ multiplications, using $n + o(n)$ additions.
   Can this be improved by either reducing the $O(\log n)$ excess multiplications or, keeping this, reducing the $(n + o(n))A$ [additions] part?
   We have some heuristic arguments to support a suspicion that the above results may be optimal in terms of order of magnitude, but this seems to be a hard problem.
\end{quote}

In this paper we answer Rabin and Winograd's conjecture in the negative by showing how to evaluate any monic polynomial of degree $n$ using just $\lfloor n/2\rfloor+1$ multiplications with \emph{rational preprocessing}.  Their question had two halves, the $O(\log n)$ excess multiplications and the $n+o(n)$ additions; we settle the first and pay on the second, using $\frac54n+O((\log n)^2)$ additions rather than $n+o(n)$, and whether $\lfloor n/2\rfloor+1$ multiplications can be combined with $n+o(n)$ additions is open.  The addition count matches the $\frac54n$ of Winograd's 1970 monic construction (Knuth's exercise~44~\cite[\S4.6.4]{knuth1997seminumerical}), the precursor of Rabin--Winograd's, which later reduced the additions to $(1+\varepsilon)n$ at the same $\frac12n+O(\log n)$ multiplications: the constant $\frac54$ survived our change of construction, while the logarithmic excess of multiplications did not.  At odd degrees the multiplication count is optimal for monic polynomials under any preprocessing whatsoever, by Motzkin's bound (for a general polynomial of odd degree $n\ge9$ our $\lfloor n/2\rfloor+2$ is one more than the complex optimum of Revah and Pan~\cite{revah1975number,pan1978complex}); at even degrees it is one more than the algebraic optimum, and we prove at degree~$6$ that the extra multiplication is necessary once the rational (single-valued) preprocessing must have no exceptional inputs, conjecturing the same at every even degree at least~$6$.  The conjecture is specific to single-valued rational preprocessing: for algebraic preprocessing Pan conjectures the opposite~\cite{pan1978complex} (\Cref{sec:open-problems}).  With fixed scalar multiplications free, the same schedules use at most $2n$ additions or subtractions uniformly and at most $\frac54n+O((\log n)^2)$ asymptotically.  The leading constant is therefore within $25\%$ of the classical $n$-addition lower bound~\cite{belaga1958some,belaga1961evaluation}, a comparison across cost models: the classical count has no free constant multiples, and synthesising ours by doubling costs about $4\%$ more additions (\Cref{sec:open-problems}).

\paragraph{Main Results.}
Section~\ref{sec:model} formalizes our model (arithmetic circuits / straight-line programs) and the notion of rational preprocessing/decoding.
Our main results are:
\begin{itemize}
   \item \textbf{Upper bound.} For every $n\ge 1$ and every field of characteristic zero, or of characteristic $p>n$, there is an evaluation scheme for monic degree-$n$ polynomials that uses $\lfloor n/2\rfloor+1$ multiplications and admits rational preprocessing.  Its addition count is at most
   $\min\{2n,\frac54n+6\lceil\log_2n\rceil^2+1\}$
   (Theorem~\ref{thm:main}).  The schedules can be arranged with
   multiplicative height at most $2\lceil\log_2n\rceil+4$ (odd $n$; $+5$
   for even $n$) at the same gate
   counts (\Cref{thm:construction-height}), and the complete
   statement for every $n\ge3$---multiplication count, decoder, and
   height---is formalized in Lean (\Cref{appendix:formalization-map}).  The
   schemes are exact, and that is where they belong: in
   floating-point arithmetic the rational keys of an arbitrary prescribed
   polynomial are large and cancel, so the schedules are numerically unstable
   there (\Cref{sec:numerical-stability} quantifies this and compares with
   Horner, Estrin, Rabin--Winograd, and Motzkin--Eve); the intended use is
   modular or exact arithmetic, as in the hashing applications below, where
   the keys are the data and the question does not arise.
   For characteristic~2, we have explicit single-circuit constructions, each
   with an explicit inverse, at every odd degree up to~21 and conjecture the
   fixed-circuit bound holds in general.
   \item \textbf{Lower bound.} Motzkin's bound shows that monic polynomials of degree $n$ need $\lceil n/2\rceil$ multiplications under any preprocessing, so our odd-degree schemes are optimal for monic polynomials.  For even degrees we prove that degree-$6$ evaluation with everywhere-defined rational decoding requires at least $4$ multiplications (Section~\ref{sec:lower})---one more than Pan's rational sextic scheme, which is undefined on a hypersurface---and conjecture that degree $2n$ requires $n+1$ under everywhere-defined rational decoding for every $n\ge3$ (at $n=2$ two multiplications suffice, by Motzkin's quartic).
      The key insight is that a polynomial automorphism of $\mathbb{C}^n$ can't have a zero Jacobian determinant, and we show any evaluation scheme with rational decoding would induce such a map if it used too few multiplications.
   \item \textbf{Injective Polynomials}. We give an ``injective'' polynomial hashing construction requiring only a single random key (Section~\ref{sec:injective}) and application-level benchmarks for hashing, data structures, and prime-field workloads (Section~\ref{sec:experiments}).
   \item \textbf{Constructions over $GF(2^n)$.}
      Our rational decoder applies in characteristic $p>n$, but it does not necessarily work in small characteristic, in particular characteristic~2, which is important in computing.
      We give fixed evaluation schemes over fields of characteristic~2 at
      every odd degree up to 21, each with an explicit inverse
      (\Cref{appendix:polynomials}), and we conjecture that such fixed
      schemes exist for all $n$ and all finite fields.
      A key insight is that we can use the bijectivity of the Frobenius map $x \mapsto x^2$.
      In the opposite direction we prove a matching lower bound in characteristic~2
      (\Cref{thm:char2-lower}): over a finite field of characteristic~2 with at least
      $2n$ elements, no chain with $n$ multiplications evaluates bijectively onto
      $\F^{2n}$ at $2n$ points, for any $n>1$ and whatever preprocessing of the
      parameters is allowed---affine, rational, or arbitrary.  Unlike \Cref{sec:lower}
      this covers every $n$, and assumes nothing about the degree.
   \item \textbf{Experiments.}
      Fast evaluation of small low-degree polynomials is important in many applications, such as cryptography~\cite{DBLP:conf/fse/CarletGPQR12} and $k$-wise independent hashing~\cite{wegman1981new, patracscu2012power}.
      We give an implementation of our method for $k$-wise independent hashing and show speedups of up to $1.8\times$ (ARM) and $2\times$ (x86) over standard Horner evaluation for $k \in \{5, \ldots, 9\}$ on modern ARM and x86 processors (Section~\ref{sec:experiments}).
\end{itemize}

\section{Polynomial Evaluation with \texorpdfstring{$\lfloor n/2\rfloor+1$}{floor(n/2)+1} Multiplications and Rational Preprocessing}
\label{sec:model}

We work over a field $\F$.
An \emph{arithmetic circuit} (or straight-line program) takes inputs $(x,\theta)\in\F\times\F^d$ and outputs a value in $\F$ using additions, subtractions, and multiplications.
We measure the primary cost of a circuit by the number of multiplications,
where a multiplication is charged whenever neither factor is a fixed element
of $\F$: products of two $x$-dependent values, of a parameter with an
$x$-dependent value such as $\theta x$, and of two parameters all count.
For the supplementary addition count, additions and subtractions are charged
separately while copying, negation, and multiplication by fixed field
constants remain free.

Fix $n\ge 1$.
An \emph{evaluation scheme} of (target) degree $n$ is a circuit $E(x;\theta)$ such that for every $\theta\in\F^d$, the function $x\mapsto E(x;\theta)$ is a polynomial of degree at most $n$.
Let $\mathsf{coeff}(\theta)\in\F^n$ denote the coefficient vector $(c_0,\dots,c_{n-1})$ of that polynomial, where we write
\[
E(x;\theta)=c_0 + c_1 x + \dots + c_{n-1}x^{n-1} + x^n.
\]
Thus $\mathsf{coeff}:\F^d\to\F^n$ is a polynomial map.

\paragraph{Rational preprocessing/decoding.}
We say $E$ has \emph{rational preprocessing} (or \emph{rational decoding}) if there is a rational map
\[
\mathsf{pre}:\F^n \dashrightarrow \F^d
\]
such that for every coefficient vector $c$ in the domain of $\mathsf{pre}$ we have
\[
E(x;\mathsf{pre}(c)) \equiv c_0 + c_1 x + \dots + c_{n-1}x^{n-1} + x^n
\]
as polynomials in $x$.
Thus $\mathsf{pre}$ is a rational right-inverse, or rational section, of
$\mathsf{coeff}$ on a Zariski-open set.
We say the preprocessing is \emph{everywhere defined}, or has \emph{no exceptional
inputs}, if the domain of $\mathsf{pre}$ is all of $\F^n$.
The distinction matters: Pan's sextic scheme (\Cref{sec:lower}) has rational
preprocessing with three multiplications, but it is undefined on a hypersurface
of coefficient vectors, whereas the schemes of this paper are everywhere defined,
and the lower bound of \Cref{sec:lower} is about everywhere-defined preprocessing.
A right-inverse alone does not imply
injectivity.  Our constructions prove more: their explicit decoder is a
polynomial left-inverse of $\mathsf{coeff}$.  Since the parameter and
coefficient spaces both have dimension $n$, the algebraic argument in
\Cref{lem:polynomial-left-inverse-automorphism} makes the two inverses agree;
the coefficient map is in fact a polynomial automorphism.

\paragraph{Hashing viewpoint.}
If we sample $\theta$ at random and evaluate $E(x;\theta)$ at an input $x$, we obtain a polynomial-hash family.
For the coefficient maps constructed here, distinct parameters induce
distinct polynomials and every monic coefficient vector occurs.  This
two-sided property---which is stronger than merely having a rational
section---is what lets preprocessing be interpreted as selecting a polynomial
from a well-defined family.

\subsection{Main Result}
\label{sec:main-theorem}

Our main contribution is an evaluation scheme with essentially the minimum possible number of multiplications while still admitting rational preprocessing.

\begin{theorem}
\label{thm:main}
For every $n\ge 1$ and every field $\F$ of characteristic zero, or of characteristic $p>n$, there is an evaluation scheme for monic degree-$n$ polynomials over $\F$ that uses $\lfloor n/2\rfloor+1$ multiplications, admits rational preprocessing, and uses at most
\[
  \min\left\{2n,\ \frac54n+6\lceil\log_2n\rceil^2+1\right\}
\]
additions or subtractions, and can be arranged with multiplicative
height at most $2\lceil\log_2 n\rceil+4$ for odd $n$ and
$2\lceil\log_2 n\rceil+5$ for even $n$
(\Cref{thm:construction-height}).
\end{theorem}

\begin{remark}[Positive characteristic]
\label{rem:char2}
The condition $p>n$ is a convenient sufficient condition, not a sharp one.
The decoder divides only by explicitly displayed integer pivots, and every
prime divisor of those pivots is at most~$n$.  Thus the same construction works in any
characteristic in which the pivots occurring in the degree-$n$ recursion are
nonzero; $p>n$ guarantees this uniformly.  In small positive characteristic
some pivots may vanish.  This is the general form of Knuth's remark that
Motzkin's quartic needs $2u_4$ invertible and Pan's sextic $6u_6$
together with the denominator of its
preprocessing~\cite[p.~494]{knuth1997seminumerical}; Belaga's normalisation
likewise divides by $n+1$~\cite{belaga1958some}.  The characteristic-$2$ circuits of
\Cref{appendix:polynomials} and \Cref{sec:ph} are decoded by field operations
and inverse Frobenius, $r\mapsto r^{1/2}$; on a fixed finite field $\F_{2^m}$
this is the polynomial function $r\mapsto r^{2^{m-1}}$, so those decoders are
everywhere-defined polynomial \emph{functions} on each such field rather than
rational maps in the algebraic-geometric sense, and for them ``rational
preprocessing'' is to be read in this way: an explicit inverse built from
field operations and $r\mapsto r^{1/2}$, valid on every perfect field of
characteristic~$2$.  A complete \emph{single-circuit} construction with
rational preprocessing covering all such cases, including characteristic~2,
remains open.
\end{remark}

\subsection{Intuition and Construction Outline}
\label{sec:main-theorem:outline}

This section gives the high-level intuition behind the construction.
The proof consists of making the invariants below precise and checking that every step preserves rational decoding.

\paragraph{From rational preprocessing to decoding coefficients.}
Fix a circuit family $E(x;\theta)$ and write $c=\mathsf{coeff}(\theta)\in\F^n$ for the output coefficient vector as in \Cref{sec:model}.
Rational preprocessing asks us to choose parameters for a desired coefficient
vector, that is, to construct a rational right-inverse of $\mathsf{coeff}$ on
a dense set.  We establish this through the stronger, easier-to-compose
invariant that the parameters can be \emph{decoded} polynomially from the
coefficients by explicit triangular relationships.  The two-sided lemma cited
above then turns that decoder into the required preprocessing map.

\paragraph{A warm-up: degree $3$ with two multiplications.}
Let $n=3$.
Compute $H_2(x)=x^2$ using one multiplication, and then define
\[
   Q_3(x;\alpha_0,\alpha_1,\alpha_2) \;=\; (x+\alpha_2)(H_2(x)+\alpha_1)+\alpha_0.
\]
This uses one additional multiplication, so two in total, matching $\lfloor 3/2\rfloor+1$.
Expanding $H_2(x)=x^2$ gives
\[
   Q_3(x)=x^3 + \alpha_2 x^2 + \alpha_1 x + (\alpha_1\alpha_2+\alpha_0),
\]
so the coefficient map is (almost) triangular:
$\alpha_2=\coeff{Q_3}{2}$ and $\alpha_1=\coeff{Q_3}{1}$ are read off directly, and then $\alpha_0=\coeff{Q_3}{0}-\alpha_1\alpha_2$.
This illustrates the general strategy: we use each multiplication to introduce (roughly) two new degrees of freedom, while keeping decoding algebraic and explicit.

\paragraph{Splitting viewpoint (``splittable pairs'').}
For larger degrees, it is convenient to construct the output polynomial as
\[
   P(x) \;=\; x\cdot T^{(1)}(x) + T^{(2)}(x),
\]
where $T^{(1)}$ and $T^{(2)}$ are degree-$(n-1)$ polynomials computed with shared intermediate values.
The final multiplication by $x$ contributes the ``$+1$'' in the multiplication count.
We call $(T^{(1)},T^{(2)})$ a \emph{splittable pair} (for degree $n$) if the coefficients of $P(x)=xT^{(1)}(x)+T^{(2)}(x)$ admit an explicit polynomial decoder for all underlying parameters.
Splittability by itself says nothing about the cost of producing the two
components: every decodable monic polynomial has a canonical compatible
split.  The stronger invariant used below is therefore a splittable pair
together with one shared circuit that jointly computes both components at
the stated cost and records the quadratic and quartic intermediates needed
by the recursion.

\paragraph{Keeping decoding stable: coefficient windows and ``compatible pairs''.}
The main technical problem is that multiplication mixes coefficients.
To control this, we maintain an invariant of the following form.
We consider a pair $(P^{(1)},P^{(2)})$ and its combination $\Phi(x)\coloneqq xP^{(1)}(x)+P^{(2)}(x)$.
We ensure that there is a \emph{window} of coefficient positions (typically a top-degree range) from which one can recover the parameters introduced so far and also recover the pair $(P^{(1)},P^{(2)})$ itself.
Pairs with this property are called \emph{compatible pairs}; the formal definition is \Cref{def:compatible-pair} in \Cref{appendix:decoder-calculus}.

Crucially, the basic operations used in the straight-line program preserve compatibility:
\begin{itemize}
   \item adding low-degree polynomials (which only affects low coefficients);
   \item multiplying by a \emph{known-power gadget} $H_{2^i}(x)$ (a fixed monic degree-$2^i$ polynomial, treated as known to the decoder),
         which shifts blocks of coefficients by $2^i$ positions without destroying recoverability of a top window;
   \item combining two compatible pairs in a block-triangular way so that the coefficient windows do not interfere.
\end{itemize}
These closure properties are what lets us scale the warm-up decoding argument beyond tiny degrees.

\paragraph{Recursion: filling coefficients with a BRW-style structure.}
At a high level, we build a decodable family $Q_{2^k-1}(x;\alpha)$ of monic degree $(2^k-1)$ polynomials using a divide-and-conquer pattern reminiscent of Rabin--Winograd / Bernstein constructions~\cite{rabin1972number}:
we keep a ladder of known-power gadgets $H_2,H_4,\dots,H_{2^{k-1}}$ (computed once and reused),
and we recursively ``fill'' disjoint coefficient windows by multiplying previously constructed pieces by the appropriate $H_{2^i}$ and adding a fresh low-degree correction.
The compatibility invariant guarantees that each recursive layer introduces new parameters that remain decodable from the top coefficients, despite coefficient mixing in intermediate multiplications.
The resulting circuits use $\Theta(2^{k-1})$ multiplications for degree $\Theta(2^k)$, i.e.\ about half the multiplications of Horner.
Concretely, the known-powers recursion is
\[
  Q_{2^k-1}=(H_{2^{k-1}}+\gamma)W+B,
\]
where $W$ and $B$ are fresh degree-$(2^{k-1}-1)$ known-powers children
computed in parallel (\Cref{alg:constr-known-2n-1}).
Monic division recovers $W$, the top remainder coefficient recovers
$\gamma$, and subtraction recovers $B$.
This binary recursion is used for the multiplication, addition, and
height bounds alike; the complete construction has logarithmic height
(\Cref{thm:construction-height}).

\paragraph{From the recursion to \Cref{thm:main}.}
The recursion above yields compatible splittable pairs together with one
costed circuit that jointly computes their two components (with a small
number of explicit cost bases).
Given such a joint realization of $(T^{(1)},T^{(2)})$ for degree $n$, the
circuit for $P(x)=xT^{(1)}(x)+T^{(2)}(x)$ uses one more multiplication than
the shared computation of $T^{(1)}$ and $T^{(2)}$.
For even degrees, we reduce to the odd case by writing
\(
   P(x)=c_0 + x\cdot Q(x)
\)
where $Q$ is a monic degree-$(n-1)$ polynomial with coefficients $(c_1,\dots,c_{n-1})$; this costs one extra multiplication by $x$.
In all cases we obtain $\lfloor n/2\rfloor+1$ multiplications overall, and the decoding invariants ensure that the induced coefficient map is generically invertible by rational functions, i.e.\ admits rational preprocessing in the sense of \Cref{sec:model}.

\subsection{\texorpdfstring{$k$}{k}-wise independent hashing and why bijectivity matters}
\label{sec:main-theorem:kwise}

\paragraph{Background: $k$-wise independence via polynomials.}
Let $\F$ be a field and fix $k\ge 1$.
A standard way to build a $k$-wise independent hash family $\F\to\F$ is to choose a random
degree-$k$ \emph{monic} polynomial and evaluate it on the input~\cite{wegman1981new}.
Concretely, for a coefficient vector $c=(c_0,\dots,c_{k-1})\in\F^k$ define
\begin{equation}
  h_c(x) \coloneqq x^k + c_{k-1}x^{k-1} + \cdots + c_1 x + c_0.
  \label{eq:kwise:monic-poly}
\end{equation}
If $c\gets\F^k$ is uniform, then the family $\{h_c\}$ is $k$-wise independent:
for any $k$ distinct inputs $x_1,\dots,x_k\in\F$, the random vector
$\bigl(h_c(x_1),\dots,h_c(x_k)\bigr)$ is uniform in $\F^k$.
Indeed,
\[
  h_c(x_j) - x_j^k = \sum_{i=0}^{k-1} c_i x_j^i,
\]
and the linear map $c\mapsto \bigl(\sum_i c_i x_j^i\bigr)_{j=1}^k$ is a Vandermonde transform,
hence invertible when the $x_j$ are distinct.

\paragraph{Why multiplications dominate in practice.}
Evaluating \eqref{eq:kwise:monic-poly} is a tight inner loop in many randomized data structures
(hash tables, sketches, filters) and in cryptographic-style workloads.
In typical finite-field implementations, additions are cheap but multiplications are expensive
because they entail modular reduction:
either integer multiplication plus reduction in $\F_p$, or carryless multiplication plus
reduction in $\F_{2^w}$.
Thus the main performance objective is often to reduce the number of field multiplications,
especially on the \emph{critical path} (dependent multiplications).

\paragraph{Classical evaluation baselines.}
The standard baseline is Horner's rule, which evaluates \eqref{eq:kwise:monic-poly} using $k-1$
multiplications and $k$ additions:
\[
  \bigl(\cdots((x+c_{k-1})x + c_{k-2})x + \cdots \bigr)x + c_0.
\]
There are two widely used alternatives:
\begin{itemize}
  \item \textbf{Estrin / Paterson--Stockmeyer style.}
  These reorganize evaluation to expose instruction-level parallelism by grouping terms,
  reducing depth to $O(\log k)$ at the cost of extra multiplications (to form powers like $x^2,x^4,\dots$).
  This can help on wide machines when throughput matters more than total work~\cite{estrin1960organization,paterson1973evaluation}.
  (The schedules of \Cref{thm:main} achieve the same logarithmic depth ---
  height at most $2\lceil\log_2k\rceil+5$, \Cref{thm:construction-height}
  --- without the extra multiplications.)
  \item \textbf{Rabin--Winograd preprocessing.}
  Rabin and Winograd~\cite{rabin1972number} showed that, with preprocessing of the coefficient
  vector $c$ using only rational operations, one can evaluate a degree-$k$ polynomial with
  about $k/2 + O(\log k)$ multiplications.
  Their work framed preprocessing in a way that is compatible with finite fields, and they
  conjectured their overhead beyond $k/2$ was essentially necessary.
\end{itemize}

\paragraph{Our setting: preprocessing as a parameterization.}
In this paper, we consider an evaluation circuit $E(x;\theta)$ whose \emph{runtime key} is a
parameter vector $\theta\in\F^d$.
For each $\theta$, the function $x\mapsto E(x;\theta)$ is a monic degree-$k$ polynomial, with
coefficient vector $\mathsf{coeff}(\theta)\in\F^k$ as in \Cref{sec:model}.
There are then two natural ways to generate a $k$-wise hash function:
\begin{enumerate}
  \item sample coefficients $c\gets\F^k$ and preprocess them into parameters
        $\theta=\mathsf{pre}(c)$, then hash by $x\mapsto E(x;\theta)$;
  \item sample parameters $\theta$ directly (for example uniformly from $\F^d$) and hash by
        $x\mapsto E(x;\theta)$.
\end{enumerate}

\paragraph{Why bijectivity/birationality is important for hashing.}
The distinction above is exactly where the ``bijective/injective'' viewpoint matters.
If $\mathsf{coeff}$ is not injective, then distinct parameters can represent the same polynomial.
This is not merely an aesthetic issue: sampling $\theta$ induces a potentially \emph{biased}
distribution on polynomials (some polynomials may have many preimages, others none),
which can break clean statements like ``the hash is $k$-wise independent'' if the key-generation
procedure is ``pick a random $\theta$''.

A rational right-inverse by itself does not rule out this bias: the map
$\mathsf{coeff}(u,v)=u$ has the polynomial section $\mathsf{pre}(c)=(c,0)$
and is nowhere injective.  What rules it out is the stronger property that
our constructions actually have: the explicit decoder is a polynomial
\emph{left}-inverse of $\mathsf{coeff}$, so $\mathsf{coeff}$ is a polynomial
automorphism of $\F^k$ (\Cref{lem:polynomial-left-inverse-automorphism}), a
bijection over every field of the stated characteristic, finite fields
included.
Operationally this gives two benefits:
\begin{itemize}
  \item \textbf{Soundness of preprocessing.}
  Given a uniformly random coefficient vector $c$ (the standard $k$-wise key), we can compute
  parameters $\theta=\mathsf{pre}(c)$ such that $E(x;\theta)\equiv h_c(x)$.
  This lets us implement $k$-wise hashing using the fast circuit while preserving the exact
  algebraic distribution that the usual Vandermonde proof assumes.
  \item \textbf{``Sample-$\theta$'' viewpoint.}
  Since $\mathsf{coeff}$ is a bijection of $\F^k$, a uniform $\theta\in\F^k$
  induces a uniform monic coefficient vector, exactly; over a finite field this
  is precisely the distribution the Vandermonde argument needs, with no generic
  or exceptional-set qualification.  (A map that is only generically injective
  would leave a bias on its exceptional set, which over a finite field is not
  negligible; \Cref{lem:polynomial-left-inverse-automorphism} is what makes
  the exact statement available.)
\end{itemize}
This is the sense in which ``fast evaluation with rational preprocessing'' becomes a
bijective/injective parameterization problem.

\paragraph{Implication of \Cref{thm:main} for $k$-wise hashing.}
Applying \Cref{thm:main} with $n=k$, we obtain an evaluation scheme for \eqref{eq:kwise:monic-poly}
that uses only $\lfloor k/2\rfloor+1$ multiplications after preprocessing the coefficients into
runtime parameters.
This improves over Horner's $k-1$ multiplications and removes the $O(\log k)$ overhead in
Rabin--Winograd while staying within the rational preprocessing model.
We evaluate practical performance trade-offs against Horner, Estrin-style evaluation, and
Rabin--Winograd in \Cref{sec:experiments:kwise}.

\subsection{Related Work}
\label{sec:related}

\paragraph{Polynomial evaluation with preprocessing.}
Horner's rule evaluates a degree-$n$ polynomial with $n$ multiplications, and Pan showed that without preprocessing this is optimal~\cite{pan1966methods}.
Motzkin~\cite{motzkin1955evaluation} introduced \emph{preconditioning}: if the same polynomial is evaluated at many points, the coefficients may be transformed for free in advance.
Motzkin~\cite{motzkin1955evaluation} and Belaga~\cite{belaga1958some,belaga1961evaluation} proved that even with preconditioning $\lfloor n/2\rfloor+1$ multiplications and $n$ additions are necessary for a general polynomial of degree $n\ge2$ (Theorems~M, A and~C in Knuth~\cite[\S4.6.4]{knuth1997seminumerical}; Belaga's note states the multiplicative bound as $\lceil n/2\rceil$, and Pan~\cite{pan1966methods} sharpened it to $\lfloor n/2\rfloor+1$);
Motzkin, Belaga, Eve and Knuth~\cite{motzkin1955evaluation,belaga1958some,eve1964evaluation,knuth1962evaluation} gave schemes with $\lceil n/2\rceil+1$ multiplications for a general polynomial of degree $n$, that is $\lceil n/2\rceil$ for a monic one (Belaga's Theorem~1, displayed as scheme~(0.5) and Theorem~3.2 of Pan's survey~\cite{pan1966methods}; Knuth's Theorem~E states the count as $\lfloor n/2\rfloor+2$, one more when $n$ is even), but their preprocessing requires finding roots of polynomials, so the transformed coefficients generally live in an algebraic extension of the field generated by the coefficients (a real extension when the coefficients are real, in Pan's scheme~(0.7)~\cite{pan1966methods}, Eve's variant~\cite{eve1964evaluation} and Knuth's Theorem~E).
These schemes attain the multiplicative lower bound for even $n$; for odd $n\ge9$ it is attained with complex parameters by Revah~\cite{revah1975number} and Pan~\cite{pan1978complex}.
Belaga already noted that the preprocessing system of his Theorem~1 is in general not solvable over the reals~\cite{belaga1958some}; the real-valued variants just cited fill that gap with algebraic preprocessing, and \Cref{thm:main} supplies rational, everywhere-defined preprocessing instead, reaching the multiplication count of Knuth's Theorem~E~\cite{knuth1962evaluation,knuth1997seminumerical} ($\lfloor n/2\rfloor+2$ for a general polynomial) without root-finding and without the shift $y=x+c$ that Theorem~E requires, although not its $n$ additions.
Pan~\cite{pan1961schemes} also gave a scheme for sextics with three multiplications whose preprocessing is rational but undefined on a hypersurface of coefficient vectors; we return to it in \Cref{sec:lower}.
Pan's survey~\cite{pan1966methods} and the monograph of B\"urgisser, Clausen and Shokrollahi~\cite[Chapter~5]{burgisser1997algebraic} give the classical picture, including the transcendence-degree argument behind the lower bounds.
Rabin and Winograd introduced \emph{rational preprocessing}~\cite{rabin1972number}, where preprocessing uses only field operations and division, obtaining $n/2+O(\log n)$ multiplications; the $O(\log n)$ term pays for computing $x^2,x^4,\ldots,x^{2^{\lfloor\log n\rfloor}}$ by repeated squaring.
Knuth's exercise~44~\cite[\S4.6.4]{knuth1997seminumerical} records Winograd's 1970 precursor of this construction for monic polynomials: $\frac12n+\lfloor\log_2 n\rfloor-1$ multiplications beyond the $\lfloor\log_2 n\rfloor$ squarings and at most $\frac54n$ additions, with parameters that are integer polynomials in the coefficients and hence valid modulo every $m$.
\Cref{thm:main} removes the logarithmic term at the same leading addition constant $\frac54$, but its decoder divides by the integer pivots of \Cref{rem:char2}, so it is not valid modulo every $m$.
Pan's resolution of the complex case~\cite{pan1978complex} explicitly sets aside evaluation over the rational numbers, referring to Rabin and Winograd; \Cref{thm:main} settles the multiplication count in that setting for monic polynomials of odd degree (for general polynomials of odd degree $n\ge9$ it is within one of the complex optimum of Revah and Pan).
A general-purpose method that instead minimises \emph{non-scalar} multiplications is Paterson--Stockmeyer~\cite{paterson1973evaluation}, and Kedlaya and Umans~\cite{kedlaya2011fast} gave asymptotically fast algorithms with preprocessing.
In cryptography, Bernstein's BRW polynomials~\cite{Bernstein:2011:RW} specialise the Rabin--Winograd recursion to message authentication: the message blocks become the coefficients and the evaluation point becomes the key, and $m$ blocks are hashed with $\lfloor m/2\rfloor$ multiplications after $\lceil\log_2 m\rceil$ squarings of the key, amortised over every message authenticated under that key; see Ghosh and Sarkar~\cite{ghosh2019evaluating} and Chakraborty et al.~\cite{chakrabortyefficient} for software and hardware evaluation.
The two schemes serve different purposes.
Rabin--Winograd preprocessing is a \emph{bijection} between coefficient vectors and parameter vectors, so a random parameter vector corresponds to a uniformly random monic polynomial and the scheme yields $k$-wise independent hashing.
BRW is only \emph{injective}: distinct messages give distinct polynomials, but not every polynomial arises, which suffices for almost-universal hashing (an ``authenticator'') but not for $k$-independence.
Our main construction is of the first kind: it keeps a bijective rational preprocessing and removes the logarithmic number of squarings, evaluating with $\lfloor n/2\rfloor+1$ multiplications while matching the logarithmic multiplicative depth of the recursion (\Cref{thm:construction-height}).
Its injective counterpart, an almost-universal hash with half the multiplications of Horner's method, is the subject of \Cref{sec:injective}, where we compare it with BRW.

\paragraph{Hashing.}
Polynomial evaluation over finite fields is a standard tool for $k$-wise independent hashing and universal hashing~\cite{wegman1981new}.
When multiplications dominate the cost (e.g.\ due to modular reduction), it is natural to seek constructions that trade fewer multiplications for other resources.
Examples include simple tabulation hashing~\cite{patracscu2012power} and expansion-based constructions~\cite{christiani2015independence}.
Another widely used universal hash is the pseudo-dot-product (NH) hash, which uses about half as many multiplications as a full dot product~\cite{winograd1968new, black1999umac}.
CLHASH~\cite{lemire2015clhash} adapts the NH construction to carryless multiplication over $\F_{2^{64}}$,
achieving very high throughput by exploiting the independence of all multiplications;
however, it requires $O(n)$ random keys to hash $n$ values, compared to $O(1)$ keys for polynomial-based methods.
We include CLHASH (denoted CLNH) in our experimental comparisons as a throughput baseline.
Recent work by Degabriele et al.~\cite{degabriele2026multivariate} explores two-level univariate/multivariate polynomial hash designs and benchmarks many variants over binary and prime fields.

\paragraph{Cryptographic workloads.}
Fast evaluation of low-degree polynomials over finite fields appears in symmetric cryptography (e.g.\ polynomial representations of S-boxes and analyses of masking schemes~\cite{DBLP:conf/fse/CarletGPQR12}) and in threshold secret sharing~\cite{shamir1979share}.
Section~\ref{sec:experiments:crypto} benchmarks our constructions on share-generation and polynomial-PRF style workloads.

\section{Injective Polynomial Hashing with Fewer Multiplications}
\label{sec:injective}

In this section we use the following recursive family.
\begin{definition}[Injective recurrence]\label{def:injective:recurrence}
Let $u,y,z$ be algebraically independent variables and define
\begin{align}
   P_0 &= z, \\
   P_i &= a_i+(b_i+y)(P_{i-1}+u),
   \qquad i\ge1. \label{eq:injective-recurrence}
\end{align}
\end{definition}
For hashing we use the recurrence with $(u,y,z)$ drawn as three independent
uniformly random field elements, the \emph{three-key family}, or the
\emph{single-key} specialization
\begin{equation}\label{eq:injective-single-key}
   y=x^3,\qquad z=x,\qquad u=x^2,
\end{equation}
that is $P_0=x$ and $P_i=a_i+(b_i+x^3)(P_{i-1}+x^2)$, which keeps a single
resident key word at the price of a weaker bound.
We prove below that distinct parameter vectors
$(a_1,b_1,\dots,a_n,b_n)$ induce distinct polynomials, both before and after
this specialization.  Neither family is bijective onto all monic
polynomials: in the single-key form $P_n$ has degree $3n+2$ but only $2n$
parameters.

\vspace{.5em}

Before proving injectivity, we note the main application: the recurrence gives an almost-universal hash family with half the multiplications of Horner's method.
We use the following standard quantitative definition.
\begin{definition}[Almost-universal hash family]\label{def:injective:universal}
   A family of hash functions $\mathcal H$ is \emph{$\varepsilon$-almost universal} if for any two distinct messages $m\ne m'$ we have
   \[
      \Pr_{h\sim\mathcal H}\!\left[h(m)=h(m')\right] \le \varepsilon.
   \]
\end{definition}

If two distinct messages induce distinct key polynomials whose \emph{difference} has total degree at most $d$, the Schwartz--Zippel lemma gives collision probability at most $d/|\F|$ over uniformly random keys.
In the three-key family $P_n$ is monic of degree $n$ in $y$ with the message-independent top part $(z+u)y^n$ (see the end of the proof of \Cref{lem:injective:coeffmap}), so the difference of two such polynomials has total degree at most $n$, and the family is $n/|\F|$-almost universal on messages of $2n$ field elements: half of Horner's bound at half of Horner's multiplications, with three resident key words.
In the single-key specialization $P_n$ has degree $3n+2$ and its three leading coefficients are again the same for every message, so the difference has degree at most $3n-1$ and the family is $(3n-1)/|\F|$-almost universal with one key word.

Notice a key difference between the bijective polynomials for $k$-wise independent hashing in the previous section, and the injective polynomials here.
In the bijective setting, the polynomial coefficients are the (random) \emph{key} and the input $x$ is the \emph{data} being hashed.
In the injective setting the roles are reversed: the stream $(a_i,b_i)$ is the \emph{data}, while the evaluation point, $x$ is the (random) \emph{key}.

\vspace{.5em}

Before proving the injectivity of our construction, we note the following construction
suggested by Daniel J. Bernstein~\cite{Bernstein:2011:RW}: a simplified version of the
Rabin--Winograd bijective polynomial hashing scheme, which he observed gives an ``authenticator''
\[
   P() = 1, \quad
   P(a_1, \ldots, a_m) = P(a_1, \ldots, a_{k-1})(x^k + a_k) + P(a_{k+1}, \ldots, a_m)
\]
where $k \le m$ is the largest power of $2$ at most $m$.
For example:
$P(a_1)=(x+a_1)+1$,
$P(a_1,a_2) = (x+a_1+1)(x^2+a_2)+1$
and $P(a_1,a_2,a_3) = (x+a_1+1)(x^2+a_2)+x+a_3+1$
and so on.

To see this gives an injective family, first note
that each polynomial (let's write $P_m = P(a_1,\dots,a_m)$ for simplicity)
is monic, since $1$ is monic and product and sums of monic polynomials are monic.
Next note that $\deg(P_0)=0$ and $\deg(P_m) = 2k-1$, where $k=k(m)$ is the largest power of 2 less than or equal to $m$ as above.
From the construction, and the fact $m-k<k$, we have:
\[
   \deg(P_m) = \deg(P_{k-1}) + k.
\]
Since $k = 2^j$, we have $k - 1 = 2^j - 1$, so the largest power of 2 $\le k-1$ is $k' = 2^{j-1} = k/2$.
Therefore $\deg(P_{k-1}) = 2(k/2) - 1 = k - 1$,
and $\deg(P_m) = 2k-1$.
Finally, the same degree separation gives a decoder.  From the coefficients of
$P_m$ in degrees $k,\ldots,2k-1$ we recover $P(a_1,\ldots,a_{k-1})$, because
only its product with $x^k$ reaches those degrees.  We recursively decode
$a_1,\ldots,a_{k-1}$.  After subtracting the $x^k$ product, the coefficient of
$x^{k-1}$ is $a_k$ plus the known leading coefficient of
$P(a_{k+1},\ldots,a_m)$ when that polynomial has degree $k-1$ (and just
$a_k$ otherwise).  Thus we recover $a_k$, subtract its product with the known
left polynomial, and recursively decode the right polynomial.  This proves
injectivity.

The ``Bernstein Rabin--Winograd'' scheme is simple to describe and prove, but
the recursive nature is a bit inconvenient for fast hashing.
Some authors~\cite{chakrabortyefficient} have proposed non-recursive variants that are more efficient to compute, but they still need to precompute $x^k$ for all powers of 2 up to $n$, which is extra accounting that our method does not require.
Bernstein himself observes that the $\lceil\log_2 m\rceil$ key squarings can be traded for $\lceil\log_2 m\rceil$ cached key powers~\cite[\S5.8]{Bernstein:2011:RW}; the three-key recurrence of \Cref{def:injective:recurrence} needs no key powers at all, and its single-key specialization \eqref{eq:injective-single-key} needs only the two cached powers $x^2,x^3$, for every message length.

\vspace{.5em}

We now prove the injectivity property used by both the one-key and two-key
specializations.

\begin{lemma}[Injectivity of the recurrence]\label{lem:injective:coeffmap}
The parameter-to-polynomial map in \Cref{def:injective:recurrence} is
injective.  Consequently, the single-variable specialization
\eqref{eq:injective-single-key} is injective as a polynomial in $x$.
\end{lemma}
\begin{proof}
   We first note the following facts about \eqref{eq:injective-recurrence}:
   \begin{enumerate}
      \item $P_n$ has the form $f_1(y) + z f_2(y) + u f_3(y)$,
         where $f_1$ has degree at most $n-1$ and $f_2$ and $f_3$ are degree $n$ polynomials in $y$.
      \item $f_1(y) = \sum_{i=1}^n a_i \prod_{j>i}(y+b_j)$.
      \item $f_2(y) = \prod_{j=1}^n (b_j + y)$,
      \item $f_3(y) = \sum_{i=1}^n \prod_{j\ge i} (b_j+y)$.
   \end{enumerate}
   Here 2. follows if we let $u=z=0$ and expand
   \[P_n=a_n+(b_n+y)P_{n-1}=a_n+(b_n+y)(a_{n-1}+(b_{n-1}+y)P_{n-2}),\]
   etc.
   The other facts follow similarly.

   To show injectivity, we reconstruct the $a$s and $b$s from the coefficients
   of $P_n$.
   It is convenient to start by extracting the $b$s.
   If $n=1$, then $f_2(y)=y+b_1$, so its constant coefficient gives $b_1$
   immediately.  Suppose henceforth that $n\ge2$.
   For this, note that
   \begin{align}
      f_2(y) &= y^n + (\sum_{i=1}^n b_i) y^{n-1} + (\sum_{i<j} b_i b_j) y^{n-2} + O(y^{n-3})
      \quad\text{and}\\
      f_3(y) &= y^{n} + (1+\sum_{i=1}^n b_i) y^{n-1} \\
             &\quad + (1 + \sum_{i=2}^n b_i + \sum_{i<j} b_i b_j) y^{n-2} + O(y^{n-3}),
      \quad\text{so}\\
      f_3(y)-f_2(y) &= y^{n-1} + (1 + \sum_{i=2}^n b_i) y^{n-2} + O(y^{n-3}).
   \end{align}
   (For $n=2$ the constant $1$ in the $y^{n-2}$ coefficient of $f_3$, and hence of $f_3-f_2$, is absent, since it comes from the $i=3$ term of $f_3$; the extraction below is unaffected.)
   We can thus extract $\sum_{i=1}^n b_i$ and $\sum_{i=2}^n b_i$, and from the difference we get $b_1$.
   Using polynomial synthetic division, we may then divide $f_2(y)$ by $(b_1+y)$ to obtain $f_2^{(2)}(y) := \prod_{j=2}^n (b_j + y)$.
   Similarly define $f_3^{(2)}(y) := f_3(y)-f_2(y)=\sum_{i=2}^n \prod_{j\ge i} (b_j+y)$.
   Then proceed by induction to extract the remaining $b$s.

   To extract the $a$s it suffices to notice that $f_1(y) = a_1 y^{n-1} + O(y^{n-2})$ so we may read $a_1$ off the leading coefficient.
   In the process of extracting the $b$s we constructed for each $i$ the polynomial $\prod_{j>i}(y+b_j)$, so we can subtract $a_1\prod_{j>1}(y+b_j)$ and proceed by induction.

   \vspace{.5em}

   The three-variable construction reduces to a \emph{single random key} $x$
   by the specialization \eqref{eq:injective-single-key}.
   The three summands $f_1(x^3)$, $x f_2(x^3)$, and $x^2 f_3(x^3)$ occupy
   distinct exponent classes modulo~$3$, so the specialization preserves injectivity.
   Its degree is $3n+2$.  The coefficients of $x^{3n+2}$, $x^{3n+1}$ and $x^{3n}$ are $1$, $1$ and $0$ for every message: the first two are the leading coefficients of $f_3$ and $f_2$, and no term of exponent $3n$ can arise, since $f_1(x^3)$ has degree at most $3n-3$ while the other two summands only produce exponents $\equiv1,2\pmod 3$.
   Hence two distinct messages give polynomials whose difference has degree at most $3n-1$, and the collision probability is at most $(3n-1)/|\F|$.

   Alternatively, retain $y$ as an independent key and set $z=x$ and $u=x^2$:
   \begin{align}
      P_0 &= x, \\
      P_{i} &= a_{i} + (b_{i}+y)(P_{i-1}+x^2),
   \end{align}
   which has total degree $n+2$ in $(x,y)$; its terms $x^2y^n$ and $xy^n$ have coefficient $1$ for every message, so differences have total degree at most $n+1$ and the collision probability is at most $(n+1)/|\F|$, comparable
   to standard Horner evaluation.
   With all three of $u$, $y$, $z$ drawn as independent uniform keys the bound
   improves further: $f_2$ and $f_3$ are monic of degree $n$, so the
   degree-$(n+1)$ part $zy^n+uy^n$ of $P_n$ is the same for every message, the
   difference of two distinct messages has total degree at most $n$, and the
   collision probability is at most $n/|\F|$ for messages of $n$ pairs, half of
   Horner's $(2n-1)/|\F|$ on the same $2n$ words, with the same three resident
   key words as the single-key form (which keeps $x$, $x^2$ and $x^3$).
\end{proof}

\subsection{Experiments}
\label{sec:injective:experiments}

The recurrence is benchmarked against Horner's rule in
\Cref{sec:experiments:injective}: with the same field arithmetic
($\mathbb F_{2^{64}}$, carryless multiplication) and the same message lengths,
hashing $2N$ words costs $N$ multiplications instead of $2N-1$, and the measured
speedup grows from $1.4\times$ at $2N=8$ to $1.8$--$2.1\times$ at $2N\ge16$ on ARM,
and from $1.1\times$ to $1.9$--$2.3\times$ on x86 (\Cref{tab:injective}).
\Cref{sec:experiments:universal} compares it further with the parallel and
tree-structured universal hashes (Horner-unrolled, BRW, CLNH).

As an implementation data point we also ported the recurrence to
SMHasher3~\cite{smhasher3}, a hash-testing framework, on an Apple M2 Pro
(ARM64), with $L=9$ independent lanes each consuming 16 bytes per step, a
ternary-tree reduction of the lane states and a finalizer that mixes in the
message length.  Over the Mersenne prime field $p=2^{61}-1$ (128-bit product
and two rounds of shift-and-add reduction) it reaches $1.59$~bytes/cycle of
bulk throughput on a 256\,KiB message and $133$~cycles per hash on short
(1--31 byte) keys.  Replacing the field multiplication by the ``MUM'' fold
$\mathrm{lo}_{64}(ab)\oplus\mathrm{hi}_{64}(ab)$ of the $128$-bit product makes
the bulk throughput $9\times$ higher ($14.4$~bytes/cycle), but that variant is
not a field operation: the fold is many-to-one, so the injectivity argument of
this section and the Schwartz--Zippel collision bound do not apply to it, and
message sets with elevated collision rates exist.  We report the number only to
locate the cost of exact field arithmetic; the guarantees of this section are
for the field version.

\paragraph{Polynomial-level comparison.}
\Cref{tab:injective:universal} times the recurrence against the other
$O(1)$-key polynomial methods, and against CLNH as the throughput ceiling of
an $O(N)$-key inner product, on messages of $2N$ words over $\F_{2^{64}}$ with
carryless multiplication, on an Apple M2 Pro and an Intel Xeon Platinum 8375C.
All methods share the same field arithmetic and the same benchmark harness
(\texttt{tools/bench/carryless\_arm.cpp}, \texttt{tools/bench/carryless.cpp}).
Sequentially, the recurrence halves the running time of Horner's rule on both
machines once $2N\ge16$, as its multiplication count predicts.  Fanning out
matters more than the count: the prefix-scan parallelization spends $3N$
multiplications and loses to the tree form of Horner on the Xeon, whereas
dealing the pairs to $L=8$ interleaved lanes keeps the count at $N+L$
(\Cref{sec:experiments:universal}).  With lanes the recurrence is the fastest
$O(1)$-key method on the M2 Pro from $2N=32$ on, by $1.2$--$1.5\times$ over the
next best and within $2\times$ of the $O(N)$-key CLNH ($1.6\times$ at $2N=256$), and on the Xeon from
$2N=128$ on, by $1.2\times$ at $2N=256$; at $32$--$64$ words on the Xeon the
tree form of Horner is $5$--$9\%$ faster, the count advantage being absorbed by
the per-multiplication cost of moving operands between integer and vector
registers in this implementation style (a vector-resident implementation of
the lanes reaches $23.8$\,GB/s on the M2 Pro, \Cref{tab:injective:adversarial}).
BRW is slower than the unrolled, scanned and laned forms on either machine; it
beats only sequential Horner from $2N=32$, the sequential recurrence from
$2N=64$, and, on the M2 Pro, the tree form of Horner up to $2N=128$.

\begin{table}[!htbp]
\centering
\caption{Universal hashing of $2N$ words over $\F_{2^{64}}$, in \textmu s per
$10^5$ hashes (mean over the fastest 100 of 200 repetitions; standard deviations
are below $5\%$ except where marked~$^{\dagger}$).  Horner: sequential, unrolled
over interleaved chains, and the tree evaluation with cached powers
(``parallel'' in \Cref{sec:experiments:universal}).  This paper: the
sequential recurrence ($N$ multiplications), the prefix-scan parallelization
($3N$), and $L=\min(N,8)$ lanes ($N+L$).  CLNH uses $O(N)$ key words; all other
methods use $O(1)$.  Best $O(1)$-key method per row in bold.}
\label{tab:injective:universal}
\resizebox{\columnwidth}{!}{%
\begin{tabular}{c|rrr|rrr|r|r}
 & \multicolumn{3}{c|}{Horner} & \multicolumn{3}{c|}{This paper} & BRW & CLNH \\
 $2N$ & Seq & Unrolled & Tree & Seq & Scan & Lanes & & \\
\hline
\multicolumn{9}{l}{\emph{Apple M2 Pro (ARM, PMULL)}} \\
  8 & 552 & 464 & 2481 & \textbf{393} & 828 & 604 & 1237 & 184 \\
 16 & 2109 & 1045 & 3463 & \textbf{991} & 1544 & 1400 & 2332 & 346 \\
 32 & 7646 & 2727 & 6607 & 4099 & 2735 & \textbf{2107} & 4481 & 1066 \\
 64 & 22976 & 6684 & 12818 & 11672 & 5162 & \textbf{4153} & 8833 & 2115 \\
128 & 64638 & 13710 & 22976 & 27706 & 10288 & \textbf{7850} & 17548 & 4657 \\
256 & 146944 & 34503 & 21157 & 71018 & 22144 & \textbf{14336} & 34879 & 8863 \\
\hline
\multicolumn{9}{l}{\emph{Intel Xeon Platinum 8375C (x86, PCLMULQDQ)}} \\
  8 & 1338 & 1134 & \textbf{1107} & 1192 & 1878 & 1836 & 4302 & 573 \\
 16 & 4560 & \textbf{1999} & 2053 & 2445 & 3168 & 3020 & 6131 & 1153 \\
 32 & 14065 & 7246 & \textbf{4067} & 6002 & 5796 & 4266 & 10350 & 2307 \\
 64 & 36279 & 9140 & \textbf{7443} & 17588 & 10835 & 8135 & 15877 & 4619 \\
128 & 80611 & 20546 & 13725 & 40697 & 20081 & \textbf{13325} & 27669 & 9244 \\
256 & 169985 & 43804 & 26990 & 86905 & 38612 & \textbf{23163} & 47747 & 18492 \\
\end{tabular}
}
\end{table}

\paragraph{Comparison with general-purpose hashes.}
We adopt the standard model for keyed hashing: the secret key material is drawn
uniformly at random and hidden from the attacker, who chooses the input
messages.  The quantity of interest is the worst case, over attacker-chosen
inputs, of the collision probability taken over the random secret.  A universal
hash bounds this by a small $\varepsilon$ for \emph{every} input pair; a
heuristic hash need not.  To compare hashes across message lengths we report
\[
   \mathrm{bits} \;=\; \min_{L}\;\log_2\frac{L}{\varepsilon(L)},
\]
where $L$ is the message length in $64$-bit words and $\varepsilon(L)$ is the
collision probability at that length, floored at $2^{-w}$ for a $w$-bit output.
Dividing by $L$ cancels the Schwartz--Zippel factor, so a degree-$L$ polynomial
hash over a field of size $q$ scores $\log_2 q$ regardless of length, and
taking the minimum keeps the score of a hash with a length-independent flaw
honest.  For the proven hashes $\varepsilon$ is the published bound; for the
heuristic hashes it is the largest collision rate we could \emph{find} by
differential search, measured over $2^{31}$ or more random secrets, so those
entries (marked~$^*$) are upper bounds on the true score: worse inputs may
exist.  We further restrict the main comparison to hashes whose per-key state
is at most eight $64$-bit words, one cache line, following Bernstein's
observation that a server handling thousands of keys cannot keep large per-key
tables hot~\cite{Bernstein:2011:RW}; the budget also makes the throughput
comparison fair, since key size is otherwise a free parameter whose limit is a
key as long as the message.  Shared public constants and per-call derivations
do not count.  Hashes with larger keys are listed below the line.

\Cref{tab:injective:adversarial} gives the result.  Within the budget, the
proven methods score $57$--$64$ bits and the heuristic methods far less:
\texttt{wyhash}, \texttt{rapidhash} and \texttt{XXH3} lose to the all-ones
difference on the two words feeding their first multiply-fold, which collides
with probability about $2^{-27}$ under a random secret, some $2^{37}$ times the
ideal; the $128$-bit \texttt{XXH3-128} fares no better, because one of its
two accumulators sees only a raw word sum that the attacker can hold fixed; \texttt{MUM} has \emph{key-free} collisions, a fixed pair of $8$-byte
messages that collides for every seed; \texttt{komihash} resisted every
differential we tried at $2^{32}$ seeds per surface, which bounds its score
below only by our search effort.  The attacks and their measurements are in
\Cref{app:adversarial}, which also records that the low $64$ bits of an $89$-bit
Mersenne residue are \emph{not} a safe output: a pair differing by $2^{64}-1$
in the last word collides with probability $2^{-25}$, so the Mersenne rows
output the full residue.

\begin{table}[!htbp]
\centering
\caption{Provable security versus throughput on an Apple M2 Pro.
\emph{Key} is the resident per-key state in $64$-bit words.  \emph{Bits} is
$\min_L\log_2(L/\varepsilon(L))$ as defined in the text; entries marked $^*$
are measured on the worst inputs we found and may be lower.  Throughput is
single-threaded, for $16$\,KB and $512$-byte messages.  Rows above the line
keep at most eight words of key.  The two ``this paper'' rows over
$\F_{2^{64}}$ are the three-key recurrence with its state kept in vector
registers (\texttt{tools/bench/adversarial/}); the framework implementation of
\Cref{tab:injective:universal}, which moves the state through general
registers, runs the same one-chain recurrence at $2.2$\,GB/s.  The other
polynomial rows use the framework implementations.  The ChainHash bounds
(\Cref{sec:ph}) are proved for a key of $W+9$ independent uniform field
elements; the implementation timed and tested here expands a $64$-bit seed
with \texttt{splitmix64}, a $2^{64}$-member subfamily that the theorems do
not cover (\Cref{rem:ph:gaps}).}
\label{tab:injective:adversarial}
\begin{tabular}{lrrrr}
\hline
hash & key & bits & GB/s (16\,KB) & GB/s (512\,B) \\
\hline
\multicolumn{5}{l}{\emph{proven, key $\le 8$ words}} \\
This paper, one chain, $\F_{2^{64}}$        & 3 & 64 &  4.1 & 10.1 \\
This paper, 8 lanes, $\F_{2^{64}}$          & 4 & 61 & 23.8 & 17.9 \\
This paper, $\F_{2^{89}-1}$ (89-bit output) & 5 & 88 &  4.4 &  3.9 \\
Horner, $\F_{2^{64}}$                       & 1 & 64 &  1.3 &  2.3 \\
Horner, unrolled                            & 1 & 64 &  5.1 &  7.4 \\
BRW~\cite{Bernstein:2011:RW}                & 1 & 63 &  6.0 &  5.3 \\
Polymur~\cite{polymur}                      & 4 & 57 & 19.7 & 16.2 \\
\multicolumn{5}{l}{\emph{heuristic, key $\le 8$ words}} \\
\texttt{wyhash} v4.3                        & 5 & 28.6$^*$ & 26.5 & 34.8 \\
\texttt{rapidhash} v1                       & 4 & 28.4$^*$ & 27.1 & 32.1 \\
\texttt{XXH3}                               & 1 & 27.3$^*$ & 38.2 & 27.2 \\
\texttt{XXH3-128} (128-bit output)          & 1 & 28.5$^*$ & 33.8 & 24.0 \\
\texttt{MUM} v3                             & 1 &  0$^*$   & 33.0 & 26.8 \\
\texttt{komihash} v5.34                     & 1 & $\ge32^*$ & 24.8 & 24.3 \\
\hline
\multicolumn{5}{l}{\emph{proven, key $> 8$ words}} \\
ChainHash, 1\,KB blocks                     & 137 & 62 & 61.7 & 40.5 \\
ChainHash, 256\,B blocks                    &  41 & 62 & 57.7 & 36.8 \\
ChainHash, 64\,B blocks                     &  17 & 62 & 26.6 & 27.4 \\
UMASH-64~\cite{umash}                       &  38 & 55 & 40.7 & 32.9 \\
UMASH-128 (128-bit output)                  &  38 & 83 & 23.9 & 19.7 \\
CLNH (fixed length)~\cite{lemire2015clhash} & $L$ & 64 & 27.3 & 26.3 \\
Multiply-shift (fixed length)               & $L{+}1$ & 64 & --- & 16.5 \\
\hline
\end{tabular}
\end{table}

\paragraph{A large-key hash built from the recurrence.}
The rows labelled ChainHash compose the recurrence with the reduction-free
inner level of the NH-based hybrids, in the same way that UMAC, VMAC and CLHASH
compose a polynomial with NH~\cite{black1999umac,lemire2015clhash}.  The name
reflects the structure: each step of the recurrence is an NH-style product into
which the running value is mixed, and that mixing is what lets one key be reused
across the whole message instead of one key word per message word.  The message
is split into blocks of $B$ bytes and each block into $S$ sub-blocks ($S=1$ for
$B\le256$, $S=2$ for $1$\,KB blocks); each sub-block is hashed by carryless NH
(one \texttt{PMULL} per $16$ bytes into two $128$-bit accumulators) and its
unreduced sum is consumed as one pair $(a_i,b_i)$ of the three-key recurrence
$P_0=z$, $P_i=a_i+(b_i+y)(P_{i-1}+u)$, one field multiplication per
sub-block; the byte length is XORed into the last $a_i$; one further key
word is added to the result modulo $2^{64}$ (the \emph{twist}: an integer
addition, carries and all); and the sum is passed through a random monic
polynomial of degree~$5$ evaluated with three multiplications,
$\lfloor5/2\rfloor+1$ as in \Cref{thm:main}, by a characteristic-$2$ circuit
whose coefficient map is a bijection of $\F^5$ (\Cref{sec:ph} gives the
explicit decoder).  With its five parameters uniform, the finalizer is a
$5$-wise independent family on distinct field inputs; for the full hash
\Cref{thm:ph:kwise} gives the qualified form: conditioned on the level-2
values of up to five messages being distinct, which fails with probability
at most $\binom{t}{2}(p+1)/2^{64}$, the outputs are independent and uniform,
so that two messages collide with probability $r+(1-r)/2^{64}$, $r$ being the
level-1/2 collision probability, rather than $1/2^{64}$.  Five-wise
independence is what linear probing provably
needs~\cite{pagh2009linear,patracscu2016kindependence}.  Both statements are
for the ideal key of $W+9$ uniform words; the implementation derives its key
from a $64$-bit seed (\Cref{rem:ph:gaps}).
The twist answers a question that $k$-wise independence never asks: over
$\F_{2^{64}}$ the $\F_2$-degree of $v\mapsto v^e$ is the number of ones in
the binary expansion of $e$, so every polynomial of degree at most $6$ is
at most quadratic in the bits of its input, with affine discrete derivatives, and the
fixed-seed keysets of SMHasher3~\cite{smhasher3} (Zeroes, Sparse,
Permutation, TwoBytes, Bitflip) are sensitive to exactly that structure: the
untwisted degree-$5$ finalizer fails $22$ of the $200$ tests, and an earlier
version of the hash used degree $7$, the first cubic option, at the price of
a fourth multiplication.  Any fixed bijection of the finalizer input
preserves the $2^{-64}$ bound and the $5$-wise independence exactly, and the
carry chain of an integer addition is not $\F_2$-affine
(\Cref{sec:ph:twist}); with the twist the degree-$5$ hash passes $200/200$,
degree $3$ with the twist still fails $17$, and whether the twist suffices
for a given test suite is an empirical question, not a theorem.  $S=2$
gives a one-block message a second message-dependent multiplication for the
same kind of reason.  The collision probability is at most $(p+2)/2^{64}$ for two
messages of at most $n$ blocks, $p=Sn$ sub-blocks: $2^{-64}$ for the block
level (the full-width XOR-universality of carryless NH, extended to last
blocks of different pair counts by the length term), $p/2^{64}$ for the
recurrence (the three-key bound above, with $p$ counting sub-blocks rather
than word pairs) and $2^{-64}$ for the finalizer (\Cref{sec:ph}).  No
heuristic mixing step is used anywhere, and both variants pass all $200$
tests of the full SMHasher3 suite, on an Apple M2 Pro and on an Intel Xeon
8375C, including a new seeded-differential test
(full-output collisions of structured pairs, complemented leading or adjacent
interior words, over $2^{24}$ random seeds, $2^{30}$ in an extended tier) that
\texttt{MUM} fails outright and \texttt{wyhash}, \texttt{rapidhash},
\texttt{XXH3} and \texttt{XXH3-128} fail in the extended tier, while \texttt{komihash},
Polymur, SipHash and the polynomial hashes pass (\Cref{app:adversarial}).
With $1$\,KB blocks and a $137$-word key it runs at $62$\,GB/s on the M2 Pro,
above every heuristic hash in the table and above UMASH, with a proof---for
the ideal key; the seeded interface that was timed and tested is a subfamily
outside the theorems (\Cref{rem:ph:gaps}); the
block size is exactly the key-budget knob, and at $64$-byte blocks, closest to
the one-cache-line budget ($17$ key words), the hybrid matches or beats the
eight-lane recurrence on long messages ($26.6$ against $23.8$\,GB/s at
$16$\,KB).  The implementation and a
bit-serial reference are in \texttt{tools/bench/chainhash/} of the repository.

\paragraph{Key expansion.}
One may ask whether the $O(B)$ key of the block level can be derived from a few
seed words, say by a $k$-wise independent polynomial.  The collision condition
of NH is affine in the key words, because the products of two key words cancel
in the difference of two block sums, so with polynomially expanded keys it is a
linear map from the seed words to the difference.  For a \emph{reduced} NH,
whose block sums are taken in $\F_{2^{64}}$, this is fatal: a difference
supported on $k+1$ positions chosen in the kernel of the $k\times(k+1)$
Vandermonde system is independent of the key, and for $k=6$ we verified that
such a pair of $38$-word blocks has the same difference under $200$ random
seeds, a key-free collision.  For the \emph{unreduced} carryless NH used here
the products are taken in $\F_2[X]$ while the expansion lives in
$\F_{2^{64}}$, and the same crafted difference only lowers the $\F_2$-rank of
the seed-to-difference map from $127$ to $63$, leaving a collision probability
of about $2^{-63}$.  We found no better attack, but also no proof; the
security of unreduced NH under $k$-wise independent keys is an open problem
(\Cref{sec:open-problems}).  UMAC's answer is to expand the key with AES,
trading the information-theoretic bound for a computational one.

\section{Lower Bound}
\label{sec:lower}

\paragraph{Straight-line programs.}
Fix a field $\mathbb{F}$.
A (polynomial) straight-line program with input $x$ and parameters $p\in\mathbb{F}^m$ is a sequence of assignments that produces intermediate values $u_0,u_1,\dots$ and finally an output $P_p(x)$, where each assignment is obtained from earlier values (and from $x$, the parameters and fixed elements of $\mathbb{F}$) using additions and multiplications by fixed elements of $\mathbb{F}$, which are free, and a bounded number of \emph{nonscalar} multiplications, which are counted: a multiplication is nonscalar when neither factor is a fixed element of $\mathbb{F}$.
Thus multiplications by fixed scalars are free, but multiplications involving $x$ or the parameters are not.
Consequently each factor of a nonscalar multiplication, and the output, is a fixed linear combination of $x$ and the outputs of the earlier nonscalar multiplications plus an affine function of the parameters (the \emph{constant slot} of that factor); the parameters enter the program only through these constant slots.
An \emph{everywhere-defined rational left inverse} of a polynomial map $F\colon\mathbb{F}^m\to\mathbb{F}^m$ is a tuple of rational functions $g_i/h_i$ ($1\le i\le m$) such that no $h_i$ has a zero in $\mathbb{F}^m$ and $g_i(F(p))=p_i\,h_i(F(p))$ for every $p\in\mathbb{F}^m$.
The preprocessing maps of \Cref{sec:model} are right inverses, not left inverses; the following standard fact bridges the two.

\begin{lemma}[everywhere-defined right inverses are left inverses]
\label{lem:right-left}
Let $\mathbb{F}$ be infinite, let $F\colon\mathbb{F}^n\to\mathbb{F}^n$ be a polynomial map, and let $R=(g_1/h_1,\dots,g_n/h_n)$ be a tuple of rational functions such that no $h_i$ has a zero in $\mathbb{F}^n$ and $F(R(y))=y$ for every $y\in\mathbb{F}^n$.
Then $R$ is an everywhere-defined rational left inverse of $F$: $g_i(F(p))=p_i\,h_i(F(p))$ for every $p\in\mathbb{F}^n$.
\end{lemma}

\begin{proof}
Let $y=(y_1,\dots,y_n)$ and $p=(p_1,\dots,p_n)$ be indeterminates and $R_i=g_i/h_i\in\mathbb{F}(y)$.
Clearing denominators in $F_k(R(y))=y_k$ gives an identity of polynomials in $y$, because it holds at every point of $\mathbb{F}^n$ and $\mathbb{F}$ is infinite; so $F_k(R_1,\dots,R_n)=y_k$ in $\mathbb{F}(y)$ for each $k$.
Hence the subfield $\mathbb{F}(R_1,\dots,R_n)\subseteq\mathbb{F}(y)$ contains $y_1,\dots,y_n$ and equals $\mathbb{F}(y)$, a field of transcendence degree $n$ over $\mathbb{F}$; since every generating set of a field extension contains a transcendence basis, the $n$ generators $R_1,\dots,R_n$ are algebraically independent over $\mathbb{F}$.
Therefore $p_i\mapsto R_i$ defines an injective $\mathbb{F}$-algebra homomorphism $\phi\colon\mathbb{F}[p]\to\mathbb{F}(y)$, which extends to the fraction field $\mathbb{F}(p)$.
The polynomial $h_i(F(p))$ is nonzero, because $h_i$ vanishes nowhere, so $R_i(F(p))=g_i(F(p))/h_i(F(p))$ is an element of $\mathbb{F}(p)$, and
\[
   \phi\bigl(R_i(F(p))\bigr)
   =\frac{g_i\bigl(F(R_1,\dots,R_n)\bigr)}{h_i\bigl(F(R_1,\dots,R_n)\bigr)}
   =\frac{g_i(y)}{h_i(y)}=R_i=\phi(p_i).
\]
By injectivity, $R_i(F(p))=p_i$ in $\mathbb{F}(p)$, i.e.\ $g_i(F(p))=p_i\,h_i(F(p))$ in $\mathbb{F}[p]$, hence at every point $p\in\mathbb{F}^n$.
\end{proof}

\begin{lemma}[$6$ parameters]
\label{lem:six-params}
Assume $\mathbb{F}$ is infinite and $\mathrm{char}(\mathbb{F})\neq 2$.
Fix any six distinct evaluation points $x_0,\dots,x_5\in\mathbb{F}$, and fix a straight-line program as above that uses at most three nonscalar multiplications, has six scalar input parameters $p\in\mathbb{F}^6$, and outputs a polynomial $P_p(x)$.
Explicitly, the program is
\[
   u_1=(Ax+a)(Bx+b),\qquad u_2=\ell_2r_2,\qquad u_3=\ell_3r_3,\qquad
   P_p=s_3u_3+s_2u_2+s_1u_1+s_0x+b_1,
\]
where $A,B,s_i$ are fixed, $\ell_i,r_i$ are affine in $x$ and the earlier $u_j$ with fixed coefficients and constant slots $a_i,b_i$, and the seven constant slots $(a,b,a_2,b_2,a_3,b_3,b_1)$---both constants of the first multiplication included---are an arbitrary affine function of $p$.
Let
\[
   F:\mathbb{F}^6\to\mathbb{F}^6,\qquad
   F(p) = \bigl(P_p(x_0),\dots,P_p(x_5)\bigr).
\]
Then $F$ admits no everywhere-defined rational left inverse; by \Cref{lem:right-left} it then admits no everywhere-defined rational right inverse either.
\end{lemma}

\paragraph{Connection to degree-$6$ decoding.}
Let $E(x;\theta)$ be an evaluation scheme for monic sextics in the model of \Cref{sec:model} with three multiplications and six parameters $\theta\in\mathbb{F}^6$, and suppose its preprocessing is rational and everywhere defined: $\mathsf{pre}\colon\mathbb{F}^6\to\mathbb{F}^6$ is a tuple of rational functions with nowhere-vanishing denominators and $\mathsf{coeff}(\mathsf{pre}(c))=c$ for every $c=(c_0,\dots,c_5)\in\mathbb{F}^6$.
Fix six distinct points $x_0,\dots,x_5$ and let $F(\theta)=(E(x_0;\theta),\dots,E(x_5;\theta))$.
Since $E(x;\theta)=\sum_{j<6}c_jx^j+x^6$ with $c=\mathsf{coeff}(\theta)$, we have $F=V\,\mathsf{coeff}+v$, where $V=(x_k^j)_{k,j<6}$ is the Vandermonde matrix, invertible because the points are distinct, and $v=(x_k^6)_{k<6}$.
Hence $R(y)=\mathsf{pre}\bigl(V^{-1}(y-v)\bigr)$ is everywhere defined and rational with $F(R(y))=y$ for every $y$, so by \Cref{lem:right-left} it is an everywhere-defined rational left inverse of $F$.
The scheme is a straight-line program of the kind in \Cref{lem:six-params} (its seven constant slots are affine in $\theta$), so the lemma says that no such $R$ exists: no three-multiplication scheme for monic sextics with six parameters has everywhere-defined rational preprocessing.

The restriction to six parameters is immaterial.
With $d<6$ parameters, $\mathsf{coeff}(\mathsf{pre}(c))=c$ holds identically as rational functions of $c$ (clear denominators and use that $\mathbb{F}$ is infinite), and differentiating gives $J_{\mathsf{coeff}}(\mathsf{pre}(c))\,D\mathsf{pre}(c)=I_6$, which is impossible because $J_{\mathsf{coeff}}$ has only $d<6$ columns.
With $d>6$ parameters, define $R$ as above (now $R\colon\mathbb{F}^6\to\mathbb{F}^d$, still with $F\circ R=\mathrm{id}$).
The reduction of \Cref{appendix:lower}, whose gauge identity \eqref{eq:lower-gauge} and factorization $F=E'\circ Q$ do not depend on the number of parameters, writes $F=E'\circ Q$ with $Q\colon\mathbb{F}^d\to\mathbb{F}^6$ polynomial and $E'\colon\mathbb{F}^6\to\mathbb{F}^6$ the evaluation map of a normal-form program whose parameters are its six slots (unless the first multiplication is scalar, in which case $F$ factors through $\mathbb{F}^5$ and the rank argument applies again).
Then $Q\circ R$ is an everywhere-defined rational map with $E'\circ(Q\circ R)=\mathrm{id}$, hence by \Cref{lem:right-left} a left inverse of $E'$; but $E'$ is itself an instance of \Cref{lem:six-params}, with constant slots $(0,a_1,a_2,b_2,a_3,b_3,b_1)$.

\paragraph{Warm-up (Motzkin, degree $4$ with $2$ multiplications).}
The analogous statement is false for two multiplications.
Motzkin's classic program
\begin{align}
   u_1 &= x (x + a_0), \\
   u_2 &= (u_1 + x + a_1)(u_1 + a_2), \\
   P &= u_2 + a_3
\end{align}
uses exactly two nonscalar multiplications and produces a monic degree-$4$
polynomial.  Moreover its decoding is rational with no exceptional inputs when
$\mathrm{char}(\mathbb{F})\neq2$: comparing coefficients gives
\[
   c_3=2a_0+1,\quad
   c_2=a_0^2+a_0+a_1+a_2,\quad
   c_1=a_0(a_1+a_2)+a_2,\quad
   c_0=a_1a_2+a_3,
\]
so $a_0=(c_3-1)/2$; then $a_1+a_2$ is determined by $c_2$, whence
$a_2=c_1-a_0(a_1+a_2)$ and $a_1$, and finally $a_3=c_0-a_1a_2$.
Every step is a polynomial expression in $c_3,\dots,c_0$ with denominators
only powers of~$2$, so the Jacobian obstruction below cannot hold at degree~$4$.
Two multiplications therefore already evaluate all monic quartics; the
obstruction we prove is specific to degree~$6$ with three multiplications.

\paragraph{Motzkin's bound, and what is left to prove.}
Under \emph{any} preprocessing, a program with $m$ multiplications can produce at
most $2m$ independent coefficients: after the normalization of \Cref{appendix:lower}
(which trades the second constant of the first multiplication for an $x$-free
correction absorbed into the later constants), the first multiplication carries one
free slot, every later multiplication two, and the final addition one, so the
coefficient vector is a fixed polynomial function of $2m$ scalars, themselves
polynomial in the parameters, and its image cannot cover an $n$-dimensional family of monic polynomials
unless $2m\ge n$ (Motzkin~\cite{motzkin1955evaluation}; Theorem~M and exercise~30
in Knuth~\cite[\S4.6.4]{knuth1997seminumerical}, whose proof Knuth credits to
Pan~\cite{pan1966methods}; see also~\cite[Chapter~5]{burgisser1997algebraic}).
Hence over any infinite field, monic polynomials of degree $n$ need $\lceil n/2\rceil$
multiplications, whatever the preprocessing.
For odd $n$ this equals the $\lfloor n/2\rfloor+1$ of \Cref{thm:main}, so the
odd-degree schedules of this paper are optimal outright.
For even $n=2m$ the bound is $m$, and it is attained: by Motzkin and
Belaga~\cite{belaga1958some,pan1966methods} with algebraic preprocessing, and at
$n=6$ even with \emph{rational} preprocessing.  Pan's scheme~\cite{pan1961schemes}
(displayed as (16) in \cite[\S4.6.4]{knuth1997seminumerical})
\[
   z=(x+\alpha_0)x+\alpha_1,\qquad w=z+x+\alpha_2,\qquad
   P=\bigl((z-x+\alpha_3)\,w+\alpha_4\bigr)\,z+\alpha_5
\]
evaluates monic sextics with three multiplications, and its parameters are
rational functions of the coefficients---but with the denominator
$27c_3-18c_5c_4+5c_5^3$, so on that hypersurface the preprocessing is undefined.
(The bounds of Paterson and Stockmeyer~\cite{paterson1973evaluation} concern the
different model in which multiplications by the coefficients are free, where
$\Theta(\sqrt n)$ nonscalar multiplications are necessary and sufficient.)
The remaining question is therefore whether the exceptional set can be removed.
The decoders of this paper have no exceptional inputs (they divide only by
integers), and we conjecture that this costs exactly one multiplication at even
degrees: \emph{for $n\ge3$, evaluating monic polynomials of degree $2n$ with
everywhere-defined rational decoding requires at least $n+1$ multiplications}.
The restriction $n\ge3$ is necessary: the quartic program above evaluates every
monic degree-$4$ polynomial with two multiplications and rational decoding with no
exceptional inputs, so the statement fails at $n=2$.
Here we prove the first open case: degree $6$ cannot be done with three
multiplications, so Pan's hypersurface cannot be avoided.

\begin{proof}[Proof sketch]
The full proof is in \Cref{appendix:lower}; it is a Jacobian obstruction.
Let $J_F(p)$ be the $6\times6$ Jacobian of $F$.  If $F$ had an
everywhere-defined rational left inverse with coordinates $g_i/h_i$, then
$g_i(F(p))=p_i\,h_i(F(p))$ identically; differentiating at a point $p_0$ with
$J_F(p_0)v=0$, $v\neq0$, leaves $v_i\,h_i(F(p_0))=0$ for all $i$, which is
impossible because no $h_i$ vanishes.  So it suffices to exhibit, for every
three-multiplication program, one parameter point with singular Jacobian.

Every such program can be brought to a normal form
$u_1=x(R_{1,0}x+a_1)$, $u_2=\ell_2r_2$, $u_3=\ell_3r_3$ with $\ell_i,r_i$
affine in $x$ and the earlier $u_j$, and
$P=s_3u_3+s_2u_2+s_1u_1+s_0x+b_1$, by the gauge identity
$(Ax+a)(Bx+b)=A\,x(Bx+b+\frac BAa)+ab$, whose $x$-free correction $ab$ is
absorbed into the later constant slots.  The six normal-form slots
$q=(a_1,a_2,b_2,a_3,b_3,b_1)$ are therefore a polynomial map $q=Q(p)$ of
degree two of the parameters, and $J_F(p)=J(Q(p))\,DQ(p)$ with $J$ the
Jacobian of the normal form.  Unless $Q$ is an explicit quadratic
automorphism of $\mathbb{F}^6$, $DQ$ is singular at an explicit point (a
kernel direction of the slot map, or the midpoint of two parameter points
with the same normal-form slots, using $\mathrm{char}(\mathbb{F})\neq2$);
when it is, a singular point of $J$ pulls back to one of $J_F$.  So it
suffices to make $J$ singular for every normal form.  The sensitivities
$\partial P/\partial(\text{slot})$ are products of multiplicands with
the adjoints $\bar u_i=\partial P/\partial u_i$.  The argument splits on the
determinant $D$ of the coefficients with which $u_2,u_1$ enter the last gate.
If $D\neq0$, the slots can be chosen so that $\ell_3$ and $r_3$ take equal
values at two evaluation points while $\bar u_1,\bar u_2$ vanish there, so
two rows of $J$ coincide.  If $D=0$, a second-difference functional over
three evaluation points that annihilates constants and $x$ annihilates all
six sensitivities, so three rows of $J$ are dependent.  Both this case
split (for every circuit in the normal form and every affine reparameterization of
its six slots; \texttt{FastPoly/LowerBound/Main.lean}) and the reduction to the
normal form (\texttt{FastPoly/LowerBound/General/}) are machine-checked in Lean:
the lemma as stated, for every program of the displayed shape and every affine
slot map $Mp+h_0$, is \texttt{no\_rationalInverse\_general}.  The end of
\Cref{appendix:lower} lists the correspondence and the one remaining modelling
step, the shape of the program display.
\end{proof}
\subsection{Beyond degree six}

We conjecture that the pattern continues: evaluating a generic monic
polynomial of degree $2n$ with everywhere-defined rational decoding requires at least $n+1$
multiplications.  The proof above suggests an approach: in the normal form,
each additional multiplication contributes two constant slots, hence two
Jacobian columns (and two further evaluation points, hence two rows), and the moment
functionals $\langle\,\cdot\,\rangle$ annihilating the partial derivatives
satisfy one quadratic (conic) consistency condition per level.  Beyond three
multiplications these conditions are no longer linear in the unknown
functional, and making the count rigorous is an open problem
(\Cref{sec:open-problems}).

\section{Experimental Evaluation}
\label{sec:experiments}

We evaluate polynomial hashing constructions over the finite field $\mathbb{F}_{2^{64}}$
in two distinct settings:
\begin{enumerate}
    \item \textbf{$k$-wise independent hashing} (Section~\ref{sec:experiments:kwise}):
          The polynomial coefficients are random \emph{keys}, and we hash a single data point $x$.
          Applications include hash tables, load balancing, and randomized algorithms
          requiring bounded independence.
    \item \textbf{Universal hashing} (Section~\ref{sec:experiments:universal}):
          The polynomial evaluation point(s) are random \emph{keys}, and we hash a
          variable-length \emph{message}.
          Applications include message authentication codes (MACs), checksums,
          and data structure fingerprinting.
\end{enumerate}
The key distinction is key size: $k$-wise hashing uses $O(k)$ keys to hash one value,
while universal hashing uses $O(1)$ keys to hash messages of any length.

\subsection{Experimental Setup}

We benchmark on two platforms:
\begin{itemize}
    \item \textbf{ARM}: Apple M2 Pro (ARM64) running macOS, using ARM NEON intrinsics
          for carryless multiplication via the \texttt{PMULL} instruction.
    \item \textbf{x86}: Intel Xeon Platinum 8375C (Ice Lake, x86-64) running Linux, using Intel intrinsics
          for carryless multiplication via the \texttt{PCLMULQDQ} instruction.
          One x86 result is from a different host: the prime-field kernels
          of \Cref{sec:experiments:crypto} report the run on an AMD EPYC
          9R14 (Zen~4, Linux) recorded in
          \texttt{tools/\allowbreak bench/\allowbreak x86\_output.txt}.
\end{itemize}
Both instructions compute the product of two 64-bit polynomials over $\mathbb{F}_2$
using dedicated hardware support.

We compare the following evaluation methods, all operating over $\mathbb{F}_{2^{64}}$:
\begin{itemize}
    \item \textbf{Horner}: Standard Horner's rule with carryless multiplication
          and full polynomial reduction after each step.
          For a monic degree-$n$ polynomial (with $n$ coefficient keys), this requires $n-1$ multiplications
          executed sequentially.
    \item \textbf{Lemire}: Horner's rule using Lemire's fast modular reduction~\cite{lemire2019fast},
          which replaces one of the three \texttt{PMULL} operations in the Barrett reduction
          with a 256-byte table lookup.
    \item \textbf{Estrin}: Estrin's parallel evaluation scheme~\cite{estrin1960organization},
          which groups coefficients into pairs and evaluates them in parallel.
          This exposes instruction-level parallelism but requires computing powers
          $x^2, x^4, \ldots$, increasing the total multiplication count.
    \item \textbf{Rabin--Winograd}: The recursive divide-and-conquer method of
          Rabin and Winograd~\cite{rabin1972number}, which splits a polynomial
          into two halves and evaluates them with shared intermediate values.
          This achieves $\lceil n/2 \rceil + O(\log n)$ multiplications.
    \item \textbf{This Paper}: Our optimized polynomial chains,
          using the minimum number of multiplications for each degree
          as described in this paper.  By \Cref{thm:construction-height}
          these schedules can moreover be arranged with multiplicative height
          $2\lceil\log_2 k\rceil+O(1)$ at the same gate counts, so they do not
          trade critical-path depth for the lower multiplication count.
\end{itemize}

\begin{sloppypar}
Each algorithm was benchmarked by hashing $10^6$ random 64-bit integers per
repetition.  The harness behind \Cref{tab:kwise_both,tab:injective,tab:injective:universal}
and the tabulation comparison (\texttt{test\_speed\_function64} in
\texttt{tools/\allowbreak bench/\allowbreak carryless\_arm.cpp} and
\texttt{tools/\allowbreak bench/\allowbreak carryless.cpp};
\texttt{tools/\allowbreak bench/\allowbreak bench\_tabrows.cpp} times every row of the tabulation comparison, the non-polynomial hashes included, under the same wrapper)
runs $200$ repetitions, sorts them, keeps the fastest $100$ and reports their
mean and population standard deviation.  Every ``mean$\pm$std'' in those
tables, and every error bar in the figures, is of that selected half: the
deviation measures the spread of the fast tail, not the sampling variability
of an ordinary mean, and we derive no confidence interval from it.  The
application benchmarks use different rules, stated with their tables: the
fastest $50$ of $100$ repetitions for the sketch kernel; for the hash-table
and filter kernels a single timed build and, for lookups and queries, the
total over $5$ passes of the query set, with no selection; and one timed loop
over the point set for the prime-field kernels.
We measure evaluation time only; key setup and preprocessing (when applicable) are excluded.
\end{sloppypar}

\paragraph{Platform scope.}
We report results for both ARM (Apple M2 Pro) and x86 (Intel Xeon Platinum
8375C; AMD EPYC 9R14 for the prime-field kernels, as noted above).
The qualitative trade-offs (critical-path structure and the cost of modular reduction)
are consistent across platforms, though quantitative speedups vary with microarchitecture.

\paragraph{Reproducibility.}
\begin{sloppypar}
The ARM benchmarks are in \texttt{tools/bench/carryless\_arm.cpp}, compiled with
\texttt{Apple clang 17.0.0} using \texttt{-O3 -std=c++17 -march=armv8-a+crypto}.
The x86 benchmarks are in \texttt{tools/bench/carryless.cpp}, compiled with
\texttt{clang 21.1.8} using \texttt{-O3 -std=c++17 -march=native}; the x86
header \texttt{injective\_hashing.h} is a line-for-line port of the ARM one,
checked bit-for-bit against it on the same inputs.
We use the same input sequence across methods and include a warmup phase before timing.
\end{sloppypar}

\paragraph{Implementation Details.}
Our ARM implementation uses several key optimizations:
(1) Keys are stored as scalar 64-bit integers rather than 128-bit SIMD vectors,
    allowing XOR operations to use the faster scalar ALU;
(2) Modular reduction uses three \texttt{PMULL} operations rather than table lookup,
    which we found to be faster on Apple Silicon;
(3) Independent multiplication chains are interleaved to exploit instruction-level parallelism.

\paragraph{Measured families.}
\label{sec:experiments:families}
\Cref{tab:families} records, for every family timed in this section, the
implementation that was run, whether the circuit it computes is displayed in
the paper, and which proof, if any, covers it.  Only the rows with a proof
carry an independence or collision guarantee; the others are timings of
search candidates.
\begin{table}[!htbp]
\centering
\caption{Correspondence between the measured families and the proofs.
         ``Circuit'' names the displayed circuit the implementation computes,
         or notes that it is not displayed.  The circuits marked ``timing
         only'' have no inverse displayed in this paper, so no guarantee is
         claimed for them beyond their multiplication count.}
\label{tab:families}
\footnotesize
\setlength{\tabcolsep}{3pt}
\begin{tabular}{p{0.95in}p{1.6in}p{1.25in}p{1.7in}}
\toprule
Family (field, degree) & Implementation & Circuit & Proof; data \\
\midrule
\raggedright $\F_{2^{64}}$, $k=3$ (\Cref{tab:kwise_both}, \Cref{sec:experiments:tabulation}) &
\raggedright \texttt{smartcl\_64::\allowbreak mult3} in \texttt{tools/\allowbreak bench/\allowbreak framework/\allowbreak fast\_hashing\_arm.h} and \texttt{fast\_hashing.h} &
\raggedright the cubic $Q_3$ of \Cref{sec:main-theorem:outline} with $H_2=x^2$ &
\raggedright triangular decoder given there, every characteristic; $10^6$ uniform inputs, uniform keys \tabularnewline
\raggedright $\F_{2^{64}}$ and $\F_{2^{61}-1}$, $k=4$ (\Cref{sec:experiments:tabulation}) &
\raggedright \texttt{quartic2\_64} (Motzkin's two-multiplication quartic), \texttt{quartic3\_64} (the three-multiplication lift $x\,Q_3(x)+c_0$) and \texttt{motzkin\_61} (Motzkin's quartic over $2^{61}-1$) in \texttt{tools/\allowbreak bench/\allowbreak framework/\allowbreak fast\_hashing\_arm.h} and \texttt{fast\_hashing.h}; \texttt{mult4} is the first on ARM and the second on x86 &
\raggedright footnote in \Cref{sec:experiments:kwise}; \Cref{sec:experiments:tabulation} &
\raggedright Motzkin over $\F_{2^{64}}$: $3$-wise only (footnote); lift: bijective, hence $4$-wise; Motzkin over $\F_{2^{61}-1}$: bijective onto the monic quartics, hence $4$-wise on $[p]$ \tabularnewline
\raggedright $\F_{2^{64}}$, $k=5$ &
\raggedright \texttt{mult5} &
\raggedright the finalizer circuit \eqref{eq:ph:chain5} &
\raggedright \Cref{lem:ph:chain}, every characteristic-$2$ field \tabularnewline
\raggedright $\F_{2^{64}}$, $k=7$ &
\raggedright ARM \texttt{mult7}, x86 \texttt{mult7\_alt2}: $y=x(x+k_0)$, $z=(x+k_1)(y+k_2)$, $t=z(z+k_3)$, $u=(x+y+t+k_4)\allowbreak(x+k_5)$, $P=u+k_6$ &
\raggedright not displayed; differs from the degree-$7$ circuit of \Cref{appendix:polynomials} &
\raggedright timing only \tabularnewline
\raggedright $\F_{2^{64}}$, $k=9$ &
\raggedright ARM \texttt{mult9}, x86 \texttt{mult9\_alt}: $y=x^2$, $z=(x+k_0)(y+k_1)$, $u=(x+k_2)(y+k_3)$, $t=(z+k_4)\allowbreak(y+z+k_5)$, $v=(t+k_6)\allowbreak(x+z+k_7)$, $P=u+v+k_8$ &
\raggedright not displayed; differs from both degree-$9$ circuits of \Cref{appendix:polynomials} &
\raggedright timing only \tabularnewline
\raggedright $\F_{2^{64}}$, degrees $7$--$21$ (nanosecond timings in \Cref{appendix:polynomials}) &
\raggedright \texttt{tools/\allowbreak bench/\allowbreak appendix\_char2\_\allowbreak circuits\_arm.cpp} &
\raggedright the displayed circuits &
\raggedright inverses for $7,15,17,19,21$; the displayed $9,11,13$ are search candidates and their certified alternatives (A.0) are not timed \tabularnewline
\raggedright $\F_{2^{89}-1}$, $k=3$--$9$ (\Cref{fig:bench_all}) &
\raggedright \texttt{smartpoly\_64} &
\raggedright not displayed &
\raggedright timing only \tabularnewline
\raggedright $\F_{2^{89}-1}$ and Goldilocks, degrees $13$--$21$ (\Cref{sec:experiments:crypto}) &
\raggedright \texttt{x2s\_mersenne\_\allowbreak chains.h}, \texttt{x2s\_goldilocks\_\allowbreak chains.h}, generated from \texttt{tools/x2s.res} &
\raggedright not displayed (the degree-$13$ chain is the degree-$13$ alternative of (A.0)) &
\raggedright timing only: no inverse over $\F_p$ (the (A.0) certificate is for characteristic~$2$); points $x_i=i$ or uniform \tabularnewline
\raggedright injective recurrence, $\F_{2^{64}}$ (\Cref{tab:injective,tab:injective:universal,tab:injective:adversarial}) &
\raggedright \texttt{tools/\allowbreak bench/\allowbreak framework/\allowbreak injective\_hashing.h}; vector-resident rows: \texttt{tools/\allowbreak bench/\allowbreak adversarial/} &
\raggedright \eqref{eq:injective-single-key} and \Cref{def:injective:recurrence} &
\raggedright \Cref{lem:injective:coeffmap} with Schwartz--Zippel; uniform random messages \tabularnewline
\raggedright ChainHash (\Cref{tab:injective:adversarial}) &
\raggedright \texttt{tools/\allowbreak bench/\allowbreak chainhash/\allowbreak chainhash.h} &
\raggedright \Cref{def:ph:hash} &
\raggedright \Cref{thm:ph:collision,thm:ph:kwise} for the ideal key; the seeded implementation is outside the theorems (\Cref{rem:ph:gaps}); SMHasher3 and \Cref{app:adversarial} \tabularnewline
\bottomrule
\end{tabular}
\end{table}

\subsection{\texorpdfstring{$k$}{k}-wise Independent Hashing}
\label{sec:experiments:kwise}

We evaluate the monic degree-$k$ polynomial family $h(x) = x^k + a_{k-1}x^{k-1} + \cdots + a_1 x + a_0$ over $\mathbb{F}_{2^{64}}$.
Here the $k$ coefficients $(a_0, \ldots, a_{k-1})$ are the random key, and we hash a single data point $x$.

Table~\ref{tab:kwise_both} shows the carryless hashing results on both platforms.
Our polynomial chains are the fastest method for every $k\ge5$ on both platforms; on ARM, Rabin--Winograd ties our method at $k=7$.
At $k=3$ the monic cubic admits no saving (two multiplications either way), and plain Horner (ARM) or Lemire's reduction (x86) is fastest.

\begin{table}[!htbp]
\centering
\caption{$k$-wise carryless hashing (\textmu s per $10^6$ hashes).
         Best method in bold for each platform.}
\label{tab:kwise_both}
\small
\begin{tabular}{c|rrrrr|rrrrr}
    & \multicolumn{5}{c|}{ARM (Apple M2 Pro)} & \multicolumn{5}{c}{x86 (Intel Xeon 8375C)} \\
    $k$ & Horner & Lemire & Estrin & R--W & Ours & Horner & Lemire & Estrin & R--W & Ours \\
    \hline
    3 & \textbf{825} & 882 & 1137 & 900 & 870 & 1930 & \textbf{1649} & 2190 & 1930 & 1750 \\
    5 & 1895 & 2149 & 2326 & 2204 & \textbf{1434} & 4816 & 4223 & 4621 & 4504 & \textbf{3090} \\
    7 & 3225 & 4115 & 3126 & \textbf{2235} & \textbf{2241} & 8535 & 7749 & 6217 & 5074 & \textbf{4880} \\
    9 & 4869 & 6694 & 4910 & 3996 & \textbf{2641} & 12817 & 12031 & 8709 & 8250 & \textbf{5952} \\
\end{tabular}
\end{table}
For $k = 9$, our method achieves a speedup of $1.84\times$ over Horner on ARM and $2.15\times$ on x86.
For smaller $k$, we use Motzkin's quartic (2 multiplications), which over $\F_{2^{64}}$ is only $3$-wise independent\footnote{Over $\F_{2^{64}}$ Motzkin's two-multiplication quartic $(y+b)(y+x+c)+d$ with $y=x(x+a)$ has $x^3$-coefficient $2a+1=1$, so its coefficients range only over the monic quartics with $e_3=1$ and the family is exactly $3$-wise independent (the remaining three coefficients are uniform); a $4$-wise quartic needs a third multiplication in characteristic~$2$, the same Frobenius effect as in \Cref{sec:ph:twist}.}, and the two-multiplication cubic of \Cref{sec:main-theorem:outline} for $k=3$ (the same count as Horner).
The speedup increases with $k$ because our polynomial chains save more multiplications
at higher degrees: for monic degree $n$, Horner uses $n-1$ multiplications while our
chains use only $\lfloor n/2\rfloor+1$; \Cref{thm:main} does not apply over
$\F_{2^{64}}$, so the count here is that of the characteristic-$2$ circuits
themselves (\Cref{tab:families}).

\paragraph{Discussion.}
The Rabin--Winograd method is strongest at $k = 7$, where its recursive structure
aligns well with the degree, tying our method on ARM.
For other values of $k$, Rabin--Winograd is slower than our method but remains competitive
with standard approaches.
Estrin's scheme provides modest speedup at higher degrees due to its parallelism,
but the additional multiplications limit its effectiveness.
Lemire's reduction is slower than Horner because, on Apple Silicon,
the \texttt{PMULL} instruction is faster than table lookup operations.

Our method achieves the best performance by combining two key insights:
(1) minimizing the number of multiplications through carefully constructed polynomial chains,
and (2) optimizing the implementation to avoid unnecessary data movement between
scalar and vector registers.

On x86 (Intel Xeon 8375C), similar trends hold: our method achieves $2.15\times$ speedup
over Horner at $k=9$ (vs.\ $1.84\times$ on ARM).
Rabin--Winograd is competitive at $k=7$ on both platforms.
Notably, Lemire's table-lookup reduction is faster than Horner on x86,
unlike ARM where \texttt{PMULL} outperforms memory access.

\paragraph{Scope of application benchmarks.}
The remaining experiments measure end-to-end kernels where hashing is only one component.
These are representative workloads rather than an exhaustive study of all data structure designs and microarchitectures.

\subsection{Application-Level Throughput: Sketch Updates}
\label{sec:experiments:sketch}

To check that these savings translate beyond microbenchmarks, we also measure a simple
CountSketch-style update kernel: for each key, compute a hash value, use it to choose a bucket
and a sign bit, and update an array of counters.
This mixes hashing with memory traffic, so speedups are expected to be smaller than in the raw
hash-evaluation benchmark.

With table size $2^{16}$ and $2^{20}$ updates (mean of the fastest $50$ of $100$ repetitions), we measure:
\begin{center}
\begin{tabular}{l|rrr|rrr}
& \multicolumn{3}{c|}{ARM} & \multicolumn{3}{c}{x86} \\
Method & Horner & R--W & Ours & Horner & R--W & Ours \\
\hline
ns/update & 5.49 & 4.96 & \textbf{4.33} & 38.3 & 24.9 & \textbf{23.7}
\end{tabular}
\end{center}
On ARM, our method achieves $1.27\times$ speedup over Horner.
On x86, the speedup is $1.62\times$, reflecting the larger benefit from reducing multiplications.
The benchmark is implemented in \texttt{tools/\allowbreak bench/\allowbreak app\_\allowbreak countsketch\_\allowbreak arm.cpp} (ARM) and
\texttt{tools/\allowbreak bench/\allowbreak app\_\allowbreak countsketch\_\allowbreak x86.cpp} (x86).

\subsection{Application-Level Throughput: Hash Tables}
\label{sec:experiments:hashtable}

Finally, we benchmark a simple linear-probing hash table,
measuring insertion throughput (table construction) and lookup throughput at load factor $70\%$.
As expected, improvements are smaller than in the microbenchmarks but still measurable.

With table size $2^{20}$ we obtain the following (insertion is a single
timed build after a $2^{12}$-key warm-up; lookups are the total over $5$
passes of the query set, reported per lookup; no selection):
\begin{center}
\begin{tabular}{l|rrr|rrr}
& \multicolumn{3}{c|}{ARM} & \multicolumn{3}{c}{x86} \\
Method & Horner & R--W & Ours & Horner & R--W & Ours \\
\hline
ns/insert & 32.5 & 28.5 & \textbf{28.5} & 37.4 & 29.9 & \textbf{24.5} \\
ns/query  & 42.7 & 41.9 & \textbf{38.6} & 46.3 & 35.4 & \textbf{31.4}
\end{tabular}
\end{center}
On ARM, our method achieves $1.14\times$ speedup for inserts and $1.11\times$ for lookups.
On x86, speedups are larger: $1.53\times$ for inserts and $1.47\times$ for lookups.
The benchmark is implemented in \texttt{tools/bench/\allowbreak app\_linearprobe\_arm.cpp} (ARM) and
\texttt{tools/bench/\allowbreak app\_linearprobe\_x86.cpp} (x86).

\subsection{Application-Level Throughput: Membership Filters}
\label{sec:experiments:filter}

We also implement a simple XOR-filter-style approximate membership structure, which builds a
small fingerprint array using a peeling algorithm and answers queries with three table reads and
two XORs.
We measure both build throughput (ns/key) and query throughput (ns/query) at load $75\%$ and table
size $2^{20}$ (build is a single timed construction including any re-seeded
peel retries; queries are the total over $5$ passes of the query set,
reported per query; no selection):
\begin{center}
\begin{tabular}{l|rrr|rrr}
& \multicolumn{3}{c|}{ARM} & \multicolumn{3}{c}{x86} \\
Method & Horner & R--W & Ours & Horner & R--W & Ours \\
\hline
ns/key (build) & 96.1 & \textbf{82.4} & 92.0 & 142.0 & 110.3 & \textbf{97.1} \\
ns/query       & 5.56 & 5.53 & \textbf{5.04} & 19.6 & 15.6 & \textbf{11.1}
\end{tabular}
\end{center}
On ARM, our method improves query throughput by $1.10\times$ over Horner, while R--W is faster at build time.
On x86, our method is fastest for both: $1.46\times$ speedup for build and $1.76\times$ for queries.
The benchmark is implemented in \texttt{tools/bench/\allowbreak app\_xorfilter\_arm.cpp} (ARM) and
\texttt{tools/bench/\allowbreak app\_xorfilter\_x86.cpp} (x86).

\subsection{Prime-Field Evaluation Kernels: Share Generation and Random-Point Evaluation}
\label{sec:experiments:crypto}

We also benchmark the ``evaluate one polynomial at many points'' inner loop that
appears in threshold secret sharing~\cite{shamir1979share} and related primitives.
These are evaluation-kernel benchmarks: the chain's keys are drawn uniformly at
random and the evaluation points are the input; sampling the polynomial,
placing a secret in it, preprocessing its coefficients into keys, and any
setup are not timed, and no secret-sharing or pseudorandom-function security
claim is made.  Each number is one timed loop over the point set, reported
per evaluation.
To model prime-field arithmetic (where a multiplication is ``full width'' regardless of whether
the input happens to be a small integer), we use the Mersenne prime field
$\F_p$ with $p = 2^{89}-1$ and implement multiplication using 128-bit integer arithmetic and
Mersenne reduction.

We compare Horner evaluation against our ``x2s'' chains (degrees $13,15,17,19,21$ from our search;
\texttt{tools/bench/framework/x2s\_mersenne\_chains.h}, generated from \texttt{tools/x2s.res}).
These are search candidates: they are not the circuits of \Cref{appendix:polynomials}, and no
inverse is displayed for them over $\F_p$ (the degree-$13$ chain coincides with the degree-$13$
alternative of \Cref{appendix:polynomials}, whose certificate is for characteristic~$2$), so
these rows time evaluation only (\Cref{tab:families}).  We time them
in two regimes:
(1) ``Share generation'': evaluation points $x_i = i$ (but treated as elements of $\F_p$, so each
multiply is a full field multiplication), and
(2) ``random-point evaluation'': evaluation points $x_i \sim \F_p$ uniformly at random,
the access pattern of a polynomial evaluated at a random point, as in polynomial
universal hashing keyed by the coefficients.

On an Apple M2 Pro, we observe consistent $2.0\times$--$2.5\times$ speedups over full-field Horner
evaluation for degrees $13$--$21$ in both regimes; see the benchmark implementation in
\texttt{tools/\allowbreak bench/\allowbreak shamir\_\allowbreak sharegen\_\allowbreak mersenne.cpp}.
Including the cost of writing shares to memory gives essentially the same speedups; see
\texttt{tools/\allowbreak bench/\allowbreak shamir\_\allowbreak sharegen\_\allowbreak mersenne\_\allowbreak store.cpp}.
The x86 rows of the two tables below, and the x86 store-to-memory range for
Goldilocks, are the run on an AMD EPYC 9R14 recorded in
\texttt{tools/bench/x86\_output.txt}, not the Xeon 8375C of the other x86
tables; the ARM rows are the Apple M2 Pro.

\begin{center}
\begin{tabular}{c|ccccc}
Degree $n$ & 13 & 15 & 17 & 19 & 21 \\
\hline
ARM: Sharegen ($x_i=i$) & 2.02$\times$ & 2.37$\times$ & 2.47$\times$ & 2.49$\times$ & 2.44$\times$ \\
ARM: Random point ($x_i \sim \F_p$) & 2.01$\times$ & 2.34$\times$ & 2.48$\times$ & 2.48$\times$ & 2.48$\times$ \\
\hline
x86 (EPYC 9R14): Sharegen ($x_i=i$) & 1.66$\times$ & 4.50$\times$ & 4.57$\times$ & 4.63$\times$ & 4.60$\times$ \\
x86 (EPYC 9R14): Random point ($x_i \sim \F_p$) & 1.66$\times$ & 4.50$\times$ & 4.57$\times$ & 4.63$\times$ & 4.60$\times$
\end{tabular}
\end{center}
On the EPYC, the speedups for degrees $15$--$21$ reach $4.5\times$--$4.6\times$, significantly
exceeding the ARM results.

\paragraph{Goldilocks prime field.}
Prime fields of 64-bit size are widely used in proof systems; a common choice is the
``Goldilocks'' prime $p = 2^{64}-2^{32}+1$.
We repeat the same benchmark in $\F_p$ and observe similar improvements:
\begin{center}
\begin{tabular}{c|ccccc}
Degree $n$ & 13 & 15 & 17 & 19 & 21 \\
\hline
ARM: Sharegen ($x_i=i$) & 1.98$\times$ & 2.23$\times$ & 2.39$\times$ & 2.33$\times$ & 2.32$\times$ \\
ARM: Random point ($x_i \sim \F_p$) & 1.96$\times$ & 2.29$\times$ & 2.35$\times$ & 2.34$\times$ & 2.34$\times$ \\
\hline
x86 (EPYC 9R14): Sharegen ($x_i=i$) & 1.27$\times$ & 1.38$\times$ & 1.42$\times$ & 1.43$\times$ & 1.37$\times$ \\
x86 (EPYC 9R14): Random point ($x_i \sim \F_p$) & 1.27$\times$ & 1.38$\times$ & 1.42$\times$ & 1.43$\times$ & 1.37$\times$
\end{tabular}
\end{center}
The benchmark is implemented in \texttt{tools/bench/\allowbreak app\_goldilocks\_stark\_eval.cpp}.
On ARM, including memory stores gives $2.19\times$--$2.36\times$ speedup;
on the EPYC, the range is $1.31\times$--$1.90\times$.
The smaller x86 gains reflect that Goldilocks multiplication (64-bit) is relatively cheaper,
reducing the benefit of fewer multiplications.
\par\noindent\texttt{tools/bench/\allowbreak app\_goldilocks\_sharegen\_store.cpp} contains the benchmark.

\subsection{Comparison with Tabulation Hashing}
\label{sec:experiments:tabulation}

Tabulation hashing~\cite{patracscu2012power} achieves 3-wise independence using
table lookups, providing excellent performance for applications that only require
low independence.
On ARM, simple tabulation (8 tables of 256 entries) achieves
1668$\pm$11\,\textmu s per $10^6$ hashes; on x86, it achieves 2244$\pm$4\,\textmu s.
Every row of the table below is timed by one driver,
\texttt{tools/bench/bench\_tabrows.cpp}, under the wrapper and batching of the
$k$-wise comparison ($10^6$ inputs, fastest $100$ of $200$ repetitions with
fresh keys, mean $\pm$ standard deviation in \textmu s per $10^6$ hashes), on the
same two machines as the $k$-wise and injective tables
(\Cref{tab:kwise_both,tab:injective}): the Apple M2 Pro (Apple clang~17,
\texttt{-O3 -march=armv8-a+crypto}) and the Intel Xeon Platinum 8375C
(clang~21.1.8, \texttt{-O3 -march=native}, pinned to eight cores).
MurmurHash3 and xxHash64 are the single-word variants (the 64-bit finalizer and
the 8-byte xxHash64 path), Dietzfelbinger is the $W=192$ Horner evaluation
described below, and the polynomial rows are the implementations of
Table~\ref{tab:kwise_both}; because the wrapper differs, they remain separate
runs from that table (x86, $k=3$: $1750$ there against $1779$\,\textmu s here).

Notably, our 3-wise independent hash function runs in 886$\pm$13\,\textmu s on
ARM---nearly $2\times$ faster than tabulation with the same independence
guarantee; on x86 (1779\,\textmu s) it is $1.26\times$ faster than tabulation.
The table lists three quartics.  Motzkin's two-multiplication quartic
$y = x(x+a)$, $P=(y+b)(y+x+c)+d$ over $\F_{2^{64}}$ is only $3$-wise (see the
footnote above) and costs 992 / 1903\,\textmu s, still below tabulation on both
platforms.  A $4$-wise quartic over $\F_{2^{64}}$ needs a third multiplication:
the lift $x\,Q_3(x)+c_0$ (\Cref{tab:families}) uses three carryless
multiplications and runs in 1428 / 2909\,\textmu s, between our $k=3$ and $k=5$
rows on both platforms.  Alternatively, the same two-multiplication Motzkin
circuit can be evaluated over the Mersenne prime $p=2^{61}-1$, where the odd
characteristic makes the key map a bijection onto the monic quartics
($b_0=(a_3-1)/2$, then $b_1,b_2,b_3$ by back-substitution), so with uniform keys
the hash is exactly $4$-wise independent on the $61$-bit universe ($64$-bit
inputs are folded modulo $p$).  Its cost, 2008 / 3199\,\textmu s, exceeds even
our $5$-wise carryless quintic ($1470$ / $3141$\,\textmu s): each of its two
multiplications is a $64\times 64\to 128$-bit integer product followed by a
fold rather than a single carryless multiply, so in the presence of
\texttt{PMULL}/\texttt{PCLMULQDQ} the third carryless multiplication is cheaper
than the odd characteristic.

We also compare against Dietzfelbinger's polynomial hashing over integers~\cite{dietzfelbinger1996universal},
which evaluates $g(x) = \sum_{i=0}^{k-1} a_i x^i \bmod 2^W$ using standard integer arithmetic
and extracts the top 64 bits as the hash.
For $k$-wise independence with 64-bit input/output, this requires $W \geq 127 + \lceil\log_2\binom{k}{2}\rceil$,
so $W=128$ suffices only for $k=2$; we use $W=192$ for $k \geq 3$.
The 192-bit arithmetic requires 3 multiplications per Horner step versus 1 carryless multiply for our method,
explaining the performance gap.

\begin{center}
\small\setlength{\tabcolsep}{4pt}
\begin{tabular}{l|rr|rr}
    & \multicolumn{2}{c|}{ARM (Apple M2 Pro)} & \multicolumn{2}{c}{x86 (Intel Xeon 8375C)} \\
    Method & Time (\textmu s) & Indep. & Time (\textmu s) & Indep. \\
    \hline
    MurmurHash3 & 589$\pm$13 & none & 854$\pm$5 & none \\
    xxHash64 & 1160$\pm$14 & none & 1888$\pm$8 & none \\
    Tabulation & 1668$\pm$11 & 3-wise & 2244$\pm$4 & 3-wise \\
    Dietzfelbinger ($k=3$) & 1563$\pm$29 & 3-wise & 2280$\pm$5 & 3-wise \\
    Dietzfelbinger ($k=5$) & 3085$\pm$24 & 5-wise & 4986$\pm$19 & 5-wise \\
    Dietzfelbinger ($k=7$) & 4893$\pm$44 & 7-wise & 7905$\pm$5 & 7-wise \\
    This Paper ($k=3$) & 886$\pm$13 & 3-wise & 1779$\pm$5 & 3-wise \\
    This Paper (Motzkin quartic, $\F_{2^{64}}$) & 992$\pm$10 & 3-wise & 1903$\pm$5 & 3-wise \\
    This Paper (quartic $x\,Q_3(x)+c_0$, $\F_{2^{64}}$) & 1428$\pm$17 & 4-wise & 2909$\pm$4 & 4-wise \\
    This Paper (Motzkin quartic, $\F_{2^{61}-1}$) & 2008$\pm$22 & 4-wise & 3199$\pm$8 & 4-wise \\
    This Paper ($k=5$) & 1470$\pm$17 & 5-wise & 3141$\pm$6 & 5-wise \\
    This Paper ($k=7$) & 2299$\pm$20 & not certified & 4968$\pm$6 & not certified \\
\end{tabular}
\end{center}
The three quartic rows are the implementations \texttt{quartic2\_64},
\texttt{quartic3\_64} and \texttt{motzkin\_61} of \Cref{tab:families};
the \texttt{mult4} of Table~\ref{tab:kwise_both} is the first of these on
ARM and the second on x86.  The $k=7$ row times the search circuit of the
$k=7$ row of \Cref{tab:families} (ARM \texttt{mult7}, x86
\texttt{mult7\_alt2}), for which no inverse is displayed: it has seven keys,
but no independence is claimed for it.

\subsection{Comparison Across Finite Fields}

\begin{figure}[!htbp]
    \centering
    \includegraphics[width=\columnwidth]{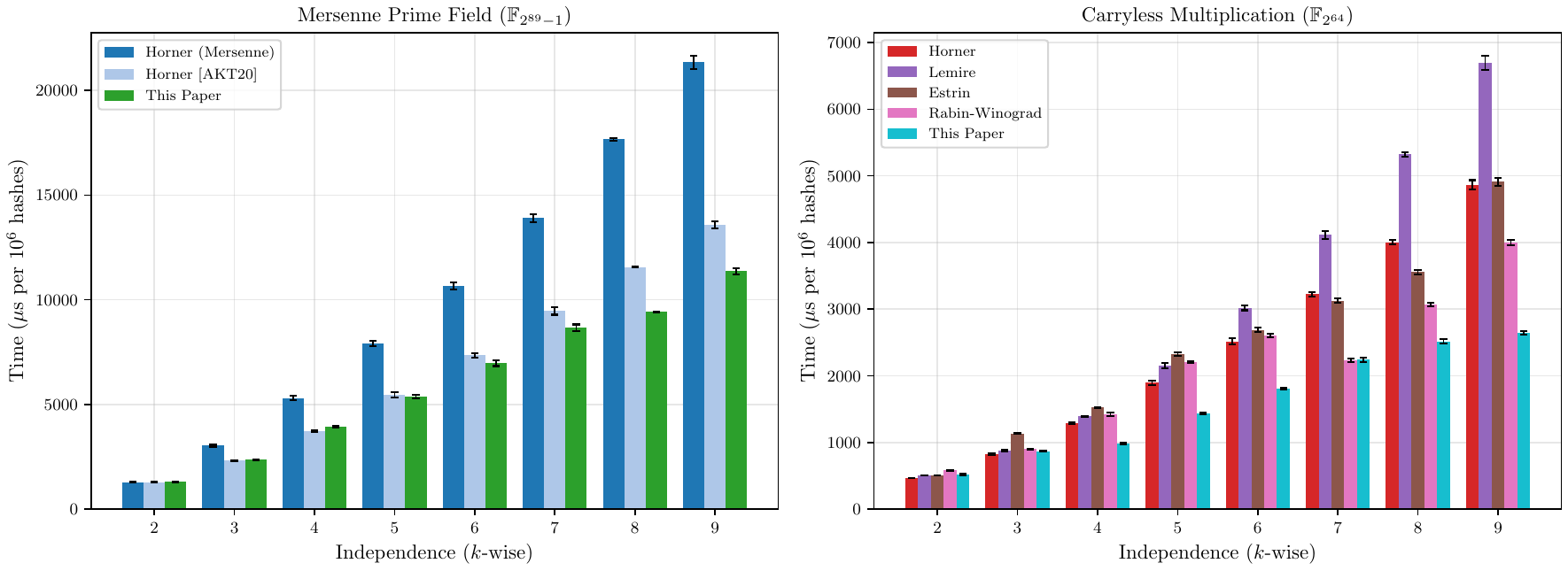}
    \caption{Comparison across finite fields: Mersenne prime $\mathbb{F}_{2^{89}-1}$ (left) vs.\
             carryless $\mathbb{F}_{2^{64}}$ (right).
             Carryless multiplication is $3$--$5\times$ faster due to
             hardware support via \texttt{PMULL}/\texttt{PCLMULQDQ}.
             Same $k$ on both panels; the vertical scales differ by
             $3$--$5\times$.  This Paper is the rightmost bar in each group.}
    \label{fig:bench_all}
\end{figure}

We also benchmark polynomial evaluation over the Mersenne prime field
$\mathbb{F}_{2^{89}-1}$, which uses 89-bit integers and standard multiplication
with modular reduction (Figure~\ref{fig:bench_all}).
For the Mersenne baseline, we use the optimized Horner implementation
from Ahle et al.~\cite{ahle2020power}, which employs fast branch-free
modular reduction and delays the final reduction to the end of the computation.

The carryless multiplication methods over $\mathbb{F}_{2^{64}}$ are consistently
$3$--$5\times$ faster than their Mersenne counterparts.
This speedup comes from hardware support for carryless multiplication (via \texttt{PMULL} on Apple M2 Pro),
which computes a 64-bit carryless product directly,
whereas Mersenne arithmetic requires multiple standard multiplications
and careful handling of the 89-bit modulus.

For $k=9$, our method over $\mathbb{F}_{2^{64}}$ (2641\,\textmu s mean) is
$4.3\times$ faster than the best Mersenne method in our benchmark (11361\,\textmu s mean).

\subsection{Injective Polynomial Hashing}
\label{sec:experiments:injective}

We briefly summarize the injective polynomial construction from Section~\ref{sec:injective}
before the comprehensive comparison in Section~\ref{sec:experiments:universal}.
The recurrence
\[
    P_0 = x, \quad P_i = a_i + (b_i + x^3)(P_{i-1} + x^2)
\]
uses a single random key $x$ (the single-key specialization~\eqref{eq:injective-single-key} of \Cref{def:injective:recurrence}, with $z=x$, $y=x^3$, $u=x^2$) to hash $2N$ message values $(a_1, b_1, \ldots, a_N, b_N)$
with $N$ multiplications---half the $2N-1$ required by Horner.

\begin{table}[!htbp]
\centering
\caption{Injective vs.\ Horner: $N$ multiplications vs.\ $2N-1$.
         Times in \textmu s per $10^6$ hashes (mean$\pm$std).}
\label{tab:injective}
\begin{tabular}{c|rrr|rrr}
    & \multicolumn{3}{c|}{ARM (Apple M2 Pro)} & \multicolumn{3}{c}{x86 (Intel Xeon 8375C)} \\
    $2N$ & Horner & Injective & Speedup & Horner & Injective & Speedup \\
    \hline
    8  & 5380$\pm$70 & 3830$\pm$50 & $1.40\times$ & 13390$\pm$20 & 11940$\pm$10 & $1.12\times$ \\
    16 & 20320$\pm$370 & 9600$\pm$110 & $2.12\times$ & 45610$\pm$30 & 24270$\pm$30 & $1.88\times$ \\
    32 & 72010$\pm$980 & 39780$\pm$500 & $1.81\times$ & 140650$\pm$30 & 60100$\pm$20 & $2.34\times$ \\
    64 & 218100$\pm$1540 & 111260$\pm$980 & $1.96\times$ & 362870$\pm$50 & 175840$\pm$50 & $2.06\times$ \\
\end{tabular}
\end{table}

The speedup approaches $2\times$ as expected from the multiplication count.
On x86, the speedup grows from $1.12\times$ at $2N=8$ to a peak of
$2.34\times$ at $2N=32$, and is $2.06\times$ at $2N=64$.
For parallel variants and comparison with other methods (CLNH, BRW, etc.),
see Section~\ref{sec:experiments:universal}.

\subsection{Universal Hashing}
\label{sec:experiments:universal}

Universal hashing uses $O(1)$ random keys to hash variable-length messages.
Unlike $k$-wise hashing where polynomial coefficients are keys,
here the evaluation point(s) are keys and message values form the polynomial.
We compare several constructions, each offering different trade-offs between
key size, multiplication count, and parallelism.

\subsubsection{Horner-based Methods}

The simplest universal hash evaluates the polynomial
$h(x) = m_0 x^{n-1} + m_1 x^{n-2} + \cdots + m_{n-1}$
where $x$ is the random key and $(m_0, \ldots, m_{n-1})$ is the message.

\paragraph{Horner (Sequential).}
Horner's rule computes $h_0 = m_0$, $h_i = h_{i-1} \cdot x + m_i$ for $i = 1, \ldots, n-1$.
This requires $n-1$ multiplications with a \emph{critical path} of $n-1$ dependent operations.
On CPUs where multiplication has multi-cycle latency, this sequential
dependency chain leaves execution units idle.

\paragraph{Horner-Unrolled.}
We process four values per iteration using Estrin-style grouping:
\begin{align}
& h x^4 + (m_i x + m_{i+1}) x^2 \nonumber\\
&\quad + (m_{i+2} x + m_{i+3}). \nonumber
\end{align}
The terms $(m_i \cdot x + m_{i+1})$ and $(m_{i+2} \cdot x + m_{i+3})$ can be computed
in parallel, exposing instruction-level parallelism (ILP) within each 4-element group.
This reduces effective critical path by $\approx 4\times$ with minimal overhead.

\paragraph{Horner-Parallel.}
Horner's recurrence $h_i = h_{i-1} \cdot x + m_i$ is a linear map with transfer function
$(c, d) = (x, m_i)$, meaning $h \mapsto h \cdot c + d$.
Writing $T_{c,d}(h)\coloneqq h\cdot c + d$, composition is
\[
   T_{c_2,d_2}\!\circ T_{c_1,d_1} = T_{c_1c_2,\; d_1c_2 + d_2}.
\]
Since $c = x$ for all steps, we precompute powers $x, x^2, x^4, \ldots$ and combine
message values in a binary tree with $O(\log n)$ depth.
This requires $O(n)$ total multiplications but only $O(\log n)$ on the critical path.

\subsubsection{Injective Polynomial (This Paper)}

Our injective construction from Section~\ref{sec:injective} uses the recurrence:
\[
P_0 = x, \quad P_i = a_i + (b_i + x^3)(P_{i-1} + x^2).
\]
The message $(a_1, b_1, \ldots, a_N, b_N)$ has $2N$ values, and the key is the single field element $x$ (with the derived
constants $x^2,x^3$) regardless of message length; the three-key family of \Cref{def:injective:recurrence} keeps three.
Each step requires one multiplication (computing $(b_i + x^3)(P_{i-1} + x^2)$),
so $N$ multiplications hash $2N$ values.
This is nearly $2\times$ fewer than Horner's $2N-1$ multiplications.

\paragraph{Injective-Parallel.}
The recurrence $P_i = a_i + c_i(P_{i-1} + x^2)$ where $c_i = b_i + x^3$ has transfer functions
$(c_i, d_i)$ with $d_i = a_i + c_i \cdot x^2$.
Unlike Horner, the coefficients $c_i$ vary, so we cannot simply precompute powers.
Instead, we compute all $(c_i, d_i)$ pairs in parallel (Phase~1: $N$ multiplications),
then perform parallel prefix reduction (Phase~2: $2N$ multiplications in $O(\log N)$ depth).
Total work is $3N$ multiplications, but critical path is only $O(\log N)$.

\paragraph{Injective-Lanes.}
The prefix scan pays for parallelism with multiplications.  Instead, deal the
$N$ pairs round-robin to $L$ independent copies of the sequential recurrence,
so that the $L$ multiplications of each round are independent, and combine the
lane results once at the end.  With a single key the lanes are combined as
$\sum_j P^{(j)}(x)\,x^{jD}$ with $D=3\lceil N/L\rceil+3$, above the degree of
every lane polynomial, so the lanes occupy disjoint degree ranges and the map
stays injective; the cost is $N+L-1$ multiplications plus $O(\log N)$ for
$x^D$, about half of parallel Horner's.  Two alternatives trade key size for a
tighter bound: an independent second key $y$ with $\sum_j P^{(j)}(x)\,y^j$
(collision probability at most $(3\lceil N/L\rceil+L+2)/|\F|$ by bivariate
Schwartz--Zippel), or an independent key per lane with
$\bigoplus_j x_j\,P^{(j)}(x_j)$ (at most $(3\lceil N/L\rceil+3)/|\F|$, the
per-lane degree, since conditioning on the other keys leaves a nonzero
polynomial in one variable, at the price of $3L$ key words).  In the last form
the multiplication by $x_j$ is essential: the final $a$-word enters every lane
with coefficient~$1$, so a plain XOR of the lanes has key-free collisions
(change the last $a$-word of two lanes by the same amount).  All three forms do
the same work per pair; we benchmark the single-key form with $L=\min(N,8)$.
With the three-key recurrence in every lane the bounds tighten to
$(\lceil N/L\rceil+L-1)/|\F|$ for a shared key triple plus one combining key
and $\lceil N/L\rceil/|\F|$ for independent triples per lane (messages of
equal length).

\subsubsection{CLNH (Multilinear Hashing)}

Carry-less NH (CLNH)~\cite{lemire2015clhash} is a multilinear hash:
\[
h = \bigoplus_{i=0}^{N-1} (k_{2i} \oplus m_{2i})(k_{2i+1} \oplus m_{2i+1})
\]
where $(k_0, \ldots, k_{2N-1})$ are random keys and $(m_0, \ldots, m_{2N-1})$ is the message.
All $N$ multiplications are \emph{completely independent}, giving $O(1)$ critical path depth.
This makes CLNH extremely fast on CPUs with multiple execution units.

\paragraph{Trade-off.}
CLNH requires $O(N)$ keys to hash an $N$-word message, compared to $O(1)$ for polynomial methods.
This is acceptable when key material can be generated once and reused (e.g., for a fixed
maximum message length), but problematic for variable-length or streaming applications.
We include CLNH as a performance baseline representing the ``best possible'' when key size
is unconstrained.

\subsubsection{BRW (Bernstein-Rabin-Winograd)}

The BRW construction~\cite{rabin1972number} uses divide-and-conquer to build a polynomial
with $O(\log N)$ depth.
For a message $(m_1, \ldots, m_n)$, the BRW polynomial is defined recursively:
\begin{align}
\text{BRW}(m_1) &= m_1 \\
\text{BRW}(m_1, m_2) &= m_1 \cdot x + m_2 \\
\text{BRW}(m_1, m_2, m_3) &= (x + m_1)(x^2 + m_2) + m_3 \\
\text{BRW}(m_1, \ldots, m_{2^k-1}) &= \text{BRW}(\text{left}) \cdot (x^{2^{k-1}} + m_{2^{k-1}}) + \text{BRW}(\text{right})
\end{align}
This achieves $\approx n/2$ multiplications with $O(\log n)$ depth, which is asymptotically optimal up to lower-order terms~\cite{rabin1972number}.

\paragraph{Practical Limitation.}
Despite theoretical optimality, BRW performs poorly in practice.
The recursive structure with irregular base cases prevents effective compiler optimization.
Function call overhead and unpredictable branching inhibit instruction scheduling,
resulting in slower execution than simpler iterative methods.

\subsubsection{c-Decimated BRW}

To improve BRW's practical performance, $c$-decimated BRW splits the message into
$c$ interleaved streams, applies BRW to each stream, and combines results:
\[
h = \text{BRW}(m_0, m_c, m_{2c}, \ldots) + x^d \cdot \text{BRW}(m_1, m_{c+1}, \ldots) + \cdots
\]
where $d$ is chosen to avoid coefficient collisions.
The $c$ independent BRW evaluations can execute in parallel, improving ILP.
We benchmark $c = 2$ and $c = 4$.

\paragraph{Result.}
Even with decimation, BRW variants remain slower than Horner-Unrolled and Injective-Parallel.
The recursive structure still prevents optimal code generation.

\subsubsection{Comprehensive Comparison}
\Cref{tab:injective:universal} in \Cref{sec:injective:experiments} compares all
methods across message sizes on both architectures; the observations below
refer to its ARM block.

\paragraph{Key Observations.}
\begin{enumerate}
    \item \textbf{CLNH is fastest} but uses $O(N)$ keys---suitable when key material is not a constraint.
    \item For \textbf{$O(1)$-key methods}:
    \begin{itemize}
        \item \textbf{Small messages ($2N \le 16$)}: Sequential Injective wins due to minimal overhead
              and half the multiplications of Horner.
        \item \textbf{Large messages ($2N \ge 32$)}: Injective-Lanes is fastest, $1.2$--$1.5\times$
              ahead of the next $O(1)$-key method (Horner-Unrolled at $2N=32$, Injective-Parallel
              at $2N=64$--$128$, Horner-Parallel at $2N=256$), $10\times$ faster than sequential
              Horner at $2N=256$, and there within $1.6\times$ of CLNH.
        \item Horner-Unrolled is a good middle ground: simple to implement, $4.3\times$ faster than Horner.
    \end{itemize}
    \item \textbf{BRW underperforms} despite its $O(\log N)$ depth.
          The recursive structure prevents effective compiler optimization.
          c-decimated BRW variants are similarly slow.
    \item \textbf{Horner-Parallel} shows non-monotonic behavior: $2N = 256$ is slightly faster than
          $2N = 128$ in wall-clock time, likely due to better instruction scheduling at that
          specific unrolled size. The $O(\log N)$ depth advantage becomes apparent at large $N$.
\end{enumerate}

\paragraph{Parallelization Strategy.}
The injective recurrence $P_i = a_i + c_i \cdot (P_{i-1} + x^2)$ where $c_i = b_i + x^3$
can be expressed as transfer functions $(c_i, d_i)$ where $d_i = a_i + c_i \cdot x^2$.
With $T_{c,d}(h)\coloneqq h\cdot c + d$, the same composition rule applies:
$T_{c_2,d_2}\!\circ T_{c_1,d_1} = T_{c_1c_2,\; d_1c_2 + d_2}$.
We compute all $(c_i, d_i)$ pairs in parallel (Phase~1), then combine them with
parallel prefix scan (Phase~2), achieving $O(\log N)$ critical path depth.

Unlike Horner where $c = x$ for all steps (allowing power precomputation),
the injective recurrence has varying $c_i$, requiring $3N$ multiplications total.
Despite $3\times$ more work than sequential ($N$ multiplications),
Injective-Parallel is $3.2\times$ faster at $2N = 256$ on the M2 Pro because the
core can execute multiple independent \texttt{PMULL} instructions per cycle;
the lane form (Injective-Lanes) keeps the count at $N+L$ and is $5.0\times$
faster than sequential.

\subsection{Implementation Insights}
\label{sec:experiments:implementation}

Our benchmarks reveal several microarchitectural insights for polynomial evaluation
on Apple M2 Pro (and, qualitatively, on other out-of-order CPUs with carryless multiplication):

\paragraph{Latency vs.\ Throughput.}
The \texttt{PMULL} instruction on the Apple M2 Pro has $\approx 3$ cycle latency but the
processor can issue multiple independent \texttt{PMULL}s per cycle.
Sequential algorithms like Horner are \emph{latency-bound}: each multiplication
depends on the previous result, leaving execution units idle.
Parallel algorithms like our prefix scan are \emph{throughput-bound}: they generate
enough independent work to keep all units busy.

For message length $2N=256$, sequential Horner has a critical path of $255$ dependent multiplications,
while Injective-Parallel has only $O(\log N)$ levels of dependent multiplications (about 8 levels for $N=128$),
plus parallel work at each level.

\paragraph{Compiler Optimization.}
The effectiveness of parallelism depends heavily on compiler optimization.
Our Injective-Parallel implementation uses simple loops that the compiler
fully unrolls and interleaves, generating dense streams of independent \texttt{PMULL}
instructions.
In contrast, recursive algorithms like BRW (Bernstein-Rabin-Winograd) suffer from
function call overhead and irregular control flow that inhibit optimization.

At $2N = 256$, Injective-Parallel ($22144$\,\textmu s per $10^5$ hashes, \Cref{tab:injective:universal}) outperforms BRW ($34879$\,\textmu s)
by $1.6\times$ despite BRW's $O(\log N)$ depth,
because the iterative structure admits better code generation.
(The depth advantage is also no longer unique: the schedules of
\Cref{thm:construction-height} achieve $O(\log N)$ multiplicative
depth at the minimum multiplication count, though the entrants benchmarked
here are the injective variants.)

\paragraph{Memory Layout.}
Storing keys as scalar 64-bit integers rather than 128-bit SIMD vectors improves
performance by $10$--$15\%$ on Apple Silicon.
This allows XOR operations to use the faster scalar ALU, and reduces register pressure
when many coefficients are live simultaneously during tree reductions.

\section{Open Problems}
\label{sec:open-problems}

\paragraph{Constructions for characteristic 2.}
Our main construction relies on the identity $(a+b)^2 = a^2 + 2ab + b^2$, which fails in characteristic 2 since $2ab = 0$.
Does one fixed circuit topology with rational preprocessing achieve
$\lfloor n/2\rfloor+1$ multiplications in characteristic~2?  We currently give
such fixed constructions, with explicit inverses, only at odd degrees up
to~$21$ (\Cref{appendix:polynomials}; \Cref{sec:ph} for degree~$5$).

\paragraph{Exceptional inputs at even degree.}
Pan's sextic scheme (\Cref{sec:lower}) reaches Motzkin's bound of $n/2$
multiplications at $n=6$ with rational preprocessing that is undefined on a
hypersurface, and \Cref{sec:lower} shows that no single-valued rational
preprocessing removes this hypersurface.
Does a generically rational scheme with $n/2$ multiplications exist for every even
$n\ge8$?  Pan~\cite{pan1978complex} credits Pan (1961)~\cite{pan1961schemes} with a
real scheme for $n=8$ with $(n+2)/2$ multiplications, that is $n/2$ for a monic
polynomial, and $n+2$ additions for almost all polynomials; we could not
determine whether its preprocessing is rational, as it is at $n=6$.  And is the
exceptional set always unavoidable for rational (single-valued) preprocessing, as we
conjecture?  For algebraic preprocessing Pan conjectures the opposite, that the
exceptional sets of his schemes can be removed at every degree, and proves it over
$\mathbb C$ for even $n\equiv2\pmod4$, $n\ge10$, at $n/2$ monic multiplications and
$n$ additions~\cite{pan1978complex}.

\paragraph{Lower bounds for characteristic 2.}
The proof in \Cref{sec:lower} uses properties specific to fields of characteristic
$\neq 2$, but a different argument applies there: over a finite field of characteristic~$2$
with at least $2n$ elements, \Cref{thm:char2-lower} rules out $n$ multiplications for
bijective evaluation at $2n$ points at every $n>1$, for every preprocessing map.  Two
questions remain.  Does the analogous bound hold for an odd number of points, that is,
for monic polynomials of odd degree?  And does it hold over small fields, where the
hypothesis $\abs{\F}\ge 2n$ fails?

\paragraph{Key expansion for carryless NH.}
The block level of ChainHash, like every NH-style hash, uses one random key
word per message word of a block.  Deriving those words from a few seed words
by a $k$-wise independent polynomial is unsafe for a reduced NH, whose block
sums are taken in $\F_{2^{64}}$: the collision condition is affine in the key
words, hence linear in the seed words, and a difference on $k+1$ positions in
the kernel of the Vandermonde system is key-independent
(\Cref{sec:injective:experiments}).  For the unreduced carryless NH the
products live in $\F_2[X]$ while the expansion lives in $\F_{2^{64}}$, and the
same construction only halves the $\F_2$-rank of the seed-to-difference map,
from $127$ to $63$.  Is unreduced carryless NH with $k$-wise independent keys
almost-XOR-universal with $\varepsilon$ close to $2^{-64}$, or is there an
attack?  A positive answer would shrink the key of ChainHash and the other NH-based hybrids to a few
words without a pseudorandom function.

\paragraph{Reduction-free universal hashing with a short key.}
The fastest proven hashes, NH and its carryless variant, owe their speed to
the absence of modular reductions: products are accumulated in $\mathbb
Z/2^{128}$ or $\F_2[X]$ and reduced once at the end, but they need one key
word per message word.  Polynomial hashes need only $O(1)$ key words but pay
a reduction per multiplication, and the fastest heuristic hashes
(\texttt{wyhash}, \texttt{MUM}, \texttt{XXH3}) try to have both at once by
folding the two halves of each $128$-bit product into one word; the fold is
not injective, and \Cref{app:adversarial} shows what that costs.  Is there an
$\varepsilon$-almost-universal family with $\varepsilon$ close to $2^{-64}$,
a key of $O(1)$ words independent of the message length, and an evaluation
that uses one $64\times64$-bit multiplication per $16$ bytes and no reduction
modulo a prime or an irreducible polynomial?  Multiplication in $\mathbb
Z/2^{w}$ has zero divisors, so the Schwartz--Zippel argument does not apply
to it directly; Dietzfelbinger's multiply-shift shows that the high bits of
products can be universal, but again with one key word per message word.

\paragraph{Unit-coefficient chains.}
The odd steps of the $T$ recursion multiply a block by the integer $k-1$:
with $H_{2D}=H^2+\Sigma H+G$ the step must use $H-\tfrac{k-1}{2}\Sigma$, and
the symmetric difference-of-squares core has $\Sigma=2U_1$.  These multiples
are free in our cost model and cost about $4\%$ extra additions if
synthesised by doubling (and nothing in characteristic~$2$), but they are
the only wire coefficients other than $\pm1$ in our chains, whereas the
classical schemes of Motzkin, Knuth--Eve, Belaga and Rabin--Winograd have
none.  An asymmetric core $H_{2D}=(H+A_1)(H+A_2)+W$ removes them for $k=3$
and halves them to $(k-1)/2$ otherwise at the same multiplication and
addition counts; we verified this by exact decoding for $n\le45$ in the
$4k+1$ family, where it changes the boundary constants of \Cref{lem:Rk2l}
and adds the prime $3$ to the exceptional set at some steps, but have not
carried it through the $8k+3$ and $8k+7$ families.  Is there a uniform
construction with $\lfloor n/2\rfloor+1$ multiplications, everywhere-defined
rational preprocessing, and all wire coefficients in $\{0,\pm1\}$?

\paragraph{Additions and height at small degrees.}
Our general construction is not optimised for additions or depth at small
degrees.  For instance, the everywhere-defined quintic
\[
 P_5=(x+\alpha_2)\bigl((x^2+\alpha_4)(x^2+x+\alpha_3)+\alpha_1\bigr)+\alpha_0,
\]
with the polynomial preprocessing $\alpha_2=a_4-1$,
$\alpha_4=a_2-\alpha_2a_3+\alpha_2^2$, $\alpha_3=a_3-\alpha_2-\alpha_4$,
$\alpha_1=a_1-\alpha_4a_3+\alpha_4^2$,
$\alpha_0=a_0-\alpha_2a_1+\alpha_2^2\alpha_4$, uses three multiplications and
six additions, where the general schedule of \Cref{sec:addition-count} uses
$A_5=8$.  An exhaustive search at $n\le7$ finds unit-coefficient chains with
$3,4,6,9$ additions for $n=3,4,5,7$ against our $3,4,8,10$, so it matches the
general construction at $n=3,4$ and saves additions at $n=5,7$.  Is there a
uniform construction with $n+O(1)$ additions at $\lfloor n/2\rfloor+1$
multiplications, and what is the minimal height?

\paragraph{Code and formalization.}
The benchmark code, the chain compilers, the characteristic-two certificates and
the Lean development referred to throughout (the directories \texttt{tools/\allowbreak bench/},
\texttt{tools/}, \texttt{char2/} and \texttt{FastPoly/}) are available at\par\noindent\coderepo.

\bibliographystyle{alpha}
\bibliography{references}  
\appendix

\section{Optimized Polynomial Circuits}\label{appendix:polynomials}

This appendix records optimized candidate circuits with four through eleven
multiplication gates.  A circuit with $m$ products below has $2m-1$ keys and
outputs a monic polynomial of degree $2m-1$.  The circuits were found by
randomized search, optimizing both the number of additions and the critical
path depth.  Explicit inverses are proved below for the displayed characteristic-$2$
circuits of degrees $7$, $15$, $17$, $19$, and $21$, and for separately displayed
alternatives of degrees $9$, $11$, and $13$.  The other search results are not used in
the proof of the main theorem.  In particular, a constant nonzero Jacobian found
by the search is only a diagnostic (and a necessary condition for a
polynomial automorphism), not a proof of bijectivity.

\subsection{Characteristic 2 (\texorpdfstring{$\F_{2^{64}}$}{GF(2\^{}64)})}

Each circuit computes a polynomial $P(x)$ over $\F_{2^{64}}$ using carryless multiplications,
parameterized by keys $a_0, a_1, \ldots$.
All additions are XOR, and each multiplication uses Galois field reduction.
Timings are medians over $10^6$ random inputs on an Apple M2 Pro using ARM NEON
\texttt{PMULL} (\texttt{tools/bench/appendix\_char2\_circuits\_arm.cpp}).

\medskip
\noindent
\begin{minipage}[t]{0.47\textwidth}
\textbf{4 products, degree 7} (7 keys)
\[
\begin{aligned}
y &= (x + 0)(x + a_0) \\
z &= (x + a_1)(y + a_2) \\
t &= (x + y + z + a_3) \\
  &\quad\cdot(x + y + z + 0) \\
u &= (x + a_4)(y + t + a_5) \\
P &= u + a_6
\end{aligned}
\]
\emph{2.3 ns, h=4, 12 XORs}
\end{minipage}
\hfill
\begin{minipage}[t]{0.47\textwidth}
\textbf{5 products, degree 9} (9 keys)
\[
\begin{aligned}
y &= (x + 0)(x + 0) \\
z &= (x + y + a_0)(x + a_1) \\
t &= (z + a_2)(y + z + a_3) \\
u &= (x + z + t + a_4) \\
  &\quad\cdot(x + y + z + a_5) \\
v &= (x + a_6)(y + a_7) \\
P &= u + v + a_8
\end{aligned}
\]
\emph{3.1 ns, h=4, 16 XORs}
\end{minipage}

\begin{lemma}[Direct inverse for the first characteristic-2 circuit]\label{lem:first-char2-circuit-inverse}
Let $\F$ be a perfect field of characteristic $2$, for example $\F_{2^{64}}$.
Fix distinct $x_0,\ldots,x_6\in\F$.
For the first circuit above, the map
\[
    (a_0,\ldots,a_6)
        \longmapsto
    \bigl(P_{a_0,\ldots,a_6}(x_0),\ldots,P_{a_0,\ldots,a_6}(x_6)\bigr)
\]
is a bijection $\F^7\to\F^7$.
\end{lemma}
\begin{proof}
Given values $v_i=P(x_i)$, first recover the coefficients $p_0,\ldots,p_6$ of the unique monic degree-$7$ polynomial
\[
    P(x)=x^7+\sum_{j=0}^6 p_jx^j
\]
with $P(x_i)=v_i$.
Equivalently, solve the Vandermonde system
\[
    \sum_{j=0}^6 p_jx_i^j = v_i+x_i^7,
    \qquad i=0,\ldots,6.
\]
The system is invertible because the $x_i$ are distinct.

We now recover the keys from the $p_j$.
Since the square map is bijective over $\F$, write $r^{1/2}$ for the unique square root of $r$.
Set
\[
\begin{aligned}
    a_4 &= p_6,\\
    b &= p_5^{1/2},\\
    a_3 &= p_4+a_4p_5,\\
    c &= \bigl(p_3+a_3b+1+a_4a_3\bigr)^{1/2},\\
    a_0 &= p_2+a_3c+a_4(c^2+a_3b+1),\\
    a_1 &= b+1+a_0,\\
    a_2 &= c+1+a_0+a_0a_1,\\
    d &= a_1a_2,\\
    a_5 &= p_1+d^2+a_3d+a_4(a_3c+a_0),\\
    a_6 &= p_0+a_4(d^2+a_3d+a_5).
\end{aligned}
\]
To verify these formulas, define
\[
    b=1+a_0+a_1,\qquad
    c=1+a_0+a_2+a_0a_1,\qquad
    d=a_1a_2.
\]
Then
\[
    x+y+z=x^3+bx^2+cx+d.
\]
Using characteristic $2$,
\[
    t=(x+y+z+a_3)(x+y+z)=(x+y+z)^2+a_3(x+y+z),
\]
and hence
\[
    y+t+a_5
    =x^6+b^2x^4+a_3x^3+(c^2+a_3b+1)x^2+(a_3c+a_0)x+(d^2+a_3d+a_5).
\]
Multiplying by $x+a_4$ and adding $a_6$ gives
\[
\begin{aligned}
    p_6 &= a_4,\\
    p_5 &= b^2,\\
    p_4 &= a_3+a_4b^2,\\
    p_3 &= c^2+a_3b+1+a_4a_3,\\
    p_2 &= a_3c+a_0+a_4(c^2+a_3b+1),\\
    p_1 &= d^2+a_3d+a_5+a_4(a_3c+a_0),\\
    p_0 &= a_4(d^2+a_3d+a_5)+a_6.
\end{aligned}
\]
The displayed recovery procedure solves these equations in order, so it recovers a unique key tuple.
Conversely, applying the recovered keys gives the same coefficients $p_j$, hence the same values at the $x_i$.
\end{proof}

\medskip
\noindent
\begin{minipage}[t]{0.47\textwidth}
\textbf{6 products, degree 11} (11 keys)
\[
\begin{aligned}
y &= (x + 0)(x + a_0) \\
z &= (x + y + a_1)(x + y + 0) \\
t &= (x + z + a_2)(z + a_3) \\
u &= (t + a_4)(x + a_5) \\
v &= (x + y + z + t + u + a_6) \\
  &\quad\cdot(y + a_7) \\
w &= (x + a_8)(y + z + t + a_9) \\
P &= z + v + w + a_{10}
\end{aligned}
\]
\emph{4.4 ns, h=5, 22 XORs}
\end{minipage}
\hfill
\begin{minipage}[t]{0.47\textwidth}
\textbf{7 products, degree 13} (13 keys)
\[
\begin{aligned}
y &= (x + 0)(x + 0) \\
z &= (y + a_0)(x + y + a_1) \\
t &= (x + z + a_2)(x + a_3) \\
u &= (x + y + a_4)(x + t + a_5) \\
v &= (y + z + a_6)(u + a_7) \\
w &= (y + z + t + a_8) \\
  &\quad\cdot(x + u + a_9) \\
s &= (x + z + u + v + w + a_{10}) \\
  &\quad\cdot(x + a_{11}) \\
P &= t + s + a_{12}
\end{aligned}
\]
\emph{5.5 ns, h=6, 26 XORs}
\end{minipage}

\medskip
\noindent
The three benchmarked circuits above are retained as search candidates.  For
the proofs, we use the following separately optimized alternatives; unlike the
candidate circuits, each comes with the explicit inverse below.
\[
\begin{array}{c|l}
9 &
\begin{aligned}
y&=x(x+a_0),& z&=x(y+a_1),& t&=(y+z+a_2)(z+a_3),\\
u&=(x+z+a_4)(t+a_5),& v&=(y+a_6)(z+a_7),& P&=u+v+a_8;
\end{aligned}\\[2ex]
11 &
\begin{aligned}
y&=x(x+a_0),& z&=(y+a_1)(x+y+a_2),& t&=x(y+a_3),\\
u&=(t+a_4)(z+a_5),&v&=(x+y+t+u+a_6)(z+t+a_7),\\
w&=(t+a_8)(y+u+a_9),&P&=v+w+a_{10};
\end{aligned}\\[2ex]
13 &
\begin{aligned}
y&=x^2,&z&=(x+y+a_{12})(y+a_{11}),&w&=(y+z+a_{10})(z+a_9),\\
v&=(y+z+a_8)(w+a_7),&u&=(z+v+a_6)(x+a_5),\\
t&=(x+y+a_4)(x+a_3),&s&=(w+t+a_2)(y+a_1),&P&=u+v+s+a_0.
\end{aligned}
\end{array}                                                     \tag{A.0}
\]
Their respective $(\text{height},\text{XOR count})$ pairs are
$(4,12)$, $(4,18)$, and $(5,21)$; these alternatives were not included in
the timing run quoted above.

\begin{lemma}[Explicit inverses for the degree-$9$ and degree-$11$ circuits]
\label{lem:char2-small-staircase-butterfly}
Over every field of characteristic $2$, the coefficient map of the displayed
degree-$9$ circuit is a polynomial bijection.  Over every perfect field of
characteristic $2$, the coefficient map of the displayed degree-$11$ circuit is
a bijection.  Consequently their evaluation maps at respectively nine and
eleven distinct points are bijections.
\end{lemma}
\begin{proof}
We give the inverse maps.  This also makes clear where perfectness is used.

For degree $9$, write
\(P=x^9+\sum_{j=0}^8c_jx^j\), and first set
\[
\begin{aligned}
 a_0&=c_8+1,\\
 a_1&=c_7+1+a_0+a_0^2,\\
 r&=c_6+1+a_0^2+a_0^3,\\
 s&=c_5+1+a_0^2+a_0^3+a_0^2a_1+a_1^2,\\
 q&=c_4+a_0^2+a_0^2r+a_0a_1^2+a_1+a_1^2.
\end{aligned}                                                    \tag{9.1}
\]
Expanding the first four products gives the constant block
\[
 r=a_2+a_3+a_4,\qquad s=a_3+a_4,\qquad q=a_2+a_3.
\]
Its inverse is
\[
 a_2=r+s,\qquad a_3=q+a_2,\qquad a_4=s+a_3.             \tag{9.2}
\]
Let $B$ be the output of the same circuit after setting
$a_5=a_6=a_7=a_8=0$, with the recovered $a_0,\ldots,a_4$ left in place,
and write $b_j=[x^j]B$.  The remaining descent is
\[
\begin{aligned}
 h&=c_3+b_3,\\
 a_7&=c_2+b_2+a_0h,\\
 a_5&=c_1+b_1+a_1h+a_0a_7,\\
 a_6&=h+a_5,\\
 a_8&=c_0+b_0+a_4a_5+a_6a_7.
\end{aligned}                                                    \tag{9.3}
\]
Indeed, after subtracting the baseline, rows $3,2,1,0$ are respectively
\[
 a_5+a_6,\quad
 a_7+a_0(a_5+a_6),\quad
 a_5+a_1(a_5+a_6)+a_0a_7,\quad
 a_8+a_4a_5+a_6a_7.
\]
Thus every step in (9.1)--(9.3) has unit slope; no root or division is used.

For degree $11$, write
\(P=x^{11}+\sum_{j=0}^{10}c_jx^j\), and put
\[
 a_0=c_{10},\qquad a_3=c_9+1,\qquad a_4=c_8+a_0,
 \qquad s=a_1+a_2,\qquad h=a_5+a_7+a_8.
\]
Define the already-known expressions
\[
\begin{aligned}
 K_7&=1+a_0^2+a_0^4+a_3,\\
 K_6&=1+a_0+a_0^3+a_0^5+a_4,\\
 K_5&=a_0+a_0^2+a_0^4a_3+a_0^2a_3+a_3.
\end{aligned}
\]
The next three rows are
\[
\begin{aligned}
 c_7&=K_7+s^2+h,\\
 c_6&=K_6+a_0s^2+(a_0+1)h,\\
 c_5&=K_5+a_0^2(s^2+h)+s+a_1^2+a_3s^2+sh+a_3h.
\end{aligned}                                                    \tag{11.1}
\]
Since the field is perfect, Frobenius is invertible.  Hence
\[
\begin{aligned}
 r&=c_7+K_7,\\
 h&=c_6+K_6+a_0r,\\
 s&=(r+h)^{1/2},\\
 a_1&=\bigl(c_5+K_5+a_0^2r+s+a_3s^2+sh+a_3h\bigr)^{1/2},\\
 a_2&=s+a_1.
\end{aligned}                                                    \tag{11.2}
\]
To expose the lower butterfly, let $B_0$ be the circuit output with
\[
 (a_5,a_6,a_7,a_8,a_9,a_{10})=(0,0,0,h,0,0).
\]
Then $a_6=c_4+[x^4]B_0$.  Recompute this baseline with the recovered $a_6$
in place and call it $B$; put $d_j=c_j+[x^j]B$.  A direct cancellation of
the two final branches gives
\[
 P+B=\kappa(t+a_4)+a_5y+a_7(x+t+a_6)+a_9(t+a_8)+a_{10},
 \quad
 \kappa=a_5(h+a_5),\quad a_8=h+a_5+a_7.              \tag{11.3}
\]
Consequently
\[
\begin{aligned}
 a_5&=d_2+a_0d_3,\\
 \kappa&=a_5(h+a_5),\\
 a_7&=d_1+a_3d_3+a_0a_5,\\
 a_9&=d_3+\kappa+a_7,\\
 a_8&=h+a_5+a_7,\\
 a_{10}&=d_0+\kappa a_4+a_7a_6+a_9a_8.
\end{aligned}                                                    \tag{11.4}
\]
This recovers all eleven keys.  The only non-polynomial-looking operations are
the two inverse-Frobenius steps in (11.2), and these are unique over a perfect
field.

In either degree, evaluation at distinct points is the coefficient map followed
by the corresponding invertible Vandermonde map.
\end{proof}

\begin{lemma}[Unitriangular inverse for the degree-$13$ circuit]
\label{lem:char2-degree13-inverse}
Over every field of characteristic $2$, the coefficient map of the displayed
degree-$13$ circuit is a polynomial bijection.  Consequently its evaluation
map at any thirteen distinct points is a bijection.
\end{lemma}
\begin{proof}
Make the invertible linear change of key coordinates
\[
\begin{aligned}
(q_0,\ldots,q_{12})={}&(a_5, a_{11}+a_{12}, a_{12}, a_9, a_8, a_{10},
 a_1,\\
&a_7, a_3, a_4, a_6, a_2, a_0),
\end{aligned}                                                    \tag{13.1}
\]
and order the coefficient rows as
\[
 (d_0,\ldots,d_{12})=(12,11,10,7,6,9,8,5,4,3,1,2,0).
                                                               \tag{13.2}
\]
After expressing the original keys through (13.1), define the explicit
baseline
\[
 K_i(q_0,\ldots,q_{i-1})
   =[x^{d_i}]P(q_0,\ldots,q_{i-1},0,\ldots,0).
                                                               \tag{13.3}
\]
The seven gate identities give the following unitriangular certificate:
\[
 [x^{d_i}]P=q_i+K_i(q_0,\ldots,q_{i-1})\qquad(0\le i\le12).
                                                               \tag{13.4}
\]
For reference, its complete pivot table is
\[
\begin{array}{c|rrrrrrrrrrrrr}
i&0&1&2&3&4&5&6&7&8&9&10&11&12\\ \hline
d_i&12&11&10&7&6&9&8&5&4&3&1&2&0\\
q_i&a_5&a_{11}{+}a_{12}&a_{12}&a_9&a_8&a_{10}&a_1&a_7&a_3&a_4&a_6&a_2&a_0
\end{array}                                                     \tag{13.5}
\]
The first three rows, for example, are
\[
 [x^{12}]P=q_0,\qquad [x^{11}]P=q_0+q_1,
 \qquad [x^{10}]P=1+q_0+q_0q_1+q_2.
\]
We record the only coupled cancellation behind the nonmonotone middle of
(13.5).  If
\[
 W_p=(A+p)B,\qquad V_t=(A+t)(W_p+\ell),\qquad
 F_{p,t}=RV_t+SW_p+E,
\]
then characteristic $2$ gives the literal identity
\[
 F_{p,t}+F_{0,0}
  =(p+t)RAB+ptRB+tR\ell+pSB.                         \tag{13.6}
\]
In the circuit take
\[
 A=z+y,\quad B=z+a_9,\quad R=x+a_5+1,\quad S=y+a_1,\quad
 p=a_{10},\quad t=a_8,\quad \ell=a_7.
\]
The row-$6$ coefficients of both $RAB$ and $SB$ are $1$, so their $p$
terms cancel: row $6$ exposes $a_8$.  Row $9$ then exposes $a_{10}$, and
row $5$ exposes $a_7$.  The remaining entries of (13.5) are the ordinary
unit offsets in unequal-degree factors; their later offsets lie strictly
below the indicated row.  This proves (13.4) without a Jacobian argument.

The inverse is now the explicit recurrence
\[
 q_i=[x^{d_i}]P+K_i(q_0,\ldots,q_{i-1})\qquad(0\le i\le12),
\]
followed by the inverse of (13.1).  It uses only addition and multiplication,
so it works over every characteristic-$2$ field.  The evaluation claim again
follows from Vandermonde interpolation.
\end{proof}

\medskip
\noindent
\begin{minipage}[t]{\textwidth}
\textbf{8 products, degree 15} (15 keys)
\[
\begin{aligned}
y &= x^2, \\
z &= (y+a_0)(x+y+a_1), \\
t &= (x+a_2)(z+a_3), \\
u &= (y+t+a_4)(z+t+a_5), \\
v &= (x+z+a_6)(z+a_7), \\
w &= (x+y+z+a_8)(y+v+a_9), \\
s &= (z+a_{10})(v+a_{11}), \\
r &= (t+a_{12})(u+a_{13}), \\
P &= w+s+r+a_{14}.
\end{aligned}
\]
\emph{$h=5$, 24 XORs}
\end{minipage}

\smallskip
\noindent
The next decoder is unitriangular: like those of degrees $19$ and $21$ below,
and unlike the degree-$7$ and degree-$17$ decoders, it requires no Frobenius
roots.

\begin{lemma}[Uniform inverse for the degree-$15$ characteristic-$2$ circuit]
\label{lem:char2-degree15-inverse}
Let $\F$ be any field of characteristic $2$.  The coefficient map of
the displayed degree-$15$ circuit is a bijection from $\F^{15}$ to the monic
degree-$15$ polynomials.  Consequently, for any fifteen distinct
$x_0,\ldots,x_{14}\in\F$, its evaluation map at those points is a bijection
$\F^{15}\to\F^{15}$.
\end{lemma}
\begin{proof}
Write $P=x^{15}+\sum_{j=0}^{14}c_jx^j$.  Replace the gate offsets by the
following invertible linear coordinates:
\[
\begin{array}{lll}
q_0=a_2, & q_1=a_0+a_1, & q_2=a_0,\\
q_3=a_3, & q_4=a_4+a_5+a_{12}, & q_5=a_5,\\
q_6=a_8+a_{10}, & q_7=a_{12}, & q_8=a_6+a_7,\\
q_9=a_{13}, & q_{10}=a_9+a_{11}, & q_{11}=a_7,\\
q_{12}=a_9+a_{10}, & q_{13}=a_9, & q_{14}=a_{14}.
\end{array}                                                       \tag{A.1}
\]
For reference, its inverse is
\[
\begin{array}{lll}
a_0=q_2, & a_1=q_1+q_2, & a_2=q_0,\\
a_3=q_3, & a_4=q_4+q_5+q_7, & a_5=q_5,\\
a_6=q_8+q_{11}, & a_7=q_{11},
  & a_8=q_6+q_{12}+q_{13},\\
a_9=q_{13}, & a_{10}=q_{12}+q_{13},
  & a_{11}=q_{10}+q_{13},\\
a_{12}=q_7, & a_{13}=q_9, & a_{14}=q_{14}.
\end{array}                                                       \tag{A.2}\label{eq:char2-15-q-inverse}
\]

Let $\mathcal P(\mathbf q)$ denote the circuit after the substitution
\eqref{eq:char2-15-q-inverse}.  For $0\le i\le14$, define the explicitly
evaluable baseline
\[
 K_i(q_0,\ldots,q_{i-1})
 =\coeff{\mathcal P(q_0,\ldots,q_{i-1},0,\ldots,0)}{14-i}.
                                                                    \tag{A.3}\label{eq:char2-15-baseline}
\]
Collecting coefficients in the nine displayed assignments gives the
unitriangular identities
\[
 \coeff{\mathcal P(\mathbf q)}{14-i}
   =q_i+K_i(q_0,\ldots,q_{i-1})
   \qquad(0\le i\le14).                                           \tag{A.4}\label{eq:char2-15-triangular}
\]
Here is a compact certificate for the collection: it records the branch in
which each new pivot first survives after the preceding rows have been
removed.
\[
\begin{gathered}
\begin{array}{c|rrrrrrrr}
\text{row}&14&13&12&11&10&9&8&7\\ \hline
\text{pivot}&q_0&q_1&q_2&q_3&q_4&q_5&q_6&q_7\\
\text{branch}&r&r&r&r&r&r&w+s&r
\end{array}\\[3pt]
\begin{array}{c|rrrrrrr}
\text{row}&6&5&4&3&2&1&0\\ \hline
\text{pivot}&q_8&q_9&q_{10}&q_{11}&q_{12}&q_{13}&q_{14}\\
\text{branch}&w+s&r&w+s&w+s&w+s&w+s&\mathrm{out}
\end{array}
\end{gathered}                                                    \tag{A.5}\label{eq:char2-15-pivot-table}
\]
The two cancellations responsible for the alternating lower rows are
\[
\begin{aligned}
u&=t^2+(y+z+a_4+a_5)t+(y+a_4)(z+a_5),\\
w+s&=(x+y+a_8+a_{10})v
 +(x+y+z+a_8)(y+a_9)+(z+a_{10})a_{11}.
\end{aligned}                                                     \tag{A.6}
\]
Thus the common $zv$ term disappears from $w+s$; the remaining products have
distinct degrees.  Descending through the rows in
\eqref{eq:char2-15-pivot-table}, the displayed
factor degrees give the named pivot with slope one, while every other term
contains only coordinates from an earlier column.  This proves
\eqref{eq:char2-15-triangular}
without any nonzero-slope or field-size assumption.

The inverse is now literal.  For $i=0,\ldots,14$, successively set
\[
 q_i=c_{14-i}+K_i(q_0,\ldots,q_{i-1}),                            \tag{A.7}\label{eq:char2-15-decoder}
\]
and then recover the original offsets from
\eqref{eq:char2-15-q-inverse}.  Equations
\eqref{eq:char2-15-triangular} and \eqref{eq:char2-15-decoder} are mutually
inverse polynomial maps, so the
coefficient map is a polynomial automorphism over every characteristic-$2$
field.  Finally, evaluation at fifteen distinct points is an invertible
Vandermonde transformation of the fifteen nonleading coefficients.
\end{proof}

\medskip
\noindent
\begin{minipage}[t]{\textwidth}
\textbf{9 products, degree 17} (17 keys)
\[
\begin{aligned}
y &= x(x+a_0),\\
z &= (x+a_1)(x+y+a_2),\\
t &= (y+a_3)(x+y+a_4),\\
u &= (y+z+a_5)(z+t+a_6),\\
v &= (x+z+a_7)(x+z+t+u+a_8),\\
h &= (y+a_9)x,\\
j &= (y+a_{10})(x+a_{11}),\\
\ell &= (t+a_{12})(h+a_{13}),\\
w &= (x+u+a_{14})(u+v+a_{15}),\\
P &= j+\ell+w+a_{16}.
\end{aligned}
\]
\emph{$h=5$, 29 XORs}
\end{minipage}

\begin{lemma}[Uniform inverse for the degree-$17$ characteristic-$2$ circuit]
\label{lem:char2-degree17-inverse}
Let $\F$ be a perfect field of characteristic $2$.  The coefficient map of
the displayed degree-$17$ circuit is a bijection from $\F^{17}$ to the monic
degree-$17$ polynomials.  Consequently, for any seventeen distinct
$x_0,\ldots,x_{16}\in\F$, its evaluation map at those points is a bijection
$\F^{17}\to\F^{17}$.
\end{lemma}
\begin{proof}
Write $[f]_i=[x^i]f$.  We first replace the gate offsets by normalized gate
coordinates
\[
\begin{aligned}
q_0&=[y]_1,
&(q_1,q_2)&=([z]_2,[z]_1),
&(q_3,q_4)&=([t]_2,[t]_1),\\
(q_5,q_6)&=([u]_4,[u]_3),
&(q_7,q_8)&=([v]_7,[v]_3),
&q_9&=[h]_1,\\
(q_{10},q_{11})&=([j]_2,[j]_1),
&(q_{12},q_{13})&=([\ell]_4,[\ell]_3),
&(q_{14},q_{15})&=([w]_{10},[w]_7),
\end{aligned}                                                     \tag{A.8}
\]
and $q_{16}=a_{16}$.  These are polynomial coordinates on the original
keys.  Indeed, the first three gates give
\[
\begin{aligned}
a_0&=q_0,\\
a_1&=q_1+q_0+1,
&a_2&=q_2+(q_0+1)a_1,\\
\sigma&=q_3+q_0^2+q_0,
&a_3&=q_4+q_0\sigma,
&a_4&=\sigma+a_3.
\end{aligned}                                                     \tag{A.9}
\]
For every remaining two-offset gate use the following unit-pivot identity.
If $A,B$ are already known and monic, $0<\deg A<\deg B$, and
$G=(A+\alpha)(B+\beta)$, then
\[
 \alpha=[G]_{\deg B}+[AB]_{\deg B},\qquad
 \beta=[G]_{\deg A}+[AB+\alpha B]_{\deg A}.                     \tag{A.10}
\]
It applies successively to $u,v,j,\ell,w$ (placing the lower-degree factor
first); also $a_9=q_9$.  Thus (A.8)--(A.10) explicitly recover every
$a_i$ from the $q_i$, without division or a nonvanishing hypothesis.

Now make the elementary coordinate change
\[
 s=q_0+q_3,\qquad r=q_5+q_7,\qquad e=q_4+q_5^2+q_5,               \tag{A.11}
\]
and order the resulting coordinates as
\[
\begin{aligned}
(z_1,\ldots,z_{17})={}&
(q_1,q_2,s,r,q_0,e,q_{14},q_6,q_5,q_{15},\\
&\hspace{34mm}q_8,q_9,q_{12},q_{13},q_{10},q_{11},q_{16}).
\end{aligned}                                                     \tag{A.12}
\]
This change is polynomially invertible:
$q_3=s+q_0$, $q_7=r+q_5$, and $q_4=e+q_5^2+q_5$.

We record the coefficient calculation in a form that is both compact and an
explicit decoder.  For each $i$, let $P_i$ be the output of the same displayed
circuit after retaining $z_1,\ldots,z_{i-1}$ and setting
$z_i,\ldots,z_{17}$ to zero; its original keys are obtained explicitly from
(A.9)--(A.12).  If $P=x^{17}+\sum_{d=0}^{16}c_dx^d$, direct collection of
the indicated rows gives
\[
\begin{array}{c|c|c@{\qquad\qquad}c|c|c}
i&d_i&c_{d_i}+[P_i]_{d_i}&i&d_i&c_{d_i}+[P_i]_{d_i}\\ \hline
1&16&z_1   &10&7&z_{10}\\
2&15&z_2   &11&6&z_{11}\\
3&13&z_3^2 &12&5&z_{12}\\
4&14&z_4   &13&4&z_{13}\\
5&12&z_5   &14&3&z_{14}\\
6&11&z_6   &15&2&z_{15}\\
7&10&z_7   &16&1&z_{16}\\
8&9 &z_8^2 &17&0&z_{17}\\
9&8 &z_9^4 &&&
\end{array}                                                       \tag{A.13}
\]
For example, the first six identities before baseline notation are
\[
\begin{aligned}
c_{16}&=q_1,\\
c_{15}&=q_2+q_1^2,\\
c_{13}&=s^2+q_1^2q_2+q_2^2+q_2,\\
c_{14}&=r+sq_1+q_1^3+q_1^2+q_2,\\
c_{12}&=q_0+rq_1^2+r+s^2q_1+sq_1^3+sq_1
          +q_1^4+q_1^2q_2+q_1^2+q_1q_2^2+q_2,\\
c_{11}&=e+r+s^2q_2+s^2+s+q_0+q_1^2q_2+q_1^2+q_1
          +q_2^3+q_2^2+q_2+1.
\end{aligned}                                                     \tag{A.14}
\]
The remaining rows of (A.13) are obtained in exactly the same way: substitute
the nine gate assignments, set the current and later coordinates to zero for
the baseline, and collect one coefficient.  Thus (A.13) is a list of literal
polynomial identities, rather than a Jacobian or finite-field test.  The
companion script \texttt{char2/verify\_n17\_uniform\_symbolic.py} (run from
the repository root as \texttt{python3 -m char2.verify\_n17\_uniform\_symbolic},
since it imports \texttt{char2/symexpr.py}) independently
expands all seventeen identities in $\F_2[z_1,\ldots,z_{17}]$.

Equation (A.13) recovers the $z_i$ in its displayed order.  All slopes are
one except for the three Frobenius pivots $z_3^2,z_8^2,z_9^4$; these have
unique inverses over a perfect field.  Equations (A.9)--(A.12) then recover
the original keys.  Conversely, applying this procedure to arbitrary
$c_0,\ldots,c_{16}$ makes the reconstructed circuit agree with those
coefficients in all seventeen rows, so the inverse is two-sided.  The final
evaluation claim follows from the invertible Vandermonde map at seventeen
distinct points.
\end{proof}

\medskip
\noindent
\begin{minipage}[t]{\textwidth}
\textbf{10 products, degree 19} (19 keys)
\[
\begin{aligned}
y &= x^2,\\
z &= (y+a_0)(x+y+a_1),\\
t &= (x+a_2)(z+a_3),\\
u &= (y+t+a_4)(z+t+a_5),\\
v &= (x+z+a_6)(z+a_7),\\
w &= (x+y+z+a_8)(y+v+a_9),\\
s &= (x+a_{10})(y+a_{11}),\\
r &= (x+a_{12})(y+a_{13}),\\
q &= (v+a_{14})(t+v+s+a_{15}),\\
\ell &= (s+a_{16})(u+w+q+a_{17}),\\
P &= r+\ell+a_{18}.
\end{aligned}
\]
\emph{$h=5$, 31 XORs}
\end{minipage}

\begin{lemma}[Uniform inverse for the degree-$19$ characteristic-$2$ circuit]
\label{lem:char2-degree19-inverse}
Let $\F$ be any field of characteristic $2$.  The coefficient map of
the displayed degree-$19$ circuit is a polynomial automorphism from $\F^{19}$
to the monic degree-$19$ polynomials.  Consequently, for any nineteen
distinct $x_0,\ldots,x_{18}\in\F$, its evaluation map at those points is a
bijection $\F^{19}\to\F^{19}$.
\end{lemma}
\begin{proof}
Write $P=x^{19}+\sum_{j=0}^{18}c_jx^j$.  Use the following polynomial key
coordinates:
\[
\begin{array}{lll}
q_0=a_{10},&q_1=a_{11},&q_2=a_{16},\\
q_3=a_2,&q_4=a_0+a_1,&q_5=a_0,\\
q_6=a_3+a_6+a_7,
 &q_7=a_{14}+a_{15}+a_8+a_3^2+a_3,&q_8=a_3,\\
q_9=a_6,&q_{10}=a_{14}+a_4+a_5,&q_{11}=a_4+a_9,\\
q_{12}=a_4+a_5,&q_{13}=a_8,&q_{14}=a_5,\\
q_{15}=a_{17},&q_{16}=a_{12},&q_{17}=a_{13},\\
q_{18}=a_{18}.&&
\end{array}                                                       \tag{A.15}\label{eq:char2-19-q-forward}
\]
This change is polynomially invertible.  Explicitly,
\[
\begin{array}{lll}
a_0=q_5,&a_1=q_4+q_5,&a_2=q_3,\\
a_3=q_8,&a_4=q_{12}+q_{14},&a_5=q_{14},\\
a_6=q_9,&a_7=q_6+q_8+q_9,&a_8=q_{13},\\
a_9=q_{11}+q_{12}+q_{14},&a_{10}=q_0,&a_{11}=q_1,\\
a_{12}=q_{16},&a_{13}=q_{17},&a_{14}=q_{10}+q_{12},\\
a_{15}=q_7+q_{10}+q_{12}+q_{13}+q_8^2+q_8,
 &a_{16}=q_2,&a_{17}=q_{15},\\
a_{18}=q_{18}.&&
\end{array}                                                       \tag{A.16}\label{eq:char2-19-q-inverse}
\]

Put
\[
 S=s+a_{16},\qquad C=u+w+q+a_{17},\qquad \ell=SC.                 \tag{A.17}\label{eq:char2-19-shell}
\]
The point of this factorization is that the top of $C$ is fixed.  Indeed,
directly from the displayed gates,
\[
\begin{aligned}
s&=x^3+q_0x^2+q_1x+q_0q_1,\\
r&=x^3+q_{16}x^2+q_{17}x+q_{16}q_{17},\\
v&=z^2+xz+(a_6+a_7)z+a_7x+a_6a_7,\\
q&=v^2+(t+s+a_{14}+a_{15})v+a_{14}(t+s+a_{15}).
\end{aligned}                                                     \tag{A.18}\label{eq:char2-19-terminal}
\]
Thus $v$ is monic of degree $8$ and $[v]_7=0$, while
$t+s+a_{14}+a_{15}$ is monic of degree $5$.  Since $u$ and $w$ have degrees
$10$ and $12$, respectively, it follows that
\[
 [C]_{16}=1,\qquad [C]_{15}=[C]_{14}=0,\qquad [C]_{13}=1.         \tag{A.19}\label{eq:char2-19-signature}
\]
This is the cubic-shell invariant: the square contributes no odd rows, the
missing subleading coefficient of $v$ kills row $14$, and the monic product
$(t+s)v$ supplies row $13$.

Now
\[
 S=x^3+q_0x^2+q_1x+(q_0q_1+q_2).
\]
Using \eqref{eq:char2-19-signature} in $P=SC+r+a_{18}$ gives the first three
decoder steps explicitly:
\[
 q_0=c_{18},\qquad q_1=c_{17},\qquad
 q_2=c_{16}+q_0q_1+1.                                            \tag{A.20}\label{eq:char2-19-shell-top}
\]
Hence $S$ is known.  The rows $19$ down to $4$ of $P$ equal those of $SC$,
because $\deg(r+a_{18})\le3$.  Top-down division by the known monic cubic
$S$ therefore recovers $[C]_{16},\ldots,[C]_1$.  At the final boundary,
$[r+a_{18}]_3=1$, so $[SC]_3=c_3+1$ and the same division recurrence recovers
$[C]_0$.  Thus the whole polynomial $C$ is known.

It remains to decode the inner crown.  Let $\mathcal C(\mathbf q)$ denote
$C$ after the substitution \eqref{eq:char2-19-q-inverse}.  For
$3\le i\le15$, define the explicit baseline
\[
 B_i(q_0,\ldots,q_{i-1})
 =\coeff{\mathcal C(q_0,\ldots,q_{i-1},0,\ldots,0)}{15-i}.
                                                                    \tag{A.21}\label{eq:char2-19-inner-baseline}
\]
Substitution in the six inner gates and coefficient collection gives the
thirteen unit-pivot identities
\[
 [C]_{15-i}=q_i+B_i(q_0,\ldots,q_{i-1})\qquad(3\le i\le15).      \tag{A.22}\label{eq:char2-19-inner-triangular}
\]
Their complete pivot table is
\[
\begin{array}{c|rrrrrrrrrrrrr}
\text{coefficient of }C&12&11&10&9&8&7&6&5&4&3&2&1&0\\ \hline
\text{pivot}&q_3&q_4&q_5&q_6&q_7&q_8&q_9&q_{10}&q_{11}&q_{12}&q_{13}&q_{14}&q_{15}.
\end{array}                                                       \tag{A.23}\label{eq:char2-19-inner-table}
\]
Equations \eqref{eq:char2-19-inner-baseline}--
\eqref{eq:char2-19-inner-triangular} are literal polynomial identities:
after the already decoded coordinates are substituted, set the later ones to
zero and collect the named row.  They give the explicit recurrence
\[
 q_i=[C]_{15-i}+B_i(q_0,\ldots,q_{i-1})\qquad(3\le i\le15).      \tag{A.24}\label{eq:char2-19-inner-decoder}
\]
The companion calculation
\texttt{char2/verify\_n19\_unitriangular\_symbolic.py} expands all thirteen
identities exactly in $\F_2[q_0,\ldots,q_{18}][x]$; its finite-field
enumeration is only a separate diagnostic.

Finally subtract the now-known shell:
\[
 P+SC=r+a_{18}
 =x^3+q_{16}x^2+q_{17}x+(q_{16}q_{17}+q_{18}).                   \tag{A.25}\label{eq:char2-19-low-tail}
\]
Rows $2,1,0$ recover $q_{16},q_{17},q_{18}$ in that order.  The inverse
coordinate change \eqref{eq:char2-19-q-inverse} then recovers every original
offset.  Every operation in this decoder is polynomial, so it is a two-sided
inverse over every characteristic-$2$ field; no root, division by a scalar, or
Jacobian inference is involved.  Finally, evaluation at nineteen distinct
points is an invertible Vandermonde transformation of the nonleading
coefficients.
\end{proof}

\medskip
\noindent
\begin{minipage}[t]{\textwidth}
\textbf{11 products, degree 21} (21 keys)
\[
\begin{aligned}
y &= x^2,\\
z &= (y+a_0)(x+y+a_1),\\
t &= (x+a_2)(z+a_3),\\
u &= (y+t+a_4)(z+t+a_5),\\
v &= (x+z+a_6)(z+a_7),\\
w &= (x+y+z+a_8)(y+v+a_9),\\
s &= (x+a_{10})(y+a_{11}),\\
r &= (x+a_{12})(y+a_{13}),\\
q &= (v+a_{14})(t+v+s+a_{15}),\\
\ell &= (s+a_{16})(u+w+q+a_{17}),\\
m &= (t+s+a_{18})(z+u+w+q+a_{19}),\\
P &= m+z+r+\ell+a_{20}.
\end{aligned}
\]
\emph{$h=5$, 39 XORs}
\end{minipage}

\begin{lemma}[Uniform inverse for the degree-$21$ characteristic-$2$ circuit]
\label{lem:char2-degree21-inverse}
Let $\F$ be any field of characteristic $2$.  The coefficient map of the
displayed degree-$21$ circuit is a polynomial automorphism from $\F^{21}$ to
the monic degree-$21$ polynomials.  Consequently its evaluation map at any
twenty-one distinct field elements is a bijection $\F^{21}\to\F^{21}$.
\end{lemma}
\begin{proof}
Write $P=x^{21}+\sum_{j=0}^{20}c_jx^j$.  The decoder uses the following
polynomial coordinates:
\[
\begin{array}{lll}
q_0=a_2,&q_1=a_0+a_1,&q_2=a_0,\\
q_3=a_3,&q_4=a_{16}+a_{18},&q_5=a_{10},\\
q_6=a_3+a_6+a_7+a_{11},&q_8=a_{11},&q_9=a_6,\\
q_{10}=a_{14}+a_4+a_5,&q_{11}=a_4+a_9,&q_{12}=a_4+a_5,\\
q_{13}=a_8,&q_{14}=a_5,&q_{15}=a_{19},\\
q_{16}=a_{18},&q_{17}=a_{17},&q_{18}=a_{12},\\
q_{19}=a_{13},&q_{20}=a_{20}.&
\end{array}                                                       \tag{A.26}\label{eq:char2-21-q-forward}
\]
The one longer coordinate is
\[
 q_7=a_{14}+a_{15}+a_8+a_3^2+a_3+a_{11}+a_{11}^2
       +a_2a_{11}+a_{10}a_{11}.                                 \tag{A.27}
\]
This change has the explicit polynomial inverse
\[
\begin{array}{lll}
a_0=q_2,&a_1=q_1+q_2,&a_2=q_0,\\
a_3=q_3,&a_4=q_{12}+q_{14},&a_5=q_{14},\\
a_6=q_9,&a_7=q_6+q_8+q_3+q_9,&a_8=q_{13},\\
a_9=q_{11}+q_{12}+q_{14},&a_{10}=q_5,&a_{11}=q_8,\\
a_{12}=q_{18},&a_{13}=q_{19},&a_{14}=q_{10}+q_{12},\\
a_{16}=q_4+q_{16},&a_{17}=q_{17},&a_{18}=q_{16},\\
a_{19}=q_{15},&a_{20}=q_{20}.&
\end{array}                                                       \tag{A.28}
\]
The omitted long row is
\[
 a_{15}=q_7+q_8+q_8^2+q_0q_8+q_5q_8+q_{10}+q_{12}+q_{13}
          +q_3^2+q_3.                                           \tag{A.29}
\]
Substitution in either direction verifies (A.26)--(A.29) without division.

Here is a compact explicit certificate for all twenty-one pivots.  Let
$\mathcal P(q_0,\ldots,q_{20})$ be the displayed circuit after substituting
(A.28)--(A.29), and define
\[
 K_i(q_0,\ldots,q_{i-1})
  =\coeff{\mathcal P(q_0,\ldots,q_{i-1},0,\ldots,0)}{20-i}.
                                                                    \tag{A.30}\label{eq:char2-21-baseline}
\]
Literal expansion in $\F_2[q_0,\ldots,q_{20}][x]$ gives
\[
 c_{20-i}=q_i+K_i(q_0,\ldots,q_{i-1})\qquad(0\le i\le20).       \tag{A.31}\label{eq:char2-21-triangular}
\]
For clarity, its first eight rows are
\[
\begin{aligned}
c_{20}&=1+q_0,\\
c_{19}&=q_0+q_1,\\
c_{18}&=1+q_2+q_0q_1,\\
c_{17}&=q_3+q_2^2+q_0q_2+q_1q_2,\\
c_{16}&=q_4+q_0^2+q_0q_3+q_0q_2^2+q_0q_1q_2,\\
c_{15}&=1+q_1+q_5,\\
c_{14}&=1+q_1+q_2+q_3+q_5+q_6+q_0q_1+q_0q_5,\\
c_{13}&=q_2+q_3+q_4+q_5+q_7+q_1^4+q_2^2+q_6^2
          +q_0q_1+q_0q_2+q_0q_5+q_1q_2+q_1q_5.
\end{aligned}                                                     \tag{A.32}
\]
The full pivot order, split across two lines for readability, is
\[
\begin{gathered}
\begin{array}{c|rrrrrrrrrrr}
\text{row}&20&19&18&17&16&15&14&13&12&11&10\\ \hline
\text{pivot}&q_0&q_1&q_2&q_3&q_4&q_5&q_6&q_7&q_8&q_9&q_{10}
\end{array}\\[1ex]
\begin{array}{c|rrrrrrrrrr}
\text{row}&9&8&7&6&5&4&3&2&1&0\\ \hline
\text{pivot}&q_{11}&q_{12}&q_{13}&q_{14}&q_{15}&q_{16}&q_{17}&q_{18}&q_{19}&q_{20}.
\end{array}
\end{gathered}                                                     \tag{A.33}
\]
Thus the decoder is the descending recurrence
\[
 q_i=c_{20-i}+K_i(q_0,\ldots,q_{i-1})\qquad(0\le i\le20).      \tag{A.34}
\]
The baseline in \eqref{eq:char2-21-baseline} is an explicit polynomial:
substitute already decoded coordinates, set the later coordinates to zero,
and collect the named row.  The companion script
\texttt{char2/verify\_n21\_unitriangular\_symbolic.py} expands every identity
in \eqref{eq:char2-21-triangular} exactly.  Its exhaustive $\F_2$ check is a
separate diagnostic and is not used in the proof.

Finally, (A.28)--(A.29) recover the original keys.  Hence the coefficient map
has a two-sided polynomial inverse over every characteristic-$2$ field; no
root extraction, scalar division, or Jacobian inference occurs.  Evaluation
at distinct points is then an invertible Vandermonde transformation.
\end{proof}

\subsection{Large-prime implementation (\texorpdfstring{$\F_{2^{89}-1}$}{GF(2\textasciicircum 89-1)})}

To implement the formulas intended for characteristic zero, we use the
Mersenne prime $M_{89}=2^{89}-1$.  The resulting computation is over the
finite field $\F_{M_{89}}$---it is not literally characteristic zero---and
uses 89-bit modular arithmetic with keys
$a_0,a_1,\ldots\in[0,M_{89})$.  Since $M_{89}$ is larger than every degree
displayed here, it lies in the large-characteristic regime covered by the
proved constructions elsewhere in the paper.  That fact does not certify
these separately optimized candidates: the search also checked a constant
nonzero Jacobian determinant, which is necessary for a polynomial inverse but
does not by itself prove one.  These candidates would require explicit
decoders before being used as bijective families.

\medskip
\noindent
\begin{minipage}[t]{0.47\textwidth}
\textbf{4 products, degree 7} (7 keys)
\[
\begin{aligned}
y &= (x + 0) \cdot (x + a_0) \\
z &= (x + a_1) \cdot (y + a_2) \\
t &= (x + z + a_3) \cdot (x + 0) \\
u &= (t + a_4) \cdot (z + a_5) \\
P &= y + u + a_6
\end{aligned}
\]
\emph{Height 4, 9 additions}
\end{minipage}
\hfill
\begin{minipage}[t]{0.47\textwidth}
\textbf{5 products, degree 9} (9 keys)
\[
\begin{aligned}
y &= (x + 0) \cdot (x + 0) \\
z &= (x + y + a_0) \cdot (y + a_1) \\
t &= (y + z + a_2) \cdot (x + a_3) \\
u &= (z + a_4) \cdot (t + a_5) \\
v &= (x + a_6) \cdot (y + a_7) \\
P &= u + v + a_8
\end{aligned}
\]
\emph{Height 4, 12 additions}
\end{minipage}

\medskip
\noindent
\begin{minipage}[t]{0.47\textwidth}
\textbf{6 products, degree 11} (11 keys)
\[
\begin{aligned}
y &= (x + 0) \cdot (x + a_0) \\
z &= (y + a_1) \cdot (x + y + a_2) \\
t &= (y + z + a_3) \cdot (x + a_4) \\
u &= (y + t + a_5) \cdot (t + a_6) \\
v &= (y + a_7) \cdot (z + a_8) \\
w &= (u + a_9) \cdot (x + 0) \\
P &= v + w + a_{10}
\end{aligned}
\]
\emph{Height 5, 15 additions}
\end{minipage}
\hfill
\begin{minipage}[t]{0.47\textwidth}
\textbf{7 products, degree 13} (13 keys)
\[
\begin{aligned}
y &= (x + 0) \cdot (x + 0) \\
z &= (y + a_0) \cdot (x + y + a_1) \\
t &= (y + z + a_2) \cdot (x + a_3) \\
u &= (y + z + a_4) \cdot (z + a_5) \\
v &= (z + t + a_6) \cdot (z + u + a_7) \\
w &= (y + a_8) \cdot (x + a_9) \\
s &= (x + y + z + t + u + w + a_{10}) \\
  &\quad\cdot(x + a_{11}) \\
P &= v + s + a_{12}
\end{aligned}
\]
\emph{Height 4, 24 additions}
\end{minipage}

\medskip
\noindent
\begin{minipage}[t]{0.47\textwidth}
\textbf{8 products, degree 15} (15 keys)
\[
\begin{aligned}
y &= (x + 0) \cdot (x + a_0) \\
z &= (y + a_1) \cdot (x + 0) \\
t &= (z + a_2) \cdot (x + y + z + a_3) \\
u &= (y + t + a_4) \cdot (z + t + a_5) \\
v &= (t + a_6) \cdot (x + t + a_7) \\
w &= (t + a_8) \cdot (x + a_9) \\
s &= (z + a_{10}) \cdot (x + y + v + a_{11}) \\
r &= (y + a_{12}) \cdot (u + v + a_{13}) \\
P &= w + s + r + a_{14}
\end{aligned}
\]
\emph{Height 5, 25 additions}
\end{minipage}
\hfill
\begin{minipage}[t]{0.47\textwidth}
\textbf{9 products, degree 17} (17 keys)
\[
\begin{aligned}
y &= (x + 0) \cdot (x + a_{16}) \\
z &= (x + y + a_{13}) \\
  &\quad\cdot(-x + y + a_{15}) \\
t &= (x + y + z + a_{11}) \\
  &\quad\cdot(-x - y + z + a_{12}) \\
u &= (y + z + a_{10}) \cdot (-y + z + a_{14}) \\
v &= (x + a_{7}) \cdot (y + a_{6}) \\
w &= (x + a_{3}) \cdot (y + a_{2}) \\
s &= (x + z + t + v + a_{5}) \\
  &\quad\cdot(x - z + t - v + a_{9}) \\
r &= (z + u + a_{4}) \cdot (-z + u + a_{8}) \\
q &= (w + s + a_{1}) \cdot (x + 0) \\
P &= r + q + a_{0}
\end{aligned}
\]
\emph{Height 5, 35 additions}
\end{minipage}

\smallskip
\noindent Unlike the searched candidates above, the degree-17 circuit is not a
search result: it is the degree-17 instance of the proved general family,
printed by \texttt{tools/polychain.py chain 17 --reduced} after a
unit-triangular key normalization that makes every gate constant a single
fresh key.  Its decoder is the paper's final decoder, and bijectivity of the
key map is covered by \Cref{cor:all-odd-decodable}; no separate
certification is needed.

\medskip
\noindent\textbf{Note:} In the large-prime implementation, the products
$(x+0)\cdot(x+0)$, i.e.\ the squaring gates $y=x^2$ of the degree-$9$ and
degree-$13$ circuits above, can be computed more efficiently than general
multiplications using $(a+b)^2=a^2+2ab+b^2\pmod{M_{89}}$; a product
$(x+0)\cdot w$ with $w\ne x$ is an ordinary multiplication.

\section{Decoder Calculus}
\label{appendix:decoder-calculus}

\paragraph{Polynomial recovery notation.}
The proofs repeatedly use the same elementary notion of dependence.  If
$Y=(y_1,\ldots,y_r)$ is a finite list of quantities, we write
\[
    z\polyfrom Y
\]
and say that $z$ is \emph{polynomially recoverable from $Y$} if there is a
polynomial $f\in\mathbb{F}[t_1,\ldots,t_r]$ with
$z=f(y_1,\ldots,y_r)$.  Equivalently, starting with the entries of $Y$, one
may obtain $z$ by finitely many additions, subtractions, multiplications, and
multiplications by fixed constants from $\mathbb{F}$.  This notation does not
assert that the recovery is efficient; the multiplication cost of the
construction is tracked separately.

We use a semicolon to separate already known data from newly observed data:
$z\polyfrom (K;Y)$ means that the entries of $K$ may also be used in the
polynomial formula.  The only composition rule we need is substitution:
\[
    z\polyfrom Y,\qquad
    y_i\polyfrom Z\ \text{for every }i
    \quad\Longrightarrow\quad
    z\polyfrom Z.
\]
The Lean formalization uses the corresponding generated-algebra membership
statement; readers need not use that formulation.

Only fixed constants from $\mathbb F$ may be used for free.  In particular,
the entries called ``known'' must come from auxiliary polynomials or from
parameters recovered at an earlier stage; data-dependent division is not
silently allowed.  This provenance convention is what makes the substitution
rule useful in a descending decoder.
The phrase ``given $K$'' can therefore be read as ordinary side information
for a conditional decoder: $K$ is fixed and available throughout that step.
No probabilistic averaging or conditional expectation is intended.

Throughout the appendix, a \emph{scalar} means a quantity independent of
$x$; it may still be one of the active parameters.  Such a scalar is not
available for free unless it is explicitly included in the given data or has
already been recovered.  Thus, for example, the shift $\delta$ in a square
gadget is an unknown scalar, whereas a fixed pivot such as $2$ or $k$ lies in
$\mathbb F$ and may be inverted when the characteristic hypothesis permits
it.  All coefficient identities below are polynomial identities in these
unknown scalars, so they remain valid in the ambient parameter polynomial
ring.

\begin{definition}[Extractable]
    We say that $(\alpha_i)_{i \in \rng{n}}$ are \emph{extractable} from a
    polynomial $P \in \mathbb{F}[\alpha_0,\ldots,\alpha_{n-1}][x]$ if
    \[
        \alpha_i\polyfrom
        \bigl((\coeff{P}{j})_{\,j\in[\deg P]}\bigr)
        \qquad(i\in\rng n).
    \]

    Given a sequence of monic polynomials $(B_k)_{k\in[t]}$, we say that the
    parameters are extractable from $P$ \emph{given $(B_k)$} if
    \[
        \alpha_i\polyfrom
        \Bigl((\coeff{B_k}{r})_{\,k\in[t],\,r\in[\deg B_k]}\ ;\
              (\coeff{P}{j})_{\,j\in[\deg P]}\Bigr)
        \qquad(i\in\rng n).
    \]
\end{definition}

\begin{definition}[Decodable]
    We say that a monic polynomial $P \in \mathbb{F}[\alpha_0, \ldots, \alpha_{n - 1}][x]$ is \emph{decodable} if $\deg P = n$ and $(\alpha_i)_{i \in \rng{n}}$ are extractable from $P$.

    We say that a monic polynomial $P \in \mathbb{F}[\alpha_0, \ldots, \alpha_{n - 1}][x]$ is \emph{decodable} given a sequence of monic polynomials, $(B_i)_{i \in [t]}$, if $\deg P = n$ and $(\alpha_i)_{i \in \rng{n}}$ are extractable from $P$ given $(B_i)_{i \in [t]}$.
\end{definition}

\begin{definition}[Algebraically splittable pair]
    For $n\in\mathbb{N}$, a pair of monic polynomials
    $(T^{(1)},T^{(2)})\in
    \mathbb{F}[\alpha_0,\ldots,\alpha_{n-1}][x]^2$ is a
    \emph{algebraically splittable pair}, or simply a \emph{splittable pair},
    for $n$ if
    \[
        \deg T^{(1)}=\deg T^{(2)}=n-1
        \quad\text{and}\quad
        xT^{(1)}+T^{(2)}\;\text{is decodable}.
    \]
\end{definition}

\begin{definition}[Joint realization and recorded byproducts]
\label{def:joint-realization}
Let $(T^{(1)},T^{(2)})$ be a pair over the parameters
$\alpha_0,\ldots,\alpha_{n-1}$.  An \emph{$m$-realization} of this pair is
one arithmetic circuit, on the common inputs
$(x,\alpha_0,\ldots,\alpha_{n-1})$, that outputs both components using
exactly $m$ multiplication gates.  If a polynomial $H$ is also an output
wire of that same circuit, we say that the realization \emph{records $H$ as
a byproduct}.  Declaring an existing wire to be an output costs no
additional gate.
\end{definition}

\begin{remark}[Why realizability is part of the induction invariant]
\label{rem:canonical-split}
The preceding algebraic notion of splittability is intentionally weaker than
joint realizability.  Indeed, every decodable monic family
\[
  P=x^n+\sum_{j=0}^{n-1}c_jx^j\qquad(n\ge2)
\]
has the canonical compatible split
\[
  T^{(1)}=x^{n-1}+(c_{n-1}-1)x^{n-2},\qquad
  T^{(2)}=x^{n-1}+\sum_{j=0}^{n-2}c_jx^j.
\]
It satisfies $xT^{(1)}+T^{(2)}=P$; the only nonconstant coefficient of the
first component is read from row $n-1$, and the coefficient of degree $j$ in
the second component is read from row $j$.  Thus it is compatible on
$\rng n$.

This observation does not give a cheap circuit for the two components:
evaluating the coefficient polynomials $c_j(\alpha)$ may require additional
multiplications.  The recursive theorem below therefore carries the stronger
object that is actually used---a compatible pair together with one
costed joint realization and its recorded power byproducts.  In particular,
the septic base is algebraically splittable by the display above; what is not
supplied is a three-multiplication joint realization suitable for the
cost-optimal recursion.
\end{remark}

\begin{lemma}[A polynomial decoder is two-sided]
\label{lem:polynomial-left-inverse-automorphism}
Let
\[
  C=(C_0,\ldots,C_{n-1}):\mathbb A^n_{\mathbb F}\longrightarrow
       \mathbb A^n_{\mathbb F}
\]
be a polynomial coefficient map.  If there is a polynomial map $D$ with
$D\circ C=\operatorname{id}$ as a polynomial identity, then $C$ is a
polynomial automorphism and $C\circ D=\operatorname{id}$ as well.
Consequently, a decodable monic degree-$n$ family in the sense above covers
every monic degree-$n$ coefficient vector and supplies polynomial (hence
rational) preprocessing.
\end{lemma}
\begin{proof}
The substitution map
\[
  C^*:\mathbb F[y_0,\ldots,y_{n-1}]\longrightarrow
       \mathbb F[\alpha_0,\ldots,\alpha_{n-1}],
  \qquad y_i\longmapsto C_i(\alpha),
\]
has a right inverse $D^*$, so it is surjective.  Let
$\iota:\mathbb F[\alpha_0,\ldots,\alpha_{n-1}]\simeq
\mathbb F[y_0,\ldots,y_{n-1}]$ be the variable-renaming isomorphism.
Then $\varphi=\iota\circ C^*$ is a surjective endomorphism of the
Noetherian ring $R=\mathbb F[y_0,\ldots,y_{n-1}]$.  Every surjective endomorphism
$\varphi$ of a Noetherian ring is injective: the ascending chain
$\ker\varphi\subseteq\ker\varphi^2\subseteq\cdots$ stabilizes, say at
$m$; if $\varphi(a)=0$, choose $b$ with $\varphi^m(b)=a$, and then
$b\in\ker\varphi^{m+1}=\ker\varphi^m$, so $a=0$.  Hence $\varphi$, and
therefore $C^*$, is also
injective and therefore is an isomorphism.  Its right inverse $D^*$ must be
its inverse, which is exactly the identity
$C\circ D=\operatorname{id}$.
\end{proof}

\begin{definition}[Polynomially recoverable polynomial]
    Let $P\in\mathbb{F}[\alpha_0,\ldots,\alpha_{m-1}][x]$ and let
    $(B_i)_{i\in[t]}$ be a sequence of polynomials in the same ring.  We say
    that $P$ is \emph{polynomially recoverable} (or \emph{derivable}) from
    $(B_i)_{i\in[t]}$ if
    \[
        \coeff{P}{j}\polyfrom
        \bigl((\coeff{B_k}{r})_{\,k\in[t],\,r\in\idx{\deg B_k}}\bigr)
        \qquad(j\in\idx{\deg P}).
    \]

    Given an auxiliary sequence $(C_\ell)_{\ell\in[s]}$, we say that $P$ is
    recoverable from $(B_i)$ \emph{given $(C_\ell)$} if, for every
    $j\in\idx{\deg P}$,
    \[
       \coeff{P}{j}\polyfrom
       \Bigl((\coeff{C_\ell}{r})_{\ell\in[s],\,r\in\idx{\deg C_\ell}}\ ;\
             (\coeff{B_k}{r})_{k\in[t],\,r\in\idx{\deg B_k}}\Bigr).
    \]
    Thus ``derivable'' in the remainder of the appendix is simply shorthand
    for this polynomial-recovery relation.
\end{definition}

\paragraph{Terminology.}
``Extractable'' refers to recovering the hidden parameter list, while
``recoverable'' (and the older shorthand ``derivable'') refers to recovering
an intermediate polynomial or one of its coefficients.  ``Compatible'' adds
the causal requirement that each coefficient be recovered only from the
appropriate high-degree part of the combined polynomial.

\begin{lemma}[Discharging conditional side information]
\label{lem:discharge-side-information}
Suppose $(\alpha_i)_{i\in\rng n}$ are extractable from $Q$ given a list of
auxiliary polynomials $B=(B_1,\ldots,B_t)$.  If $Q$ and every $B_i$ are
polynomially recoverable from $P$ given a remaining side list $K$, then the
$\alpha_i$ are extractable from $P$ given $K$.
\end{lemma}
\begin{proof}
The conditional decoder writes every $\alpha_i$ as a polynomial in the
coefficients of $Q$ and the $B_j$'s.  Substitute the assumed polynomial
formulas for all of those coefficients in terms of $(K;P)$.  The result is a
polynomial recovery formula for $\alpha_i$ from $(K;P)$, with no occurrence
of the discharged list $B$.
\end{proof}

\begin{corollary}[Recovery composes]\label{lem:extractable-via-derivable}
    If $(\alpha_i)_{i\in\rng{n}}$ are extractable from a polynomial $Q$ given
    $(B_k)_{k\in[t]}$, and $Q$ is polynomially recoverable from a polynomial
    $P$ given $(B_k)_{k\in[t]}$, then $(\alpha_i)_{i\in\rng{n}}$ are
    extractable from $P$ given $(B_k)_{k\in[t]}$.
\end{corollary}
\begin{proof}
    Apply \Cref{lem:discharge-side-information} with the remaining side list
    $K=(B_k)_{k\in[t]}$: $Q$ is recoverable by hypothesis, and every $B_k$
    is recoverable trivially from that same side list.
\end{proof}

\begin{definition}[Compatible pair]\label{def:compatible-pair}
    Let $n\in\mathbb{N}$ and put $\Phi=xP^{(1)}+P^{(2)}$.  We say that a
    pair of monic polynomials $(P^{(1)},P^{(2)})$ is a \emph{compatible pair}
    on a set of indices $G\subseteq\idx n$ if:
    \begin{enumerate}
        \item $\deg P^{(1)} = \deg P^{(2)} = n$.
        \item For every $j\in\rng n$,
        \begin{align}
            \coeff{P^{(1)}}{j}
              &\polyfrom
              \bigl(\coeff{\Phi}{i}:i\in G,\ i\ge j+1\bigr),\\
            \coeff{P^{(2)}}{j}
              &\polyfrom
              \bigl(\coeff{\Phi}{i}:i\in G,\ i\ge j\bigr).
        \end{align}
    \end{enumerate}

    Given an auxiliary sequence of monic polynomials $(B_k)_{k\in[t]}$, we
    say that the pair is compatible on $G$ \emph{given $(B_k)$} if the same
    two relations hold with the coefficients of the $B_k$'s included among
    the known inputs.
\end{definition}

When the pair is used as a splittable pair for a degree-$n$ polynomial, its
two components have degree $n-1$.  In that application the natural window is
\(
  \idx{n-1}=\rng n,
\)
because the combined polynomial \(xP^{(1)}+P^{(2)}\) has degree \(n\) and its
leading coefficient is fixed by monicity.  We freely use either notation below
depending on whether we are emphasizing the component degree or the target
degree.

\begin{figure}[H]
  \centering
  \begin{minipage}[b]{0.32\textwidth}\centering
\begin{tikzpicture}[x=4.2mm, y=4.2mm, font=\scriptsize]
  \fill[fill=white] (0,0) rectangle (1,-1);
  \fill[fill=white] (1,0) rectangle (2,-1);
  \fill[fill=white] (2,0) rectangle (3,-1);
  \fill[fill=white] (3,0) rectangle (4,-1);
  \fill[fill=blue!70!black] (4,0) rectangle (5,-1);
  \fill[fill=white] (5,0) rectangle (6,-1);
  \fill[fill=blue!70!black] (6,0) rectangle (7,-1);
  \node[anchor=east] at (-0.15,-0.5) {$x^{6}$};
  \fill[fill=white] (0,-1) rectangle (1,-2);
  \fill[fill=white] (1,-1) rectangle (2,-2);
  \fill[fill=white] (2,-1) rectangle (3,-2);
  \fill[fill=white] (3,-1) rectangle (4,-2);
  \fill[fill=orange!35] (4,-1) rectangle (5,-2);
  \fill[fill=blue!70!black] (5,-1) rectangle (6,-2);
  \fill[fill=orange!35] (6,-1) rectangle (7,-2);
  \node[anchor=east] at (-0.15,-1.5) {$x^{5}$};
  \fill[fill=white] (0,-2) rectangle (1,-3);
  \fill[fill=white] (1,-2) rectangle (2,-3);
  \fill[fill=blue!70!black] (2,-2) rectangle (3,-3);
  \fill[fill=blue!70!black] (3,-2) rectangle (4,-3);
  \fill[fill=orange!35] (4,-2) rectangle (5,-3);
  \fill[fill=orange!35] (5,-2) rectangle (6,-3);
  \fill[fill=orange!35] (6,-2) rectangle (7,-3);
  \node[anchor=east] at (-0.15,-2.5) {$x^{4}$};
  \fill[fill=white] (0,-3) rectangle (1,-4);
  \fill[fill=blue!70!black] (1,-3) rectangle (2,-4);
  \fill[fill=orange!35] (2,-3) rectangle (3,-4);
  \fill[fill=orange!35] (3,-3) rectangle (4,-4);
  \fill[fill=orange!35] (4,-3) rectangle (5,-4);
  \fill[fill=orange!35] (5,-3) rectangle (6,-4);
  \fill[fill=orange!35] (6,-3) rectangle (7,-4);
  \node[anchor=east] at (-0.15,-3.5) {$x^{3}$};
  \fill[fill=white] (0,-4) rectangle (1,-5);
  \fill[fill=orange!35] (1,-4) rectangle (2,-5);
  \fill[fill=orange!35] (2,-4) rectangle (3,-5);
  \fill[fill=orange!35] (3,-4) rectangle (4,-5);
  \fill[fill=orange!35] (4,-4) rectangle (5,-5);
  \fill[fill=orange!35] (5,-4) rectangle (6,-5);
  \fill[fill=orange!35] (6,-4) rectangle (7,-5);
  \node[anchor=east] at (-0.15,-4.5) {$x^{2}$};
  \fill[fill=white] (0,-5) rectangle (1,-6);
  \fill[fill=orange!35] (1,-5) rectangle (2,-6);
  \fill[fill=orange!35] (2,-5) rectangle (3,-6);
  \fill[fill=orange!35] (3,-5) rectangle (4,-6);
  \fill[fill=orange!35] (4,-5) rectangle (5,-6);
  \fill[fill=orange!35] (5,-5) rectangle (6,-6);
  \fill[fill=orange!35] (6,-5) rectangle (7,-6);
  \node[anchor=east] at (-0.15,-5.5) {$x^{1}$};
  \fill[fill=blue!70!black] (0,-6) rectangle (1,-7);
  \fill[fill=orange!35] (1,-6) rectangle (2,-7);
  \fill[fill=orange!35] (2,-6) rectangle (3,-7);
  \fill[fill=white] (3,-6) rectangle (4,-7);
  \fill[fill=orange!35] (4,-6) rectangle (5,-7);
  \fill[fill=orange!35] (5,-6) rectangle (6,-7);
  \fill[fill=white] (6,-6) rectangle (7,-7);
  \node[anchor=east] at (-0.15,-6.5) {$x^{0}$};
  \draw[gray!60, very thin] (0,0) grid (7,-7);
  \node[anchor=west, rotate=90] at (0.5,0.15) {$\alpha_{0}$};
  \node[anchor=west, rotate=90] at (1.5,0.15) {$\alpha_{1}$};
  \node[anchor=west, rotate=90] at (2.5,0.15) {$\alpha_{2}$};
  \node[anchor=west, rotate=90] at (3.5,0.15) {$\alpha_{3}$};
  \node[anchor=west, rotate=90] at (4.5,0.15) {$\alpha_{4}$};
  \node[anchor=west, rotate=90] at (5.5,0.15) {$\alpha_{5}$};
  \node[anchor=west, rotate=90] at (6.5,0.15) {$\alpha_{6}$};
\end{tikzpicture}\end{minipage}\hfill
  \begin{minipage}[b]{0.62\textwidth}\centering\input{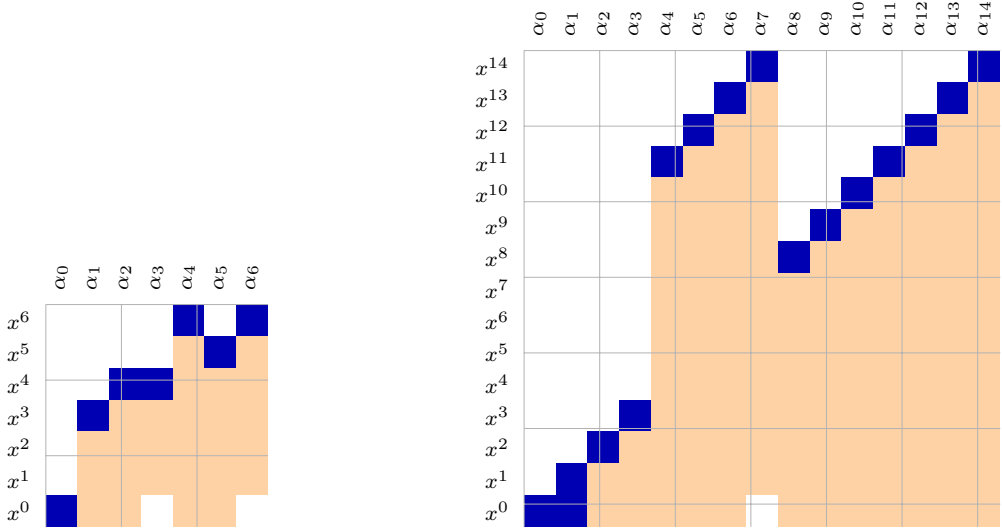}\end{minipage}
  \caption{Sparsity pattern of the Jacobian $\partial\,\coeff{P_n}{d}/\partial\alpha_i$ for the septic base $P_7$ (left) and the special case $P_{15}$ (right); rows are degrees $d=n-1,\ldots,0$ and columns are parameters.
  Blue (dark) cells are nonzero field constants, orange (light) cells are non-constant polynomials in the parameters, white cells are zero.
  Decoding proceeds top-down: each row exposes one fresh parameter with a constant slope once the parameters of the rows above are known.
  Where a row contains two constant cells (e.g.\ $\coeff{P_7}{6}$, which involves $\alpha_4$ and $\alpha_6$), the decoder first recovers an internal quantity of the construction (here $z_2=\alpha_4+1+\alpha_6$) and separates the two parameters at a lower row; this is the composition structure that the compatible-pair lemmas make precise.}
  \label{fig:jacobian-pattern}
\end{figure}

\Cref{fig:jacobian-pattern} is only a visualization of the explicit pivot order.  We do
not infer invertibility from a Jacobian determinant; every inverse used below
is supplied by the coefficient recurrences themselves.

\paragraph{Constructive viewpoint.}
Compatibility is a \emph{causal} recovery condition.  At cutoff $j$, the
decoder for $[x^j]P^{(1)}$ may use only coefficients of
$\Phi=xP^{(1)}+P^{(2)}$ in degrees at least $j+1$, and the decoder for
$[x^j]P^{(2)}$ may use only degrees at least $j$.  In particular, mere
recoverability of the pair from the whole window is not sufficient.  The
closure proofs below preserve this cutoff condition because all their
reconstructions run in descending degree order.

Thus a proof of compatibility must provide a cutoff-respecting decoder, not
merely recover both components from the entire window.  We will usually give
that decoder as a descending pivot table.  The following stronger form is
tailored to the $T_{k,2^l}$ remainder recursion.

\begin{definition}[Coefficient-triangular remainder pair]
\label{def:coefficient-triangular}
Let $d\ge0$, let $\alpha_0,\ldots,\alpha_{d-1}$ be the active parameters, and
let $K$ be the quantities already known to the decoder.  A pair
$(R^{(1)},R^{(2)})$ is \emph{coefficient-triangular} if, for every
$j\in\rng d$,
\begin{align}
 [x^j]R^{(1)}
   &\polyfrom (K;\alpha_{j+1},\ldots,\alpha_{d-1}),&
 [x^j]R^{(2)}
   &\polyfrom (K;\alpha_j,\ldots,\alpha_{d-1}),
   \label{eq:filtered-remainder}\\
 [x^j](xR^{(1)}+R^{(2)})
   &=\lambda_j\alpha_j+F_j,
   &F_j&\polyfrom (K;\alpha_{j+1},\ldots,\alpha_{d-1}).
   \label{eq:filtered-pivot}
\end{align}
Every pivot slope $\lambda_j$ is required to lie in $\mathbb F^\times$.
Coefficients in degrees at least $d$ are required to be polynomially
recoverable from $K$ alone.
\end{definition}

\begin{lemma}[A triangular remainder pair is causal]
\label{lem:triangular-implies-compatible}
Let $H,\widetilde H$ be auxiliary monic polynomials of degree $h$, let
$N=kh$ and $d=(k-1)h$, and set
\[
 T^{(1)}=H^k+R^{(1)},\qquad
 T^{(2)}=\widetilde H^k+R^{(2)}.
\]
If $(R^{(1)},R^{(2)})$ is coefficient-triangular and has degree at most $d$,
then the parameters are recovered from
$D=xR^{(1)}+R^{(2)}$ by descending induction, and
$(T^{(1)},T^{(2)})$ is compatible on $\rng d$ given
$(H,\widetilde H)$ and the other auxiliary data.
\end{lemma}
\begin{proof}
Equation~\eqref{eq:filtered-pivot} is a descending decoder: after
$\alpha_{j+1},\ldots,\alpha_{d-1}$ have been recovered, divide by the fixed
nonzero scalar $\lambda_j$ to recover $\alpha_j$.  The support conditions in
\eqref{eq:filtered-remainder} say that $[x^j]R^{(1)}$ uses only the already
recovered parameters with index at least $j+1$, while $[x^j]R^{(2)}$ may also
use $\alpha_j$.  Coefficients in degrees at least $d$ are known from $K$.
Finally
\[
 D=xT^{(1)}+T^{(2)}-\bigl(xH^k+\widetilde H^k\bigr),
\]
so the same cutoff statements hold with the coefficients of
$xT^{(1)}+T^{(2)}$.  This is precisely compatibility.
\end{proof}

\begin{lemma}[Shifting a triangular remainder block]
\label{lem:triangular-shift}
Let $(A^{(1)},A^{(2)})$ be coefficient-triangular for active
parameters $\eta_0,\ldots,\eta_{e-1}$, given known data $K$.
Let $L_1,L_2$ be monic polynomials of the same degree $h$.  Replace $\eta_i$
by a parameter $\alpha_{h+i}$.  The coefficients of $L_1,L_2$ may either be
included in the known data $K$, or be polynomially recoverable from
\[
       (K;\zeta_1,\ldots,\zeta_s),
\]
where every $\zeta_u$ has been assigned a global parameter index at least
$h+e$.  Thus the $\zeta$'s belong to a block higher than the block being
shifted.  Then the pair

\[
       (L_1A^{(1)},L_2A^{(2)})
\]

has the triangular support conditions on the shifted block
$h,\ldots,h+e-1$, and its combined pivot at $h+i$ is the old pivot at
$i$.  More precisely, coefficients below $h+e$ have no active parameter
of index smaller than their degree (with the usual one-degree offset for
the first component), and

\[
 [x^{h+i}](xL_1A^{(1)}+L_2A^{(2)})
   =\lambda_i\alpha_{h+i}+F_{h+i},
   \qquad F_{h+i}\polyfrom
      (K;\zeta_1,\ldots,\zeta_s,\alpha_{h+i+1},\ldots).
\]

The assertion remains true after adding polynomials whose coefficients are
polynomially recoverable from $(K;\zeta_1,\ldots,\zeta_s)$, provided their
active parameters also have global indices at least $h+e$.
\end{lemma}
\begin{proof}
Write $A^{(1)}_q=[x^q]A^{(1)}$ and similarly for $A^{(2)}$.  By
coefficient-triangularity, $A^{(1)}_q$ is polynomially recoverable from
$(K;\eta_{q+1},\ldots,\eta_{e-1})$, while $A^{(2)}_q$ is polynomially
recoverable from $(K;\eta_q,\ldots,\eta_{e-1})$.
In a coefficient of $L_rA^{(r)}$, a summand has the form
$L_{r,t}A^{(r)}_q$ with $t+q=j$ and $0\le t\le h$.  Thus a parameter
$\eta_i$ occurring in the first (respectively second) component can occur
only when $j\le h+i-1$ (respectively $j\le h+i$), which is exactly the
required support after reindexing.
For the coefficient of degree $h+i$ in the combined polynomial, the
summands with $t=h$ are
$A^{(1)}_{i-1}+A^{(2)}_i$, because both factors are monic.  All summands
with $t<h$ involve only $A^{(1)}_q$ with $q>i-1$ or $A^{(2)}_q$ with
$q>i$, hence only parameters of index strictly larger than $i$.  The old
    pivot formula therefore gives the displayed new pivot with the same
    $\lambda_i$.  Substituting the formulas for the coefficients of
    $L_1,L_2$ introduces only the higher parameters $\zeta_u$, whose global
    indices are at least $h+e$ and hence are strictly larger than every
    index in the shifted block.  The same observation applies to the allowed
    auxiliary additions, so neither the support nor the pivot is changed.

The support assertion is only needed on the shifted block
$h,\ldots,h+e-1$.  Rows below $h$ can of course contain the low
coefficients of the inner pair (the multiplier $L_r$ need not be a
monomial); this is not a violation, since every active parameter there has
index at least $h$.  Once the shifted block has been decoded, those low
rows are known polynomials and can be subtracted when the remaining
low-degree rows are processed.
\end{proof}

\begin{lemma}[Concatenating triangular blocks]
\label{lem:triangular-block-concatenation}
Let the active parameters be partitioned into consecutive blocks
\[
  I_r=\{b_r,\ldots,b_{r+1}-1\},\qquad
  0=b_0<b_1<\cdots<b_s=d.
\]
Suppose that, for every $r$, the coefficient and pivot formulas on the rows
$j\in I_r$ have been proved with a finite side list $Y_r$, in the form
\[
\begin{aligned}
 [x^j]R^{(1)}&\polyfrom
   (Y_r;\alpha_{j+1},\ldots,\alpha_{b_{r+1}-1}),\\
 [x^j]R^{(2)}&\polyfrom
   (Y_r;\alpha_j,\ldots,\alpha_{b_{r+1}-1}),\\
 [x^j](xR^{(1)}+R^{(2)})
   &=\lambda_j\alpha_j+F_j,
 \qquad F_j\polyfrom
   (Y_r;\alpha_{j+1},\ldots,\alpha_{b_{r+1}-1}),
\end{aligned}
\]
where each $\lambda_j\in\mathbb F^\times$, and suppose moreover that every
entry of $Y_r$ is polynomially recoverable from
\[
       (K;\alpha_{b_{r+1}},\ldots,\alpha_{d-1}).
\]
Then the whole pair is coefficient-triangular for
$(\alpha_0,\ldots,\alpha_{d-1})$ given $K$.

In particular, a quantity which has been decoded in a higher block may be
treated as known while proving a lower block: substituting its polynomial
formula only introduces parameters with strictly larger global indices.
\end{lemma}
\begin{proof}
Substitute the displayed formulas for the entries of $Y_r$.  For a row
$j\in I_r$, all parameters occurring in the substituted side list have
index at least $b_{r+1}>j$, while the explicitly displayed parameters have
index at least $j$ (and at least $j+1$ in the first component and in the
remainder term $F_j$).  The three resulting formulas are therefore exactly
the three formulas in \Cref{def:coefficient-triangular}.  Processing the
blocks in the order $I_{s-1},\ldots,I_0$ gives the descending decoder.
\end{proof}

\paragraph{How the tables compose.}
The closure lemmas below prove their own cutoffs by descending induction.
Within the $T$ recursion, \Cref{lem:triangular-shift} transports a block
through overlapping monic factors, and
\Cref{lem:triangular-block-concatenation} joins consecutive blocks.  After an
intermediate polynomial has been recovered, all later uses of it are ordinary
substitution as in \Cref{lem:extractable-via-derivable}; no separate causal
composition principle is used.

\begin{figure}[H]
  \centering
  \resizebox{\textwidth}{!}{
\begin{tikzpicture}[x=1cm,y=1cm,>=Latex]
  \node[fp panel title] at (0,4.1) {(a) Shift through a monic factor};
  \node[fp active, minimum width=42mm] (old) at (2.15,2.55)
    {old pivots $\eta_{e-1},\ldots,\eta_0$};
  \node[fp known, minimum width=24mm] (factor) at (2.15,3.42)
    {$L=x^h+\cdots$};
  \node[fp label, right=1mm of factor] {leading coefficient $1$};
  \node[fp active, minimum width=42mm] (shifted) at (1.55,1.05)
    {same pivots $\alpha_{h+e-1},\ldots,\alpha_h$};
  \node[fp known, minimum width=16mm, right=0pt of shifted]
    {rows $<h$};
  \draw[fp decode] (old.south) -- node[fp label, right]
    {$\times L$: shift by $h$} (shifted.north);
  \node[fp label, anchor=west, text=fpBlue!80!black] at (0,0.22)
    {monicity preserves every pivot slope $\lambda_i$};

  \node[fp panel title] at (7.15,4.1) {(b) Concatenate descending blocks};
  \draw[fp decode] (7.15,3.42) -- node[fp label, above]
    {decoding order} (13.55,3.42);
  \node[fp active, minimum width=22mm, anchor=west] (tail) at (7.15,2.55)
    {new tail};
  \node[fp second, minimum width=27mm, anchor=west] (inner) at (9.35,2.55)
    {shifted recursion};
  \node[fp recursive, minimum width=18mm, anchor=west] (low) at (12.05,2.55)
    {low block};
  \draw[fp dependence] (tail.south) to[bend right=12]
    node[fp label, below] {decoded side data} (inner.south);
  \draw[fp dependence] (inner.south) to[bend right=12] (low.south);
  \node[fp label] at (8.25,1.18)
    {$I_{s-1}$ decoded first};
  \node[fp label] at (10.7,1.18)
    {then $I_{s-2},\ldots$};
  \node[fp label] at (12.95,1.18)
    {finally $I_0$};
  \node[fp label, anchor=west, text=fpBlue!80!black] at (7.15,0.22)
    {higher-block formulas may be substituted into every lower block};
\end{tikzpicture}}
  \caption{The two structural moves behind the stage tables (styles as
  keyed in \Cref{fig:decoder-language}).
  Multiplication by a monic degree-$h$ factor shifts an inner triangular
  block by $h$ rows without changing its pivot slopes; lower rows may contain
  that block but are processed only after it is known.  Consecutive blocks
  are then decoded from high degree to low degree, with each completed block
  becoming legitimate side information for the next.  This is the preceding
  shifting and concatenation calculus in pictures.}
  \label{fig:triangular-blocks}
\end{figure}
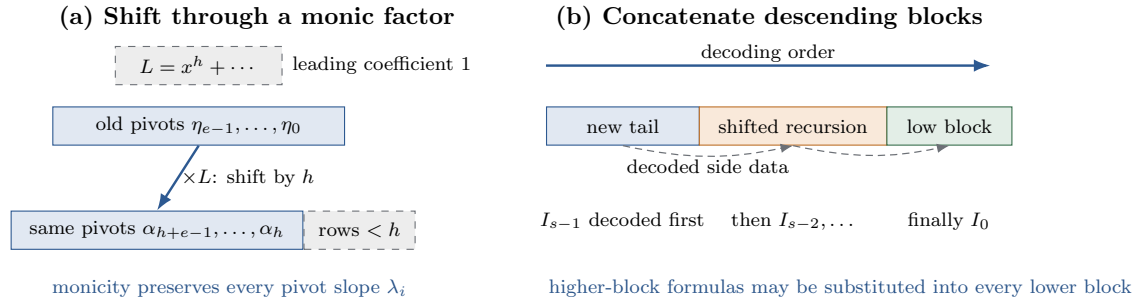

\paragraph{Stage-table convention.}
Every stage table is read from its highest row downward.  A line with row set
$I$, active block $(\alpha_j)_{j\in I}$, and slope $\lambda$ abbreviates the
three assertions, for each $j\in I$,
\[
\begin{aligned}
 [x^j]R^{(1)}&\polyfrom(K;\alpha_{>j}),&
 [x^j]R^{(2)}&\polyfrom(K;\alpha_{\ge j}),\\
 [x^j](xR^{(1)}+R^{(2)})
   &=\lambda\alpha_j+F_j(K;\alpha_{>j}).
\end{aligned}
\]
The source column identifies the term producing the pivot.  The text beside
the table verifies that all other occurrences lie in lower rows or involve
only higher parameters.  Once the listed row sets partition the parameter
range, \Cref{lem:triangular-block-concatenation} turns the table into the
coefficient-triangular invariant.  Thus the tables are compact decoder
proofs, not coefficient-pattern heuristics; \Cref{fig:triangular-blocks}
draws the two moves.

\subsection{Causal closure and top-window calculus}

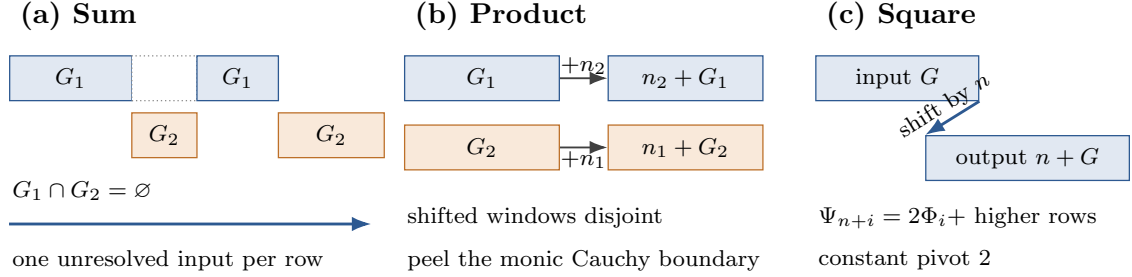
\begin{figure}[H]
  \centering
  \resizebox{\textwidth}{!}{
\begin{tikzpicture}[x=1cm,y=1cm,>=Latex]
  \node[fp panel title] at (0,3.2) {(a) Sum};
  \node[fp active, minimum width=15mm, anchor=west] at (0,2.43) {$G_1$};
  \node[fp unread, minimum width=8mm, anchor=west] at (1.5,2.43) {};
  \node[fp active, minimum width=10mm, anchor=west] at (2.3,2.43) {$G_1$};
  \node[fp second, minimum width=8mm, anchor=west] at (1.5,1.73) {$G_2$};
  \node[fp second, minimum width=13mm, anchor=west] at (3.3,1.73) {$G_2$};
  \node[fp label, anchor=west] at (0,1.05) {$G_1\cap G_2=\varnothing$};
  \draw[fp decode] (0,0.64) -- (4.45,0.64);
  \node[fp label, anchor=west] at (0,0.2) {one unresolved input per row};

  \node[fp panel title] at (4.85,3.2) {(b) Product};
  \node[fp active, minimum width=19mm, anchor=west] (g1) at (4.85,2.43) {$G_1$};
  \node[fp second, minimum width=19mm, anchor=west] (g2) at (4.85,1.58) {$G_2$};
  \node[fp active, minimum width=19mm, anchor=west] (sg1) at (7.35,2.43) {$n_2+G_1$};
  \node[fp second, minimum width=19mm, anchor=west] (sg2) at (7.35,1.58) {$n_1+G_2$};
  \draw[fp flow] (g1.east) -- node[fp label, above] {$+n_2$} (sg1.west);
  \draw[fp flow] (g2.east) -- node[fp label, below] {$+n_1$} (sg2.west);
  \node[fp label, anchor=west] at (4.85,0.72) {shifted windows disjoint};
  \node[fp label, anchor=west] at (4.85,0.2) {peel the monic Cauchy boundary};

  \node[fp panel title] at (9.9,3.2) {(c) Square};
  \node[fp active, minimum width=20mm, anchor=west] (gin) at (9.9,2.43) {input $G$};
  \node[fp active, minimum width=25mm, anchor=west] (gout) at (11.25,1.45) {output $n+G$};
  \draw[fp decode] (gin.south east) -- node[fp label, above, sloped]
    {shift by $n$} (gout.north west);
  \node[fp label, anchor=west] at (9.9,0.72) {$\Psi_{n+i}=2\Phi_i+$ higher rows};
  \node[fp label, anchor=west] at (9.9,0.2) {constant pivot $2$};
\end{tikzpicture}}
  \caption{Window geometry of the three causal closure operations.  A sum
  uses disjoint observed rows; a product translates each input window by the
  degree of the other monic factor; and a square translates the window by
  the common degree and exposes the old row with slope~$2$.  In every case
  the output descent encounters at most one unresolved input row.}
  \label{fig:causal-closures}
\end{figure}

The three closure operations used later have the same shape.  Their window
conditions ensure that each observed row belongs to at most one unresolved
input block; monicity then supplies a constant pivot.
\[
\begin{array}{c|c|c}
  \text{operation}&\text{output window}&\text{descending step}\\ \hline
  \text{sum}&G_1\cup G_2&\text{subtract the other input row}\\
  \text{product}&(n_2+G_1)\cup(n_1+G_2)
    &\text{peel the monic Cauchy boundary}\\
  \text{square}&n+G&\Psi_{n+i}=2\Phi_i+\text{higher terms}
\end{array}
\]
The proofs below establish the exact first- and second-component cutoffs; the
table is only a map of those proofs, and \Cref{fig:causal-closures} pictures
the closure moves.

\begin{lemma}[Additivity of Compatible Pairs]
    Let $(P^{(1)}_1, P^{(2)}_1), (P^{(1)}_2, P^{(2)}_2) \in \mathbb{F}[x]^2$ be compatible pairs on $G_1 \subseteq \idx{n_1}$ and $G_2 \subseteq \idx{n_2}$ given $(B_k)_{k \in [t]}$, respectively, with $\deg P^{(1)}_1 = n_1$ and $\deg P^{(1)}_2 = n_2$.
    Assume $G_1 \cap G_2 = \emptyset$. If $n_1 \ne n_2$, define
    \[
        (P^{(1)},P^{(2)}) \coloneqq (P^{(1)}_1 + P^{(1)}_2,\; P^{(2)}_1 + P^{(2)}_2).
    \]
    If instead $n_1 = n_2$, define
    \[
        (P^{(1)},P^{(2)}) \coloneqq \bigl(P^{(1)}_1 + P^{(1)}_2 - x^{n_1},\; P^{(2)}_1 + P^{(2)}_2 - x^{n_1}\bigr).
    \]
    Then $(P^{(1)}, P^{(2)})$ is a compatible pair on $G_1 \cup G_2$ given $(B_k)_{k \in [t]}$, and $P^{(1)}_1, P^{(2)}_1, P^{(1)}_2, P^{(2)}_2$ are all derivable from $(P^{(1)}, P^{(2)}, (B_k)_{k \in [t]})$.
\end{lemma}
\begin{proof}
    For $i\in\{1,2\}$ define $\Phi_i(x)\coloneqq xP^{(1)}_i(x)+P^{(2)}_i(x)$, and adopt the convention $\coeff{P^{(1)}_i}{-1}=0$.
    Define
    \[
        \Delta(x)\coloneqq
        \begin{cases}
            0 & \text{if } n_1\ne n_2,\\
            x^{n_1+1}+x^{n_1} & \text{if } n_1=n_2.
        \end{cases}
    \]
    Then the combined polynomial for $(P^{(1)},P^{(2)})$ satisfies
    \[
        xP^{(1)}(x)+P^{(2)}(x)=\Phi_1(x)+\Phi_2(x)-\Delta(x).
    \]

    Set $G\coloneqq G_1\cup G_2$. From $\bigl(\coeff{xP^{(1)}+P^{(2)}}{d}\bigr)_{d\in G}$ we can recover $\bigl(\coeff{\Phi_1+\Phi_2}{d}\bigr)_{d\in G}$ by adding the known coefficients of $\Delta$.
    We now show that from the coefficients $\bigl(\coeff{\Phi_1+\Phi_2}{d}\bigr)_{d\in G}$ (and the coefficients of $(B_k)_{k\in[t]}$) we can recover $\bigl(\coeff{\Phi_1}{d}\bigr)_{d\in G_1}$ and $\bigl(\coeff{\Phi_2}{d}\bigr)_{d\in G_2}$.
    We proceed by descending induction on $d$, starting above
    $\max(n_1,n_2)+1$.  Initialize
    \[
      \coeff{\Phi_i}{d}=0\quad(d>n_i+1),\qquad
      \coeff{\Phi_i}{n_i+1}=1.
    \]
    At a row $d\le n_i$ outside $G_i$, compatibility of the $i$th input pair
    expresses $\coeff{P^{(1)}_i}{d-1}$ in terms of its higher
    $\Phi_i$-rows (with $\coeff{P^{(1)}_i}{-1}=0$).  It does the same for
    $\coeff{P^{(2)}_i}{d}$ when $d<n_i$, while for $d=n_i$ that coefficient
    is the known leading~$1$.  Hence it determines
    \[
      \coeff{\Phi_i}{d}
       =\coeff{P^{(1)}_i}{d-1}+\coeff{P^{(2)}_i}{d}.
    \]
    The leading coefficient at $d=n_i+1$ and the zero rows above it are
    supplied by monicity, so no unobserved row is used.

    If $d\in G_1$ (so $d\notin G_2$), then $d\in G$ and we can compute
    \[
        \coeff{\Phi_1}{d}=\coeff{\Phi_1+\Phi_2}{d}-\coeff{\Phi_2}{d}.
    \]
    The case $d\in G_2$ is symmetric.
    Finally, if $d\notin G_1\cup G_2$ then both $d\notin G_1$ and $d\notin G_2$, so compatibility of each pair determines $\coeff{\Phi_1}{d}$ and $\coeff{\Phi_2}{d}$ from higher coefficients and $(B_k)_{k\in[t]}$ without using $\coeff{\Phi_1+\Phi_2}{d}$.

    This completes the induction and thus recovers the required coefficients of $\Phi_1$ on $G_1$ and $\Phi_2$ on $G_2$.
    Applying compatibility of each original pair, we can recover all coefficients of $P^{(1)}_1,P^{(2)}_1$ from $(\coeff{\Phi_1}{d})_{d\in G_1}$ and $(B_k)_{k\in[t]}$, and similarly for $P^{(1)}_2,P^{(2)}_2$.
    Therefore $P^{(1)}_1,P^{(2)}_1,P^{(1)}_2,P^{(2)}_2$ are derivable from $(P^{(1)},P^{(2)},(B_k)_{k\in[t]})$.
    We record the information cutoff explicitly.  The induction just given has
    the following invariant: for every $r$, each input coefficient
    $\coeff{\Phi_i}{r}$ that is used is a polynomial in the auxiliary data and
    in output coefficients $\coeff{\Phi}{s}$ with $s\in G$ and $s\ge r$.
    Indeed, at a row $r\in G_i$ we read the output row $r$, while at a row
    outside $G_i$ the compatibility formula for the $i$th input uses only
    rows strictly above $r$; the latter are handled recursively.  The known
    correction $\Delta$ does not change this invariant.

    To recover $\coeff{P^{(1)}}{j}$, apply the two input compatibility formulas
    only at rows $r\ge j+1$.  The invariant then uses output rows $s\ge r\ge
    j+1$.  For $\coeff{P^{(2)}}{j}$ the corresponding rows satisfy $r\ge j$,
    and hence use only output rows $s\ge j$.  Thus the output pair is
    compatible on $G$ with exactly the two cutoffs in
    \Cref{def:compatible-pair}, not merely recoverable from the whole window.
\end{proof}

\begin{lemma}[Multiplicativity of Compatible Pairs]
    Let $(P^{(1)}_1, P^{(2)}_1), (P^{(1)}_2, P^{(2)}_2) \in \mathbb{F}[x]^2$ be compatible pairs on $G_1 \subseteq \idx{n_1}$ and $G_2 \subseteq \idx{n_2}$ given $(B_k)_{k \in [t]}$, respectively, with $\deg P^{(1)}_1 = n_1$ and $\deg P^{(1)}_2 = n_2$.
    If $(n_2 + G_1) \cap (n_1 + G_2) = \emptyset$ then $(P^{(1)}_1 \cdot P^{(1)}_2, P^{(2)}_1 \cdot P^{(2)}_2)$ is a compatible pair on $(n_2 + G_1) \cup (n_1 + G_2)$ given $(B_k)_{k \in [t]}$, and $P^{(1)}_1, P^{(2)}_1, P^{(1)}_2, P^{(2)}_2$ are all derivable from $(P^{(1)}_1 \cdot P^{(1)}_2, P^{(2)}_1 \cdot P^{(2)}_2, (B_k)_{k \in [t]})$.
\end{lemma}
\begin{proof}
    Define $\Phi_i(x)\coloneqq xP^{(1)}_i(x)+P^{(2)}_i(x)$ for $i\in\{1,2\}$ and
    \[
        \Phi(x)\coloneqq x\bigl(P^{(1)}_1(x)P^{(1)}_2(x)\bigr)+P^{(2)}_1(x)P^{(2)}_2(x).
    \]
    Write $N\coloneqq n_1+n_2$, and adopt the convention $\coeff{P}{r}=0$ for $r<0$.
    Without loss of generality assume $n_1\ge n_2$ (otherwise swap the roles of the pairs).

    Put $W\coloneqq(n_2+G_1)\cup(n_1+G_2)$.  The high-degree part of the
    induction is a simultaneous descent on $j=0,\ldots,n_2$.  At stage $j$,
    assume that $\coeff{\Phi_1}{n_1-r}$ and $\coeff{\Phi_2}{n_2-r}$ have
    already been recovered for $r<j$.  Compatibility of the input pairs then
    gives all coefficients of $P^{(1)}_1$ in degrees at least $n_1-j$ and
    of $P^{(2)}_1$ in degrees strictly greater than $n_1-j$, and the
    analogous coefficients of the second pair.

    For $0\le j\le n_2$, the Cauchy product and monicity give
    \begin{align}
      \coeff{\Phi}{N-j}
        &=\coeff{\Phi_1}{n_1-j}+\coeff{\Phi_2}{n_2-j}
          -\mathbf 1_{\{j=0\}} \notag\\
        &\quad+
        \sum_{\substack{i_1\in\rng{n_1},\,i_2\in\rng{n_2}\\
                         i_1+i_2=N-j-1}}
          \coeff{P^{(1)}_1}{i_1}\coeff{P^{(1)}_2}{i_2}\notag\\
        &\quad+
        \sum_{\substack{i_1\in\rng{n_1},\,i_2\in\rng{n_2}\\
                         i_1+i_2=N-j}}
          \coeff{P^{(2)}_1}{i_1}\coeff{P^{(2)}_2}{i_2}.
      \label{eq:multiplicativity-high-row}
    \end{align}
    Every coefficient in the two sums has already been determined at this
    stage: in the first sum its degrees are at least $n_i-j$, and in the
    second they are at least $n_i-j+1$.  If $N-j\notin W$, then neither
    $n_1-j\in G_1$ nor $n_2-j\in G_2$, so both current $\Phi_i$-coefficients
    are obtained from their own higher rows by compatibility.  If
    $N-j\in W$, at least one of these two indices is in its input window,
    and the shifted-window disjointness says that they are not both in their
    windows.  Subtracting the two sums in \eqref{eq:multiplicativity-high-row}
    (and adding $\mathbf 1_{\{j=0\}}$) therefore gives the current windowed
    coefficient after the other current coefficient has been obtained by
    compatibility.  This completes the descent through
    $\coeff{\Phi_2}{0}$ and $\coeff{\Phi_1}{n_1-n_2}$.

    If $n_1>n_2$, there is a remaining low tail of the first pair.  Suppose
    that all rows of $\Phi_1$ strictly above degree~$a$, and all rows of
    $\Phi_2$, have been recovered, and descend on
    $a=n_1-n_2-1,\ldots,0$.  The second pair is now completely known.  By
    monicity and the Cauchy product,
    \begin{align}
      \coeff{\Phi}{n_2+a}
        &=\coeff{\Phi_1}{a}
          +\sum_{\substack{r+s=n_2+a-1\\s<n_2}}
             \coeff{P^{(1)}_1}{r}\coeff{P^{(1)}_2}{s}
          +\sum_{\substack{r+s=n_2+a\\s<n_2}}
             \coeff{P^{(2)}_1}{r}\coeff{P^{(2)}_2}{s}.
      \label{eq:multiplicativity-low-row}
    \end{align}
    Every coefficient of $P_1^{(1)}$ or $P_1^{(2)}$ in the sums has degree
    strictly larger than the current degree $a-1$ or $a$, respectively, and
    is consequently already known.  If $a\notin G_1$, compatibility supplies
    $\coeff{\Phi_1}{a}$ from higher rows; if $a\in G_1$, the coefficient on
    the left is available at the window row $n_2+a\in W$, and
    \eqref{eq:multiplicativity-low-row} solves for $\coeff{\Phi_1}{a}$.
    (For these low rows $n_2+a<n_1$, so they cannot simultaneously belong to
    $n_1+G_2$.)  Thus all coefficients of $\Phi_1$ on $G_1$ and of
    $\Phi_2$ on $G_2$ are recovered.

    Applying compatibility of the original pairs now recovers all four
    component polynomials, proving derivability.
    We spell out the cutoff, since it is stronger than global recoverability.
    The induction invariant says that an input row
    $\coeff{\Phi_1}{n_1-r}$ (respectively
    $\coeff{\Phi_2}{n_2-r}$) is a polynomial in output rows of degree at least
    $N-r$.  Equivalently, an input row of degree $a$ in the first factor is
    recovered using output rows of degree at least $n_2+a$, and an input row
    of degree $b$ in the second factor using rows of degree at least $n_1+b$.
    (Rows outside the displayed windows are supplied by monicity, zero
    coefficients, or the input compatibility recurrence, and obey the same
    lower bound.)

    A coefficient of degree $q$ in $P^{(1)}_1P^{(1)}_2$ is a sum of terms with
    factor degrees $a+b=q$.  The first factor coefficient is recovered from
    rows at least $n_2+a+1=q+1+(n_2-b)\ge q+1$, and the second from rows at
    least $n_1+b+1=q+1+(n_1-a)\ge q+1$.  Thus the first output component uses
    only output combined coefficients in degrees at least $q+1$.  For a
    coefficient of degree $q$ in $P^{(2)}_1P^{(2)}_2$, the same calculation has
    no extra $+1$: the two thresholds are $n_2+a\ge q$ and $n_1+b\ge q$.
    Consequently the product pair is compatible on
    $(n_2+G_1)\cup(n_1+G_2)$ with the required first- and second-component
    cutoffs.
\end{proof}

\begin{figure}[H]
  \centering
  \resizebox{\textwidth}{!}{
\begin{tikzpicture}[x=1cm,y=1cm,>=Latex]
  \node[fp panel title] at (0,4.65) {(a) Monic division};
  \node[fp active, minimum width=32mm, anchor=west] (f) at (0,3.95) {top rows of $F=QD+R$};
  \node[fp known, minimum width=19mm, anchor=west] (d) at (3.45,3.95) {$D=x^d+\cdots$};
  \node[fp pivot, minimum width=29mm, anchor=west] (q) at (1.25,3.08) {$q_t,q_{t-1},\ldots$};
  \draw[fp decode] (f.south) -- node[fp label, left] {leading $1$} (q.north);
  \node[fp label, anchor=west] at (0,2.52) {each row exposes the next coefficient of $Q$};

  \node[fp panel title] at (7.05,4.65) {(b) Monic root};
  \node[fp active, minimum width=34mm, anchor=west] (pow) at (7.05,3.95) {top $n$ rows of $P^m$};
  \node[fp pivot, minimum width=28mm, anchor=west] (proot) at (8.35,3.08) {$p_{n-1},\ldots,p_0$};
  \draw[fp decode] (pow.south) -- node[fp label, left] {pivot $m$} (proot.north);
  \node[fp label, anchor=west] at (7.05,2.52) {$[x^{nm-s}]P^m=mp_{n-s}+$ earlier terms};

  \node[fp panel title] at (0,1.78) {(c) One boundary error};
  \node[fp active, minimum width=31mm, anchor=west] at (0,1.08) {clean top window};
  \node[fp seam, minimum width=9mm, anchor=west] at (3.1,1.08) {$E_b$};
  \node[fp unread, minimum width=17mm, anchor=west] at (4.0,1.08) {lower error};
  \draw[fp decode] (0,0.45) -- node[fp label, above]
    {subtract the named seam, then use (b)} (5.7,0.45);

  \node[fp panel title] at (7.05,1.78) {(d) Square gadget};
  \node[fp active, minimum width=27mm, anchor=west] (sq) at (7.05,1.08) {recover monic $S$};
  \node[fp seam, minimum width=11mm, anchor=west] (delta) at (9.75,1.08) {$\delta$};
  \node[fp recursive, minimum width=21mm, anchor=west] (low) at (10.85,1.08) {low remainder};
  \draw[fp decode] (sq.south) -- ++(0,-.55);
  \draw[fp decode] (delta.south) -- ++(0,-.55);
  \draw[fp decode] (low.south) -- ++(0,-.55);
  \node[fp label, anchor=west] at (7.05,0.14)
    {$xS^2+(S+\delta)^2+E$: peel $S$, then $\delta$, then $E$};
\end{tikzpicture}}
  \caption{The top-window calculus used in the remaining constructions.
  Monicity makes division triangular; taking a monic $m$th root has constant
  pivot~$m$; a lower-degree perturbation changes only the explicitly marked
  boundary row; and the square gadget is decoded as three consecutive
  shells.  In particular, panels (a)--(c) compose into the relative shell
  $Y=MS^2+E$: divide the top window by the known monic $M$, subtract the one
  named seam of $E$, and take the monic square root.  The proofs below are
  the coefficient recurrences represented by these four glyphs.}
  \label{fig:top-window-calculus}
\end{figure}
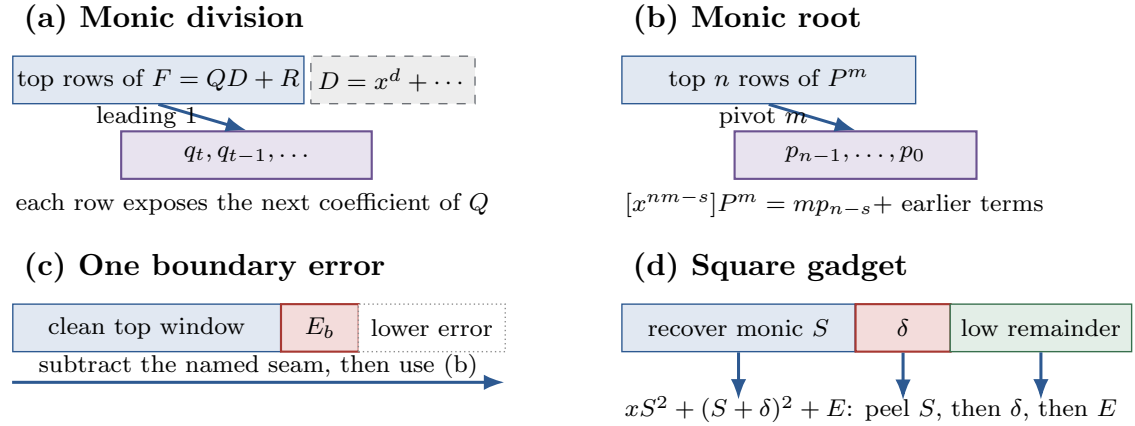

\Cref{fig:top-window-calculus} summarizes the operations of this calculus.

\begin{lemma}[Division by a monic polynomial]\label{lem:monic-division}
    Let $D,F\in\mathbb{F}[x]$ with $D$ monic of degree $d\ge 1$ and $\deg F=n\ge d$.
    Let $Q,R\in\mathbb{F}[x]$ be the unique polynomials with
    \[
      F=QD+R,\qquad \deg R<d.
    \]
    Then $Q$ and $R$ are derivable from $(F,D)$.
\end{lemma}
\begin{proof}
    Write $q\coloneqq n-d$ so $\deg Q=q$.
    Since $D$ is monic, the coefficient of $x^{n}$ in $F$ equals the coefficient of $x^{q}$ in $Q$.
    More generally, a descending induction on $t=q,q-1,\ldots,0$ recovers $\coeff{Q}{t}$:
    after the coefficients $\coeff{Q}{q},\ldots,\coeff{Q}{t+1}$ are known, the coefficient of $x^{d+t}$ in $F$
    equals $\coeff{Q}{t}$ plus a polynomial in the already-known coefficients of $Q$ and the coefficients of $D$
    (this is the Cauchy product for $QD$, isolating the term where $D$ contributes degree $d$).
    Thus $Q$ is derivable from $(F,D)$, and then $R=F-QD$ is derivable as well.
\end{proof}

\begin{lemma}[Recovering a monic polynomial from a power]\label{lem:monic-from-power}
    Let $P\in\mathbb{F}[x]$ be monic of degree $n$, and let $m>1$ be an integer that is invertible in $\mathbb{F}$.
    Then $P$ is derivable from $P^m$.
\end{lemma}
\begin{proof}
    For $1\le s\le n$, the coefficient $\coeff{P^m}{nm-s}$ equals
    \[
        m\,\coeff{P}{n-s}
    \]
    plus a polynomial in $\coeff{P}{n-1},\ldots,\coeff{P}{n-s+1}$ (this is the Cauchy product, isolating the terms where exactly one factor contributes degree $n-s$).
    Since $m$ is invertible, a descending induction on $s$ recovers all coefficients of $P$ from $P^m$, proving derivability.
\end{proof}

\begin{lemma}[Decoding the square gadget $xS^2+(S+\delta)^2$]\label{lem:square-gadget}
    Assume $\mathrm{char}(\mathbb{F})\ne 2$.
    Let $S$ be monic of degree $d\ge 1$ and let $\delta$ be an unknown
    $x$-independent scalar.
    Define
    \[
      G(x)\coloneqq x\,S(x)^2 + \bigl(S(x)+\delta\bigr)^2.
    \]
    Then $S$ and $\delta$ are derivable from $G$.
\end{lemma}
\begin{proof}
    Write $A(x)\coloneqq S(x)^2$.
    Since $\deg S=d$, we have $\deg A=2d$, and for every $i>d$ the scalar shift does not affect degree $i$, i.e.
    \[
      [x^i]\bigl(S+\delta\bigr)^2=[x^i]S^2=[x^i]A.
    \]
    Therefore, for $i=d+1,\ldots,2d+1$,
    \[
      [x^i]G = [x^i]\bigl(xA\bigr) + [x^i]A = [x^{i-1}]A + [x^i]A.
    \]
    Since $A$ is monic, $[x^{2d}]A=1$, so descending induction on $i=2d+1,2d,\ldots,d+1$ recovers $[x^j]A$ for all $j=d,\ldots,2d$ from the coefficients of $G$.
    By \Cref{lem:monic-from-power} (with $m=2$), this recovers $S$ from $A=S^2$.

    Finally, compute $(S+\delta)^2 = G - xS^2$.
    Since $[x^d](S+\delta)^2 = [x^d]S^2 + 2\delta$ and $\mathrm{char}(\mathbb{F})\ne 2$, we can solve
    \[
      \delta=\frac{[x^d]\bigl((S+\delta)^2-S^2\bigr)}{2},
    \]
    so $\delta$ is derivable once $S$ is.
\end{proof}

\begin{lemma}[Square gadget with a boundary error]
\label{lem:square-gadget-boundary}
Assume $\operatorname{char}(\mathbb F)\ne2$.  Let $S$ be monic of degree
$d\ge1$, let $\delta$ be an unknown $x$-independent scalar, and let $E$ have
degree at most $d$.  If
\[
 W=xS^2+(S+\delta)^2+E,
\]
then $S$ and $\delta$ are derivable from
$\bigl(W,[x^d]E\bigr)$.
\end{lemma}
\begin{proof}
The error does not meet degrees $d+1,\ldots,2d+1$, so the descending
recurrence in the proof of \Cref{lem:square-gadget} recovers the top half of
$S^2$ and hence $S$.  At the remaining boundary degree,
\[
 [x^d]\bigl(W-xS^2-S^2\bigr)=2\delta+[x^d]E.
\]
Subtracting the supplied boundary coefficient and dividing by~$2$ recovers
$\delta$.
\end{proof}

\begin{lemma}[Causal Cauchy transport through a monic factor]
\label{lem:monic-cauchy-transport}
Let $A$ be monic of degree $a$ and let $\deg B\le b$.  With coefficients
outside the natural range interpreted as zero, for every $0\le r\le b$,
\begin{equation}
  [x^{a+r}](AB)=[x^r]B+
    \sum_{s=r+1}^{b}[x^{a+r-s}]A\,[x^s]B.
  \tag{monic-Cauchy}\label{eq:monic-Cauchy}
\end{equation}
Thus row $a+r$ introduces $[x^r]B$ with unit slope and otherwise uses only
higher coefficients of $B$.  In particular, if $b\ge1$, then
\begin{equation}
  [x^{a+b}](AB)=[x^b]B,
  \qquad
  [x^{a+b-1}](AB)=[x^{b-1}]B+[x^{a-1}]A\,[x^b]B.
  \tag{monic-top-two}\label{eq:monic-top-two}
\end{equation}
\end{lemma}
\begin{proof}
In the Cauchy product for degree $a+r$, terms with index below $r$ in $B$
would require a coefficient of $A$ above degree $a$, and terms above $b$
vanish.  The term with index $r$ has coefficient $[x^a]A=1$, which gives
\eqref{eq:monic-Cauchy}.  Taking $r=b$ and $r=b-1$ gives
\eqref{eq:monic-top-two}.
\end{proof}

\begin{lemma}[Top-window division by a monic factor]\label{lem:peel-monic-factor}
    Let $A,B,E\in\mathbb{F}[x]$ with $A$ monic of degree $a$ and $\deg B\le b$.
    Fix $t$ with $1\le t\le b+1$ and define $P\coloneqq AB+E$.
    If $\deg E\le a+b-t$, then the coefficients $\coeff{B}{b},\coeff{B}{b-1},\ldots,\coeff{B}{b-(t-1)}$ are derivable from the coefficients of $A$ and the coefficients $\coeff{P}{a+b},\coeff{P}{a+b-1},\ldots,\coeff{P}{a+b-(t-1)}$.
\end{lemma}
\begin{proof}
    Adopt the convention $\coeff{Q}{r}=0$ for $r<0$.
    Since $\deg E\le a+b-t$, we have $\coeff{P}{a+b-s}=\coeff{AB}{a+b-s}$ for $s=0,1,\ldots,t-1$.
    For each such $s$, apply \Cref{lem:monic-cauchy-transport} with
    $r=b-s$.  Its causal formula gives
    \[
        \coeff{AB}{a+b-s}=\coeff{B}{b-s}+G_s\!\bigl(\coeff{B}{b},\ldots,\coeff{B}{b-s+1},\coeff{A}{a-1},\ldots,\coeff{A}{a-s}\bigr)
    \]
    for some polynomial $G_s$.
    A descending induction on $s$ therefore recovers $\coeff{B}{b-s}$ from $\coeff{P}{a+b-s}$ and previously recovered higher coefficients of $B$, proving the claim.
\end{proof}

\begin{lemma}[Relative square shell]\label{lem:relative-square-shell}
    Assume $\operatorname{char}(\mathbb F)\ne2$.  Let $M,S,E,Y\in\mathbb
    F[x]$, where $M$ and $S$ are monic of degrees $m$ and $d\ge1$, and
    \[
       Y=M S^2+E,\qquad \deg E\le m+d.
    \]
    Then $S$ is derivable from
    \(
      \bigl(Y,M,[x^{m+d}]E\bigr).
    \)
    More precisely, for every $j<d$, the coefficient $[x^j]S$ uses only
    coefficients $[x^r]Y$ with $r\ge m+d+j$ (besides $M$ and the supplied
    seam value).
\end{lemma}
\begin{proof}
    Descend on $i=2d-1,2d-2,\ldots,d$.  Monicity of $M$ gives
    \[
      [x^{m+i}]Y=[x^i]S^2+
        F_i\bigl(M;[x^{i+1}]S^2,\ldots,[x^{2d}]S^2\bigr)
        +[x^{m+i}]E
    \]
    for a polynomial $F_i$ coming from the remaining Cauchy terms.  The error
    term vanishes when $i>d$; at $i=d$ it is the one supplied boundary
    coefficient.  Starting from the known monic coefficient
    $[x^{2d}]S^2=1$, this recurrence recovers
    $[x^{2d-1}]S^2,\ldots,[x^d]S^2$.  These are exactly the top coefficients
    used by the descending monic-square recursion in
    \Cref{lem:monic-from-power}, so division by the fixed pivot~$2$ recovers
    $S$.  In this combined descent, $[x^j]S$ starts with
    $[x^{d+j}]S^2$, which starts with row $m+d+j$ of $Y$; every other term
    comes from a strictly higher row.  This proves the stated cutoff as well.
\end{proof}

\begin{lemma}[Scalar shift from a square boundary coefficient]\label{lem:scalar-shift-square}
    Assume $\mathrm{char}(\mathbb{F})\ne 2$.
    Let $H,M\in\mathbb{F}[x]$ be monic with $\deg H=d\ge 1$ and $\deg M=e$.
    Let $\delta$ be an unknown $x$-independent scalar and let
    $\lambda\in\mathbb{F}^\times$.
    Define $S\coloneqq H+\delta$ and $P\coloneqq \lambda\,S^2M+E$ for some $E\in\mathbb{F}[x]$ with $\deg E\le d+e-1$.
    Then $\delta$ is extractable from $P$ given $(H,M)$.
\end{lemma}
\begin{proof}
    Since $\deg E\le d+e-1$, we have $\coeff{P}{d+e}=\lambda\,\coeff{S^2M}{d+e}$.
    Expanding $S^2=H^2+2\delta H+\delta^2$ gives
    \[
      S^2M = H^2M + 2\delta\,HM + \delta^2M.
    \]
    The term $\delta^2M$ has degree $e<d+e$, so it does not contribute to $\coeff{S^2M}{d+e}$.
    Moreover, $HM$ is monic of degree $d+e$, so $\coeff{HM}{d+e}=1$ and therefore
    \[
      \coeff{S^2M}{d+e}=\coeff{H^2M}{d+e}+2\delta.
    \]
    Since $\mathrm{char}(\mathbb{F})\ne 2$ and $\lambda\ne 0$, we can solve
    \[
      \delta=\frac{\coeff{P}{d+e}-\lambda\,\coeff{H^2M}{d+e}}{2\lambda},
    \]
    which is a polynomial expression in the coefficients of $P,H,M$.
\end{proof}

\begin{lemma}[Square closure for compatible pairs]\label{lem:compatible-power}
    Assume $\mathrm{char}(\mathbb{F})\ne 2$.
    Let $n\ge1$.
    Let $(P^{(1)}, P^{(2)}) \in \mathbb{F}[x]^2$ be a compatible pair on $G \subseteq \idx{n}$ given $(B_k)_{k \in [t]}$, with $\deg P^{(1)}=\deg P^{(2)}=n$.
    Then $\bigl((P^{(1)})^2, (P^{(2)})^2\bigr)$ is a compatible pair on $n + G$ given $(B_k)_{k \in [t]}$, and $P^{(1)}$ and $P^{(2)}$ are derivable from $\bigl((P^{(1)})^2, (P^{(2)})^2\bigr)$.
\end{lemma}
\begin{proof}
    By \Cref{lem:monic-from-power} with $m=2$, each of $P^{(1)}$ and $P^{(2)}$ is derivable from its square (using $\mathrm{char}(\mathbb{F})\ne 2$ to divide by $2$).

    It remains to prove compatibility of $\bigl((P^{(1)})^2,(P^{(2)})^2\bigr)$ on $n+G$ given $(B_k)_{k\in[t]}$.
    Define the combined polynomials
    \[
        \Phi(x)\coloneqq xP^{(1)}(x)+P^{(2)}(x)
        \qquad\text{and}\qquad
        \Psi(x)\coloneqq x\,(P^{(1)}(x))^2 + (P^{(2)}(x))^2.
    \]

    \paragraph{Step 1: Recover $\bigl(\coeff{\Phi}{i}\bigr)_{i\in G}$ from $\bigl(\coeff{\Psi}{n+i}\bigr)_{i\in G}$.}
    Fix $i\in\idx{n}$.
    For $1\le i\le n-1$, expanding coefficients of squares shows that
    \[
        \coeff{\Psi}{n+i}
        = 2\,\coeff{\Phi}{i} + F_i,
    \]
    where $F_i$ is a polynomial in the coefficients $\coeff{P^{(1)}}{r}$ with $r>i-1$ and $\coeff{P^{(2)}}{r}$ with $r>i$.
    (Indeed, in $\coeff{(P^{(1)})^2}{n+i-1}$ the two ``boundary'' terms $(r,s)=(n,i-1)$ and $(i-1,n)$ contribute $2\,\coeff{P^{(1)}}{i-1}$, and all other Cauchy-product summands involve only degrees $>i-1$; similarly $\coeff{(P^{(2)})^2}{n+i}$ has boundary contribution $2\,\coeff{P^{(2)}}{i}$ and all other summands use only degrees $>i$.)

    The endpoint cases are similar: for $i=n$ we have
    \[
        \coeff{\Psi}{2n}
        = \coeff{(P^{(1)})^2}{2n-1}+\coeff{(P^{(2)})^2}{2n}
        = 2\,\coeff{P^{(1)}}{n-1}+1
        = 2\,\coeff{\Phi}{n}-1,
    \]
    and for $i=0$ we have
    \[
        \coeff{\Psi}{n}
        = \coeff{(P^{(1)})^2}{n-1}+\coeff{(P^{(2)})^2}{n}
        = 2\,\coeff{\Phi}{0} + F_0,
    \]
    where, writing $a_r=[x^r]P^{(1)}$ and $b_r=[x^r]P^{(2)}$,
    \[
      F_0=
        \sum_{\substack{r+s=n-1\\0\le r,s<n}}a_ra_s
        +\sum_{\substack{r+s=n\\0\le r,s<n\\r,s\ne0}}b_rb_s .
    \]
    Thus $F_0$ may involve $a_0$ (through the first sum), but it does not
    involve $b_0=\coeff{\Phi}{0}$, which is the coefficient being solved for.
    This is not circular: compatibility recovers $a_0$ from the coefficients
    of $\Phi$ in degrees at least~$1$, all of which have already been
    recovered in the descending induction before the $i=0$ row is used.

    By compatibility of $(P^{(1)},P^{(2)})$ on $G$, these ``higher-degree'' coefficients are determined by $\bigl(\coeff{\Phi}{d}\bigr)_{d\in G,\; d>i}$ and the $(B_k)$-coefficients.
    Thus, descending induction on $i$ recovers $\coeff{\Phi}{i}$ for all $i\in G$ from $\bigl(\coeff{\Psi}{n+i}\bigr)_{i\in G}$ and $(B_k)_{k\in[t]}$ (dividing by $2$ in $\mathbb{F}$ when needed, and using the explicit constant $-1$ term for $i=n$).

    \paragraph{Step 2: Recover $P^{(1)}$ and $P^{(2)}$, then their squares.}
    Applying compatibility of $(P^{(1)},P^{(2)})$ on $G$ now derives $P^{(1)}$ and $P^{(2)}$ from $\bigl(\coeff{\Phi}{i}\bigr)_{i\in G}$ and $(B_k)_{k\in[t]}$.
    Squaring then derives $(P^{(1)})^2$ and $(P^{(2)})^2$.

    This reconstruction respects the cutoffs for the squared pair.  In a
    summand of $[x^j](P^{(1)})^2$, both factor indices are at least
    $j-n$.  Recovering a first-component factor of degree $r$ uses only
    $\Phi$-rows at least $r+1$, so every such summand uses $\Phi$-rows at
    least $j-n+1$ and hence, by Step~1, $\Psi$-rows at least $j+1$.  For
    $[x^j](P^{(2)})^2$, the corresponding bound is $j-n$ and therefore the
    required $\Psi$-rows are at least $j$.  If $j<n$, these lower bounds are
    below the bottom of the shifted window and all rows of $n+G$ are already
    above them, so the same conclusion holds.  Therefore
    $\bigl((P^{(1)})^2,(P^{(2)})^2\bigr)$ is compatible on $n+G$ given
    $(B_k)_{k\in[t]}$.
\end{proof}

\section{Large-Characteristic Construction Details}
\label{appendix:constructions}


This appendix provides costed compatible joint realizations of the
splittable-pair constructions for all odd degrees except~$7$; for $n=7$ we
use a separate decodable septic base.  Algebraic splittability itself is not
exceptional at degree~$7$; the missing object is a cost-optimal joint
realization, as explained in \Cref{rem:canonical-split}.
The results hold over fields of characteristic zero and, for target degree
$n$, over fields of characteristic $p>n$.
The resulting degree-$n$ circuit uses exactly
$\lfloor n/2\rfloor+1$ multiplications and, after sharing affine forms,
at most
\[
   \min\!\left\{2n,\ \frac54n+6\lceil\log_2n\rceil^2+1\right\}
\]
additions or subtractions, with multiplicative height
$2\lceil\log_2 n\rceil+4$ for odd $n$ ($+5$ for even $n$, whose lift
adds one product), using the binary known-powers recursion of
\Cref{sec:peeled-Q}.  Thus the construction has leading arithmetic
cost $n/2$ multiplications and $5n/4$ additions on a critical path of
logarithmic length.  The preceding
decoder-calculus appendix supplies the common recovery language.  This appendix
certifies the construction, the multiplication count, and the logarithmic
height; the separate addition-accounting appendix proves the two stated
addition bounds for the same polynomial family.

\paragraph{Characteristic boundary.}
The recovery and circuit interfaces are characteristic-free.  The formulas
below are not: square peels and scalar pivots require~$2$ and several small
integers to be units.  A characteristic-$2$ treatment can reuse those
interfaces, but needs a separate recursive family rather than exceptional
branches of the $T_{k,2^l}$ recurrence.  Every helper that divides by~$2$
states that hypothesis explicitly.

\paragraph{Proof architecture.}
Although the formulas are recursive, the proof uses one decoder pattern
throughout:
\[
\begin{aligned}
  \text{construction identity}
    &\Longrightarrow \text{degree and support bounds}\\
    &\Longrightarrow \text{descending pivot table}
     \Longrightarrow \text{polynomial decoder}.
\end{aligned}
\]
The decoder calculus of \Cref{appendix:decoder-calculus} records polynomial
recovery and its causal version.  The fill construction supplies the known-powers gadgets
$Q_{2^s-1}$; the depth-balanced peeled variant of \Cref{sec:peeled-Q}
computes the same family at the identical ledger with height $s$.  The central $T_{k,2^l}$ lemma proves one
coefficient-triangular remainder invariant by three stage tables (even,
shared odd base, and ordinary odd).  The remaining lemmas attach small outer
gadgets and close a strong induction on the target degree.  Decoder summaries
after the longer proofs collect their steps; the proofs themselves establish
every pivot and information cutoff.

\paragraph{Construction atlas.}
The appendix can be read as the following small collection of circuit moves.
Each entry says both what the construction computes and how its inverse exposes
the hidden block.  On a first reading, this atlas together with the decoder
figures gives the logical spine; the intervening coefficient calculations
certify the stated arrows.
\begin{multicols}{2}
\small
\raggedcolumns
\noindent\textbf{Compatible closures.}
Sum, multiply, or square two pairs; descend through the translated windows
and peel the unique unresolved row.  These are the composition rules.

\smallskip\noindent\textbf{Fill $A_l$.}
Attach monic heads above and below a pair; decode upper head
$\to$ shifted input window $\to$ lower head.  Iterating this move gives the
Mersenne gadgets $Q_{2^s-1}$.

\smallskip\noindent\textbf{Peeled $Q^{\mathrm{peel}}$.}
Two half-size children under one keyed glue factor; decode by peeling
against the known monic $H$, reading $\gamma$ at the top row, and
subtracting.  Same ledger as the fill route, height $k$.

\smallskip\noindent\textbf{The $T_{k,2^l}$ recursion.}
Halve the exponent and attach a fresh tail; decode tail pivots
$\to$ shifted recursive block $\to$ low $Q$-block.  This is the central
compatible-pair construction.

\smallskip\noindent\textbf{The $Q_{4k+1}$ crown.}
Place a $T_{2k,2}$ pair under an affine crown; read five crown pivots and
then run the recovered $T$ decoder.  This handles degrees $1\pmod4$.

\smallskip\noindent\textbf{The barred gadget $\bar Q_{8k+7}$.}
Place a $T_{k,8}$ pair inside one level-$4$ fill; decode its top scalar/block
pivots $\to$ two seam rows $\to$ inner $T$ block $\to$ low rows.

\smallskip\noindent\textbf{Outer induction.}
Wrap a smaller pair in one or two difference-of-squares shells; peel the
shells, decode the smaller pair and its recorded powers, and only then decode
the auxiliary gadgets.

\smallskip\noindent\textbf{Finite bases.}
The fused circuits at $7,15,27,31$ use the same top-down shell and pivot
rules; they bridge the degrees at which a generic branch would miss the
optimal multiplication count.
\end{multicols}

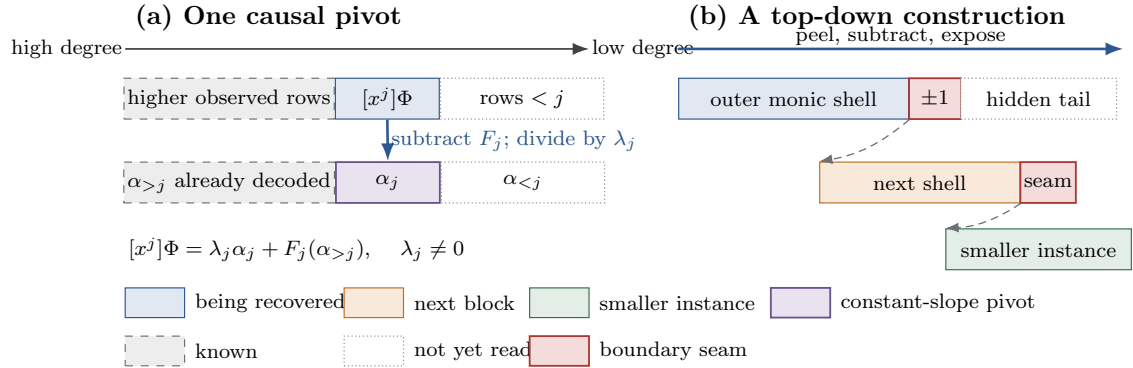
\begin{figure}[H]
  \centering
  \resizebox{\textwidth}{!}{
\begin{tikzpicture}[x=1cm,y=1cm,>=Latex]
  \node[fp panel title] at (0,4.05) {(a) One causal pivot};
  \draw[fp flow] (0,3.62) -- (6.25,3.62);
  \node[fp label, anchor=east] at (0,3.62) {high degree};
  \node[fp label, anchor=west] at (6.25,3.62) {low degree};

  \node[fp known, minimum width=28mm] (seen) at (1.4,2.95)
    {higher observed rows};
  \node[fp active, minimum width=14mm, right=0pt of seen] (row)
    {$[x^j]\Phi$};
  \node[fp unread, minimum width=22mm, right=0pt of row] (unread)
    {rows $<j$};

  \node[fp known, minimum width=28mm] (highpar) at (1.4,1.82)
    {$\alpha_{>j}$ already decoded};
  \node[fp pivot, minimum width=14mm, right=0pt of highpar] (par)
    {$\alpha_j$};
  \node[fp unread, minimum width=22mm, right=0pt of par]
    {$\alpha_{<j}$};
  \draw[fp decode] (row.south) -- node[fp label, right, text=fpBlue!80!black]
    {subtract $F_j$; divide by $\lambda_j$} (par.north);
  \node[fp label, anchor=west] at (0,0.92)
    {$[x^j]\Phi=\lambda_j\alpha_j+F_j(\alpha_{>j})$,
     \quad$\lambda_j\ne0$};

  \node[fp panel title] at (7.45,4.05) {(b) A top-down construction};
  \draw[fp decode] (7.45,3.62) -- node[fp label, above]
    {peel, subtract, expose} (13.45,3.62);

  \node[fp active, minimum width=31mm, anchor=west] (outer) at (7.45,2.95)
    {outer monic shell};
  \node[fp seam, minimum width=7mm, anchor=west] at (10.55,2.95)
    {$\pm1$};
  \node[fp unread, minimum width=21mm, anchor=west] at (11.25,2.95)
    {hidden tail};

  \node[fp second, minimum width=27mm, anchor=west] (inner) at (9.35,1.82)
    {next shell};
  \node[fp seam, minimum width=7mm, anchor=west] at (12.05,1.82)
    {seam};

  \node[fp recursive, minimum width=25mm, anchor=west] (rec) at (11.05,0.92)
    {smaller instance};
  \draw[fp dependence] (outer.south east) to[bend left=15] (inner.north west);
  \draw[fp dependence] (inner.south east) to[bend left=15] (rec.north west);

  \node[fp active, minimum width=8mm, minimum height=4mm] at (0.4,0.2) {};
  \node[fp label, anchor=west] at (0.9,0.2) {being recovered};
  \node[fp second, minimum width=8mm, minimum height=4mm] at (3.35,0.2) {};
  \node[fp label, anchor=west] at (3.85,0.2) {next block};
  \node[fp recursive, minimum width=8mm, minimum height=4mm] at (5.85,0.2) {};
  \node[fp label, anchor=west] at (6.35,0.2) {smaller instance};
  \node[fp pivot, minimum width=8mm, minimum height=4mm] at (9.1,0.2) {};
  \node[fp label, anchor=west] at (9.6,0.2) {constant-slope pivot};
  \node[fp known, minimum width=8mm, minimum height=4mm] at (0.4,-0.45) {};
  \node[fp label, anchor=west] at (0.9,-0.45) {known};
  \node[fp unread, minimum width=8mm, minimum height=4mm] at (3.35,-0.45) {};
  \node[fp label, anchor=west] at (3.85,-0.45) {not yet read};
  \node[fp seam, minimum width=8mm, minimum height=4mm] at (5.85,-0.45) {};
  \node[fp label, anchor=west] at (6.35,-0.45) {boundary seam};
\end{tikzpicture}}
  \caption{The visual language used in this appendix.  Coefficient degree
  decreases from left to right, which is also the order in which a decoder
  runs.  A causal pivot reads the current observed row and quantities decoded
  strictly earlier, but never a lower row.  Larger constructions are drawn as
  nested monic shells: peel the highest shell, subtract it (including any
  displayed boundary correction), and continue with the newly exposed
  block.  Colours distinguish blocks, while labels and line styles carry the
  same information in monochrome.}
  \label{fig:decoder-language}
\end{figure}

For concision, call a field $d$-\emph{admissible} if it has characteristic
zero or characteristic $p>d$.  This is a convenient sufficient condition,
not the exact one (\Cref{rem:char2}).  Every decoder displayed in this
appendix is a composition of unit pivots, monic divisions, and affine
pivots, root extractions and block solves whose slopes, root indices and
determinants are fixed integers; it is therefore defined, and inverts the
construction, over every field in which those integers are units.
Replaying the routing of \Cref{alg:final-construction} lists them exactly.
Along the recursive calls made for degree~$n$, every integer divided by is
one of
\begin{equation}
\label{eq:bad-primes}
  2,\qquad 2k,\qquad k\ (k\text{ even})\ \text{or}\ k(k-1)\ (k\text{ odd}),
  \qquad k^2,
\end{equation}
for an internal index $k$, and these occur as follows.
\begin{itemize}
  \item $2$: every monic square root, square gadget and scalar shift
    (\Cref{lem:monic-from-power,lem:square-gadget-boundary,lem:scalar-shift-square}),
    the septic base (\Cref{lem:septic-base}) and the finite bases $15,27,31$
    (\Cref{lem:special-cases-splittable}).
  \item $2k$: the five crown slopes $2k,2k,-2k,k,k$ of the $4k+1$ branch
    (\Cref{lem:4k+1-splittable}) and of $Q_{4k+1}(x,H_2)$
    (\Cref{lem:Q4k+1-from-H2}), and the slope $M=2k$ of a perturbed
    $T_{2k,2^l}$ call (\Cref{lem:causal-perturbed-T}).
  \item $k$ or $k(k-1)$: the $T$ recursion at index $k$
    (\Cref{lem:Rk2l,lem:Rk2l-leading-coeff}), followed down its halving
    chain $k\mapsto\lfloor k/2\rfloor$.  For even $k$ the stage table has
    slopes $-k$ and $k/2$ together with one square root and one scalar
    shift; for odd $k$ it has slopes $-k(k-1)$, $-(k-1)$ and $(k-1)/2$; the
    leading coefficient $\gamma_k$ divides $k(k-1)$ in both cases.
  \item $k^2$: the block determinant $\det M=-k^2$ and the two slopes $k$
    of the barred gadget $\bar Q_{8k+7}$ (\Cref{lem:barQ8k+7}).
\end{itemize}
The seam corrections $\gamma_m$,
$\binom m2$, $\tau=k(k-1)(k-2)/3=2\binom k3$ and $\theta_m$ are integers
that are subtracted, never divided by, and the fill gadgets $A_{2^l}$, the
known-powers gadgets $Q_{2^t-1}$, the gadgets
$Q_1,Q_3,Q_7$, the degree-$3$ base and the even lift use only unit pivots
and monic division.  Let $\mathrm{Bad}(n)$ be the set of primes dividing at
least one integer in \eqref{eq:bad-primes} along the routing for degree
$n$.  Then the construction and the decoder of $P_n$ are valid over every
field in which the elements of $\mathrm{Bad}(n)$ are units, i.e.\ whose
characteristic is not in $\mathrm{Bad}(n)$.  Every internal index $k$ in
\eqref{eq:bad-primes} is at most $(n-1)/4$, so every odd element of
$\mathrm{Bad}(n)$ is at most $(n-1)/4$ and $n$-admissibility is sufficient.
The script \texttt{tools/bad\_primes.py} replays the routing and prints
the pivot list; for instance
\[
\begin{aligned}
  &\mathrm{Bad}(5)=\mathrm{Bad}(9)=\mathrm{Bad}(15)=\mathrm{Bad}(17)
    =\mathrm{Bad}(31)=\{2\},\\
  &\mathrm{Bad}(13)=\mathrm{Bad}(27)=\{2,3\},\qquad
  \mathrm{Bad}(21)=\{2,5\},\qquad
  \mathrm{Bad}(29)=\{2,3,7\}.
\end{aligned}
\]
Thus $\mathrm{Bad}(n)=\{2\}$ for $n\in\{5,\dots,12\}$ and, e.g., for
$n=31,\,47,\,79$.  As a check, the reference decoder
\texttt{tools/polychain.py} completes its encode--decode round trip over
$\mathrm{GF}(p)$ for every odd $n\le75$ (and selected $n\le127$) and
every prime $5\le p\le31$ with $p\notin\mathrm{Bad}(n)$, including many
pairs with $p<n$; for $p\in\mathrm{Bad}(n)$ the displayed decoder divides
by zero.  (At $p=3$ that implementation additionally evaluates $\tau$ by a
division, which the paper's decoder does not need.)
No characteristic restriction beyond \eqref{eq:bad-primes} is hidden in
the recursion.

\subsection{Known-powers and fill gadgets}
\label{sec:peeled-Q}

We first construct $Q_{2^k-1}$ from a tower of known monic polynomials
$H_2,\ldots,H_{2^{k-1}}$, with $\deg H_{2^i}=2^i$.
Its two half-size children run in parallel and are joined by one product.
This is the known-powers construction used throughout the paper and Lean.
We then define the fill gadget $A_{2^l}$, which uses these $Q$-blocks to
extend a compatible input pair.  The known-powers decoder is independent
of the fill decoder, so no mutual recursion is needed.

\begin{algorithm}[H]
  \caption{Known-powers construction (binary recursion)}\label{alg:constr-known-2n-1}
  \begin{algorithmic}
    \Require $k\ge1$, known monic powers $(H_2,\ldots,H_{2^{k-1}})$,
      and parameters $(\alpha_0,\ldots,\alpha_{2^k-2})$
    \If{$k=1$}
      \State $Q_1[\alpha_0]=x+\alpha_0$.
    \ElsIf{$k=2$}
      \State $Q_3[\alpha_0,\alpha_1,\alpha_2]=(x+\alpha_2)(H_2+\alpha_1)+\alpha_0$.
    \Else
      \State Set $h=2^{k-1}$ and $\gamma=\alpha_0$.
      \State Compute $W=Q_{h-1}[\alpha_1,\ldots,\alpha_{h-1}]$ and
        $B=Q_{h-1}[\alpha_h,\ldots,\alpha_{2h-2}]$ in parallel.
      \State $Q_{2^k-1}=(H_h+\gamma)W+B$.
    \EndIf
  \end{algorithmic}
\end{algorithm}

Thus, for $k\ge3$, the recursion is
\begin{equation}
  Q_{2^k-1}[\gamma,\vec\alpha,\vec\beta]
  =\bigl(H_{2^{k-1}}+\gamma\bigr)Q_{2^{k-1}-1}[\vec\alpha]
    +Q_{2^{k-1}-1}[\vec\beta],
  \label{eq:peeled-Q}
\end{equation}
where the two child blocks are fresh and have length $2^{k-1}-1$ each.
All $Q$-blocks below use this definition, including those inside a fill,
a tower update, or one of the special degree-$15,27,31$ constructions.

We first record the two direct decoder bases.

\begin{lemma}
    $Q_{1}[\alpha_0](x)$ is decodable.
\end{lemma}
\begin{proof}
    By definition $Q_1(x)=x+\alpha_0$, so $\alpha_0=[x^0]Q_1$ is extractable from $Q_1$.
\end{proof}

\begin{lemma}
    $Q_{3}[\alpha_0, \alpha_1, \alpha_2](x, H_2)$ is decodable given $H_2$.
\end{lemma}
\begin{proof}
    By definition, $Q_3(x,H_2)=(x+\alpha_2)(H_2+\alpha_1)+\alpha_0$ with $H_2$ monic of degree 2.
    Writing $H_2(x)=x^2 + h_1 x + h_0$, a direct expansion gives
    \begin{align}
        [x^2]Q_3 &= h_1 + \alpha_2, \\
        [x^1]Q_3 &= h_0 + \alpha_1 + h_1\alpha_2, \\
        [x^0]Q_3 &= \alpha_0 + h_0\alpha_2 + \alpha_1\alpha_2.
    \end{align}
    Hence
    \begin{align}
        \alpha_2 &= [x^2]Q_3 - [x^1]H_2, \\
        \alpha_1 &= [x^1]Q_3 - [x^0]H_2 - \alpha_2\,[x^1]H_2, \\
        \alpha_0 &= [x^0]Q_3 - \alpha_2\bigl([x^0]H_2 + \alpha_1\bigr),
    \end{align}
    which are polynomials in the coefficients of $Q_3$ and $H_2$. Thus $Q_3$ is decodable given $H_2$.
\end{proof}

\begin{lemma}[The known-powers gadget is decodable]
\label{lem:peeled-Q-decodable}
For every $k\ge1$, $Q_{2^k-1}$ is monic of degree $2^k-1$ and is decodable
given $H_2,\ldots,H_{2^{k-1}}$.
\end{lemma}
\begin{proof}
The bases were established above.  For $k\ge3$, write $h=2^{k-1}$ and
$m=h-1$.  By induction both children $W,B$ in \eqref{eq:peeled-Q} are
monic of degree $m$, so the output is monic of degree $h+m$.
Since $\deg(\gamma W+B)\le m<h$, monic division by the known $H_h$
(\Cref{lem:monic-division}) returns quotient $W$ and remainder
$R=\gamma W+B$.  The coefficient $[x^m]R=\gamma+1$ recovers $\gamma$;
then $B=R-\gamma W$, and we decode both children recursively.
These steps are polynomial in the supplied coefficients.

Conversely, starting with any monic polynomial of degree $h+m$, division
by $H_h$ gives a monic quotient of degree $m$ and a remainder of degree
at most $m$.  Setting $\gamma=[x^m]R-1$ makes $B=R-\gamma W$ monic of
degree $m$.  The child inverses therefore apply, and re-expansion returns
the input.  This proves both directions of the decoder.
\end{proof}

\begin{lemma}[Coefficient pivots of the known-powers gadgets]
\label{lem:Q-unitriangular}
Let $q=2^k-1$ and write $Q_q=x^q+\sum_{j<q}q_jx^j$.
There is an explicit row ordering $(\eta_0,\ldots,\eta_{q-1})$ of its
parameters such that
\[
  q_j=\eta_j+f_j(\eta_{j+1},\ldots,\eta_{q-1})
  \qquad(0\le j<q).
  \tag{Q-tri}
\]
The coefficients of $f_j$ depend polynomially only on the supplied
known-power coefficients.  For $k=1,2$, the row order is the raw
parameter order.  For $k\ge3$, put $h=2^{k-1}$ and $m=h-1$:
the first $m$ slots are the $B$ child's row order, slot $m$ is $\gamma$,
and slots $h,\ldots,2h-2$ are the $W$ child's row order.
\end{lemma}
\begin{proof}
The base coefficient formulas have the claimed unit slopes.
For the induction step, the coefficients in rows $h,\ldots,h+m-1$
come only from $H_hW$.  Descending monic multiplication gives
$q_{h+j}=[x^j]W$ plus a polynomial in higher coefficients of $W$.
The induction hypothesis for $W$ therefore gives the high-block pivots.
In row $m$, both children have leading coefficient one, so
$q_m=\gamma+1+[x^m](H_hW)$; all parameters of $W$ are in higher rows.
Finally, each row $j<m$ is $[x^j]B+[x^j]((H_h+\gamma)W)$.
The second term now depends only on higher-row parameters, and the
induction hypothesis for $B$ gives the remaining unit pivots.
The two child blocks and the glue slot are disjoint and exhaust the raw
parameters, so this recursively defined row order is a permutation.
\end{proof}

\begin{algorithm}[H]
  \caption{Decoder summary for $Q_{2^k-1}$ from known powers}\label{alg:decode-Q-2kminus1}
  \begin{algorithmic}
    \Require $k\ge1$, known powers $(H_2,\ldots,H_{2^{k-1}})$, and
      $Q$ produced by \Cref{alg:constr-known-2n-1}
    \Ensure the raw parameter block $(\alpha_0,\ldots,\alpha_{2^k-2})$
    \If{$k=1$}
      \State Output $\alpha_0=[x^0]Q$.
    \ElsIf{$k=2$}
      \State Apply the explicit $Q_3$ coefficient formulas above.
    \Else
      \State Set $h=2^{k-1}$ and divide $Q$ by the known monic $H_h$,
        obtaining quotient $W$ and remainder $R$.
      \State Set $\alpha_0=[x^{h-1}]R-1$ and $B=R-\alpha_0W$.
      \State Recursively decode $W$ to recover $(\alpha_1,\ldots,\alpha_{h-1})$.
      \State Recursively decode $B$ to recover $(\alpha_h,\ldots,\alpha_{2h-2})$.
    \EndIf
  \end{algorithmic}
\end{algorithm}

\begin{figure}[H]
  \centering
  \resizebox{\textwidth}{!}{
\begin{tikzpicture}[x=1cm,y=1cm,>=Latex]
  \node[fp panel title] at (0,4.0) {(a) Binary known-powers recursion};
  \node[fp active, minimum width=42mm] (qroot) at (2.6,3.3)
    {$\bigl(H_{2^{k-1}}+\gamma\bigr)\cdot W+B$};
  \node[fp recursive, minimum width=26mm] (wch) at (0.9,2.35)
    {$W=Q_{2^{k-1}-1}[\vec\alpha]$};
  \node[fp recursive, minimum width=26mm] (bch) at (4.3,2.35)
    {$B=Q_{2^{k-1}-1}[\vec\beta]$};
  \draw[fp flow] (wch.north) -- ([xshift=-12mm]qroot.south);
  \draw[fp flow] (bch.north) -- ([xshift=12mm]qroot.south);
  \node[fp label, anchor=west, text=fpGray] at (6.6,2.8)
    {\eqref{eq:peeled-Q}: both children in parallel\\
     under one keyed glue factor;\\
     one new product; height at most $k$};
  \node[fp panel title] at (0,1.75) {(b) Decoder of \Cref{lem:peeled-Q-decodable}};
  \draw[fp decode] (0,1.12) -- node[fp label, above]
    {descending coefficient degree} (13.4,1.12);
  \node[fp active, text width=33mm, anchor=west] (topw) at (0,0.45)
    {top window $=H_{2^{k-1}}\cdot W$;\\divide by the monic $H$};
  \node[fp pivot, text width=26mm, anchor=west] (grow) at (3.65,0.45)
    {$[x^{m}]$ residual $=\gamma+1$;\\one unit pivot};
  \node[fp second, text width=24mm, anchor=west] (brow) at (6.6,0.45)
    {$B=R-\gamma W$\\by subtraction};
  \node[fp recursive, text width=32mm, anchor=west] (rec) at (9.35,0.45)
    {recurse on $W$ and $B$\\(both children monic)};
  \draw[fp dependence] (topw.south east) to[bend right=10] (grow.south west);
  \draw[fp dependence] (grow.south east) to[bend right=10] (brow.south west);
  \draw[fp dependence] (brow.south east) to[bend right=10] (rec.south west);
\end{tikzpicture}}
  \caption{The known-powers gadget.  (a) Two half-size children run in
  parallel under one keyed factor, using one new product.
  (b) Monic division recovers $W$; the top remainder coefficient reveals
  $\gamma$ with slope one; subtraction recovers $B$.  Both children then
  recurse.  The exact gate counts and height are
  \Cref{lem:fill-Q-count,lem:peeled-Q-count}.}
  \label{fig:peeled-Q}
\end{figure}

\paragraph{The fill construction.}
The level-$l$ fill takes a compatible degree-$n$ pair and known powers
$H_2,\ldots,H_{2^l}$, and outputs a monic polynomial of degree
$n+2^{l+1}-1$.  Its decoder recovers the input pair and all fresh
$\alpha$- and $\beta$-parameters.

\begin{figure}[H]
  \centering
  \resizebox{\textwidth}{!}{
\begin{tikzpicture}[x=1cm,y=1cm,>=Latex]
  \node[fp panel title] at (0,5.05) {(a) One fill level, $D=2^l$};
  \node[fp recursive, minimum width=27mm] (input) at (1.45,4.18)
    {input pair\\$(S_D^{(1)},S_D^{(2)})$};
  \node[fp known, minimum width=33mm] (heads) at (4.65,4.18)
    {heads $H_D+Q_{D/2-1}$\\and $H_D+\beta_D$};
  \node[fp second, minimum width=31mm] (mid) at (8.2,4.18)
    {multiply; add\\a fresh low $Q$-block};
  \node[fp active, minimum width=30mm] (lower) at (11.75,4.18)
    {intermediate pair\\at level $l-1$};
  \draw[fp flow] (input) -- (heads);
  \draw[fp flow] (heads) -- (mid);
  \draw[fp flow] (mid) -- (lower);
  \node[fp label, below=2mm of lower] {recurse, then output
    $P=(x+\beta_0)A^{(1)}+A^{(2)}$};

  \node[fp panel title] at (0,2.95) {(b) The three output pivot bands};
  \draw[fp decode] (0,2.5) -- node[fp label, above] {descending coefficient degree}
    (13.55,2.5);
  \node[fp active, minimum width=38mm, anchor=west] (betas) at (0,1.78)
    {high $\beta$-block\\unit pivots in $Q$ row order};
  \node[fp recursive, minimum width=47mm, anchor=west] (old) at (3.8,1.78)
    {input-pair pivots shifted by $2D-2$\\with their original slopes};
  \node[fp second, minimum width=46mm, anchor=west] (alphas) at (8.5,1.78)
    {low $\alpha$-block\\unit pivots in $Q$ row order};
  \draw[fp dependence] (betas.south east) to[bend right=10] (old.south west);
  \draw[fp dependence] (old.south east) to[bend right=10] (alphas.south west);

  \node[draw=black!50, line width=.45pt, inner sep=2pt, font=\scriptsize,
    align=center, minimum width=105mm, minimum height=6mm] at (6.7,0.52)
    {proof order:
     $Q_1\longrightarrow Q_3\longrightarrow Q_7\longrightarrow\cdots$,
     \quad then $F_1\longrightarrow F_2\longrightarrow F_3\longrightarrow\cdots$};
  \node[fp label, anchor=west, text=fpBlue!80!black] at (0,-0.08)
    {The binary known-powers decoder is independent of the fill induction.};
\end{tikzpicture}}
  \caption{Forward and inverse views of the fill construction.  One level
  multiplies the input pair by monic heads, adds a fresh low gadget, and
  delegates to the preceding fill level.  In the inverse direction its
  coefficients separate into three descending pivot bands: new top
  $\beta$-parameters, the shifted input-pair pivots, and new low
  $\alpha$-parameters.  The bottom line displays the proof order:
  the known-powers decoder is proved independently, before the fill induction.
  Within each $Q$-block, parameters are read in the row order of
  \Cref{lem:Q-unitriangular}.}
  \label{fig:fill-slots}
\end{figure}

\begin{algorithm}[H]
    \caption{High-level fill construction}\label{alg:constr-fill}
    \begin{algorithmic}
        \If{$l = 1$}
            \Statex $A^{(1)}_{2}[\alpha_0, \alpha_1, \beta_2, \beta_1](S^{(1)}_2, (x, H_2)) = (H_2 + \beta_1) S^{(1)}_2 + \alpha_1$
            \Statex $A^{(2)}_{2}[\alpha_0, \alpha_1, \beta_2, \beta_1](S^{(2)}_2, (x, H_2)) = (H_2 + \beta_2) S^{(2)}_2 + \alpha_0$
        \ElsIf{$l = 2$}
            \Statex $S^{(1)}_{2} = (H_{4} + \beta_3) S^{(1)}_{4} + Q_{3}[\alpha_3, \alpha_4, \alpha_5](x, H_2)$
            \Statex $A^{(1)}_{4}[\alpha_0, \ldots, \alpha_5, \beta_3, \ldots, \beta_1](S^{(1)}_{4}, (x, H_2, H_4))$
            \Statex \hspace{\algorithmicindent}$= A^{(1)}_{2}[\alpha_0, \alpha_1, \beta_2, \beta_1](S^{(1)}_{2}, (x, H_2))$
            \Statex $S^{(2)}_{2} = (H_{4} + \beta_4) S^{(2)}_{4} + \alpha_2$
            \Statex $A^{(2)}_{4}[\alpha_0, \ldots, \alpha_5, \beta_4, \ldots, \beta_1](S^{(2)}_{4}, (x, H_2, H_{4}))$
            \Statex \hspace{\algorithmicindent}$= A^{(2)}_{2}[\alpha_0, \alpha_1, \beta_2, \beta_1](S^{(2)}_{2}, (x, H_2))$
        \Else
            \Statex $S^{(1)}_{2^{l - 1}} = (H_{2^{l}} + Q_{2^{l - 1} - 1}[\beta_{2^{l} - 1}, \ldots, \beta_{2^{l - 1} + 1}](x, H_2, \ldots, H_{2^{l - 2}})) S^{(1)}_{2^l}$
            \Statex \hspace{\algorithmicindent}$+\; Q_{2^{l} - 1}[\alpha_{2^{l} - 1}, \ldots, \alpha_{2^{l + 1} - 3}](x, H_2, \ldots, H_{2^{l - 1}})$
            \Statex $A^{(1)}_{2^l}[\alpha_0, \ldots, \alpha_{2^{l + 1} - 3}, \beta_{2^l - 1}, \ldots, \beta_{1}](S^{(1)}_{2^l}, (x, H_2, \ldots, H_{2^{l}}))$
            \Statex \hspace{\algorithmicindent}$= A^{(1)}_{2^{l - 1}}[\alpha_0, \ldots, \alpha_{2^l - 3}, \beta_{2^{l - 1} - 1}, \ldots, \beta_{1}](S^{(1)}_{2^{l - 1}}, (x, H_2, \ldots, H_{2^{l - 1}}))$
            \Statex $S^{(2)}_{2^{l - 1}} = (H_{2^{l}} + \beta_{2^{l}}) S^{(2)}_{2^l} + \alpha_{2^l - 2}$
            \Statex $A^{(2)}_{2^l}[\alpha_0, \ldots, \alpha_{2^{l + 1} - 3}, \beta_{2^l}, \ldots, \beta_{1}](S^{(2)}_{2^l}, (x, H_2, \ldots, H_{2^{l}}))$
            \Statex \hspace{\algorithmicindent}$= A^{(2)}_{2^{l - 1}}[\alpha_0, \ldots, \alpha_{2^l - 3}, \beta_{2^{l - 1}}, \ldots, \beta_{1}](S^{(2)}_{2^{l - 1}}, (x, H_2, \ldots, H_{2^{l - 1}}))$
        \EndIf
        \Statex $A_{2^l}[\alpha_0, \ldots, \alpha_{2^{l + 1} - 3}, \beta_{2^l}, \ldots, \beta_{0}](S^{(1)}_{2^l}, S^{(2)}_{2^l}, (x, H_2, \ldots, H_{2^{l}})) = (x + \beta_{0}) A^{(1)}_{2^l} + A^{(2)}_{2^l}$
    \end{algorithmic}
\end{algorithm}

The construction, \Cref{alg:constr-fill}, proceeds by recursively building intermediate pairs and finally outputting
$P=(x+\beta_0)A^{(1)}_{2^l}+A^{(2)}_{2^l}$.
We start by isolating this final step.

\begin{lemma}[$(x+\alpha)$-extraction]\label{lem:x-alpha-extraction}
    Let $(T^{(1)}, T^{(2)})$ be a compatible pair on
    $G \subseteq \idx{n}$ given $(B_i)_{i \in [t]}$, with
    $\deg T^{(1)}=\deg T^{(2)}=n$.
    Define $P[\alpha](x)=(x+\alpha)\,T^{(1)}(x)+T^{(2)}(x)$.
    If $G\subseteq\rng n$, then: (1) $\alpha$ is extractable from $P$
    given $(B_i)_{i\in[t]}$; and (2) $T^{(1)}$ and $T^{(2)}$ are
    derivable from $\bigl(P,(B_i)_{i\in[t]}\bigr)$.
    More generally, for an arbitrary $G\subseteq\idx n$, conclusion~(2)
    still holds whenever $\alpha$ has already been derived.
\end{lemma}
\begin{proof}
    First suppose $G \subseteq \rng{n}$.  Compatibility for $j=n-1$
    then expresses $\coeff{T^{(1)}}{n-1}$ as a polynomial in the
    $(B_i)$-coefficients (the tuple
    $\bigl(\coeff{xT^{(1)}+T^{(2)}}{i}\bigr)_{i\in G,\, i\ge n}$ is
    empty).
    Moreover, since $T^{(1)}$ and $T^{(2)}$ are monic of degree $n$,
    \[
        \coeff{P}{n}
            = \coeff{(x+\alpha)T^{(1)}}{n} + \coeff{T^{(2)}}{n}
            = \alpha\,\coeff{T^{(1)}}{n} + \coeff{T^{(1)}}{n-1} + 1
            = \alpha + \coeff{T^{(1)}}{n-1} + 1.
    \]
    Hence $\alpha=\coeff{P}{n}-\coeff{T^{(1)}}{n-1}-1$ is extractable from $P$ given $(B_i)_{i\in[t]}$.

    For the second conclusion, no restriction beyond $G\subseteq\idx n$
    is needed once $\alpha$ is known.  Define
    $\Phi(x)\coloneqq xT^{(1)}(x)+T^{(2)}(x)$. Since
    $P=\Phi+\alpha T^{(1)}$, we have
    \[
        \coeff{\Phi}{k}=\coeff{P}{k}-\alpha\,\coeff{T^{(1)}}{k}\qquad(k\ge 0).
    \]
    We recover the coefficients of $T^{(1)}$ and $T^{(2)}$ by descending induction on $j=n-1,n-2,\ldots,0$.
    The leading coefficients in degree $n$ are already known to be one.
    The induction hypothesis is that $\coeff{T^{(1)}}{r}$ and
    $\coeff{T^{(2)}}{r}$ are known for all $r>j$, and thus
    $\coeff{\Phi}{i}=\coeff{P}{i}-\alpha\coeff{T^{(1)}}{i}$ is known for
    all $i\in G$ with $i>j$.  This includes the possible top row $i=n$.
    By compatibility, $\coeff{T^{(1)}}{j}$ is a polynomial in $\bigl(\coeff{\Phi}{i}\bigr)_{i\in G,\,i\ge j+1}$ and the $(B_i)$-coefficients; since $i\ge j+1$ implies $i>j$, these $\Phi$-coefficients are already known by the induction hypothesis.
    Once $\coeff{T^{(1)}}{j}$ is known, we can compute $\coeff{\Phi}{j}=\coeff{P}{j}-\alpha\,\coeff{T^{(1)}}{j}$, and then compatibility expresses $\coeff{T^{(2)}}{j}$ as a polynomial in $\bigl(\coeff{\Phi}{i}\bigr)_{i\in G,\,i\ge j}$ and the $(B_i)$-coefficients.
    This completes the induction and shows that $T^{(1)}$ and $T^{(2)}$ are derivable from $\bigl(P,(B_i)_{i\in[t]}\bigr)$.
\end{proof}

\begin{lemma}[Aux-head compatibility for degree $2$]\label{lem:compatible-auxhead-deg2}
    Let $H_2\in\mathbb{F}[x]$ be monic of degree $2$, and let $b_1,b_2$ be scalars.
    Define $F_1\coloneqq H_2+b_1$ and $F_2\coloneqq H_2+b_2$.
    Then $(F_1,F_2)$ is a compatible pair on $\{0,1\}$ given $(H_2)$.
\end{lemma}
\begin{proof}
    Define $\Phi\coloneqq xF_1+F_2$.
    Since $F_1$ and $F_2$ are monic of degree $2$, their leading coefficients are known.
    Also $\coeff{F_1}{1}=\coeff{F_2}{1}=\coeff{H_2}{1}$ is derivable from $H_2$.

    Finally, using that $\coeff{\Phi}{0}=\coeff{F_2}{0}$ and $\coeff{\Phi}{1}=\coeff{F_1}{0}+\coeff{F_2}{1}$, we can derive
    \[
        \coeff{F_2}{0}=\coeff{\Phi}{0},
        \qquad
        \coeff{F_1}{0}=\coeff{\Phi}{1}-\coeff{H_2}{1}.
    \]
    The unknown constant coefficient of $F_1$ is therefore read from row~$1$,
    while that of $F_2$ is read from row~$0$; their degree-one coefficients
    and all higher coefficients are supplied by $H_2$.  Hence the first
    component uses only rows at least $j+1$ and the second only rows at least
    $j$, which witnesses compatibility on $\{0,1\}$ given $(H_2)$.
\end{proof}

\begin{lemma}[A trivial compatible pair for adding constants]\label{lem:compatible-monic-plus-constants}
    Let $n\ge 2$ and let $u,v$ be scalars.
    Then the pair $\bigl(x^n+u,\;x^n+v\bigr)$ is compatible on $\{0,1\}$.
\end{lemma}
\begin{proof}
    Define $\Phi\coloneqq x(x^n+u)+(x^n+v)=x^{n+1}+x^n+ux+v$.
    Since $n\ge 2$, we have $\coeff{\Phi}{0}=v$ and $\coeff{\Phi}{1}=u$.
    Thus the only unknown coefficient of the first component, $u$ in degree
    zero, is read from row $1$, and the only unknown coefficient of the second
    component, $v$ in degree zero, is read from row $0$; every other
    coefficient is fixed.  These are exactly the two compatibility cutoffs.
\end{proof}

\noindent
In later constructions we repeatedly meet pairs of the form $(H+Q,\;H+b)$ where $H$ is a known monic
polynomial, $Q$ is an \emph{unknown} lower-degree monic polynomial, and $b$ is an unknown scalar shift.
The next lemma records that such a pair is compatible on the low-degree window that contains all
coefficients of $Q$: the map $\Phi=x(H+Q)+(H+b)$ is block-triangular, so one can peel off $Q$ (and $b$)
from low-degree coefficients of $\Phi$ given $H$.

\begin{lemma}[A block-triangular compatible pair]\label{lem:compatible-aux-add-left}
    Let $H\in\mathbb{F}[x]$ be monic of degree $m\ge 1$, let $Q\in\mathbb{F}[x]$ be monic of degree $q<m$, and let $b$ be a scalar.
    Define $P^{(1)}\coloneqq H+Q$ and $P^{(2)}\coloneqq H+b$.
    Then $(P^{(1)},P^{(2)})$ is a compatible pair on $\idx{q}$ given $(H)$.
\end{lemma}
\begin{proof}
    Define $\Phi\coloneqq xP^{(1)}+P^{(2)}$ and note that
    \[
      \Phi = x(H+Q) + (H+b) = (xH+H) + (xQ+b).
    \]
    Since $H$ is given, the polynomial $xH+H$ is derivable.
    Therefore the polynomial
    \[
      \Xi \coloneqq \Phi-(xH+H)=xQ+b
    \]
    is derivable from $\Phi$ given $(H)$.
    In particular, $b=\coeff{\Xi}{0}$ is derivable.

    Next, for $1\le i\le q$ we have $\coeff{\Xi}{i}=\coeff{Q}{i-1}$, so the coefficients $\coeff{Q}{0},\ldots,\coeff{Q}{q-1}$ are derivable from $\bigl(\coeff{\Phi}{i}\bigr)_{i\in\idx{q}}$ given $(H)$.
    Together with monicity $\coeff{Q}{q}=1$, this derives $Q$, hence also $P^{(1)}=H+Q$ and $P^{(2)}=H+b$.

    It remains to verify the coefficient-dependency condition in the definition of compatibility.
    Since $q<m$, we have $\coeff{P^{(1)}}{j}=\coeff{P^{(2)}}{j}=\coeff{H}{j}$ for all $j>q$, and these coefficients are derivable from $(H)$ alone.
    For $j\le q$, each coefficient of $Q$ (and hence of $P^{(1)}$) is derivable from the coefficients $\bigl(\coeff{\Phi}{i}\bigr)_{i\in\idx{q},\, i\ge j+1}$ and $(H)$ by the explicit reconstruction above (since $\coeff{Q}{j}$ depends only on $\coeff{\Xi}{j+1}$ when $j<q$, and $\coeff{Q}{q}=1$).
    Similarly, $\coeff{P^{(2)}}{0}$ is derivable from $\coeff{\Phi}{0}$ and $(H)$ via $b$, and $\coeff{P^{(2)}}{j}=\coeff{H}{j}$ for $j\ge 1$.
    This gives the required coefficient polynomials in the definition of compatibility on $\idx{q}$ given $(H)$.
\end{proof}

\begin{lemma}[Padding a low pair]\label{lem:compatible-low-padding}
Let $N\ge r\ge1$, let $Q$ be monic of degree $r-1$, and let $b$ be a
scalar.  Then
\[
  (x^N+Q,\ x^N+b)
\]
is compatible on $\rng r$.
\end{lemma}
\begin{proof}
After subtracting the known terms $x^{N+1}+x^N$ from its combined
polynomial, the window $\rng r$ is exactly the corresponding window of
$xQ+b$.  Degree zero gives $b$, and degree $j+1$ gives $[x^j]Q$ for
$0\le j<r-1$; the remaining coefficient of $Q$ is its known leading~$1$.
Moreover $[x^j]Q$ is read in degree $j+1$ and $b$ in degree zero, so these
formulas satisfy the two compatibility cutoffs.
\end{proof}

\begin{lemma}\label{lem:fill-correctness}
    Let $L\ge 1$ and let $H_2, \ldots, H_{2^L}$ be a sequence of monic polynomials with $\deg H_{2^i} = 2^i$ for $1 \le i \le L$.
    Fix $l$ with $1\le l\le L$.
    (We allow extra known powers $H_{2^{l+1}},\ldots,H_{2^L}$ since later arguments may carry more auxiliary data than the particular fill level $l$ needs.)
    Assume that $(S^{(1)}_{2^l}, S^{(2)}_{2^l})$ are a compatible pair on $G \subseteq \rng{n - 2^l}$ given $(H_2, \ldots, H_{2^L})$ with $n = \deg S^{(1)}_{2^l} = \deg S^{(2)}_{2^l} \ge 2^l$.
    Apply the level-$l$ fill with fresh parameter blocks as follows:
    \[
      \begin{aligned}
        P\coloneqq{}&A_{2^l}
          [\alpha_0,\ldots,\alpha_{2^{l+1}-3},
           \beta_{2^l},\ldots,\beta_0]\\
        &\bigl(S^{(1)}_{2^l},S^{(2)}_{2^l},
               (x,H_2,\ldots,H_{2^l})\bigr).
      \end{aligned}
    \]
    Then the monic polynomial $P$
    satisfies that $\alpha_0, \ldots, \alpha_{2^{l + 1} - 3}, \beta_{2^l}, \ldots, \beta_{0}$ are extractable from $P$ given $(H_2, \ldots, H_{2^L})$, and that $S^{(1)}_{2^l}$ and $S^{(2)}_{2^l}$ are derivable from $(P, H_2, \ldots, H_{2^L})$.
\end{lemma}
\begin{proof}
    We induct on $l$, uniformly for every $L\ge l$ and every permitted
    input pair.  Every known-powers child is already decodable by
    \Cref{lem:peeled-Q-decodable}; only the fill level decreases in this induction.

    Base case $l=1$: The output is
    \[
        P=(x+\beta_0)\bigl((H_2+\beta_1)S^{(1)}_2+\alpha_1\bigr)+\bigl((H_2+\beta_2)S^{(2)}_2+\alpha_0\bigr).
    \]
    We claim that $(A^{(1)}_2,A^{(2)}_2)$ is a compatible pair on
    \[
        G' \coloneqq (2+G)\cup(n+\{0,1\})\cup\{0,1\}\subseteq\rng{n+2}
    \]
    given $(H_2,\ldots,H_{2^L})$.
    Indeed, by \Cref{lem:compatible-auxhead-deg2} the degree-$2$ pair $\bigl(H_2+\beta_1,\;H_2+\beta_2\bigr)$ is compatible on $\{0,1\}$ given $(H_2)$.
    Since $G\subseteq\rng{n-2}$, the shifted windows $(2+G)$ and $(n+\{0,1\})$ are disjoint, so the Multiplicativity lemma implies that $\bigl((H_2+\beta_1)S^{(1)}_2,\;(H_2+\beta_2)S^{(2)}_2\bigr)$ is compatible on $(2+G)\cup(n+\{0,1\})$ given $(H_2,\ldots,H_{2^L})$.
    Moreover, by \Cref{lem:compatible-monic-plus-constants} (applied with $n\gets n+2$), the pair $(x^{n+2}+\alpha_1,\;x^{n+2}+\alpha_0)$ is compatible on $\{0,1\}$, and this window is disjoint from $(2+G)\cup(n+\{0,1\})$.
    Therefore, applying the Additivity lemma in the equal-degree case yields that $\bigl(A^{(1)}_2,A^{(2)}_2\bigr)=\bigl((H_2+\beta_1)S^{(1)}_2+\alpha_1,\;(H_2+\beta_2)S^{(2)}_2+\alpha_0\bigr)$ is compatible on $G'$ given $(H_2,\ldots,H_{2^L})$.
    Now we may apply \Cref{lem:x-alpha-extraction} to $P=(x+\beta_0)A^{(1)}_2+A^{(2)}_2$ to extract $\beta_0$ and derive $A^{(1)}_2,A^{(2)}_2$.

    Since $(S^{(1)}_2,S^{(2)}_2)$ is compatible on $G\subseteq\rng{n-2}$ given $(H_2,\ldots,H_{2^L})$, its coefficients of degrees $n-1$ and $n-2$ are derivable from the given data alone.
    Comparing the coefficient of $x^{n}$ in $A^{(1)}_2=(H_2+\beta_1)S^{(1)}_2+\alpha_1$ then extracts $\beta_1$, and similarly extracts $\beta_2$ from $A^{(2)}_2=(H_2+\beta_2)S^{(2)}_2+\alpha_0$.
    Since $H_2+\beta_1$ and $H_2+\beta_2$ are monic, \Cref{lem:monic-division} applied to
    \(
      A^{(1)}_2=(H_2+\beta_1)S^{(1)}_2+\alpha_1
    \)
    and
    \(
      A^{(2)}_2=(H_2+\beta_2)S^{(2)}_2+\alpha_0
    \)
    derives $S^{(1)}_2,S^{(2)}_2$ and the remainders $\alpha_1,\alpha_0$, completing the base case.

    Induction step: Assume the claim holds for all smaller levels, and fix $l\ge 2$.
    The recursive equalities in \Cref{alg:constr-fill} show that $P$ is obtained by applying the level-$(l-1)$ fill construction to the intermediate pair $(S^{(1)}_{2^{l-1}},S^{(2)}_{2^{l-1}})$ (with parameters $\alpha_0,\ldots,\alpha_{2^l-3}$ and $\beta_{2^{l-1}},\ldots,\beta_0$).

    We first check that this intermediate pair satisfies the assumptions needed to apply the induction hypothesis at level $l-1$.
    Write $n\coloneqq\deg S^{(1)}_{2^l}=\deg S^{(2)}_{2^l}$, so $\deg S^{(1)}_{2^{l-1}}=\deg S^{(2)}_{2^{l-1}}=n+2^l$.

    If $l=2$, \Cref{alg:constr-fill} defines
    \[
      S^{(1)}_{2}=(H_{4}+\beta_3)\,S^{(1)}_{4}+Q_{3}[\alpha_3,\alpha_4,\alpha_5](x,H_2),
      \qquad
      S^{(2)}_{2}=(H_{4}+\beta_4)\,S^{(2)}_{4}+\alpha_2.
    \]
    Since $H_4$ is given, subtracting $xH_4+H_4$ from $x(H_4+\beta_3)+(H_4+\beta_4)$ isolates $\beta_3x+\beta_4$, so $\bigl(H_4+\beta_3,\;H_4+\beta_4\bigr)$ is compatible on $\{0,1\}$ given $(H_4)$.
    Together with compatibility of $(S^{(1)}_{4},S^{(2)}_{4})$ on $G\subseteq\rng{n-4}$, the Multiplicativity lemma shows that the product pair
    \(
      \bigl((H_{4}+\beta_3)S^{(1)}_{4},\;(H_{4}+\beta_4)S^{(2)}_{4}\bigr)
    \)
    is compatible on $(4+G)\cup(n+\{0,1\})$.
    Put $N=n+4$.  By \Cref{lem:compatible-low-padding}, the padded low pair
    $(x^N+Q_3,x^N+\alpha_2)$ is compatible on $\rng4$.  Its window is
    disjoint from the product window, so Additivity (with the compensating
    $-x^N$ in its equal-degree case) shows that
    $(S^{(1)}_2,S^{(2)}_2)$ is compatible on
    \[
      \rng4\cup(4+G)\cup(n+\{0,1\})\subseteq\rng{(n+4)-2}.
    \]

    Now assume $l\ge 3$.
    Let
    \[
      \begin{aligned}
      Q_{\mathrm{low}}\coloneqq{}&
        Q_{2^{l-1}-1}[\beta_{2^l-1},\ldots,\beta_{2^{l-1}+1}]
          (x,H_2,\ldots,H_{2^{l-2}}),\\
      Q_{\mathrm{high}}\coloneqq{}&
        Q_{2^l-1}[\alpha_{2^l-1},\ldots,\alpha_{2^{l+1}-3}]
          (x,H_2,\ldots,H_{2^{l-1}}).
      \end{aligned}
    \]
    By \Cref{alg:constr-fill}, the intermediate pair is defined by
    \[
      S^{(1)}_{2^{l-1}}=(H_{2^l}+Q_{\mathrm{low}})\,S^{(1)}_{2^l}+Q_{\mathrm{high}},
      \qquad
      S^{(2)}_{2^{l-1}}=(H_{2^l}+\beta_{2^l})\,S^{(2)}_{2^l}+\alpha_{2^l-2}.
    \]
    By \Cref{lem:compatible-aux-add-left}, the pair $\bigl(H_{2^l}+Q_{\mathrm{low}},\;H_{2^l}+\beta_{2^l}\bigr)$ is compatible on $\idx{2^{l-1}-1}$ given $(H_{2^l})$.
    Since $(S^{(1)}_{2^l},S^{(2)}_{2^l})$ is compatible on $G$ and $G\subseteq\rng{n-2^l}$, we have
    \[
      (n+\idx{2^{l-1}-1})\cap(2^l+G)=\emptyset,
    \]
    and therefore the Multiplicativity lemma implies that the product pair
    \[
      \left(
      \begin{aligned}
        &(H_{2^l}+Q_{\mathrm{low}})S^{(1)}_{2^l},\\[-2pt]
        &(H_{2^l}+\beta_{2^l})S^{(2)}_{2^l}
      \end{aligned}
      \right)
    \]
    is compatible on $(2^l+G)\cup(n+\idx{2^{l-1}-1})$ given $(H_2,\ldots,H_{2^L})$.
    Put $N=n+2^l$.  By \Cref{lem:compatible-low-padding}, the padded pair
    $(x^N+Q_{\mathrm{high}},x^N+\alpha_{2^l-2})$ is compatible on
    $\rng{2^l}$.  This is disjoint from the displayed product window, so a
    further application of Additivity shows that the intermediate pair is
    compatible on
    \[
      G_{\mathrm{int}}\coloneqq \rng{2^l}\cup(2^l+G)\cup(n+\idx{2^{l-1}-1})
      \subseteq \rng{(n+2^l)-2^{l-1}}
    \]
    given $(H_2,\ldots,H_{2^L})$.

    By the induction hypothesis at level $l-1$, the parameters
    $\alpha_0,\ldots,\alpha_{2^l-3}$ and
    $\beta_{2^{l-1}},\ldots,\beta_0$ are extractable from $P$ given the
    powers.  The same hypothesis derives
    $(S^{(1)}_{2^{l-1}},S^{(2)}_{2^{l-1}})$ from $P$ and those powers.

    If $l=2$, we now recover $(S^{(1)}_{4},S^{(2)}_{4})$ and the remaining parameters as follows.
    Since $(S^{(1)}_{4},S^{(2)}_{4})$ is compatible on $G\subseteq\rng{n-4}$, its top four coefficients are derivable from the given power data alone.
    Comparing the coefficient of $x^{n}$ in $S^{(1)}_{2}=(H_{4}+\beta_3)S^{(1)}_{4}+Q_3$ then extracts $\beta_3$ (the $Q_3$ term does not contribute to degree $n$), and similarly extracts $\beta_4$ from $S^{(2)}_{2}=(H_{4}+\beta_4)S^{(2)}_{4}+\alpha_2$.
    Applying \Cref{lem:monic-division} to $S^{(1)}_{2}$ and $H_4+\beta_3$ derives $S^{(1)}_{4}$ and the remainder $Q_3$, and applying it to $S^{(2)}_{2}$ and $H_4+\beta_4$ derives $S^{(2)}_{4}$ and the remainder $\alpha_2$.
    Finally, the earlier $Q_3$ lemma decodes $Q_3$ to extract $\alpha_3,\alpha_4,\alpha_5$.
    This completes the induction step for $l=2$.

    Now assume $l\ge 3$.
    It remains to recover the two blocks
    \[
      \beta_{2^l},\beta_{2^l-1},\ldots,\beta_{2^{l-1}+1}
      \quad\text{and}\quad
      \alpha_{2^l-2},\alpha_{2^l-1},\ldots,\alpha_{2^{l+1}-3},
    \]
    and to derive $(S^{(1)}_{2^l},S^{(2)}_{2^l})$.
    Define
    \[
      \Phi\coloneqq xS^{(1)}_{2^{l-1}}+S^{(2)}_{2^{l-1}}.
    \]
    Using the displayed definitions and $\deg Q_{\mathrm{high}}<2^l$, the polynomial $xQ_{\mathrm{high}}+\alpha_{2^l-2}$ contributes only to degrees $\le 2^l$, and its contribution in degree $2^l$ is the known constant $1$ (since $Q_{\mathrm{high}}$ is monic of degree $2^l-1$).
    Therefore, from $\Phi$ we can derive the combined product polynomial
    \[
      \begin{aligned}
      \Phi_{\mathrm{prod}}\coloneqq{}&\Phi-x^{2^l}\\
        ={}&x\bigl((H_{2^l}+Q_{\mathrm{low}})S^{(1)}_{2^l}\bigr)
           +(H_{2^l}+\beta_{2^l})S^{(2)}_{2^l}\\
        &\quad +(\text{terms supported in degrees }<2^l).
      \end{aligned}
    \]
    In particular, the coefficients of $\Phi_{\mathrm{prod}}$ in the window $(2^l+G)\cup(n+\idx{2^{l-1}-1})$ agree with those of
    \(x\bigl((H_{2^l}+Q_{\mathrm{low}})S^{(1)}_{2^l}\bigr) + (H_{2^l}+\beta_{2^l})S^{(2)}_{2^l}\).
    Since the corresponding product pair is compatible on this window, we can derive
    \(
      (H_{2^l}+Q_{\mathrm{low}})S^{(1)}_{2^l}
    \)
    and
    \(
      (H_{2^l}+\beta_{2^l})S^{(2)}_{2^l}
    \),
    and then the Multiplicativity lemma derives the factors $(H_{2^l}+Q_{\mathrm{low}},H_{2^l}+\beta_{2^l})$ and $(S^{(1)}_{2^l},S^{(2)}_{2^l})$ from these products given $(H_2,\ldots,H_{2^L})$.

    With these polynomials derived, we can extract $\beta_{2^l}$ from the constant term of $H_{2^l}+\beta_{2^l}$.
    Also, applying \Cref{lem:monic-division} to $S^{(1)}_{2^{l-1}}$ and the monic divisor $H_{2^l}+Q_{\mathrm{low}}$ derives the remainder $Q_{\mathrm{high}}$, and applying it to $S^{(2)}_{2^{l-1}}$ and $H_{2^l}+\beta_{2^l}$ derives the remainder $\alpha_{2^l-2}$.

    Finally, to extract the remaining $\beta$- and $\alpha$-blocks, observe that $Q_{\mathrm{low}}$ and $Q_{\mathrm{high}}$ are the known-powers gadgets with exponents $s=l-1$ and $s=l$, respectively.  Both are decodable by \Cref{lem:peeled-Q-decodable}.  Hence they are decodable given the relevant known powers.
    Hence the remaining $\beta$-block is extractable from $Q_{\mathrm{low}}$, and the remaining $\alpha$-block is extractable from $Q_{\mathrm{high}}$.
    This completes the induction on the fill level.
\end{proof}

\begin{algorithm}[H]
  \caption{Decoder summary for the fill construction $A_{2^l}$}\label{alg:decode-fill}
  \begin{algorithmic}
    \Require level $l\ge 1$, known powers $(H_2,\ldots,H_{2^L})$ with $L\ge l$, and $P=A_{2^l}(\cdot)$ as in \Cref{lem:fill-correctness} (including the cutoff-respecting decoder witnessing compatibility of the input pair on $G\subseteq\rng{n-2^l}$)
    \Ensure the parameter blocks $(\alpha_0,\ldots,\alpha_{2^{l+1}-3},\beta_{2^l},\ldots,\beta_0)$ and the input pair $(S^{(1)}_{2^l},S^{(2)}_{2^l})$

    \If{$l=1$}
      \State Use the explicit compatibility window $G'$ from the base case in \Cref{lem:fill-correctness} and apply \Cref{lem:x-alpha-extraction} to extract $\beta_0$ and derive $(A^{(1)}_2,A^{(2)}_2)$.
      \State Extract $\beta_1,\beta_2$ from the coefficient of $x^n$ in $A^{(1)}_2,A^{(2)}_2$ (using that the relevant top coefficients of $(S^{(1)}_2,S^{(2)}_2)$ are derivable from the auxiliary data by compatibility).
      \State Apply \Cref{lem:monic-division} to recover $S^{(1)}_2,S^{(2)}_2$ and the remainders $\alpha_1,\alpha_0$.
    \ElsIf{$l=2$}
      \State First run the $l=1$ decoder on $P$ to derive $(S^{(1)}_2,S^{(2)}_2)$ and extract $(\alpha_0,\alpha_1,\beta_0,\beta_1,\beta_2)$.
      \State Undo the $l=2$ fill step as in \Cref{lem:fill-correctness}: extract $\beta_3,\beta_4$ from the coefficient of $x^n$ in the equations
      \(S^{(1)}_2=(H_4+\beta_3)S^{(1)}_4+Q_3\) and \(S^{(2)}_2=(H_4+\beta_4)S^{(2)}_4+\alpha_2\),
      then divide by the monic divisors to recover $(S^{(1)}_4,S^{(2)}_4)$ and $(Q_3,\alpha_2)$.
      \State Decode $Q_3$ by the explicit $Q_3$ lemma to extract $(\alpha_3,\alpha_4,\alpha_5)$.
    \Else \Comment{$l\ge 3$}
      \State First run the decoder recursively at level $l-1$ on $P$ to derive the intermediate pair $(S^{(1)}_{2^{l-1}},S^{(2)}_{2^{l-1}})$ and extract the lower blocks $(\alpha_0,\ldots,\alpha_{2^l-3},\beta_{2^{l-1}},\ldots,\beta_0)$.
      \State Form the combined polynomial $\Phi=xS^{(1)}_{2^{l-1}}+S^{(2)}_{2^{l-1}}$ and subtract the known $x^{2^l}$ contribution (coming from monicity of the degree-$(2^l-1)$ $Q_{\mathrm{high}}$ term) to obtain $\Phi_{\mathrm{prod}}$ as in \Cref{lem:fill-correctness}.
      \State Using compatibility on the disjoint window $(2^l+G)\cup(n+\idx{2^{l-1}-1})$, recover the product pair
      \(
        \bigl((H_{2^l}+Q_{\mathrm{low}})S^{(1)}_{2^l},\,(H_{2^l}+\beta_{2^l})S^{(2)}_{2^l}\bigr)
      \)
      and then (by the multiplicativity reconstruction) derive the factors $(H_{2^l}+Q_{\mathrm{low}},H_{2^l}+\beta_{2^l})$ and $(S^{(1)}_{2^l},S^{(2)}_{2^l})$.
      \State Extract $\beta_{2^l}$ from the constant term of $H_{2^l}+\beta_{2^l}$.
      \State Apply \Cref{lem:monic-division} to recover the remainders $Q_{\mathrm{high}}$ and $\alpha_{2^l-2}$ from the defining equations for $S^{(1)}_{2^{l-1}}$ and $S^{(2)}_{2^{l-1}}$.
      \State Decode $Q_{\mathrm{low}}$ and $Q_{\mathrm{high}}$ by \Cref{alg:decode-Q-2kminus1}, extracting the remaining $\beta$- and $\alpha$-blocks.
    \EndIf
  \end{algorithmic}
\end{algorithm}

\begin{lemma}[Costs of the fill and Mersenne gadgets]\label{lem:fill-Q-count}
    For all $l\ge 2$ and $k\ge 2$:
    \begin{align}
        \mathcal{A}(A^{(1)}_{2^l}) + \mathcal{A}(A^{(2)}_{2^l})
            &= 5(2^{l - 1} + 2^{l - 2}) - 4 \\
        \mathcal{M}(A^{(1)}_{2^l}) + \mathcal{M}(A^{(2)}_{2^l})
            &= 2^l + 2^{l - 1} - 1 \\
        \mathcal{A}(A_{2^l}) &= 5(2^{l - 1} + 2^{l - 2}) - 2 \\
        \mathcal{M}(A_{2^l}) &= 2^l + 2^{l - 1} \\
        \mathcal{A}(Q_{2^k - 1}) &= 5 \cdot 2^{k - 2} - 2 \\
        \mathcal{M}(Q_{2^k - 1}) &= 2^{k - 1} - 1
    \end{align}
\end{lemma}
\begin{proof}
Count each shared subexpression once.  Write $q_k^A,q_k^M$ for the
addition and multiplication counts of $Q_{2^k-1}$, and $a_l,m_l$ for
the joint counts of the fill pair.  The binary construction gives
\begin{equation}
  q_k^A=2q_{k-1}^A+2,\qquad q_k^M=2q_{k-1}^M+1
  \quad(k\ge3),
  \label{eq:count-q-rec}
\end{equation}
with $(q_2^A,q_2^M)=(3,1)$.  Induction immediately gives
$q_k^A=5\cdot2^{k-2}-2$ and $q_k^M=2^{k-1}-1$.
In particular $Q_7$ uses eight additions and three products.

The level-two fill pair has $(a_2,m_2)=(11,5)$.  For $l\ge3$,
\Cref{alg:constr-fill} adds two products, four additions, and its two
known-powers children:
\begin{equation}
  a_l=a_{l-1}+4+q_{l-1}^A+q_l^A,\qquad
  m_l=m_{l-1}+2+q_{l-1}^M+q_l^M.
  \label{eq:count-a-rec}
\end{equation}
Substituting the already-proved $q$ counts and inducting on $l$ gives
\[
  a_l=5(2^{l-1}+2^{l-2})-4,\qquad
  m_l=2^l+2^{l-1}-1.
\]
Finally, $(x+\beta_0)A^{(1)}_{2^l}+A^{(2)}_{2^l}$ adds two additions
and one product to these joint pair counts.
\end{proof}

\begin{lemma}[Height of the known-powers gadget]
\label{lem:peeled-Q-count}
For $k\ge2$, the circuit of \Cref{alg:constr-known-2n-1} has height at
most $k$ whenever $h(H_{2^i})\le i$; its height is exactly $k$ when
$h(H_{2^i})=i$.  Its gate counts are those of \Cref{lem:fill-Q-count}.
\end{lemma}
\begin{proof}
The base $Q_3$ has one product above $H_2$, hence height at most two.
At each binary node the two children run in parallel; the height is
\[
 \max\{1+\max(h(H_{2^{k-1}}),h(W)),\,h(B)\}\le k
\]
by induction.  When the tower heights are exact, the path through
$H_{2^{k-1}}$ attains $k$.  The linear base $Q_1$ has height zero.
\end{proof}

\begin{theorem}[Multiplicative height of the construction]
\label{thm:construction-height}
For every $n\ge3$, the complete construction computes a decodable monic
degree-$n$ family with exactly the multiplication count of
\Cref{thm:construction-count}, an addition count no larger than that of
\Cref{thm:construction-addition-count}, and height at most
\[
  2\lceil\log_2n\rceil+4 \quad (n\text{ odd}),\qquad
  2\lceil\log_2n\rceil+5 \quad (n\text{ even}).
\]
\end{theorem}
\begin{proof}
Let $h(U)$ denote the maximum number of nonscalar multiplications on
an input-to-$U$ path; inputs have height zero, additions take the maximum
of their operand heights, and a product adds one to that maximum.
The known-powers decoder and coefficient pivots are
\Cref{lem:peeled-Q-decodable,lem:Q-unitriangular}; their exact gate counts
are \Cref{lem:fill-Q-count}.  We prove the height bound by a simultaneous
ledger induction over the towers, the $T$ recursion, the fills, and the
complete degree-$n$ branches.  Every component uses the single
known-powers construction of \Cref{alg:constr-known-2n-1}.

\emph{Towers and gadgets.}  Write $h(U)$ for the height of a wire.  We
claim, by simultaneous induction on the level,
\[
  h(H_{2^i})=i,
  \qquad
  h\bigl(Q_{2^t-1}\bigr)\le t .
\]
Every displayed power step forms $H_{2^{\ell+1}}$ with one product whose
operands are $H_{2^\ell}$ (height $\ell$) and blocks
$H_{2^{\ell-1}}+Q_{2^{\ell-1}-1}$,
$H_{2^{\ell-2}}+Q_{2^{\ell-2}-1}$,
$Q_{2^{\ell-2}-1}$ of height at most $\ell-1$, so
$h(H_{2^{\ell+1}})=\ell+1$; and \eqref{eq:peeled-Q} gives
$h(Q_{2^t-1})
 =1+\max\bigl(h(H_{2^{t-1}}),h(Q_{2^{t-1}-1})\bigr)=t$
for $t\ge2$, from the base $h(Q_3)=2$; the linear base has height zero.

\emph{The $T$ recursion.}  For $T_{k,2^\ell}$ put
$\lambda=\ell+\lceil\log_2k\rceil$ and let $s(k)$ be the number of odd
values exceeding $1$ in the halving chain of $k$, so
$s(k)\le\lceil\log_2k\rceil=\lambda-\ell$.  Then
\[
  h\bigl(T^{(1)}_{k,2^\ell}\bigr),\,h\bigl(T^{(2)}_{k,2^\ell}\bigr)
  \;\le\;\lambda+s(k).
\]
The base $k=1$ is the tower value $H_{2^\lambda}$ at height $\lambda$; an
even step recurses at $(k/2,\ell+1)$ with the same $\lambda$ after one
tower step; an odd step multiplies the recursive pair by factors of height
at most $\ell$ (a tower level plus peeled blocks of level $<\ell$) and adds
$Q_{2^\ell-1}$ of height $\ell$, contributing exactly the
$+1$ that separates $s(k)$ from $s\bigl(\tfrac{k-1}2\bigr)$.

\emph{Fills and the odd-degree family.}  The fill
$A_{2^\ell}(S^{(1)},S^{(2)})$ is a chain of $\ell+1$ products whose factors
carry towers and known-powers blocks of level at most $\ell$, so
\[
  h(A_{2^\ell})\le\max\bigl(h(S^{(1)}),h(S^{(2)}),\ell\bigr)+\ell+1.
\]
The
family $Q_{2^{l+1}k+2^l-1}$ ($k\ge1$) runs $T_{2k,2^l}$ and finishes with
$A_{2^{l-1}}$ (for $l=1$, with a single $(x+\beta_0)$ extraction), hence
has height at most
\[
  \bigl(\lambda+s(2k)\bigr)+l+1
  \;\le\;
  2\lambda+1
  \;\le\;
  2\lceil\log_2(m+1)\rceil+1
\]
for its degree $m$: here $s(2k)\le\lambda-l$ absorbs the fill length, and
$\lambda=\lceil\log_2(2^l\cdot 2k)\rceil$ with $2^l\cdot2k\le m+1$,
so $\lambda\le\lceil\log_2(m+1)\rceil$ by monotonicity.  The same bound covers the
good-polynomial gadgets: for $\deg\equiv1\pmod4$ they \emph{are} this
family at $l=1$; $\deg\equiv3\pmod8$ adds one product; and the
$8k+7$ fallback synthesizes one keyed $H_8$ (height $\le4$), runs
$T_{k,8}$, and finishes with the constant-length fill $A_4$, all within
$2\lceil\log_2(m+1)\rceil+4$.

\emph{Branches.}  Let $B(L)$ bound the height of every costed pair of
degree $m$ with $\lceil\log_2m\rceil\le L$, and take
$B(L)=2L+3$.  The $4k+1$ family is one extraction over $T_{2k,2}$:
$1+(\lambda+s)\le2\lambda\le2L$.  The $8k+3$ and $8k+7$ steps add one
square-difference on top of odd-degree or good-polynomial blocks of degree
$m\le(n-1)/2$---since odd $n\ge3$ is not a power of two,
$\lceil\log_2(m+1)\rceil\le L-1$, so these blocks have height at most
$2(L-1)+4=2L+2$---and of a recursive pair of degree at most $(n-1)/4$
(height $\le B(L-2)=2L-1$); hence the new pair has height at most
$2L+3=B(L)$.  The special cases $15,27,31$ are finite, use the same peeled
blocks, and satisfy the bound by direct inspection of their displayed
circuits.  The final combination
$P_n=x\cdot T^{(1)}+T^{(2)}$ adds one product:
$B(L)+1=2\lceil\log_2n\rceil+4$.  For even $n$ the lift from degree
$n-1$ adds one further product on top of the odd-degree schedule, giving
$2\lceil\log_2(n-1)\rceil+5\le2\lceil\log_2n\rceil+5$; this is the
constant $2\lceil\log_2 n\rceil+4+[n\text{ even}]$ of the machine-checked
statement \texttt{polynomial\_height} (\texttt{FastPoly/HeightFinal.lean}),
which covers every $n\ge3$.
\end{proof}

\begin{remark}
The Lean statement \texttt{odd\_realizable\_pairs} carries the decoder,
exact joint multiplication count, and pair-height bound
$2\lceil\log_2n\rceil+3$ for the very circuit it constructs, together
with $h(H_2)\le1$ and $h(H_4)\le2$.
The final assembly in \texttt{FastPoly/HeightFinal.lean} proves
\texttt{odd\_polynomial\_height} and \texttt{polynomial\_height} for
this construction: the complete polynomial has exactly
$\lfloor n/2\rfloor+1$ products and the height bound stated above.
\end{remark}

\subsection{The $T$ recursion and the $4k+1$ family}
\label{sec:splittable-4k+1}
This section defines the recursive pair $T_{k,2^l}$, proves its exact cost,
and proves the coefficient-triangular remainder lemma on which all later uses
of $T$ rely.  It then adds five outer pivots to obtain the splittable
degree-$(4k+1)$ family.  \Cref{fig:T-tower} pictures the tower of known
powers that the recursion consumes.

\begin{figure}[H]
  \centering
  \resizebox{\textwidth}{!}{
\begin{tikzpicture}[>=Latex, font=\small,
    pow/.style={fp active, rounded corners, minimum width=22mm,
      minimum height=7mm, font=\small},
    gad/.style={fp second, rounded corners, font=\footnotesize,
      align=center, inner sep=2pt}]
  \node[pow] (H2) at (0,0) {$H_2 = x^2+\dots$};
  \node[pow] (H4) at (0,1.7) {$H_4 = H_2^2+\dots$};
  \node[pow] (H8) at (0,3.4) {$H_8 = H_4^2+\dots$};
  \node[pow] (H16) at (0,5.1) {$H_{16} = H_8^2+\dots$};
  \node at (0,6.1) {$\vdots$};

  \draw[fp flow] (H2) -- node[right, xshift=1mm, font=\footnotesize]
    {$(H_2+S_1)(H_2-S_1)+S_2$} (H4);
  \draw[fp flow] (H4) -- node[right, xshift=1mm, font=\footnotesize]
    {$(H_4+S_1)(H_4-S_1)+S_2$} (H8);
  \draw[fp flow] (H8) -- node[right, xshift=1mm, font=\footnotesize]
    {$\bigl((H_8+S_1)+S_2\bigr)\bigl((H_8+S_1)-S_2\bigr)+S_3$} (H16);

  \node[gad, left=12mm of H4] (g4) {$S_1 = x+a,\; S_2=e$\\ (even step, $l{=}1$)};
  \node[gad, left=12mm of H8] (g8) {$S_1 = H_2+Q_1,\; S_2 = Q_1$\\ (even step, $l{=}2$)};
  \node[gad, left=12mm of H16] (g16) {$S_1=H_4+Q_3,\; S_2=H_2+Q_1,\; S_3=Q_1$\\ (odd step, $l{=}3$)};
  \draw[fp dependence] (g4) -- ($(H2)!0.5!(H4)$);
  \draw[fp dependence] (g8) -- ($(H4)!0.5!(H8)$);
  \draw[fp dependence] (g16) -- ($(H8)!0.5!(H16)$);

  \node[font=\footnotesize, text=black!85, align=left, text width=118mm,
    anchor=north west] at (-4.6,-0.7)
    {One product per branch per rung; only the two identified bases share
     products.  The parameters inside $S_1,S_2,S_3$ are fresh and sit
     strictly below the leading term, so the top of $T$ equals $H^{\,k}$ and
     the corrections are read off top-down.};
\end{tikzpicture}}
  \caption{The tower of known powers used by the recursion $T_{k,2^l}$ (\Cref{alg:constr-Tk2l,alg:constr-Tk2l-base}).
  In each branch, a rung uses one product
  $(H+S_1)(H-S_1)+S_2$ (even step) or
  $((H+S_1)+S_2)((H+S_1)-S_2)+S_3$ (odd step), producing a monic
  $H_{2^{l+1}}=H_{2^l}^2+\text{lower terms}$; those lower terms carry fresh
  parameters through small $Q$-gadgets.  The index $k$ halves at every rung:
  at level $l$ the recursion computes
  $T_{k,2^l}=H_{2^l}^{\,k}+(\text{terms of degree}\le 2^l(k-1))$, and the
  base $T_{1,2^L}=H_{2^L}$ costs no product.
  The two branches normally use parallel products.  Product sharing occurs
  only in the explicitly identified $l=1$ even base and the immediately
  following odd $l=2$ base, where the required scalar-difference invariant
  has already been established.}
  \label{fig:T-tower}
\end{figure}

We use named parameter blocks in the recursive definition.  If $\mathsf B$
is a block of $q$ fresh parameters, $Q_q[\mathsf B]$ denotes the corresponding
known-powers gadget, with its required power arguments suppressed.  A display
$[\mathsf B_1\mid\cdots\mid\mathsf B_s]$ assigns
$\alpha_0,\ldots,\alpha_{d-1}$ to the blocks from left to right.  The sizes
shown below sum to $d=(k-1)2^l$, so this notation is only a readable
reindexing of one consecutive parameter list.  Recursive $T$ calls inherit
$x$ and the existing power tower; their displays show only the fresh block
and the newly formed top power(s).

\begin{algorithm}[H]
    \caption{Recursive construction of $T_{k,D}$, ordinary branches
    ($D=2^l$; $D\ge4$ in the even branch and $D\ge8$ in the odd branch).}
    \label{alg:constr-Tk2l}
    \begin{algorithmic}
        \State Put $H=H_D$ and $\widetilde H=\widetilde H_D$.
        \If{$k=1$}
            \State $T^{(1)}_{1,D}=H$, \qquad $T^{(2)}_{1,D}=\widetilde H$.
        \ElsIf{$k=2m$}
            \State Put $r=D/2$, $b=(k-2)D$, and allocate
            $[\mathsf I_b\mid\sigma\mid\mathsf B^-_{r-1}
              \mid\delta\mid\mathsf B^+_{r-1}]$.
            \State $U_1=H_r+Q_{r-1}[\mathsf B^+]$,
            $V_1=Q_{r-1}[\mathsf B^-]$,
            $U_2=H_r+\delta$, $V_2=\sigma$.
            \State $H_{2D}=H^2-U_1^2+V_1$, \qquad
            $\widetilde H_{2D}=\widetilde H^2-U_2^2+V_2$.
            \State $T^{(1)}_{k,D}=T^{(1)}_{m,2D}[\mathsf I](H_{2D})$,
            \qquad
            $T^{(2)}_{k,D}=T^{(2)}_{m,2D}[\mathsf I]
              (H_{2D},\widetilde H_{2D})$.
        \Else \Comment{$k=2m+1$}
            \State Put $r=D/2$, $q=D/4$, $b=(k-2)D$, and allocate
            \Statex \hspace{\algorithmicindent}$[\alpha_0\mid\mathsf B^{\rm low}_{D-1}
              \mid\mathsf I_{b-D}\mid\zeta\mid\mathsf B^-_{q-1}
              \mid\varepsilon\mid\mathsf B^0_{q-1}
              \mid\delta\mid\mathsf B^+_{r-1}]$.
            \State $U_1=H_r+Q_{r-1}[\mathsf B^+]$,
            $V_1=H_q+Q_{q-1}[\mathsf B^0]$,
            $W_1=Q_{q-1}[\mathsf B^-]$.
            \State $U_2=H_r+\delta$, $V_2=H_q+\varepsilon$,
            $W_2=\zeta$.
            \State $H_{2D}=(H+U_1)^2-V_1^2+W_1$, \qquad
            $\widetilde H_{2D}=(\widetilde H+U_2)^2-V_2^2+W_2$.
            \State $T^{(1)}_{k,D}=(H-(k-1)U_1)
              T^{(1)}_{m,2D}[\mathsf I](H_{2D})
              +Q_{D-1}[\mathsf B^{\rm low}]$.
            \State $T^{(2)}_{k,D}=(\widetilde H-(k-1)U_2)
              T^{(2)}_{m,2D}[\mathsf I](H_{2D},\widetilde H_{2D})
              +\alpha_0$.
        \EndIf
    \end{algorithmic}
\end{algorithm}

The two shared-product bases are invoked only when
$\widetilde H_D-H_D$ is scalar.  The even $D=2$ base preserves that
difference at degree four, which is exactly the hypothesis needed by the
following odd $D=4$ base.

\begin{algorithm}[H]
    \caption{Shared-product bases for $T_{k,D}$ ($D=2$ with $k$ even,
    and $D=4$ with $k$ odd).}
    \label{alg:constr-Tk2l-base}
      \begin{algorithmic}
          \If{$k=2m$ and $D=2$}
              \State Allocate $[\mathsf I_{2k-4}\mid e\mid a]$ and set
              $H_4=H_2^2-(x+a)^2+e$.
              \State Retain the supplied scalar shift
              $\rho=\widetilde H_2-H_2$ and set
              $\widetilde H_4=H_4+\rho$.
              \Comment{do not recompute the difference}
              \State $T^{(1)}_{k,2}=T^{(1)}_{m,4}[\mathsf I](H_4)$,
              \qquad
              $T^{(2)}_{k,2}=T^{(2)}_{m,4}[\mathsf I]
                 (H_4,\widetilde H_4)$.
          \Else \Comment{$k=2m+1\ge3$ and $D=4$}
              \State Allocate
              $[\alpha_0\mid\mathsf B^{\rm low}_3
                \mid\mathsf I_{4(k-3)}\mid z\mid w\mid v\mid u]$.
              \State $U_1=H_2+x+u$, $V_1=x+v$, $W_1=w$, and
              $H_8=(H_4+U_1)^2-V_1^2+W_1$.
              \State Retain $\rho=\widetilde H_4-H_4$.  For the proof put
              $U_2=U_1-\rho$, $V_2=V_1$, $W_2=W_1+z$, but compute only
              $\widetilde H_8=H_8+z$.
              \State Put $F_1=H_4-(k-1)U_1$ and use the shared affine form
              $F_2=F_1+k\rho
                    =\widetilde H_4-(k-1)U_2$.
              \State $T^{(1)}_{k,4}=F_1
                T^{(1)}_{m,8}[\mathsf I](H_8)
                +Q_3[\mathsf B^{\rm low}]$.
              \State $T^{(2)}_{k,4}=F_2
                T^{(2)}_{m,8}[\mathsf I](H_8,\widetilde H_8)
                +\alpha_0$.
          \EndIf
      \end{algorithmic}
  \end{algorithm}
\begin{remark}[Why the shared odd base is admissible]
    The condition that $\tilde H_4-H_4$ is scalar says exactly that the two
    quartics have the same nonconstant coefficients.  The difference may be
    an active parameter; once $H_4$ and $\tilde H_4$ are supplied to the
    recursive call, that scalar is recoverable from their constant
    coefficients and belongs to the given data.  It is not an additional
    condition on the target polynomial.  The shared odd $l=2$ branch is invoked
    only immediately after the even $l=1$ branch, which constructs
    \[
       \tilde H_4-H_4=\tilde H_2-H_2.
    \]
    Thus the recursion establishes the condition before it uses it.  A
    standalone odd $T_{k,4}$ call with arbitrary quartic inputs is not covered
    by this shared-base definition (it would require a separate two-product
    variant).  No such call occurs in the exact-count construction.
\end{remark}

\begin{lemma}[Exact cost of the $T$ recursion]\label{lem:T-multiplication-count}
    Let $\mu(k,l)$ be the number of multiplication gates used to compute the
    pair $(T^{(1)}_{k,2^l},T^{(2)}_{k,2^l})$ when
    $(x,H_2,\ldots,H_{2^l},\tilde H_{2^l})$ is already available.  For every
    call occurring in the construction---including the shared $l=1$ step and,
    when reached from it, the shared odd $l=2$ step---we have
    \[
       \mu(k,l)=(k-1)2^{l-1}.
    \]
\end{lemma}
\begin{proof}
    Additions, scalar shifts, and multiplication by the displayed integers are
    free in this count; the latter can be implemented by repeated additions.
    The case $k=1$ uses no multiplication and agrees with the formula.

    Suppose first that $k$ is even.  If $l=1$, the shared base computes $H_4$
    with one product and obtains $\tilde H_4=H_4+(\tilde H_2-H_2)$ without a
    second product.  If $l\ge2$, each of the two
    $Q_{2^{l-1}-1}$ blocks costs $2^{l-2}-1$ multiplications by
    \Cref{lem:fill-Q-count}, and the two new degree-$2^{l+1}$ powers cost one
    product each.  Thus in both cases the even recurrence is
    \[
       \mu(k,l)=\mu(k/2,l+1)+2^{l-1}.
    \]

    Now suppose that $k\ge3$ is odd.  In the shared $l=2$ base there is one
    product for the common octic core, one for $Q_3$, and two products for the
    final recursive factors.  The shift $\tilde H_8=H_8+z$ costs
    nothing, so the overhead is $4=2^l$.  For $l\ge3$, the four embedded
    Mersenne gadgets have costs
    \[
       (2^{l-2}-1),\quad (2^{l-3}-1),\quad
       (2^{l-3}-1),\quad (2^{l-1}-1).
    \]
    The two new powers and the two final recursive factors cost four more
    products.  Their sum is exactly $2^l$.  Hence the odd recurrence is
    \[
       \mu(k,l)=\mu((k-1)/2,l+1)+2^l.
    \]

    Strong induction on $k$ now closes the count.  In the even case,
    \[
      \mu(k,l)=\left(\frac{k}{2}-1\right)2^l+2^{l-1}
               =(k-1)2^{l-1},
    \]
    while in the odd case,
    \[
      \mu(k,l)=\left(\frac{k-1}{2}-1\right)2^l+2^l
               =(k-1)2^{l-1}.
    \]
\end{proof}

\begin{lemma}[Cubic degree loss in an odd $T$ step]
\label{lem:odd-T-cubic-loss}
Let $k\ge3$, let $D=2r$ with $r\ge1$, and let $H,U$ be monic of
degrees $D,r$.  Put
\[
  c_k=\binom{k}{2},\qquad \tau_k=2\binom{k}{3}
      =\frac{k(k-1)(k-2)}3,
\]
and
\[
  E(H,U)=\bigl(H-(k-1)U\bigr)(H+U)^{k-1}-H^k
          +c_kU^2H^{k-2}.
\]
Then
\[
  \deg E(H,U)\le (k-2)D+r,
  \qquad
  [x^{(k-2)D+r}]E(H,U)=-\tau_k.
\]
\end{lemma}
\begin{proof}
The binomial theorem gives
\[
  \bigl(H-(k-1)U\bigr)(H+U)^{k-1}-H^k
  =\sum_{t=1}^{k}
    \left(\binom{k-1}{t}-(k-1)\binom{k-1}{t-1}\right)
    H^{k-t}U^t,
\]
where $\binom{k-1}{k}=0$.  The coefficients for $t=1,2,3$ are
$0,-c_k,-\tau_k$.  The added quadratic term cancels $t=2$.
The $t=3$ term has degree $(k-3)D+3r=(k-2)D+r$, while every
$t\ge4$ term has degree $kD-tr\le(k-2)D$.  Monicity of $H,U$
therefore gives both assertions.
\end{proof}

\begin{lemma}[Boundary transport of a remainder pair]
\label{lem:seam-transport}
Let $A_1,A_2$ be monic of the same degree $f\ge0$ and put
$\varphi=[x^{f-1}]A_1$ (so $\varphi=0$ when $f=0$).  Let $e\ge0$ and let
$B_1,B_2$ have degree at most $e$ with
\[
  [x^{e}]B_1=[x^{e}]B_2=\ell,\qquad [x^{e-1}]B_1=a,
\]
where a coefficient of negative index is read as zero.  Then the shifted
combination $P=xA_1B_1+A_2B_2$ has degree at most $f+e+1$, and its two top
rows are
\begin{equation}
  [x^{f+e+1}]P=\ell,\qquad
  [x^{f+e}]P=a+(\varphi+1)\ell.
  \tag{seam-transport}\label{eq:seam-transport}
\end{equation}
In particular both values are determined by the interface
$(\ell,a,\varphi)$ alone: they involve neither $[x^{e-1}]B_2$, nor any
lower coefficient of $B_1,B_2$, nor any coefficient of $A_2$ below the
leading one.
\end{lemma}
\begin{proof}
Both products have degree at most $f+e$, so $\deg P\le f+e+1$,
\[
  [x^{f+e+1}]P=[x^{f+e}](A_1B_1),
  \qquad
  [x^{f+e}]P=[x^{f+e-1}](A_1B_1)+[x^{f+e}](A_2B_2).
\]
If $e\ge1$, the top-two specialization \eqref{eq:monic-top-two} of
\Cref{lem:monic-cauchy-transport} gives $[x^{f+e}](A_iB_i)=[x^e]B_i=\ell$
for $i=1,2$ and
$[x^{f+e-1}](A_1B_1)=[x^{e-1}]B_1+[x^{f-1}]A_1\,[x^{e}]B_1=a+\varphi\ell$.
If $e=0$, then $B_1=B_2=\ell$ are scalars, $a=0$, and the same three
values are read off directly from $A_i\ell$.  Adding the contributions
gives \eqref{eq:seam-transport}; the final sentence records which
coefficients the two formulas use.
\end{proof}

\begin{lemma}\label{lem:Rk2l}
    Let $l\ge 2$ and $k\ge 1$, and supply monic polynomials
    $H_2,\ldots,H_{2^l},\tilde H_{2^l}$ of their indicated degrees.  In the
    exceptional case where $l=2$ and $k\ge3$ is odd, assume that
    $\tilde H_4-H_4$ is a scalar and use the shared-product base in
    \Cref{alg:constr-Tk2l-base}.
    Assume that $\mathbb F$ is $(k2^l)$-admissible.
    Define the remainder polynomials by
    \[
      R^{(1)}_{k,2^l}\coloneqq T^{(1)}_{k,2^l}(x,H_2,\ldots,H_{2^l})-H_{2^l}^k,
      \qquad
      R^{(2)}_{k,2^l}\coloneqq T^{(2)}_{k,2^l}(x,H_2,\ldots,H_{2^l},\tilde H_{2^l})-\tilde H_{2^l}^k.
    \]
    Then:
    \begin{enumerate}
        \item By definition,
        \(
          T^{(1)}_{k,2^l}=H_{2^l}^k+R^{(1)}_{k,2^l}
        \)
        and
        \(
          T^{(2)}_{k,2^l}=\tilde H_{2^l}^k+R^{(2)}_{k,2^l}.
        \)
        \item If $k=1$, both remainders vanish.  If $k\ge2$, then
        $\deg R^{(1)}_{k,2^l}=\deg R^{(2)}_{k,2^l}=(k-1)2^l$.
        \item With $d=(k-1)2^l$, the remainder pair is
        coefficient-triangular in the sense of
        \Cref{def:coefficient-triangular}, in the row order specified by
        the stage tables below (with each known-powers block ordered as in
        \Cref{lem:Q-unitriangular}).  Hence the raw parameters
        $\alpha_0,\ldots,\alpha_{d-1}$ are extractable from
        $xR^{(1)}_{k,2^l}+R^{(2)}_{k,2^l}$, and
        $(T^{(1)}_{k,2^l},T^{(2)}_{k,2^l})$ is compatible on $\rng d$,
        given $(H_2,\ldots,H_{2^l},\tilde H_{2^l})$.
        \item If $k\ge2$, put $D=2^l$ and retain the monic
        degree-$D/2$ auxiliary polynomials $S^{(i)}_1$ from the defining
        branch.  With $H_1=H_D$, $H_2=\widetilde H_D$,
        $\sigma_i=[x^{D/2-1}]S^{(i)}_1$, and
        $h_i=[x^{D-1}]H_i$, one has
        \[
          [x^d]R^{(i)}_{k,D}=-\gamma_k,
          \qquad
          [x^{d-1}]R^{(i)}_{k,D}
            =-\gamma_k\bigl(2\sigma_i+(k-2)h_i\bigr),
        \]
        where
        $\gamma_k=k/2$ for even $k$ and
        $\gamma_k=k(k-1)/2$ for odd $k$.
    \end{enumerate}
\end{lemma}
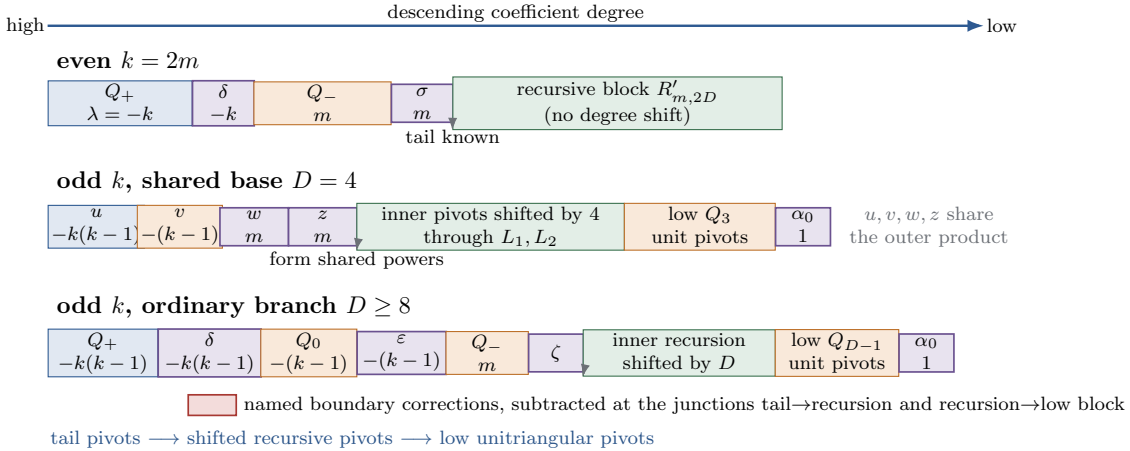
\begin{figure}[H]
  \centering
  \resizebox{\textwidth}{!}{
\begin{tikzpicture}[x=1cm,y=1cm,>=Latex]
  \draw[fp decode] (0,6.15) -- node[fp label, above] {descending coefficient degree}
    (13.65,6.15);
  \node[fp label, anchor=east] at (0,6.15) {high};
  \node[fp label, anchor=west] at (13.65,6.15) {low};

  \node[fp panel title] at (0,5.65) {even $k=2m$};
  \node[fp active, minimum width=21mm, anchor=west] (ep) at (0,5.02)
    {$Q_+$\\$\lambda=-k$};
  \node[fp pivot, minimum width=9mm, anchor=west] (ed) at (2.1,5.02)
    {$\delta$\\$-k$};
  \node[fp second, minimum width=20mm, anchor=west] (em) at (3.0,5.02)
    {$Q_-$\\$m$};
  \node[fp pivot, minimum width=9mm, anchor=west] (es) at (5.0,5.02)
    {$\sigma$\\$m$};
  \node[fp recursive, minimum width=48mm, anchor=west] (er) at (5.9,5.02)
    {recursive block $R'_{m,2D}$\\(no degree shift)};
  \draw[fp dependence] (es.south east) to[bend right=12]
    node[fp label, below] {tail known} (er.south west);

  \node[fp panel title] at (0,3.87) {odd $k$, shared base $D=4$};
  \node[fp active, minimum width=13mm, anchor=west] (ou) at (0,3.22)
    {$u$\\$-k(k-1)$};
  \node[fp second, minimum width=12mm, anchor=west] (ov) at (1.3,3.22)
    {$v$\\$-(k-1)$};
  \node[fp pivot, minimum width=10mm, anchor=west] (ow) at (2.5,3.22)
    {$w$\\$m$};
  \node[fp pivot, minimum width=10mm, anchor=west] (oz) at (3.5,3.22)
    {$z$\\$m$};
  \node[fp recursive, minimum width=39mm, anchor=west] (orr) at (4.5,3.22)
    {inner pivots shifted by $4$\\through $L_1,L_2$};
  \node[fp second, minimum width=22mm, anchor=west] (oq) at (8.4,3.22)
    {low $Q_3$\\unit pivots};
  \node[fp pivot, minimum width=8mm, anchor=west] at (10.6,3.22)
    {$\alpha_0$\\$1$};
  \draw[fp dependence] (oz.south east) to[bend right=12]
    node[fp label, below] {form shared powers} (orr.south west);
  \node[fp label, anchor=west, text=fpGray] at (11.65,3.22)
    {$u,v,w,z$ share\\the outer product};

  \node[fp panel title] at (0,2.05) {odd $k$, ordinary branch $D\ge8$};
  \node[fp active, minimum width=16mm, anchor=west] (qp) at (0,1.38)
    {$Q_+$\\$-k(k-1)$};
  \node[fp pivot, minimum width=15mm, anchor=west] (delta) at (1.6,1.38)
    {$\delta$\\$-k(k-1)$};
  \node[fp second, minimum width=14mm, anchor=west] (qz) at (3.1,1.38)
    {$Q_0$\\$-(k-1)$};
  \node[fp pivot, minimum width=13mm, anchor=west] (eps) at (4.5,1.38)
    {$\varepsilon$\\$-(k-1)$};
  \node[fp second, minimum width=12mm, anchor=west] (qm) at (5.8,1.38)
    {$Q_-$\\$m$};
  \node[fp pivot, minimum width=8mm, anchor=west] (zeta) at (7.0,1.38)
    {$\zeta$};
  \node[fp recursive, minimum width=28mm, anchor=west] (rr) at (7.8,1.38)
    {inner recursion\\shifted by $D$};
  \node[fp second, minimum width=18mm, anchor=west] (lq) at (10.6,1.38)
    {low $Q_{D-1}$\\unit pivots};
  \node[fp pivot, minimum width=8mm, anchor=west] at (12.4,1.38)
    {$\alpha_0$\\$1$};
  \node[fp seam, minimum width=7mm, minimum height=3mm] at (2.4,0.62) {};
  \node[fp label, anchor=west] at (2.82,0.62)
    {named boundary corrections, subtracted at the junctions
     tail$\to$recursion and recursion$\to$low block};
  \draw[fp dependence] (zeta.south east) to[bend right=10] (rr.south west);
  \node[fp label, anchor=west, text=fpBlue!80!black] at (0,0.12)
    {tail pivots $\longrightarrow$ shifted recursive pivots $\longrightarrow$ low unitriangular pivots};
\end{tikzpicture}}
  \caption{Decoder map for the three branches of the central
  $T_{k,2^l}$ remainder lemma.  Each coloured segment is a consecutive row
  interval; its label is the active parameter block and its constant pivot
  slope.  In every branch the fresh tail is decoded first.  The resulting
  powers and monic factors then expose the recursive block, after which the
  low unitriangular $Q$-block is read.  The red markers are the
  parameter-free boundary rows that must be subtracted before crossing into
  the next segment.}
  \label{fig:Rk2l-stages}
\end{figure}
\begin{proof}
    \emph{Reading guide.}
    The algebra below certifies the three coloured rows of
    \Cref{fig:Rk2l-stages}.  Every branch follows the same four-step template:
    derive one exact remainder identity; identify its highest nonzero term;
    verify the tail pivots and their two seam rows; then transport the recursive
    block and append the low unitriangular block.  On a first reading, one may
    follow the displayed identity, stage table, and final concatenation in each
    branch.  The paragraphs headed ``seam rows'' justify precisely the red
    boundary markers in the figure.  \Cref{alg:decode-Rk2l} restates the same
    argument as an executable decoder.

    Assertion (1) is the definition.  We prove (2--4) simultaneously by
    induction on $k$.  Carrying the two coefficients in (4) is essential:
    they are the parameter-free corrections at the boundaries between two
    consecutive decoder blocks.
    For $k=1$, we have $T^{(1)}_{1,2^l}=H_{2^l}$ and $T^{(2)}_{1,2^l}=\tilde H_{2^l}$ by the defining algorithm, hence $R^{(1)}_{1,2^l}=R^{(2)}_{1,2^l}=0$ and the claims are trivial.

    We may therefore assume $k\ge2$ below.  For the remainder of the proof
    put
    \[
       D=2^l,\qquad d=(k-1)D.
    \]
    In an even step the recursive call has size
    $(k/2)(2D)=kD$, and in an odd step it has size
    $((k-1)/2)(2D)=(k-1)D<kD$.  Thus $kD$-admissibility supplies the
    induction hypothesis in both cases.
    In the odd branch with $l=2$ this automatically means $k\ge3$, so all
    indices in the shared-base formulas are defined.

      Within each known-powers block, parameter pivots use the recursive
      row order of \Cref{lem:Q-unitriangular}, rather than the raw order in
      \Cref{alg:constr-known-2n-1}.  The stage tables first recover factor
      coefficients; the known-powers decoder then recovers the raw block.
      Row permutations are confined to their displayed blocks and do not
      change any coefficient cutoff.  In the proof and stage tables below,
      $\alpha_j$ denotes the parameter assigned to row $j$ by this blockwise
      permutation; the construction and final decoder retain the raw order.

      Each branch is now proved in the same order: one exact remainder
      identity, its degree/support bounds, its tail stage table, and finally
      the recursive and low blocks.  Every product transport below is an
      instance of the causal monic Cauchy rule
      \eqref{eq:monic-Cauchy}; together with
      \Cref{lem:Q-unitriangular}, it says that the current row has unit slope
      in the current factor coefficient and otherwise uses only higher rows.
      Every seam correction is an instance of \Cref{lem:seam-transport}:
      the two rows at which a decoder block ends receive the top two
      coefficients of the block below, carried across the monic factor that
      multiplies it and across the shift $x$ of the combined polynomial.
      The inner remainder pair is carried this way in all three branches,
      with the interface $(\ell,a,\varphi)$ fixed once and for all in the
      next paragraph; the even branch carries one further pair, its
      quadratic binomial term.

      \smallskip\noindent\emph{Inner boundary values.}
      Put $m=\lfloor k/2\rfloor$ and $h=[x^{D-1}]H_D$, recall
      \[
        \gamma_t=\begin{cases}t/2&t\text{ even},\\ t(t-1)/2&t\text{ odd},\end{cases}
        \qquad \gamma_1=0,
      \]
      and define the three fixed seam values
      \[
        \ell_m=-\gamma_m,\qquad
        a_m=-2\gamma_m\bigl(1+(m-1)h\bigr),\qquad
        \theta_m=a_m+(h+1)\ell_m+\mathbf 1_{\{k=3\}}.
        \tag{inner-boundary}\label{eq:inner-boundary}
      \]
      Let $R'^{(i)}=R^{(i)}_{m,2D}$ be the inner remainder pair of the
      recursive call at $(m,l+1)$ and put $e=(m-1)2D$.  We claim
      \[
        [x^{e}]R'^{(1)}=[x^{e}]R'^{(2)}=\ell_m,
        \qquad
        [x^{e-1}]R'^{(1)}=a_m .
        \tag{inner-top-two}\label{eq:inner-top-two}
      \]
      If $m=1$ both inner remainders vanish and $\gamma_1=0$, so there is
      nothing to prove.  If $m\ge2$, assertion~(4) of the induction
      hypothesis at $(m,l+1)$ gives $[x^e]R'^{(i)}=-\gamma_m$ and
      $[x^{e-1}]R'^{(1)}=-\gamma_m\bigl(2\sigma'_1+(m-2)h'_1\bigr)$, where
      $\sigma'_1=[x^{D-1}]S'^{(1)}_1$ and $h'_1=[x^{2D-1}]H_{2D}$ refer to
      the inner call.  Since $l+1\ge3$, that call is an ordinary branch of
      \Cref{alg:constr-Tk2l}, so $S'^{(1)}_1=H_D+Q_{D-1}[\mathsf B^+]$ with
      $Q_{D-1}$ monic by \Cref{lem:Q-unitriangular}, whence
      $\sigma'_1=h+1$.  In each of the three branches below, $H_{2D}$ is
      $H_D^2$ or $(H_D+U_1)^2$ plus terms of degree at most $D$, where
      $\deg U_1\le D/2<D-1$; hence $h'_1=2h$.  Substituting,
      $[x^{e-1}]R'^{(1)}=-\gamma_m\bigl(2(h+1)+2(m-2)h\bigr)=a_m$.
      The second inner branch has the same leading coefficient $\ell_m$;
      its next coefficient is never needed, because by
      \Cref{lem:seam-transport} it does not reach a seam row.

      In all three tables, $K$ consists only of field constants and the
      supplied power coefficients (including the tilded top power).  We use
      the stage-table convention above with
      $\alpha_{>j}=(\alpha_{j+1},\ldots,\alpha_{d-1})$.  In particular,
      quantities decoded in a higher row are substituted; they are not added
      to $K$ as free side information.

      \paragraph{Even $k$.}
      Let $k=2m$, put $b=(k-2)D$ and $r=D/2$, and write
      \[
        U=S^{(1)}_1=H_r+Q_+,\quad V=S^{(1)}_2=Q_-,\quad
        \widetilde U=S^{(2)}_1=H_r+\delta,\quad
        \widetilde V=S^{(2)}_2=\sigma.
      \]
      Here $Q_+$ and $Q_-$ have degree $r-1$, while $\delta$ and
      $\sigma$ are scalars.  In the formulas below $H_1=H_D$ and
      $H_2=\widetilde H_D$, and

      \[
      E_1=-U^2+V,\qquad E_2=-\widetilde U^2+\sigma,\qquad
      R'_i=R^{(i)}_{m,2D}.
      \]
      \smallskip\noindent\emph{Identity, degree, and top boundary.}
      The tail parameters have the order
      \[
        (\sigma,\eta^-_0,\ldots,\eta^-_{r-2},\delta,
          \eta^+_0,\ldots,\eta^+_{r-2})
        =(\alpha_{b},\ldots,\alpha_{b+D-1}),
        \tag{R-even-block}
      \]
      where the $\eta$'s are the row-ordered pivot parameters of the two
      $Q$-polynomials.
      The recursive definition and the binomial theorem give the single
      identity
      \[
        R^{(i)}_{k,D}=mE_iH_i^{k-2}+\widehat R_i+R'_i,
        \tag{R-even-exp}
      \]
      with $\deg\widehat R_i,\deg R'_i\le b$.  Explicitly,
      $\widehat R_i=\sum_{q=2}^{m}\binom mq E_i^qH_i^{k-2q}$;
      terms with $q\ge3$ have degree at most $b-D$.

      The summand $-m(S^{(i)}_1)^2H_i^{k-2}$ has degree $d$ and
      leading coefficient $-m$.  The $S^{(i)}_2$ companion has degree at
      most $b+r-1$, while $\widehat R_i$ and $R'_i$ have degree at most
      $b$.  Hence the remainder has exact degree $d$, and its top two
      coefficients are
      \begin{equation}
        [x^d]R^{(i)}_{k,D}=-m,\qquad
        [x^{d-1}]R^{(i)}_{k,D}
          =-m\bigl(2\sigma_i+(k-2)h_i\bigr),
        \tag{R-top-two-even}\label{R-top-two-even}
      \end{equation}
      where $\sigma_i=[x^{r-1}]S^{(i)}_1$ and
      $h_i=[x^{D-1}]H_i$.  This proves assertions~(2) and~(4) for the
      even branch.

      \smallskip\noindent\emph{Tail pivots and support.}
      Applying \eqref{eq:monic-Cauchy} to the first-order terms gives the complete tail
      table
      \[
      \begin{array}{c|c|c|c}
        \text{rows }j&\text{active block}&\lambda_j&\text{source}\\ \hline
        b+D-1,\ldots,b+r+1
          &\alpha_{b+r+1},\ldots,\alpha_{b+D-1}&-2m=-k&-xU^2H_1^{k-2}\\
        b+r&\alpha_{b+r}=\delta&-2m=-k&-\widetilde U^2H_2^{k-2}\\
        b+r-1,\ldots,b+1
          &\alpha_{b+1},\ldots,\alpha_{b+r-1}&m&xVH_1^{k-2}\\
        b&\alpha_b=\sigma&m&\sigma H_2^{k-2}\\
        b-1,\ldots,0&\text{recursive block}&\text{inner pivot}&R'_1,R'_2
      \end{array}
      \tag{R-even-table}
      \]
      For example, the pivot parameter $\eta^+_s$ first occurs as
      $2\eta^+_s$ plus higher-row terms in $U^2$ at degree $r+s$; the $x$-shift and the monic
      factor $H_1^{k-2}$ put it in row $b+r+s+1$, with slope $-2m$.
      The other rows follow from the same Cauchy-product calculation.  In a
      first-branch row $j$ the displayed occurrence lies in
      $[x^{j-1}]R^{(1)}$, and replacing any monic leading term by a lower
      coefficient can only lower its degree; hence that parameter occurs in
      the first component only through degree $j-1$.  The analogous
      second-branch occurrence lies in $[x^j]R^{(2)}$ and nowhere above
      degree $j$.  This proves the two component-support inequalities for all
      four tail blocks, not just the combined pivot formulas.  At the
      two bottom tail rows one must retain the boundary error

      \[
        E^{\rm bd}=x\widehat R_1+\widehat R_2+xR'_1+R'_2.
      \]
      \smallskip\noindent\emph{Seam rows.}
      Let $s=[x^{r-1}]U$; here $h=[x^{D-1}]H_1$ is the value in
      \eqref{eq:inner-boundary}.  The two seam rows $b+1$ and $b$ of
      $E^{\rm bd}$ come from two instances of \Cref{lem:seam-transport},
      both landing on these rows because $f+e=b$ in each:
      \begin{itemize}
        \item the quadratic binomial term
        $\binom m2E_i^2H_i^{k-4}$ (for $m=1$ the sum $\widehat R_i$ is
        empty and $\binom m2=0$), with
        $(A_i,B_i,f,e)=\bigl(H_i^{k-4},\binom m2E_i^2,(k-4)D,2D\bigr)$:
        here $\varphi=(k-4)h$; $\ell=\binom m2$ because
        $E_i^2=U_i^4+\cdots$ is monic of degree $2D$; and
        $a=\binom m2\,[x^{2D-1}]U^4=4s\binom m2$, because $U^2V$ has degree
        at most $3r-1<2D-1$;
        \item the recursive pair, with $(A_i,B_i,f,e)=(1,R'_i,0,b)$: here
        $\varphi=0$ and $(\ell,a)=(\ell_m,a_m)$ by \eqref{eq:inner-top-two},
        since $e=(m-1)2D=b$ in the even branch.
      \end{itemize}
      The $q\ge3$ terms of $\widehat R_i$ have degree at most $b-D<b-1$ and
      meet neither row.  Adding the two instances of
      \eqref{eq:seam-transport} gives
      \[
        [x^{b+1}]E^{\rm bd}=\binom m2-\gamma_m,\qquad
        [x^{b}]E^{\rm bd}
        =\binom m2\bigl(4s+(k-4)h+1\bigr)+a_m+\ell_m,
        \tag{R-even-boundary-values}\label{eq:R-even-boundary-values}
      \]
      where $\ell_m=-\gamma_m$ and
      $a_m=-2\gamma_m(1+(m-1)h)$ as in \eqref{eq:inner-boundary} (with both
      zero for $m=1$).
      These are constants once the already exposed tail is substituted.
      \Cref{lem:seam-transport} also shows explicitly why no inner parameter
      is present in either boundary equation: the recursive pair enters only
      through its interface $(\ell_m,a_m,\varphi)$, and each of these values
      is fixed by \eqref{eq:inner-top-two}.
      Subtracting them gives the pivots in the last two tail rows.  Every
      other summand in a displayed row either has lower degree or contains a
      strictly higher coefficient of the relevant $Q$-polynomial.  This proves
      the support and pivot assertions in the tail.

      \smallskip\noindent\emph{Recursive block and concatenation.}
      Once this tail is known, the new powers $H_{2D}$ and
      $\widetilde H_{2D}$ are known as well.  The remaining parameters are the
      unshifted inner block $\alpha_0,\ldots,\alpha_{b-1}$.  Since the even
      recursion is the direct pair $(R'_1,R'_2)$ plus known tail terms, the
      induction hypothesis applies at the same rows $0,\ldots,b-1$ (there is
      no degree shift in the even branch).  The side data
      $H_{2D},\widetilde H_{2D}$ and the coefficients of the other displayed
      terms are polynomially recoverable from
      $(K;\alpha_b,\ldots,\alpha_{d-1})$.  Consequently the tail rows and the
      inner rows are glued by
      \Cref{lem:triangular-block-concatenation}; in particular, treating the
      newly formed powers as known here is a substitution of higher-block
      formulas, not an enlargement of the cutoff.  Finally the top-two coefficient
      calculation in \eqref{R-top-two-even} shows that all four coefficients
      in degrees $d$ and $d-1$ are parameter-free (the top coefficients of
      the displayed $S_1$ blocks are fixed by monicity and the known powers).

      \paragraph{Odd $k$, shared $l=2$ base.}
      Put $D=4$, $d_0=4(k-2)$, $d=4(k-1)$, $m=(k-1)/2$, and
      $\rho=\widetilde H_4-H_4$; since both quartics are supplied,
      $\rho\polyfrom K$ and is an $x$-independent known scalar.
      The base definitions give
      \[
      \begin{aligned}
        S^{(1)}_1&=H_2+x+u,&
        S^{(1)}_2&=x+v,&
        S^{(1)}_3&=w,\\
        S^{(2)}_1&=S^{(1)}_1-\rho,&
        S^{(2)}_2&=S^{(1)}_2,&
        S^{(2)}_3&=w+z,
      \end{aligned}
      \]
      with $(z,w,v,u)=(\alpha_{d_0},\alpha_{d_0+1},
      \alpha_{d_0+2},\alpha_{d_0+3})$.  The shared identities
      $\widetilde H_4+S^{(2)}_1=H_4+S^{(1)}_1$ and
      $\widetilde H_8=H_8+z$ show that both branches use the same product.
      Here $h=[x^3]H_4$, and $\ell_m,a_m,\theta_m$ are the inner boundary
      values \eqref{eq:inner-boundary}; put
      \[
        \tau=\frac{k(k-1)(k-2)}3 .
      \]
      \smallskip\noindent\emph{Tail pivots, seam rows, and top boundary.}
      The factors $F_1,F_2$ of \Cref{alg:constr-Tk2l-base} are monic of
      degree $4$ with $[x^3]F_1=h$, because $\deg U_1=2$.  The inner pair
      at $(m,3)$ has degree at most $e=8(m-1)=d_0-4$ and the top two
      coefficients \eqref{eq:inner-top-two}.  \Cref{lem:seam-transport}
      with $(A_1,A_2,B_1,B_2,f,e)=(F_1,F_2,R'^{(1)},R'^{(2)},4,d_0-4)$ and
      $\varphi=h$ shows that the product $xF_1R'^{(1)}+F_2R'^{(2)}$
      contributes exactly $\ell_m$ in the $w$ row $d_0+1$ and
      $a_m+(h+1)\ell_m$ in the $z$ row $d_0$, and nothing above.  The low
      term $xQ_3$ contributes its known leading coefficient $1$ in row $4$,
      which is the $z$ row exactly when $k=3$; this is the indicator in
      $\theta_m$.
      After subtracting the terms depending on already exposed parameters,
      the first row therefore has the fixed correction $-m-\tau$, the $w$
      row has the fixed recursive correction $\ell_m$, and the $z$ row has
      the fixed correction $\theta_m$.  Direct expansion of the four
      boundary rows then gives
      \[
      \begin{array}{c|c|c}
        \text{row }j&\text{parameter}&\lambda_j\\ \hline
        d-1&u&-k(k-1)\\
        d-2&v&-(k-1)\\
        d-3&w&m\\
        d-4=d_0&z&m
      \end{array}
      \tag{R-odd-base-table}
      \]
      after these corrections and the preceding rows have been substituted.
      To verify the table symbolically, inspect the highest row in
      $D_R=xR^{(1)}+R^{(2)}$ in which each scalar can occur.  The scalar $u$
      is the constant coefficient of the monic quadratic $S^{(1)}_1$ (and of
      $S^{(2)}_1$ up to the known shift $\rho$); in the principal term
      $-\binom{k}{2}(S^{(1)}_1)^2H_4^{k-2}$ its first-branch occurrence is
      shifted by $x$ to row $d-1$, with slope
      $-2\binom{k}{2}=-k(k-1)$.  Every occurrence of $u$ in the second branch
      or in a term containing at least three copies of $S_1$ is at least one
      row lower.  Similarly, $v$ first occurs through the cross term in
      $-m(S^{(1)}_2)^2$, in row $d-2$ with slope $-2m$; $w$ first occurs in
      the first-branch linear $mS^{(1)}_3$ term, in row $d-3$ with slope $m$;
      and $z$ first occurs in the corresponding second-branch term, in row
      $d-4$ with slope $m$.  The monic factors multiplying these terms can
      contribute only higher already-known coefficients by
      \eqref{eq:monic-Cauchy}.
      This degree inspection also proves the component-support condition: in
      the first component
      a parameter assigned to row $j$ occurs only through degree $j-1$, and
      in the second only through degree $j$.  Thus the four displayed slopes
      are constant and nonzero under the stated admissibility hypothesis.
      The same four-row expansion also proves the strengthened boundary
      assertion: with $\sigma_i=[x^{1}]S^{(i)}_1$ and
      $h_i=[x^3]H_i$,
      \[
        [x^d]R^{(i)}_{k,4}=-\frac{k(k-1)}2,\qquad
        [x^{d-1}]R^{(i)}_{k,4}
          =-\frac{k(k-1)}2\bigl(2\sigma_i+(k-2)h_i\bigr).
      \]
      These are parameter-free because the constant shifts in the displayed
      $S_1$ blocks do not change $\sigma_i$.
      The first coefficient is nonzero by admissibility, and therefore also
      proves $\deg R^{(i)}_{k,4}=d$ in the shared-base branch.
      \smallskip\noindent\emph{Shifted recursive and low blocks.}
      After the tail is recovered, the inner pair is multiplied by the monic
      degree-four factors
      $H_4-(k-1)S^{(1)}_1$ and
      $\widetilde H_4-(k-1)S^{(2)}_1$.  At row $4$, the term
      $[x^4](xQ_3)=1$ is a known boundary contribution.  If $k>3$ it is
      subtracted before the first shifted pivot in
      \Cref{lem:triangular-shift}; if $k=3$ the shifted interval is empty,
      and the same contribution is the final tail boundary already recorded by the
      $\mathbf 1_{\{k=3\}}$ term in $\theta_m$.  The shifted induction
      hypothesis therefore applies to the rows $4,\ldots,d_0-1$, including
      the overlap of the two factors.  The coefficients of those factors are
      polynomially recoverable from
      $(K;\alpha_{d_0},\ldots,\alpha_{d-1})$, so
      \Cref{lem:triangular-shift} and
      \Cref{lem:triangular-block-concatenation} justify the substitution of
      the decoded tail.  The low rows are the unitriangular $Q_3$ block and
      the scalar $\alpha_0$; \Cref{lem:Q-unitriangular} supplies the pivots on
      rows $1,2,3,0$.  This proves the shared-base case without a
      separate composition argument.

      \paragraph{Odd $k$, $l\ge3$.}
      Put $r=D/2$, $q=D/4$, $b=(k-2)D$, $m=(k-1)/2$, and
      $c=k(k-1)/2$.
      Abbreviate
      \[
        U=H_r+Q_+,\quad V=H_q+Q_0,\quad W=Q_-,
        \qquad
        \widetilde U=H_r+\delta,\quad
        \widetilde V=H_q+\varepsilon,\quad
        \widetilde W=\zeta.
      \]
      The tail order is
      \[
        (\zeta,Q_-\text{-block},\varepsilon,Q_0\text{-block},
          \delta,Q_+\text{-block})
        =(\alpha_{b},\ldots,\alpha_{b+D-1}),
        \tag{R-odd-block}
      \]
      with block lengths $1,q-1,1,q-1,1,r-1$.  Set
      \[
      \begin{aligned}
        L_1&=H-(k-1)U,&
        L_2&=\widetilde H-(k-1)\widetilde U,\\
        K_1&=L_1(H+U)^{k-3},&
        K_2&=L_2(\widetilde H+\widetilde U)^{k-3},\\
        G_1&=-V^2+W,&
        G_2&=-\widetilde V^2+\zeta.
      \end{aligned}
      \]
      \smallskip\noindent\emph{Identity and error separation.}
      The factors $L_i$ are monic of degree $D$ and the factors $K_i$ are
      monic of degree $b$.  An exact binomial expansion
      of the odd recursion gives
      \begin{align}
        R^{(1)}_{k,D}&=-cU^2H^{k-2}+mG_1K_1
             +L_1R'^{(1)}+Q_{D-1}+E_1,\\
        R^{(2)}_{k,D}&=-c\widetilde U^2\widetilde H^{k-2}+mG_2K_2
             +L_2R'^{(2)}+\alpha_0+E_2,
        \tag{R-odd-block-exp}
      \end{align}
      where $R'$ is the inner remainder pair at $(m,l+1)$.  We split the
      error, since a single coarse degree bound does not prove the required
      support at the two adjacent windows.  With
      $(H_1,U_1)=(H,U)$ and $(H_2,U_2)=(\widetilde H,\widetilde U)$, set
      \[
      \begin{aligned}
      E_i^{U}&\coloneqq
        L_i(H_i+U_i)^{k-1}-H_i^k+cU_i^2H_i^{k-2},\\
      E_i^{G}&\coloneqq
        L_i\sum_{q=2}^{m}\binom mqG_i^q
                       (H_i+U_i)^{k-1-2q}.
      \end{aligned}
      \]
      (An empty sum is zero.)  Then $E_i=E_i^U+E_i^G$.  Put
      $\tau=2\binom{k}{3}=k(k-1)(k-2)/3$.
      \Cref{lem:odd-T-cubic-loss}, applied to $(H_i,U_i)$, gives the first
      two bounds below.  The $q$th summand of $E_i^G$ has degree at most

      \[
        D+qr+(k-1-2q)D
          =kD-\frac{3qD}{2}\le b-D\qquad(q\ge2),
      \]
      and therefore
      \[
        \deg E_i^U\le b+r=d-r,\qquad
        [x^{b+r}]E_i^U=-\tau,\qquad
        \deg E_i^G\le b-D.
        \tag{R-odd-error-bound}\label{eq:R-odd-error-bound}
      \]
      The coefficient at degree $b+r$ is parameter-free.  Hence $xE_1^U$
      meets the tail first at row $b+r+1$
      and $E_2^U$ first at row $b+r$, with a fixed correction in either
      case; every lower coefficient of $E_i^U$ depends only on the
      $Q_+$/$\delta$ block, which has already been exposed when that row is
      processed.  More precisely, using the coefficient of $x^s$ in $Q_+$
      in a term of $E_1^U$ lowers its degree by at least $r-s$, so its
      occurrence in $xE_1^U$ is in a row at most $b+s+1$, strictly below its
      pivot row $b+r+s+1$.  Using the scalar $\delta$ instead of the leading
      term of $\widetilde U$ lowers the degree by $r$, so every occurrence of
      $\delta$ in $E_2^U$ has degree at most $b$, below its pivot row $b+r$.
      Finally, the last bound in \eqref{eq:R-odd-error-bound} puts every
      $G$-error below the entire tail window.  These are the precise
      degree-loss statements used in the stage table.

      \smallskip\noindent\emph{Degree and top boundary.}
      In particular, the principal term
      $-cU_i^2H_i^{k-2}$ has degree $d$ and leading coefficient $-c$;
      the $mG_iK_i$ and $E_i^U$ terms have degree at most $d-r$, the
      product $L_iR'^{(i)}$ has degree at most $b$, and the remaining terms
      are lower still.  Since $c$ is invertible under the admissibility
      hypothesis, this proves the exact-degree assertion in the ordinary odd
      branch.

      The same degree estimate gives the other part of the strengthened
      induction invariant.  The first displayed correction is the only term
      that can reach degrees $d$ and $d-1$; its top two coefficients are
      $-1$ and $-(2\sigma_i+(k-2)h_i)$, where
      $\sigma_i=[x^{r-1}]S^{(i)}_1$ and
      $h_i=[x^{D-1}]H_i$.  Consequently
      \begin{equation}
        [x^d]R^{(i)}_{k,D}=-c,\qquad
        [x^{d-1}]R^{(i)}_{k,D}
          =-c\bigl(2\sigma_i+(k-2)h_i\bigr).
        \tag{R-top-two-odd}\label{R-top-two-odd}
      \end{equation}
      For the shared $D=4$ base the same statement follows from the four-row
      expansion above (the error terms have degree at most $d-2$).

      \smallskip\noindent\emph{Tail pivots and support.}
      The resulting stage table is
      \[
      \begin{array}{c|c|c|c}
        \text{rows }j&\text{active block}&\lambda_j&\text{source}\\ \hline
        d-1,\ldots,b+r+2&Q_+\text{-block}&-2c=-k(k-1)&-cxU^2H^{k-2}\\
        b+r+1&\text{lowest $Q_+$ coefficient}&-2c=-k(k-1)&-cxU^2H^{k-2}\\
        b+r&\delta&-2c=-k(k-1)&-c\widetilde U^2\widetilde H^{k-2}\\
        b+r-1,\ldots,b+q+1&Q_0\text{-block}&-2m=-(k-1)&mxG_1K_1\\
        b+q&\varepsilon&-2m=-(k-1)&mG_2K_2\\
        b+q-1,\ldots,b+1&Q_-\text{-block}&m&mxWK_1\\
        b&\zeta&m&mG_2K_2
      \end{array}
      \tag{R-odd-table}\label{eq:R-odd-table}
      \]
      The source column refers to the combined polynomial
      $D_R=xR^{(1)}+R^{(2)}$.  Here is the support check, separated by
      source.  In the first source, \eqref{eq:monic-Cauchy} isolates the coefficient of
      $U$ (respectively $\widetilde U$) at the current row and all other
      summands use a strictly higher coefficient.  Since $U$ has the
      coefficient-triangular $Q_+$ block and $\widetilde U$ has only the
      scalar $\delta$, this gives the first three rows of the table.  The
      term $E^U_i$ has the same dependence: its top coefficient is the fixed
      $-\tau$, and replacing a leading coefficient of $U_i$ by its scalar
      term loses $r$ degrees, so $\delta$ cannot occur before the $\delta$
      row.  The term $E^G_i$ is below $b-D$ and hence cannot meet any tail
      row.  In the second source, \eqref{eq:monic-Cauchy} applied to $G_iK_i$ first
      exposes the coefficient of $V$ or $W$ at the current row; the
      coefficients of $K_i$ and all remaining summands depend only on the
      already exposed $Q_+$ and $\delta$ blocks.  This gives the four
      stage--2 rows and their slopes.  As in the even case, a first-branch
      pivot in row $j$ is an occurrence in $[x^{j-1}]R^{(1)}$ and no occurrence
      of that active parameter can lie higher, while a second-branch pivot is
      an occurrence in $[x^j]R^{(2)}$ and none can lie higher.  The degree
      bounds for $E^U,E^G$ just proved cover the only terms not visible in
      those two Cauchy-product calculations.  Thus, after substitution, every
      row in the table has exactly the stage-table support form, with the
      displayed nonzero pivot.

      \smallskip\noindent\emph{Seam rows.}
      At row $b+r+1$ the stage-$2$ leading term and
      the cubic term in $E_1$ contribute the fixed correction
      $-m-\tau$, where $\tau=k(k-1)(k-2)/3$.  At rows $b+1$ and $b$ the
      inner pair is carried across $L_1,L_2$.  These factors are monic of
      degree $D$ with $[x^{D-1}]L_1=h$, since $\deg U=r<D-1$; the inner
      pair has degree at most $e=(m-1)2D=b-D$ and the top two coefficients
      \eqref{eq:inner-top-two}.  \Cref{lem:seam-transport} with
      $(A_1,A_2,B_1,B_2,f,e)=(L_1,L_2,R'^{(1)},R'^{(2)},D,b-D)$ and
      $\varphi=h$ therefore shows that $xL_1R'^{(1)}+L_2R'^{(2)}$
      contributes $\ell_m$ at row $b+1$ and $a_m+(h+1)\ell_m$ at row $b$,
      and nothing above row $b+1$ (when $m=1$, equivalently $k=3$, both
      inner remainders vanish and $\ell_m=a_m=0$).  The latter row also
      receives the known leading coefficient of $xQ_{D-1}$ exactly when
      $k=3$, giving $\theta_m$ as in \eqref{eq:inner-boundary}.  A
      corresponding fixed contribution from the same
      stage--$2$/cubic terms can occur in the $\delta$ row $b+r$; it is
      absorbed into the known correction term $F$ there and does not change
      the pivot slope.  After these corrections and the values of the
      already exposed higher blocks have been subtracted, every row in the
      table is exactly the indicated affine pivot.  In particular, the scalar
      $\varepsilon$ is solved before the $Q_-$ ($S_3$) window.

      \smallskip\noindent\emph{Shifted recursive and low blocks.}
      After the tail is known, apply \Cref{lem:triangular-shift} with shift
      $D$ to $(L_1R'^{(1)},L_2R'^{(2)})$ and the induction hypothesis.  The
      inner block has length $b-D$, and every coefficient of $L_i$ is
      polynomially recoverable from
      $(K;\alpha_b,\ldots,\alpha_{d-1})$.  The higher-block clause in
      \Cref{lem:triangular-shift} therefore moves the inner pivots, including
      the overlap of the two factors, to rows $D,\ldots,b-1$ without changing
      their cutoffs.  The principal and error terms outside the recursive
      product now depend only on the decoded tail block, so they are the
      permitted higher-block additions in \Cref{lem:triangular-shift} and are
      subtracted row by row.  At row $D$, $xQ_{D-1}$ contributes its known leading
      coefficient $1$, which is subtracted before the first shifted pivot
      (when $k=3$, this is the boundary row $b=D$ already accounted for).
      Below row $D$, the remaining active block is
      $Q_{D-1}[\alpha_1,\ldots,\alpha_{D-1}]$ together with $\alpha_0$;
      \Cref{lem:Q-unitriangular} gives unit pivots on rows
      $1,\ldots,D-1,0$.  The block-concatenation lemma now glues the tail,
      shifted inner block, and low block, proving the support and pivot
      conditions
      for every row: the three intervals are respectively
      $b,\ldots,d-1$, $D,\ldots,b-1$, and $0,\ldots,D-1$
      (with an empty middle interval when $k=3$).  The top-boundary assertion
      follows from the simultaneous
      two-term coefficient calculation above.  This completes the odd case.

      In each of the even, shared-base, and ordinary odd branches, the glued
      stage table is precisely a coefficient-triangular certificate.
      Applying \Cref{lem:triangular-implies-compatible} gives the extraction
      and causal compatibility claimed in~(3); the simultaneous top-two
      calculation in the same branch gives~(4).  This completes the
      induction and the proof of (2--4).

\end{proof}

\begin{algorithm}[H]
  \caption{Decoder summary for $xR^{(1)}_{k,2^l}+R^{(2)}_{k,2^l}$}\label{alg:decode-Rk2l}
  \begin{algorithmic}
    \Require integers $l\ge 2$, $k\ge 1$, known powers $(H_2,\ldots,H_{2^l})$ and the shifted power $\tilde H_{2^l}$, and $P=xR^{(1)}_{k,2^l}+R^{(2)}_{k,2^l}$; if $l=2$ and $k\ge3$ is odd, require $\tilde H_4-H_4$ to have degree zero in $x$
    \Ensure the parameter block $\alpha_0,\ldots,\alpha_{(k-1)2^l-1}$ (and the derived inputs for the recursive call at $(k',l+1)$, $k'\in\{k/2,(k-1)/2\}$)

    \If{$k=1$}
      \State Output the empty parameter list.
    \ElsIf{$k$ is even}
      \State Use the top-degree window to peel the monic factor in the first-order correction term, recovering $S^{(1)}_1$ (via \Cref{lem:peel-monic-factor} and \Cref{lem:monic-from-power} with $m=2$).
      \State Recover the scalar shift in $S^{(2)}_1=H_{2^{l-1}}+\delta$ from the boundary coefficient of the $(S^{(2)}_1)^2$ branch using \Cref{lem:scalar-shift-square}.
      \State Recover the coefficients of $S^{(1)}_2$ in descending order; the descent ends at degree $(k-2)2^l+1$, where the parameter-free boundary correction in \eqref{eq:R-even-boundary-values} is subtracted.  Decode its embedded $Q$ block.
      \State Recover the scalar $S^{(2)}_2$ from the following boundary row (it enters affinely with slope $\tfrac{k}{2}$).
      \State Derive $H_{2^{l+1}}$ and $\tilde H_{2^{l+1}}$ from the recovered $(S^{(1)}_1,S^{(1)}_2,S^{(2)}_1,S^{(2)}_2)$.
      \State Isolate the recursive combined remainder polynomial $xR^{(1)}_{k/2,2^{l+1}}+R^{(2)}_{k/2,2^{l+1}}$ and recurse at parameters $(k/2,l+1)$.
    \ElsIf{$k\ge3$ is odd and $l=2$}
      \State Recover $u,v,w,z$ from the four descending affine pivots in the shared-base part of the proof of \Cref{lem:Rk2l}.
      \State Derive $H_8$, $\tilde H_8=H_8+z$, and the two degree-$4$ factors multiplying the recursive pair.
      \State Apply the shift-$4$ triangular recurrence to the overlapping monic-factor products, invoking the inner decoder at $((k-1)/2,3)$; then subtract the recovered inner contribution and decode the low $Q_3\mathbin{\oplus}\alpha_0$ block.
    \Else \Comment{$k$ odd and $l\ge3$; $Q_\pm,Q_0$ are the $Q_{r-1}[\mathsf B^\pm]$, $Q_{q-1}[\mathsf B^0]$ blocks of \Cref{alg:constr-Tk2l}}
      \State Read the $Q_+$ block and then the scalar $\delta$ from the first three row blocks of \eqref{eq:R-odd-table}, subtracting the fixed correction $-m-\tau$ at their common boundary.
      \State Read the $Q_0$ block and $\varepsilon$, followed by the $Q_-$ block and $\zeta$, from the remaining four row blocks.  At the last two rows subtract the parameter-free inner boundary coefficients $\ell_m$ and $\theta_m$ of \eqref{eq:inner-boundary} (carried across $L_1,L_2$ by \Cref{lem:seam-transport}).
      \State Form the two monic factors $L_1,L_2$.  Descend through the overlapping polynomial $xL_1R'^{(1)}+L_2R'^{(2)}$ by the shifted triangular recurrence of \Cref{lem:triangular-shift}, invoking the inner decoder at $((k-1)/2,l+1)$; do not divide the two overlapping products separately.
      \State Subtract the recovered inner contribution and decode the low $Q_{2^l-1}\mathbin{\oplus}\alpha_0$ block.
    \EndIf
  \end{algorithmic}
\end{algorithm}

\begin{lemma}[Leading coefficients of the remainder polynomials]\label{lem:Rk2l-leading-coeff}
    Assume the hypotheses in \Cref{lem:Rk2l}, and let $k\ge2$ and
    $D=2^l$.  Put $d=(k-1)D$ and
    \[
      \gamma_k\coloneqq
      \begin{cases}k/2&k\text{ even},\\ k(k-1)/2&k\text{ odd},\end{cases}
      \qquad
      \sigma_i\coloneqq [x^{D/2-1}]S^{(i)}_1,\quad
      h_i\coloneqq [x^{D-1}]H_i,
    \]
    where $H_1=H_D$ and $H_2=\widetilde H_D$.  Then, for $i=1,2$,
    \[
      [x^d]R^{(i)}_{k,D}=-\gamma_k,
      \qquad
      [x^{d-1}]R^{(i)}_{k,D}
        =-\gamma_k\bigl(2\sigma_i+(k-2)h_i\bigr).
      \tag{R-top-two}
    \]
    In particular, the leading-coefficient formula used below is the first
    equality.
\end{lemma}
\begin{proof}
    This is assertion~(4) of \Cref{lem:Rk2l}, recorded separately for later
    reference.  Its simultaneous induction proof is the top-two coefficient
    calculation in the even, shared-base, and ordinary odd branches there.
\end{proof}

\begin{lemma}[Parameter-free top boundary]\label{lem:Rk2l-top-boundary}
    Under the hypotheses of \Cref{lem:Rk2l}, assume $k\ge2$ and put
    $d=(k-1)2^l$.
    All four coefficients
    \[
      [x^d]R^{(1)}_{k,2^l},\quad [x^{d-1}]R^{(1)}_{k,2^l},\quad
      [x^d]R^{(2)}_{k,2^l},\quad [x^{d-1}]R^{(2)}_{k,2^l}
    \]
    are derivable from the given-power data alone; in particular, they are
    independent of the parameter block of the $T_{k,2^l}$ call.
\end{lemma}
\begin{proof}
    By \Cref{lem:Rk2l-leading-coeff}, these coefficients are determined by
    the top two coefficients of $H_i$ and $S^{(i)}_1$.  In every even-$k$
    call (including $l=2$), and in every odd-$k$ call with $l\ge3$,
    $S^{(1)}_1=H_{2^{l-1}}+Q$ with $Q$ monic of degree
    $2^{l-1}-1$, while $S^{(2)}_1=H_{2^{l-1}}+$ a scalar.  Hence their top
    two coefficients are fixed by the given powers and monicity.  In the
    shared $l=2$ base, $S^{(1)}_1=H_2+x+u$ and
    $S^{(2)}_1=S^{(1)}_1-\rho$; again $u$ and $\rho$ affect only the constant
    term.  Substitution in (R-top-two) proves the claim.
\end{proof}

\begin{lemma}[Causal substitution for a perturbed top power]
\label{lem:causal-perturbed-T}
Let $D=2^l\ge4$, put $r=D/2$, and let $M=2k\ge2$ and $N=MD$.
Fix monic lower powers $H_2,\ldots,H_{D/2}$ as given data.  Let $H=H_D$
be a given monic power, let $Q$ be monic of degree $r-1$,
and let $\delta$ be an unknown $x$-independent scalar.  Set
\[
  \widehat H=H+Q,\qquad \widetilde H=\widehat H+\delta .
\]
Let
\[
  S^{(1)}=T^{(1)}_{M,D}(\widehat H),\qquad
  S^{(2)}=T^{(2)}_{M,D}(\widehat H,\widetilde H),\qquad
  \Psi=xS^{(1)}+S^{(2)} .
\]
Assume that $\mathbb F$ is $N$-admissible.  Then
$(S^{(1)},S^{(2)})$ is compatible on $\rng{N-r}$ given the original
known powers, and $\widehat H,\widetilde H$ are polynomially recoverable
from those powers and the corresponding coefficients of $\Psi$.
\end{lemma}
\begin{proof}
Because $M$ is even, this call is not in the exceptional shared odd base.
Thus $N=MD$-admissibility lets us apply
\Cref{lem:Rk2l,lem:Rk2l-top-boundary} at $(M,l)$.
Put $d=N-D$ and write
\[
  Q=x^{r-1}+\sum_{q=0}^{r-2}q_qx^q .
\]
The active quantities occur in the following descending row blocks:
\[
\begin{array}{c|c|c}
  \text{rows of }\Psi&\text{new quantity}&\text{pivot slope}\\ \hline
  d+r-1,\ldots,d+1&q_{r-2},\ldots,q_0&M\\
  d&\delta&M\\
  d-1,\ldots,0&\text{parameters of }R_{M,D}&
       \text{slopes from \Cref{lem:Rk2l}}
\end{array}
\tag{perturbed-T-table}
\]
We justify every row and its cutoff.

First, the binomial expansion and $\deg Q=r-1<D$ give
\[
 [x^{d+q}]\widehat H^M
   =M q_q+F_q(q_{q+1},\ldots,q_{r-2};H)
   \qquad(0\le q\le r-2),
\tag{perturbed-power-pivot}
\]
because a term containing two copies of $Q$ has degree at most $d-2$.
The quantity $q_q$ cannot occur one degree higher, in either
$\widehat H^M$ or $\widetilde H^M$.  The remainders do not meet these rows,
except that $xR^{(1)}$ contributes the fixed coefficient
$[x^d]R^{(1)}=-M/2$ in the last row $d+1$.  Consequently row $d+q+1$ of
$\Psi$ has the first pivot in the table, plus a polynomial in the already
recovered $q_{q+1},\ldots,q_{r-2}$ and the original known powers.  Reading
$q=r-2,r-3,\ldots,0$ therefore recovers all coefficients of $Q$, and hence
$\widehat H$.

At row $d$, a scalar perturbation contributes
\[
 [x^d]\widetilde H^M=[x^d]\widehat H^M+M\delta .
\]
The remaining boundary terms $[x^{d-1}]R^{(1)}$ and
$[x^d]R^{(2)}$ are parameter-free by
\Cref{lem:Rk2l-top-boundary}.  More explicitly, their formulas use only
$[x^{D-1}]\widehat H=[x^{D-1}]\widetilde H=[x^{D-1}]H$ and the fixed top
two coefficients of the auxiliary $S_1$ blocks.  Thus row $d$ has slope
$M$ in $\delta$ and recovers it without circular side information.

After $Q$ and $\delta$ have been recovered, subtract the two powers:
\[
 \Psi-\bigl(x\widehat H^M+\widetilde H^M\bigr)
       =xR^{(1)}+R^{(2)} .
\]
The coefficient-triangular conclusion of \Cref{lem:Rk2l} supplies the last
block of the table.

It remains only to check that this decoder is causal.  A coefficient $q_u$
can occur in either perturbed power only in degrees at most $d+u$; it is
recovered in row $d+u+1$.  The scalar $\delta$ can occur only in the second
component and only in degrees at most $d$; it is recovered in row $d$.
Any occurrence of these quantities in a lower remainder coefficient is
therefore already known before that row is processed.  The last block then
has exactly the two cutoffs asserted by \Cref{lem:Rk2l}: row $j+1$ suffices
for the first component and row $j$ for the second.  Finally, coefficients
in degrees at least $N-r$ agree with the known polynomial $H^M$, while the
coefficient in degree $N-r-1=d+r-1$ is fixed by the monicity of $Q$.
This proves compatibility on $\rng{N-r}$ and also recovers
$\widehat H=H+Q$ and $\widetilde H=\widehat H+\delta$.
\end{proof}

\begin{lemma}[The $4k+1$ family is splittable]\label{lem:4k+1-splittable}
    Let $k\ge1$ and assume that $\mathbb F$ is $(4k+1)$-admissible.
    Define the quadratic base
    \[
      H_2[\alpha_{4k-1},\alpha_{4k}](x)\coloneqq (x+\alpha_{4k})x+\alpha_{4k-1},
      \qquad
      \tilde H_2\coloneqq H_2+\alpha_{4k-2}.
    \]
    Let
    \[
      T^{(1)}_{4k+1}\coloneqq T^{(1)}_{2k,2}[\alpha_0,\ldots,\alpha_{4k-3}](x,H_2),
      \qquad
      T^{(2)}_{4k+1}\coloneqq T^{(2)}_{2k,2}[\alpha_0,\ldots,\alpha_{4k-3}](x,H_2,\tilde H_2),
    \]
    and set $P_{4k+1}\coloneqq xT^{(1)}_{4k+1}+T^{(2)}_{4k+1}$.
    Then $P_{4k+1}$ is decodable, and the pair
    $(T^{(1)}_{4k+1},T^{(2)}_{4k+1})$ is compatible on
    $\rng{4k+1}$ with no auxiliary data.  In particular it is a splittable
    pair for $4k+1$.  Moreover, the displayed construction is a joint
    $2k$-realization and records its monic $H_2$ and $H_4$.
\end{lemma}

\Cref{fig:4k1-crown} pictures the five-pivot crown over the $T_{2k,2}$ call.

\begin{figure}[H]
  \centering
  \resizebox{\textwidth}{!}{
\begin{tikzpicture}[x=1cm,y=1cm,>=Latex]
  \node[fp panel title] at (0,5.05) {(a) Forward construction};
  \node[fp active, minimum width=28mm] (h2) at (1.55,4.22)
    {$H_2=(x+b)x+c$\\one product};
  \node[fp second, minimum width=37mm] (h4) at (5.2,4.22)
    {$H_4=H_2^2-(x+a)^2+e$\\one shared product};
  \node[fp pivot, minimum width=26mm] (ht) at (8.72,4.22)
    {$\widetilde H_4=H_4+\rho$\\no product};
  \node[fp recursive, minimum width=27mm] (t) at (11.75,4.22)
    {$(U,V)=T_{k,4}$\\recursive core};
  \draw[fp flow] (h2) -- (h4);
  \draw[fp flow] (h4) -- (ht);
  \draw[fp flow] (ht) -- (t);
  \node[fp label, below=2mm of t] {$P_{4k+1}=xU+V$};

  \node[fp panel title] at (0,3.05) {(b) Inverse coefficient crown};
  \draw[fp decode] (0,2.62) -- node[fp label, above]
    {top five rows, then the $T$ remainder} (13.55,2.62);
  \node[fp active, minimum width=15mm, anchor=west] (b) at (0,1.92)
    {$b$\\$2k$};
  \node[fp active, minimum width=15mm, anchor=west] (c) at (1.5,1.92)
    {$c$\\$2k$};
  \node[fp second, minimum width=15mm, anchor=west] (a) at (3.0,1.92)
    {$a$\\$-2k$};
  \node[fp second, minimum width=15mm, anchor=west] (e) at (4.5,1.92)
    {$e$\\$k$};
  \node[fp pivot, minimum width=15mm, anchor=west] (rho) at (6.0,1.92)
    {$\rho$\\$k$};
  \node[fp recursive, minimum width=52mm, anchor=west] (inner) at (7.5,1.92)
    {$x(U-H_4^k)+(V-\widetilde H_4^k)$\\apply the $(R^{(1)}_{k,4},R^{(2)}_{k,4})$ decoder};

  \node[fp known, minimum width=25mm] (dh2) at (1.5,0.72) {reconstruct $H_2$};
  \node[fp known, minimum width=25mm] (dh4) at (4.5,0.72) {reconstruct $H_4$};
  \node[fp known, minimum width=24mm] (dht) at (7.15,0.72) {form $\widetilde H_4$};
  \draw[fp dependence] (b.south) -- (dh2.north west);
  \draw[fp dependence] (c.south) -- (dh2.north east);
  \draw[fp dependence] (a.south) -- (dh4.north west);
  \draw[fp dependence] (e.south) -- (dh4.north east);
  \draw[fp dependence] (rho.south) -- (dht.north);
  \draw[fp dependence] (dh2.east) -- (dh4.west);
  \draw[fp dependence] (dh4.east) -- (dht.west);
  \draw[fp dependence] (dht.east) -- (inner.south west);
  \node[fp label, anchor=west, text=fpBlue!80!black] at (9.05,0.72)
    {subtract the recovered powers\\before entering the inner rows};
\end{tikzpicture}}
  \caption{The $4k+1$ construction and its inverse crown.  Forward, the
  quadratic and quartic powers are shared by the two branches of the
  $T_{k,4}$ call.  Backward, the top five coefficients expose
  $b,c,a,e,\rho$ with the displayed constant slopes.  These reconstruct
  $H_2,H_4,\widetilde H_4$ before the decoder subtracts their $k$th powers
  and enters the coefficient-triangular remainder block.}
  \label{fig:4k1-crown}
\end{figure}
\begin{proof}
    Put
    \[
      b=\alpha_{4k},\quad c=\alpha_{4k-1},\quad
      \rho=\alpha_{4k-2},\quad a=\alpha_{4k-3},\quad
      e=\alpha_{4k-4}.
    \]
    The shared $l=1$ step is equivalently
    \[
      H_2=x^2+bx+c,\qquad
      H_4=H_2^2-(x+a)^2+e,\qquad
      \tilde H_4=H_4+\rho,
    \]
    followed by the call
    \[
      (U,V)=T_{k,4}[\alpha_0,\ldots,\alpha_{4k-5}]
         (x,H_2,H_4,\tilde H_4),\qquad P_{4k+1}=xU+V.
    \]
    Write $H_4=x^4+h_3x^3+h_2x^2+h_1x+h_0$.  Direct expansion gives
    \[
      h_3=2b,\quad h_2=b^2+2c-1,\quad
      h_1=2bc-2a,\quad h_0=c^2-a^2+e.
    \]

    Here $\tilde H_4-H_4=\rho$ is a scalar, and
    $(4k+1)$-admissibility implies $4k$-admissibility.  Hence all hypotheses
    of \Cref{lem:Rk2l} hold at $(k,l)=(k,2)$, and
    \[
      U=H_4^k+A,\qquad V=(H_4+\rho)^k+B,\qquad
      \deg A,\deg B\le4(k-1).
    \]
    (For $k=1$ both remainders vanish; for $k\ge2$ both inequalities are
    equalities.)
    Hence $xA+B$ has degree at most $4k-3$.  Expanding the top five
    coefficients of $P_{4k+1}$ and substituting previously recovered values
    gives the triangular pivots
    \[
    \begin{array}{c|c|c}
      \text{coefficient}&\text{new parameter}&\text{slope}\\ \hline
      \coeff{P_{4k+1}}{4k}&b&2k\\
      \coeff{P_{4k+1}}{4k-1}&c&2k\\
      \coeff{P_{4k+1}}{4k-2}&a&-2k\\
      \coeff{P_{4k+1}}{4k-3}&e&k\\
      \coeff{P_{4k+1}}{4k-4}&\rho&k
    \end{array}
    \]
    In the fourth and fifth rows the possible boundary contributions from
    $A,B$ vanish when $k=1$ and, when $k\ge2$, are parameter-free and hence
    subtractable by \Cref{lem:Rk2l-top-boundary}.  By
    $(4k+1)$-admissibility, every displayed
    slope is invertible.  Thus $H_2,H_4,\tilde H_4$ and all
    five outer parameters are derivable from $P_{4k+1}$.

    We now check the cutoffs directly.  The coefficients of $U$ above
    degree $4(k-1)$ and of $V$ above the same boundary are coefficients of
    the known powers.  At the boundary, the parameter-free top-boundary
    assertion \Cref{lem:Rk2l-top-boundary} supplies the one remaining
    correction.  Descending from the top five rows therefore derives
    $U_j$ from rows of $P_{4k+1}$ at least $j+1$ and $V_j$ from rows at least
    $j$, for all $j\ge4(k-1)$.  After the five outer parameters and the
    powers have been substituted, the residual
    \[
      x(U-H_4^k)+\bigl(V-\tilde H_4^k\bigr)=xA+B
    \]
    is coefficient-triangular on rows below $4(k-1)$ by
    \Cref{lem:Rk2l}(3).  Every outer parameter, and hence every coefficient
    of these powers, was recovered from rows at least $4(k-1)$.  For an inner
    row $j<4(k-1)$ this is at least $j+1$.  Substituting the top-block formulas
    into the triangular decoder therefore preserves both cutoffs: residual
    rows at least $j+1$ recover $A_j$, and residual rows at least $j$ recover
    $B_j$.  Hence $A,B$, and
    therefore $U,V$, satisfy the two compatibility cutoffs on the whole
    window.  If $k\ge2$, the same argument extracts
    $\alpha_0,\ldots,\alpha_{4k-5}$; for $k=1$ that internal block is empty.
    This proves the decoder and compatibility claims.  For the realization,
    the quadratic $H_2$ costs one product, and the shared
    $T_{2k,2}$ call costs $\mu(2k,1)=2k-1$ products by
    \Cref{lem:T-multiplication-count}.  The same call outputs both components
    and forms $H_4$ as an intermediate wire.  Thus the single shared circuit
    has $2k$ products and records both $H_2$ and $H_4$.
\end{proof}

\begin{lemma}[A decodable $Q_{4k+1}$ given $H_2$]\label{lem:Q4k+1-from-H2}
    Let $k\ge 1$, assume that $\mathbb F$ is $(4k+1)$-admissible, and let
    $H_2\in\mathbb{F}[x]$ be monic of degree $2$.

    Define
    \[
      \hat H_2 \coloneqq H_2+\alpha_{4k-1},
      \qquad
      \tilde H_2 \coloneqq \hat H_2+\alpha_{4k-2},
    \]
    and let
    \[
      S^{(1)}_2 \coloneqq T^{(1)}_{2k,2}[\alpha_0,\ldots,\alpha_{4k-3}](x,\hat H_2),
      \qquad
      S^{(2)}_2 \coloneqq T^{(2)}_{2k,2}[\alpha_0,\ldots,\alpha_{4k-3}](x,\hat H_2,\tilde H_2).
    \]
    Finally define
    \[
      Q_{4k+1}[\alpha_0,\ldots,\alpha_{4k}](x,H_2)\coloneqq (x+\alpha_{4k})\,S^{(1)}_2(x)+S^{(2)}_2(x).
    \]
    Then $Q_{4k+1}$ is decodable given $H_2$, and the polynomial $H_4$ computed in the internal even-$k$, $l=1$ step is derivable from $Q_{4k+1}$ given $H_2$.
\end{lemma}
\begin{proof}
    Write $H_2=x^2+bx+c_0$ and put
    \[
      \gamma=\alpha_{4k-1},\quad \rho=\alpha_{4k-2},\quad
      a=\alpha_{4k-3},\quad e=\alpha_{4k-4},\quad
      \beta=\alpha_{4k}.
    \]
    The shared base computes
    \[
      H=H_2+\gamma,\qquad
      H_4=H^2-(x+a)^2+e,\qquad
      \tilde H_4=H_4+\rho,
    \]
    and then
    \[
      (S^{(1)}_2,S^{(2)}_2)
       =T_{k,4}[\alpha_0,\ldots,\alpha_{4k-5}]
          (x,H,H_4,\tilde H_4).
    \]
    Since $\tilde H_4-H_4=\rho$ and the field is $4k$-admissible, the
    hypotheses of \Cref{lem:Rk2l} hold for this $T_{k,4}$ call.
    As in the preceding proof, the top-boundary lemma and a direct Cauchy-product
    expansion give the descending affine pivots
    \[
    \begin{array}{c|c|c}
      \coeff{Q_{4k+1}}{4k}&\beta&1\\
      \coeff{Q_{4k+1}}{4k-1}&\gamma&2k\\
      \coeff{Q_{4k+1}}{4k-2}&a&-2k\\
      \coeff{Q_{4k+1}}{4k-3}&e&k\\
      \coeff{Q_{4k+1}}{4k-4}&\rho&k
    \end{array}
    \]
    Every entry not containing the displayed new parameter is a polynomial in
    the coefficients of the given $H_2$ and in parameters recovered in earlier
    rows.  Hence all five parameters, and therefore $H,H_4,\tilde H_4$, are
    derivable.

    Given $H,H_4,\tilde H_4$, \Cref{lem:Rk2l}(3) says directly that
    $(S^{(1)}_2,S^{(2)}_2)$ is compatible on its remainder window.  The five
    top rows have already recovered those three polynomials and $\beta$, so
    the known-shift clause of \Cref{lem:x-alpha-extraction} applies to
    $Q_{4k+1}=(x+\beta)S^{(1)}_2+S^{(2)}_2$ and derives the pair.  This clause
    permits the full top row in the compatibility window; it does not try to
    recover $\beta$ from the generic top-row formula.
    If $k\ge2$, subtracting $H_4^k$ and $\tilde H_4^k$ and applying
    \Cref{lem:Rk2l}(3) at $(k,2)$ extracts
    $\alpha_0,\ldots,\alpha_{4k-5}$; for $k=1$ this block is empty.  Thus
    $Q_{4k+1}$ is decodable
    given $H_2$, and the already recovered $H_4$ is the required byproduct.
\end{proof}

\subsection{Odd-degree gadgets from known powers}
\begin{algorithm}[H]
    \caption{Odd-degree gadget $Q_{2^{l+1}k+(2^l-1)}$ from the known powers
        $(H_2,\ldots,H_{2^l})$, for $k\ge1$ and $l\ge2$.}
    \label{alg:constr-Q-odd}
      \begin{algorithmic}
          \State $\hat{H}_{2^l} = H_{2^l} + Q_{2^{l - 1} - 1}[\alpha_{2^{l + 1} k - 1}, \ldots, \alpha_{2^{l + 1}k + 2^{l - 1} - 3}](x, H_2, \ldots, H_{2^{l - 2}})$
          \State $S^{(1)}_{2^l} = T^{(1)}_{2 k, 2^{l}}[\alpha_{2^{l} - 2}, \ldots, \alpha_{2^{l + 1}k - 3}](x, H_2, \ldots, \hat{H}_{2^l})$
          \State $S^{(2)}_{2^l} = T^{(2)}_{2 k, 2^{l}}[\alpha_{2^{l} - 2}, \ldots, \alpha_{2^{l + 1}k - 3}](x, H_2, \ldots, \hat{H}_{2^l}, \hat{H}_{2^l} + \alpha_{2^{l + 1}k - 2})$
          \State $Q_{2^{l + 1}k + (2^l - 1)}[\alpha_0, \ldots, \alpha_{2^{l + 1}k + 2^l - 2}](x, H_2, \ldots, H_{2^l})$
          \Statex \hspace{\algorithmicindent}$= A_{2^{l-1}}[\alpha_0, \ldots, \alpha_{2^l - 3}, \alpha_{2^{l + 1}k + 2^{l - 1} - 2}, \ldots, \alpha_{2^{l + 1}k + 2^l - 2}](S^{(1)}_{2^l}, S^{(2)}_{2^l}, (x, H_2, \ldots, H_{2^{l-1}}))$
    \end{algorithmic}
\end{algorithm}

\begin{lemma}\label{lem:Q-odd-degree-with-powers}
    Let $k\ge 1$ and $l\ge 2$, put
    $d=2^{l+1}k+(2^l-1)$, and assume that $\mathbb F$ is $d$-admissible.
    Then \Cref{alg:constr-Q-odd} outputs a polynomial $Q_{2^{l+1}k+(2^l-1)}$ that is decodable given $(H_2,\ldots,H_{2^l})$.
\end{lemma}
\begin{proof}
    Let $n\coloneqq\deg S^{(1)}_{2^l}=\deg S^{(2)}_{2^l}=2^{l+1}k$.
    Since $n<d$, $d$-admissibility implies $n$-admissibility.  The internal
    call has parameters $(M,D)=(2k,2^l)$, so its size is exactly $MD=n$;
    moreover $M$ is even, and therefore the exceptional shared odd base is
    irrelevant.  Consequently \Cref{lem:causal-perturbed-T} applies to this
    call.
    The internal $T_{2k,2^l}$ call that defines $(S^{(1)}_{2^l},S^{(2)}_{2^l})$ uses the perturbed power
    $\hat H_{2^l}=H_{2^l}+Q_{2^{l-1}-1}$ (with $\deg Q_{2^{l-1}-1}=2^{l-1}-1$) and its scalar shift
    $\tilde H_{2^l}=\hat H_{2^l}+\alpha_{2^{l+1}k-2}$.
    Apply \Cref{lem:causal-perturbed-T} to this internal call, with
    $D=2^l$, $M=2k$, and $r=2^{l-1}$.  It gives, without treating the
    perturbed power as an already-known constant, that
    \[
      (S^{(1)}_{2^l},S^{(2)}_{2^l})
      \text{ is compatible on }G=\rng{n-r}
    \]
    given $(H_2,\ldots,H_{2^l})$.  The same lemma gives a cutoff-respecting
    recovery of $\hat H_{2^l}$ and $\tilde H_{2^l}$ from the coefficients of
    $xS^{(1)}_{2^l}+S^{(2)}_{2^l}$ and the given powers.
    Thus the internal call is handled directly by
    \Cref{lem:causal-perturbed-T}; no compatibility assertion is imported
    from the recursive program without its cutoff proof.

      Applying \Cref{lem:fill-correctness} at level $l-1$ to the outer $A_{2^{l-1}}$ call shows that the parameters appearing explicitly in that call are extractable from $Q_{2^{l+1}k+(2^l-1)}$ given $(H_2,\ldots,H_{2^l})$, and that $(S^{(1)}_{2^l},S^{(2)}_{2^l})$ is derivable from this given power data.

    It remains to extract the parameters inside the definitions of
    $\hat H_{2^l}$ and the internal $T$ call.  Since the outer fill decoder
    derives $(S^{(1)}_{2^l},S^{(2)}_{2^l})$, it also derives their combined
    polynomial.  The recovery part of \Cref{lem:causal-perturbed-T} therefore
    gives $\hat H_{2^l}$ and $\tilde H_{2^l}$ directly.  Their difference
    extracts the scalar
    $\alpha_{2^{l+1}k-2}=\tilde H_{2^l}-\hat H_{2^l}$.

    Next, from the derived $\hat H_{2^l}$ we obtain the polynomial
    \[
      Q_{2^{l-1}-1}=\hat H_{2^l}-H_{2^l}.
    \]
    This is exactly the $Q_{2^{l-1}-1}$ instance in the definition of $\hat H_{2^l}$, so decodability of $Q_{2^{l-1}-1}$ given $(H_2,\ldots,H_{2^{l-2}})$ extracts the corresponding parameter block.

    Finally, we isolate the combined remainder polynomial for the $T_{2k,2^l}$ call:
    \[
      xR^{(1)}_{2k,2^l}+R^{(2)}_{2k,2^l}
        = \bigl(xS^{(1)}_{2^l}+S^{(2)}_{2^l}\bigr) - \bigl(x\hat H_{2^l}^{2k}+\tilde H_{2^l}^{2k}\bigr).
    \]
    By \Cref{lem:Rk2l}(3), the remaining block of parameters inside $S^{(1)}_{2^l}$ and $S^{(2)}_{2^l}$ is extractable from this polynomial given $(H_2,\ldots,H_{2^{l-1}},\hat H_{2^l},\tilde H_{2^l})$, and all these polynomials are derivable from the data we already have.
    Applying \Cref{lem:discharge-side-information} substitutes those derived
    powers into the conditional remainder decoder.
    Therefore all parameters are extractable from $Q_{2^{l+1}k+(2^l-1)}$ given $(H_2,\ldots,H_{2^l})$, proving that the output is decodable.
\end{proof}

\subsection{Induction steps}

The $8k+3$ induction step and the special case $n=31$ require a small auxiliary ingredient: a degree-$15$ polynomial that is decodable given only $(H_2,H_4)$.
We define the explicit barred gadgets needed when the available power list
does not contain the next power.  In the induction steps below this choice is
abbreviated by the selected-gadget notation $\mathcal Q_d$.
\Cref{fig:odd-gadgets} summarizes both gadget families and their decoders.

\begin{figure}[H]
  \centering
  \resizebox{\textwidth}{!}{
\begin{tikzpicture}[x=1cm,y=1cm,>=Latex]
  \node[fp panel title] at (0,5.05) {(a) Two ways to obtain an odd auxiliary gadget};
  \node[fp known, minimum width=28mm] (powers) at (1.45,4.18)
    {given powers\\$H_2,\ldots,H_D$};
  \node[fp active, minimum width=32mm] (perturb) at (4.55,4.18)
    {perturb $H_D$ by\\a smaller $Q$-gadget};
  \node[fp recursive, minimum width=25mm] (teven) at (7.65,4.18)
    {even call\\$T_{2k,D}$};
  \node[fp second, minimum width=24mm] (fill) at (10.35,4.18)
    {outer fill\\$A_{D/2}$};
  \node[fp active, minimum width=16mm] (qout) at (12.65,4.18)
    {$Q_d$};
  \draw[fp flow] (powers) -- (perturb);
  \draw[fp flow] (perturb) -- (teven);
  \draw[fp flow] (teven) -- (fill);
  \draw[fp flow] (fill) -- (qout);

  \node[fp known, minimum width=25mm] (h24) at (1.45,3.05)
    {only $H_2,H_4$\\are given};
  \node[fp active, minimum width=30mm] (h8) at (4.55,3.05)
    {construct $H_8$\\with one product};
  \node[fp recursive, minimum width=25mm] (t8) at (7.65,3.05)
    {$T_{k,8}$\\($k=1$ is direct)};
  \node[fp second, minimum width=24mm] (a4) at (10.35,3.05)
    {fixed outer\\fill $A_4$};
  \node[fp active, minimum width=16mm] (barq) at (12.65,3.05)
    {$\bar Q$};
  \draw[fp flow] (h24) -- (h8);
  \draw[fp flow] (h8) -- (t8);
  \draw[fp flow] (t8) -- (a4);
  \draw[fp flow] (a4) -- (barq);

  \node[fp panel title] at (0,2.4) {(b) Top-down decoder of the barred gadget};
  \draw[fp decode] (0,1.75) -- node[fp label, above] {descending coefficient degree}
    (13.55,1.75);
  \node[fp active, minimum width=24mm, anchor=west] at (0,1.08)
    {$b_0,b_1,b_2$\\three unit pivots};
  \node[fp pivot, minimum width=34mm, anchor=west] at (2.4,1.08)
    {$(b_3,b_4,u,v)$\\explicit $4\times4$ block};
  \node[fp pivot, minimum width=19mm, anchor=west] at (5.8,1.08)
    {$w,\rho$\\slopes $k$};
  \node[fp recursive, minimum width=27mm, anchor=west] at (7.7,1.08)
    {internal $R_{k,8}$\\shifted pivots};
  \node[fp second, minimum width=31mm, anchor=west] at (10.4,1.08)
    {remaining $A_4$ slots\\unit pivots};
  \node[fp label, anchor=west] at (2.45,0.42)
    {$\det M=-k^2$ (so $-1$ for $\bar Q_{15}$, the case $k=1$)};
  \node[fp label, anchor=west, text=fpBlue!80!black] at (0,0.02)
    {The matrix inverse is displayed explicitly; no data-dependent division or Jacobian argument is used.};
\end{tikzpicture}}
  \caption{Ordinary and barred odd-degree gadgets.  When the full power
  list is available, a smaller $Q$ perturbs the top power, an even $T$ call
  supplies a compatible pair, and a fill call closes the gadget.  When only
  $(H_2,H_4)$ is available, the barred construction first synthesizes
  $H_8$ and uses a fixed $A_4$ shell.  Its decoder consists of three unit
  pivots, one explicit constant-determinant block solve, the two power
  shifts, the internal remainder decoder when that block is nonempty, and
  the low fill slots.}
  \label{fig:odd-gadgets}
\end{figure}

\begin{lemma}[A decodable $\bar Q_{8k+7}$ given $(H_2,H_4)$]\label{lem:barQ8k+7}
    Let $k\ge 1$, assume that $\mathbb F$ is $(8k+7)$-admissible, and let
    $H_2,H_4\in\mathbb{F}[x]$ be monic with $\deg H_2=2$ and $\deg H_4=4$.

    Define
    \[
      H_8 \coloneqq \bigl(H_4+(x+\alpha_1)\bigr)\bigl(H_4+(H_2+\alpha_2)\bigr)+\alpha_0,
      \qquad
      \tilde H_8 \coloneqq H_8+\alpha_3.
    \]
    When $k=1$, the interval $\alpha_4,\ldots,\alpha_{8k-5}$ below is empty,
    and the $T$ base is $T^{(1)}_{1,8}=H_8$,
    $T^{(2)}_{1,8}=\widetilde H_8$.
    Let
    \[
      \begin{aligned}
      S^{(1)}&\coloneqq
        T^{(1)}_{k,8}[\alpha_4,\ldots,\alpha_{8k-5}]
          (x,H_2,H_4,H_8),\\
      S^{(2)}&\coloneqq
        T^{(2)}_{k,8}[\alpha_4,\ldots,\alpha_{8k-5}]
          (x,H_2,H_4,H_8,\widetilde H_8),
      \end{aligned}
    \]
    and define
    \[
      \begin{aligned}
      &\bar Q_{8k+7}[\alpha_0,\ldots,\alpha_{8k+6}](x,H_2,H_4)\\
      &\qquad\coloneqq
      A_{4}[\alpha_{8k-4},\ldots,\alpha_{8k+1},
            \alpha_{8k+2},\ldots,\alpha_{8k+6}]
           (S^{(1)},S^{(2)},(x,H_2,H_4)).
      \end{aligned}
    \]
    Then $\bar Q_{8k+7}$ is decodable given $(H_2,H_4)$.
\end{lemma}
\begin{proof}
    Put $N=8k$ and write
    \[
      \begin{aligned}
      H_2&=x^2+r_1x+r_0,&
      H_4&=x^4+s_3x^3+s_2x^2+s_1x+s_0,\\
      u&=\alpha_1,&v&=\alpha_2,&w&=\alpha_0,&\rho&=\alpha_3.
      \end{aligned}
    \]
    If $k=1$, the defining base of the $T$ recursion gives
    $S^{(1)}=H_8$, $S^{(2)}=\widetilde H_8$, and both internal remainders
    vanish.  If $k\ge2$, then $N<8k+7$, so the admissibility hypothesis
    implies $N$-admissibility and
    \Cref{lem:Rk2l,lem:Rk2l-top-boundary} apply to the internal $T_{k,8}$
    call.  Its level is $l=3$, so it does not use the exceptional shared odd
    base.
    For the outer $A_4$ call put
    \[
      b_i=\alpha_{N+2+i}\ (0\le i\le4),\qquad
      a_i=\alpha_{N-4+i}\ (0\le i\le5).
    \]
    Thus the six $a_i$'s are the consecutive parameters in the $Q_3$ and
    additive slots, while the five $b_i$'s are the scalar slots.  More
    explicitly, the outer expression is
    \[
    \begin{aligned}
      U_0&=(H_4+b_3)S^{(1)}+(x+a_5)(H_2+a_4)+a_3,\\
      V_0&=(H_4+b_4)S^{(2)}+a_2,\\
      C_1&=(H_2+b_1)U_0+a_1,
      &C_2&=(H_2+b_2)V_0+a_0,\\
      \bar Q_{N+7}&=(x+b_0)C_1+C_2.
    \end{aligned}
    \]
    (The notation in this display is just a relabelling of the eleven
    parameters $\alpha_{N-4},\ldots,\alpha_{N+6}$ used by $A_4$.)

    We use the following top-coefficient notation.  If $X$ has degree $N$, let
    \[
      \widehat X(t)=x^{-N}X(x)\big|_{x=t^{-1}},
      \qquad
      p_r=[x^{N+7-r}]\bar Q_{N+7}\quad(0\le r\le7).
    \]
    Equality modulo $t^8$ below means equality of the top eight coefficients.
    For $k=1$ the two $S$-polynomials are exactly $H_8$ and
    $\widetilde H_8$; for $k\ge2$ their remainders have degree $N-8$ by
    \Cref{lem:Rk2l}.  In either case, the coefficients of
    $\widehat S^{(1)}$ and $\widehat S^{(2)}$ through $t^7$ are those of
    $H_8^k$ and $\widetilde H_8^k$, respectively.

    The first three rows have unit pivots:
    \[
      \begin{array}{c|ccc}
        \text{coefficient of }\bar Q_{N+7}&[x^{N+6}]&[x^{N+5}]&[x^{N+4}]\\ \hline
        \text{new parameter}&b_0&b_1&b_2\\
        \text{slope}&1&1&1
      \end{array}
    \]
    All entries to the right of a pivot in this table involve only parameters
    already recovered.  This follows directly by expanding the monic factors
    in the displayed definition of $\bar Q_{N+7}$.

    We next solve simultaneously for $(b_3,b_4,u,v)$.  For a compact
    verification of the four pivots, put
    \[
      c_j\coloneqq [x^{N-j}]
        \bigl((H_4+x)(H_4+H_2)\bigr)^k
        \quad(0\le j\le3).
    \]
    These are known from $H_2,H_4$ (and $c_1=2ks_3$).  The coefficients of
    $S^{(1)}$ and $S^{(2)}$ in the first eight degrees below their leading
    term are those of the corresponding powers.  In the normalized variable
    $t=x^{-1}$, the second branch carries an additional factor $t$ because
    it has degree $N+6$ rather than $N+7$; retaining this shift is essential
    in the convolution below.  Since $u$ and $v$ first
    enter $H_8$ four degrees below its leading term, the four rows
    $p_4,p_5,p_6,p_7$ are affine in $(b_3,b_4,u,v)$, where
    $p_r=[x^{N+7-r}]\bar Q_{N+7}$.  After the first three pivots, direct
    convolution gives
    \[
      \begin{pmatrix}p_4\\p_5\\p_6\\p_7\end{pmatrix}
       =\begin{pmatrix}p^0_4\\p^0_5\\p^0_6\\p^0_7\end{pmatrix}
        +M\begin{pmatrix}b_3\\b_4\\u\\v\end{pmatrix},
    \]
    with
    \[
    \begin{gathered}
      A_1=b_0+r_1+c_1,\qquad D=r_1+c_1,\\
      C=b_0(c_1+r_1)+c_2+r_1c_1+r_0+b_1,\qquad
      F=c_2+r_1c_1+r_0+b_2,\\
      E=b_0(c_2+r_1c_1+r_0+b_1)+c_3+r_1c_2
        +(r_0+b_1)c_1,\\
      L=b_0+2r_1+(2k-1)s_3+1,
    \end{gathered}
    \]
    and
    \[
      M=
      \begin{pmatrix}
        1&0&k&k\\
        A_1&1&k(A_1+1)&k(A_1+1)\\
        C&D&k(C+D)&k(C+D-1)\\
        E&F&k(E+F-1)&k(E+F-L)
      \end{pmatrix}.
    \]
    Replacing the last two columns by
    $C_3-kC_1-kC_2$ and $C_4-kC_1-kC_2$ gives
    $(0,0,0,-k)^{\mathsf T}$ and $(0,0,-k,-kL)^{\mathsf T}$.
    Since the upper $2\times2$ block in the first two columns has determinant
    one, this proves $\det M=-k^2$.  Thus this block is
    inverted by the explicit decoder
    \[
      \begin{pmatrix}b_3\\b_4\\u\\v\end{pmatrix}
        =-\frac1{k^2}\operatorname{adj}(M)
          \left(
            \begin{pmatrix}p_4\\p_5\\p_6\\p_7\end{pmatrix}
            -\begin{pmatrix}p^0_4\\p^0_5\\p^0_6\\p^0_7\end{pmatrix}
          \right).
    \]
    The only division is by the fixed field constant $k^2$, which is
    invertible under $(8k+7)$-admissibility; there is no data-dependent
    division.

    It remains to recover $w$ and $\rho$, and then, when $k\ge2$, the
    internal $T$ block.  For $k\ge2$, the next two coefficients include the
    parameter-free boundary rows of the two internal remainders; these are
    known by \Cref{lem:Rk2l-leading-coeff,lem:Rk2l-top-boundary}.  For $k=1$
    the remainders vanish.  Thus in both cases every contribution other than
    the indicated scalar shifts is already known after the preceding steps.
    Indeed, writing
    $H_8=H_8^0+w$, the first term of $(H_8^0+w)^k$ containing $w$ is
    $kw(H_8^0)^{k-1}$, of degree $N-8$; multiplication by the monic
    degree-$7$ factor $A$ places it in degree $N-1$.  Terms containing two
    copies of $w$ are at least eight degrees lower.  Likewise the first term
    containing $\rho$ in $(H_8+\rho)^k$ is
    $k\rho H_8^{k-1}$, and the monic degree-$6$ factor $B$ places it in
    degree $N-2$.  Therefore
    \[
      [x^{N-1}]\bar Q_{N+7}=kw+K_{N-1},\qquad
      [x^{N-2}]\bar Q_{N+7}=k\rho+K_{N-2}(w),
    \]
    for known polynomials $K_{N-1},K_{N-2}$.  Thus $w$ and then $\rho$ are
    recovered (the first boundary coefficient of $H_8^k$ contributes $kw$;
    the second branch contributes $k(w+\rho)$ one row later).
    Consequently $H_8$ and $\widetilde H_8$ are now known.

    Subtract the known main terms and set
    \[
      A=(x+b_0)(H_2+b_1)(H_4+b_3),\qquad
      B=(H_2+b_2)(H_4+b_4).
    \]
    Then
    \[
      \bar Q_{N+7}-AH_8^k-B\widetilde H_8^k
       =AR^{(1)}_{k,8}+BR^{(2)}_{k,8}+F,\qquad \deg F\le6,
    \]
    where
    \[
      F=(x+b_0)\bigl((H_2+b_1)Q_3+a_1\bigr)+(H_2+b_2)a_2+a_0.
    \]
    If $k=1$, then $R^{(1)}_{1,8}=R^{(2)}_{1,8}=0$ and the internal parameter
    interval $\alpha_4,\ldots,\alpha_{N-5}$ is empty, so the displayed
    residual is simply $F$.  Suppose now that $k\ge2$.  Since $A$ and $B$
    are monic of degrees $7$ and $6$, respectively, the
    coefficient in degree $j+6$ of the first two terms is
    \[
      [x^j]\bigl(xR^{(1)}_{k,8}+R^{(2)}_{k,8}\bigr)
      +\text{a polynomial in higher coefficients of }R^{(1)},R^{(2)}.
    \]
    Descending through degrees $N-3,N-4,\ldots,6$ therefore transfers the
    triangular decoder of \Cref{lem:Rk2l}(3) to the internal parameter block
    $\alpha_4,\ldots,\alpha_{N-5}$.  This explicitly handles the overlap of
    the two multiplied remainder polynomials; no fill-compatibility lemma is
    invoked.  At the final row (degree $6$, corresponding to the inner
    coefficient of degree $0$), the low term has only its known leading
    contribution $[x^6]F=1$; subtracting this constant leaves the same
    triangular pivot.  The unknown coefficients of $F$ start in degree at
    most $5$.

    (For $k=1$ this entire internal descent is void.)  After reconstructing
    $S^{(1)},S^{(2)}$, compute the low residual $F$.
    Its coefficients in degrees $5,4,3,2,1,0$ have the unit-triangular pivot
    table
    \[
      \begin{array}{c|cccccc}
        \text{degree}&5&4&3&2&1&0\\ \hline
        \text{new parameter}&a_5&a_4&a_3&a_2&a_1&a_0\\
        \text{slope}&1&1&1&1&1&1.
      \end{array}
    \]
    (The coefficient in degree $6$ is the known monic leading term.)  A direct
    expansion of $Q_3=(x+a_5)(H_2+a_4)+a_3$ verifies the table, so these six
    parameters are recovered in descending order.  Together with the already
    recovered internal block and $(w,u,v,\rho)$ this is every parameter in
    $\bar Q_{N+7}$, proving decodability given $(H_2,H_4)$.
\end{proof}

\begin{corollary}[The degree-$15$ barred gadget]\label{lem:barQ15}
    The polynomial $\bar Q_{15}$ is decodable given $(H_2,H_4)$ over every
    $15$-admissible field.
\end{corollary}
\begin{proof}
    Specialize \Cref{lem:barQ8k+7} to $k=1$.
\end{proof}

\begin{lemma}[Odd-degree gadgets from $(H_2,H_4)$]\label{lem:odd-gadgets-H2H4}
    Let $H_2,H_4\in\mathbb{F}[x]$ be monic with $\deg H_2=2$ and $\deg H_4=4$.
    For each degree $n$ asserted below, assume that $\mathbb F$ is
    $n$-admissible.
    Then:
    \begin{enumerate}
        \item If $n\equiv 1\pmod 4$ and $n\ge 5$, then $Q_n$ is decodable given $H_2$ by \Cref{lem:Q4k+1-from-H2}.
        \item The base degrees $n=3$ and $n=7$ are the explicit $Q_3$ and
        $Q_7$ constructions above.  If
        $n\equiv 3\pmod 8$ and $n\ge 11$, then $Q_n$ is decodable given
        $(H_2,H_4)$ by the known-powers construction with $l=2$ (i.e.
        \Cref{alg:constr-known-2n-1} and \Cref{lem:Q-odd-degree-with-powers}
        specialized to $l=2$).
        \item If $n\equiv 7\pmod 8$ and $n\ge 15$, then $\bar Q_n$ is
        decodable given $(H_2,H_4)$ by \Cref{lem:barQ8k+7}; its $k=1$
        specialization is the degree-$15$ gadget
        \Cref{lem:barQ15}.
    \end{enumerate}
\end{lemma}
\begin{proof}
    For $n\equiv 1\pmod 4$, write $n=4k+1$ and apply \Cref{lem:Q4k+1-from-H2}.

    The cases $n=3,7$ are the explicit $Q_3,Q_7$ constructions.  For
    $n\equiv 3\pmod 8$
    with $n\ge11$, write $n=2^{l+1}k+(2^l-1)$ with $l=2$ and apply the
    construction \Cref{alg:constr-known-2n-1} together with
    \Cref{lem:Q-odd-degree-with-powers} (specialized to $l=2$).

    For $n\equiv 7\pmod 8$ and $n\ge 15$, write $n=8k+7$ with $k\ge1$
    and apply \Cref{lem:barQ8k+7}.
\end{proof}

\paragraph{Selected odd gadget notation.}
In the two final induction steps, the degree of an auxiliary odd gadget is
not always of the Mersenne form for which the required higher powers have
already been built.  To avoid hiding this distinction in the algorithm
captions, write $\mathcal Q_d$ for the following degree-$d$ gadget (with a
fresh, consecutive parameter block):
\[
\mathcal Q_d\coloneqq
\begin{cases}
 Q_1, & d=1,\\
 Q_3, & d=3,\\
 Q_7, & d=7,\\
 Q_d, & d\equiv1\pmod4,\\
 Q_d, & d\equiv3\pmod8,\\
 \bar Q_d, & d\equiv7\pmod8\text{ and }d\ge15.
\end{cases}
\]
Here the first two non-base lines use respectively
\Cref{lem:Q4k+1-from-H2} and the known-powers construction with
$(H_2,H_4)$; the first three lines are the explicit $Q_1,Q_3,Q_7$ bases,
and the final line uses the barred construction
\Cref{lem:barQ15,lem:barQ8k+7}.  If additional powers are available, the
standard known-powers $Q_d$ may be used instead of $\mathcal Q_d$.
In every case $\mathcal Q_d$ is monic, decodable from the displayed known
powers, and has the cost $\lfloor d/2\rfloor$ recorded below.

\begin{lemma}[Exact costs of the auxiliary odd gadgets]
\label{lem:odd-gadgets-count}
    Assume the indicated known powers are already available.
    \begin{enumerate}
      \item $\mathcal Q_{4k+1}(x,H_2)=Q_{4k+1}(x,H_2)$ uses $2k$
      multiplications.
      \item For $l\ge2$, the known-powers construction
      $Q_{2^{l+1}k+(2^l-1)}$ uses
      \[
        2^l k+2^{l-1}-1
        =\left\lfloor\frac{2^{l+1}k+(2^l-1)}2\right\rfloor
      \]
      multiplications.
      \item $\mathcal Q_{8k+7}(x,H_2,H_4)=\bar Q_{8k+7}(x,H_2,H_4)$ uses
      $4k+3$ multiplications; this includes $\bar Q_{15}$ when $k=1$.
    \end{enumerate}
    Consequently every auxiliary odd-degree $Q$-gadget used in the two final
    induction steps has cost $\lfloor d/2\rfloor$, where $d$ is its degree.
\end{lemma}
\begin{proof}
    The first construction consists of the call $T_{2k,2}$, whose cost is
    $2k-1$ by \Cref{lem:T-multiplication-count}, followed by the one product
    $(x+\beta)T^{(1)}$.

    For the second construction, forming
    $\hat H_{2^l}=H_{2^l}+Q_{2^{l-1}-1}$ costs
    $2^{l-2}-1$ multiplications.  The internal $T_{2k,2^l}$ call costs
    $(2k-1)2^{l-1}$, and the outer $A_{2^{l-1}}$ call costs
    $3\cdot2^{l-2}$ (the formula in \Cref{lem:fill-Q-count}, with the direct
    $A_2$ base when $l=2$).  Their sum is
    \[
      (2^{l-2}-1)+(2k-1)2^{l-1}+3\cdot2^{l-2}
      =2^lk+2^{l-1}-1.
    \]

    Finally, $\bar Q_{8k+7}$ uses one product to form $H_8$, then
    $4(k-1)$ products for $T_{k,8}$, and six products for $A_4$.  Thus its
    cost is $1+4(k-1)+6=4k+3$.  The remaining auxiliary cases
    $Q_{2^r-1}$ have cost $2^{r-1}-1=\lfloor(2^r-1)/2\rfloor$ by
    \Cref{lem:fill-Q-count}; together these cases cover the gadgets listed in
    \Cref{lem:odd-gadgets-H2H4} (including $Q_1,Q_3,Q_7$).
\end{proof}

\subsubsection{The $8k+7$ step}

\Cref{fig:odd-induction-steps} pictures the two induction steps.

\begin{figure}[H]
  \centering
  \resizebox{\textwidth}{!}{
\begin{tikzpicture}[x=1cm,y=1cm,>=Latex]
  \draw[fp decode] (0,5.2) -- node[fp label, above] {descending coefficient degree}
    (13.55,5.2);
  \node[fp label, anchor=east] at (0,5.2) {high};
  \node[fp label, anchor=west] at (13.55,5.2) {low};

  \node[fp panel title] at (0,4.67) {(a) The $8k+7$ step: two nested square shells};
  \node[fp active, minimum width=49mm, anchor=west] (s3) at (0,3.92)
    {$xS_3^2+(S_3+b)^2$\\recover $S_3,b$};
  \node[fp seam, minimum width=8mm, anchor=west] at (4.9,3.92) {$-1$};
  \node[fp second, minimum width=36mm, anchor=west] (s2) at (5.7,3.92)
    {$-xS_2^2-(S_2+a)^2$\\recover $S_2,a$};
  \node[fp seam, minimum width=8mm, anchor=west] at (9.3,3.92) {$+1$};
  \node[fp recursive, minimum width=30mm, anchor=west] (small7) at (10.1,3.92)
    {$P_{2k+1}$\\recurse};
  \draw[fp dependence] (s3.south east) to[bend right=10] (s2.south west);
  \draw[fp dependence] (s2.south east) to[bend right=10] (small7.south west);
  \node[fp panel title] at (0,2.95) {(b) The $8k+3$ step: shell, squared recursion, delayed low gadget};
  \node[fp active, minimum width=45mm, anchor=west] (top3) at (0,2.2)
    {$xS_2^2+(S_2+a)^2$\\recover $S_2,a$};
  \node[fp seam, minimum width=8mm, anchor=west] at (4.5,2.2) {$-1$};
  \node[fp recursive, minimum width=45mm, anchor=west] (sqrec) at (5.3,2.2)
    {$-x(T^{(1)})^2-(T^{(2)})^2$\\recover squares; take monic roots};
  \node[fp second, minimum width=30mm, anchor=west] (low3) at (9.8,2.2)
    {$xS_3+\alpha_0$\\low residual};
  \draw[fp dependence] (top3.south east) to[bend right=10] (sqrec.south west);
  \draw[fp dependence] (sqrec.south east) to[bend right=10] (low3.south west);
  \node[fp known, minimum width=37mm] (powers) at (7.4,0.78)
    {recurse on $P_{2k+1}$\\and reconstruct $H_2,H_4$};
  \draw[fp decode] (sqrec.south) -- (powers.north);
  \draw[fp dependence] (powers.east) -- (low3.south west);
  \node[fp label, anchor=west, text=fpBlue!80!black] at (0,0.02)
    {Both steps recover auxiliary polynomials before invoking their conditional parameter decoders.};
\end{tikzpicture}}
  \caption{The two recursive odd-degree steps.  For $8k+7$, two square
  shells are peeled in succession and leave the smaller polynomial
  $P_{2k+1}$.  For $8k+3$, the outer shell exposes a compatible squared
  smaller pair; its monic square roots recover the smaller instance, whose
  decoder reconstructs the powers needed by the low auxiliary gadget.  The
  red cells are the fixed leading-coefficient corrections at shared
  boundaries.}
  \label{fig:odd-induction-steps}
\end{figure}

\begin{algorithm}[H]
    \caption{
        Extending a compatible joint realization from $2k+1$ to $8k+7$.
        The odd-degree auxiliary calls below use the selected gadgets
        $\mathcal Q_d$ of \Cref{lem:odd-gadgets-H2H4}; they require only the
        displayed $H_2,H_4$ (and only $H_2$ in the $d\equiv1\pmod4$ case).
    }
    \label{alg:constr-8k+7}

    \begin{algorithmic}
        \State $S^{(1)}_1 = T^{(1)}_{2k + 1}[\alpha_{0}, \ldots, \alpha_{2k}](x)$ \Comment{The smaller construction records the $H_2,H_4$ used below.}
        \State $S^{(1)}_2 = \mathcal Q_{2k + 1}[\alpha_{2k + 1}, \ldots, \alpha_{4k + 1}](x, H_2(x), H_4(x))$
        \State $S^{(1)}_3 = \mathcal Q_{4k + 3}[\alpha_{4k + 2}, \ldots, \alpha_{8k + 4}](x, H_2(x), H_4(x))$
        \State Put $s=\alpha_{8k+5}$, $d=\alpha_{8k+6}$ and, for the
        decoder, $a=(s-d)/2$, $b=(s+d)/2$.
        \State $T^{(1)}_{8k + 7}[\alpha_0, \ldots, \alpha_{8k + 6}](x)
            = (S^{(1)}_3 + S^{(1)}_2)(S^{(1)}_3 - S^{(1)}_2) + S^{(1)}_1
            = (S^{(1)}_3)^2 - (S^{(1)}_2)^2 + S^{(1)}_1$
        \\
        \State $S^{(2)}_1 = T^{(2)}_{2k + 1}[\alpha_{0}, \ldots, \alpha_{2k}](x)$
        \State For the identities put $S^{(2)}_2=S^{(1)}_2+a$ and
        $S^{(2)}_3=S^{(1)}_3+b$; the circuit does not materialize these two
        shifted polynomials.
        \State $T^{(2)}_{8k + 7}[\alpha_0, \ldots, \alpha_{8k + 6}](x)
            = \bigl((S^{(1)}_3+S^{(1)}_2)+s\bigr)
              \bigl((S^{(1)}_3-S^{(1)}_2)+d\bigr)+S^{(2)}_1
            = (S^{(2)}_3 + S^{(2)}_2)(S^{(2)}_3 - S^{(2)}_2) + S^{(2)}_1
            = (S^{(2)}_3)^2 - (S^{(2)}_2)^2 + S^{(2)}_1$

        \State \begin{align}
            P_{8k + 7}[\alpha_0, \ldots, \alpha_{8k + 6}](x)
            &= x T^{(1)}_{8k + 7} + T^{(2)}_{8k + 7}
            \\&=x(S^{(1)}_3)^2+(S^{(1)}_3+b)^2
                 -x(S^{(1)}_2)^2\notag\\
            &\quad -(S^{(1)}_2+a)^2
                 +xS^{(1)}_1+S^{(2)}_1
        \end{align}
    \end{algorithmic}
\end{algorithm}

\begin{lemma}[The $8k+7$ induction step is decodable (and preserves compatibility)]\label{lem:8k+7-splittable}
    Let $k\ge2$ and assume that $\mathbb{F}$ is $(8k+7)$-admissible.
    Suppose that $P_{2k+1}=xT^{(1)}_{2k+1}+T^{(2)}_{2k+1}$ is decodable, and that the auxiliary polynomials $S^{(1)}_2,S^{(1)}_3$ in \Cref{alg:constr-8k+7} are decodable given the relevant known-power data.
    Assume moreover that the underlying pair
    $(T^{(1)}_{2k+1},T^{(2)}_{2k+1})$ is compatible on some window $G$
    with no auxiliary data, and that its construction records the monic
    polynomials $H_2,H_4$ used by the two auxiliary gadgets as polynomial
    byproducts of the smaller parameter block.
    Define
    \[
      G_{8k+7}\coloneqq \{2k+1,\ldots,4k+3\}\cup\{4k+3,\ldots,8k+6\}\cup G.
    \]
    Then the polynomial $P_{8k+7}$ produced by the construction is decodable,
    and the output pair $(T^{(1)}_{8k+7},T^{(2)}_{8k+7})$ is compatible on
    $G_{8k+7}$ with no auxiliary data.  If the smaller pair has a joint
    $m$-realization recording $H_2,H_4$, then the displayed construction has
    a joint $(m+3k+3)$-realization and records the same two powers.
\end{lemma}
\begin{proof}
    Write $S_2\coloneqq S^{(1)}_2$ and
    $S_3\coloneqq S^{(1)}_3$ for brevity.  The two scalar parameter
    coordinates are $s=\alpha_{8k+5}$ and $d=\alpha_{8k+6}$; put
    $a=(s-d)/2$ and $b=(s+d)/2$, so that $a+b=s$ and $b-a=d$.
    This is an invertible fixed linear change of coordinates because the
    admissibility hypothesis makes $2$ invertible.

    Since $S_3=\mathcal Q_{4k+3}$ is monic of degree $4k+3$, the polynomial $xS_3^2+(S_3+b)^2$ has degree $8k+7$.
    All remaining summands in the explicit formula for $P_{8k+7}$ have degree at most $4k+3$:
    indeed, $S_2=\mathcal Q_{2k+1}$ has degree $2k+1$, so $xS_2^2$ has degree $4k+3$ and $(S_2+a)^2$ has degree $4k+2$, while $xS^{(1)}_1+S^{(2)}_1=P_{2k+1}$ has degree $2k+1$.
    Therefore we may write
    \[
        P_{8k+7} = \underbrace{xS_3^2 + (S_3+b)^2}_{\text{square gadget at degree }4k+3}
            \;+\; E_3
    \]
    where $\deg E_3\le 4k+3$.
    The boundary coefficient of this error is fixed:
    $[x^{4k+3}]E_3=-1$.  Therefore
    \Cref{lem:square-gadget-boundary}, with this supplied boundary
    coefficient, derives $S_3$ and $b$ from $P_{8k+7}$.

    Subtracting the derived square-gadget polynomial $xS_3^2+(S_3+b)^2$ from $P_{8k+7}$ leaves
    \[
        P' \coloneqq -xS_2^2 - (S_2+a)^2 + P_{2k+1}.
    \]
    Equivalently, $-P' = xS_2^2 + (S_2+a)^2 + E_2$ with $E_2=-P_{2k+1}$ and $\deg E_2\le 2k+1=\deg S_2$.
    Its boundary coefficient is again fixed, $[x^{2k+1}]E_2=-1$.
    Applying \Cref{lem:square-gadget-boundary}, with this boundary
    coefficient, derives $S_2$ and $a$ from $P'$, and by
    construction both polynomials are now available without using their
    internal known-power data.

    Finally,
    \[
        P_{2k+1} = P' + xS_2^2 + (S_2+a)^2
    \]
    is derivable from $P_{8k+7}$, so decodability of $P_{2k+1}$ extracts all
    parameters of the $2k+1$ block.  We can then recompute the recorded
    byproducts $H_2,H_4$ of that smaller construction.  Only at this point do
    we invoke decodability of the auxiliary gadgets: from the already derived
    polynomials $S_2,S_3$ and their now-known powers,
    \Cref{lem:discharge-side-information} extracts their parameter blocks;
    the reconstructed $H_2,H_4$ are thereby discharged rather than silently
    treated as fixed auxiliary data.
    Together with $s=a+b$ and $d=b-a$, this extracts
    $\alpha_0,\ldots,\alpha_{8k+6}$ and proves decodability.

    For compatibility we retain the row cutoffs in those two peelings.  If
    $S$ is monic of degree $e$, the descending recurrence in
    \Cref{lem:square-gadget-boundary} first reads the coefficient
    $[x^q]S$ in row $e+q+1$; the scalar shift is read in row $e$.
    Consequently $[x^j]S^2$ is recovered using only rows at least $j+1$,
    while $[x^j](S+\delta)^2$ uses only rows at least $j$.  Subtracting a
    recovered square gadget at row $j$ also uses only rows at least $j$, so
    the same statement remains true for the second peeling and for the
    residual $P_{2k+1}$.

    Now apply the unconditional compatibility hypothesis for the smaller
    pair to that residual.  It recovers
    $[x^j]T^{(1)}_{2k+1}$ from output rows at least $j+1$ and
    $[x^j]T^{(2)}_{2k+1}$ from rows at least $j$.  Substitution in
    \[
      T^{(1)}_{8k+7}=S_3^2-S_2^2+T^{(1)}_{2k+1},\qquad
      T^{(2)}_{8k+7}=(S_3+b)^2-(S_2+a)^2+T^{(2)}_{2k+1}
    \]
    gives exactly the two required cutoffs on $G_{8k+7}$.  No coefficients
    of the internal powers used to define $S_2,S_3$ occur in this
    reconstruction, so the output compatibility is unconditional.

    Finally run the smaller $m$-realization once.  The two selected auxiliary
    gadgets cost $k$ and $2k+1$ products by
    \Cref{lem:odd-gadgets-count}; they reuse the recorded $H_2,H_4$ wires.
    The two displayed difference-of-squares outputs cost one product each.
    Hence the joint cost is $m+k+(2k+1)+2=m+3k+3$, and the inherited power
    wires remain recorded outputs.
\end{proof}

\begin{example}[Worked example: $k=7$ (constructing $63$)]
    Instantiating the construction with $k=7$ gives $2k+1=15$, $4k+3=31$, and $8k+7=63$.
    Let $(T^{(1)}_{15},T^{(2)}_{15})$ be the compatible seven-product joint
    realization from \Cref{sec:special-cases}, which records $H_2,H_4$, and
    let $\bar Q_{15}$ and
    $\bar Q_{31}$ be the selected degree-$15$ and degree-$31$ gadgets from
    the convention above.
    Put $s\coloneqq\alpha_{61}$, $d\coloneqq\alpha_{62}$ and
    $a=(s-d)/2$, $b=(s+d)/2$.  The construction specializes to
    \[
      \begin{aligned}
        T^{(1)}_{63}
          &=\bar Q_{31}^2-\bar Q_{15}^2+T^{(1)}_{15},\\
        T^{(2)}_{63}
          &=(\bar Q_{31}+\bar Q_{15}+s)
            (\bar Q_{31}-\bar Q_{15}+d)+T^{(2)}_{15}\\
          &=(\bar Q_{31}+b)^2-(\bar Q_{15}+a)^2+T^{(2)}_{15}.
      \end{aligned}
    \]
    and hence
    \[
        P_{63}=xT^{(1)}_{63}+T^{(2)}_{63}
        = x\bar Q_{31}^2 + (\bar Q_{31}+b)^2
          -x\bar Q_{15}^2-(\bar Q_{15}+a)^2+P_{15}.
    \]
    The parameter blocks are
    \[
      \begin{array}{c|c}
        \alpha_0,\ldots,\alpha_{14}&(T^{(1)}_{15},T^{(2)}_{15})\\
        \alpha_{15},\ldots,\alpha_{29}&\bar Q_{15}\\
        \alpha_{30},\ldots,\alpha_{60}&\bar Q_{31}\\
        \alpha_{61},\alpha_{62}&s,d.
      \end{array}
    \]
\end{example}

\subsubsection{The $8k+3$ step}
\begin{algorithm}[H]
    \caption{
        Extending a compatible joint realization from $2k+1$ to $8k+3$.
        The call to $\mathcal Q_{2k-1}$ below uses the selected gadget of
        \Cref{lem:odd-gadgets-H2H4}: it needs only the $H_2,H_4$ made
        derivable by the preceding two calls (with the explicit $Q_3,Q_7$
        bases in the small cases).
    }
    \label{alg:constr-8k+3}

    \begin{algorithmic}
        \State $S^{(1)}_1 = T^{(1)}_{2k + 1}[\alpha_{2k}, \ldots, \alpha_{4k}](x)$ \Comment{The smaller construction records $H_2(x)$.}
        \State $S^{(1)}_2 = Q_{4k + 1}[\alpha_{4k + 2}, \ldots, \alpha_{8k + 2}](x, H_2(x))$ \Comment{This call records the quartic $H_4(x)$.}
        \If{$k > 1$}
            \State $S^{(1)}_3 = \mathcal Q_{2k - 1}[\alpha_{1}, \ldots, \alpha_{2k - 1}](x, H_2(x), H_4(x))$
        \Else
            \State $S^{(1)}_3 = \alpha_1$
        \EndIf
        \State $T^{(1)}_{8k + 3}[\alpha_0, \ldots, \alpha_{8k + 2}](x)
            = (S^{(1)}_2 + S^{(1)}_1)(S^{(1)}_2 - S^{(1)}_1) + S^{(1)}_3
            = (S^{(1)}_2)^2 - (S^{(1)}_1)^2 + S^{(1)}_3$
        \\
        \State $S^{(2)}_1 = T^{(2)}_{2k + 1}[\alpha_{2k}, \ldots, \alpha_{4k}](x)$
        \State $S^{(2)}_2 = S^{(1)}_2 + \alpha_{4k + 1}$
        \State $S^{(2)}_3 = \alpha_{0}$
        \State $T^{(2)}_{8k + 3}[\alpha_0, \ldots, \alpha_{8k + 2}](x)
            = (S^{(2)}_2 + S^{(2)}_1)(S^{(2)}_2 - S^{(2)}_1) + S^{(2)}_3
            = (S^{(2)}_2)^2 - (S^{(2)}_1)^2 + S^{(2)}_3$

        \State \begin{align}
            P_{8k + 3}[\alpha_0, \ldots, \alpha_{8k + 2}](x)
            &= x T^{(1)}_{8k + 3} + T^{(2)}_{8k + 3}
            \\&=x(S^{(1)}_2)^2+(S^{(1)}_2+\alpha_{4k+1})^2
                 -x(S^{(1)}_1)^2\notag\\
            &\quad -(S^{(2)}_1)^2+xS^{(1)}_3+S^{(2)}_3
        \end{align}
    \end{algorithmic}
\end{algorithm}

\begin{lemma}[The $8k+3$ induction step is decodable (and preserves compatibility)]\label{lem:8k+3-splittable}
    Let $k\ge1$ and assume that $\mathbb{F}$ is $(8k+3)$-admissible.
    Suppose that $P_{2k+1}=xT^{(1)}_{2k+1}+T^{(2)}_{2k+1}$ is decodable,
    and that the underlying pair
    $(T^{(1)}_{2k+1},T^{(2)}_{2k+1})$ is compatible on some $G$ with no
    auxiliary data.  Assume that the smaller construction records its monic
    quadratic $H_2$ as a polynomial byproduct.  Assume also that the auxiliary polynomials
    $S^{(1)}_2$ and $S^{(1)}_3$ in \Cref{alg:constr-8k+3} are decodable once
    their indicated known powers are supplied.
    Define
    \[
      G_{8k+3}\coloneqq
      \{4k+1,\ldots,8k+2\}\cup (2k+G)\cup \rng{2k+1}.
    \]
    Then the polynomial $P_{8k+3}$ produced by the construction is decodable,
    and the output pair $(T^{(1)}_{8k+3},T^{(2)}_{8k+3})$ is compatible on
    $G_{8k+3}$ with no auxiliary data.  If the smaller pair has a joint
    $m$-realization recording $H_2$, then the displayed construction has a
    joint $(m+3k+1)$-realization and records both $H_2$ and the quartic $H_4$
    formed by its $Q_{4k+1}$ call.
\end{lemma}
\begin{proof}
    Write $S_2\coloneqq S^{(1)}_2$, $S_3\coloneqq S^{(1)}_3$, and $a\coloneqq \alpha_{4k+1}$ for the scalar shift.

    From the explicit formula in the algorithm we have
    \[
        P_{8k+3} = \underbrace{xS_2^2 + (S_2+a)^2}_{\text{square gadget at degree }4k+1} + E,
    \]
    where $\deg E\le 4k+1$ (all remaining terms involve only the $2k+1$-block and $xS_3+\alpha_0$).
    The boundary coefficient of this error is fixed:
    $[x^{4k+1}]E=-1$.  Therefore
    \Cref{lem:square-gadget-boundary}, with this supplied boundary
    coefficient, derives $S_2$ and $a$ from $P_{8k+3}$.

    Subtracting the derived square-gadget polynomial
    $xS_2^2+(S_2+a)^2$ yields
    \[
      P'=-\Psi+xS_3+\alpha_0,
      \qquad
      \Psi\coloneqq x(S^{(1)}_1)^2+(S^{(2)}_1)^2,
    \]
    where $S^{(i)}_1=T^{(i)}_{2k+1}$.  The last two terms have degree at
    most $2k$.  Hence $[x^d]\Psi=-[x^d]P'$ for $d>2k$.  At $d=2k$ the
    only correction is the fixed leading coefficient of $xS_3$: it is zero
    for $k=1$, and one for $k>1$.  Thus every coefficient of $\Psi$ in the
    window $2k+G$ is recoverable from $P'$ without knowing any coefficient of
    the internal powers.

    The unconditional compatibility hypothesis and
    \Cref{lem:compatible-power} show that
    $((S^{(1)}_1)^2,(S^{(2)}_1)^2)$ is compatible on $2k+G$, again with no
    auxiliary data.  Its decoder therefore recovers both squares from the
    displayed $\Psi$-window.  Taking the descending monic square roots
    recovers $S^{(1)}_1,S^{(2)}_1$, and hence
    $P_{2k+1}=xS^{(1)}_1+S^{(2)}_1$.  Decodability of $P_{2k+1}$ now extracts
    its parameter block, after which the recorded polynomial $H_2$ can be
    recomputed.

    Finally, with $(S^{(1)}_1,S^{(2)}_1)$ derived, we can isolate
    \[
        xS_3+\alpha_0 = P' + x(S^{(1)}_1)^2 + (S^{(2)}_1)^2.
    \]
    The constant term of $xS_3+\alpha_0$ equals $\alpha_0$, so $\alpha_0$ is extractable.
    Subtracting $\alpha_0$ then derives $xS_3$, and hence $S_3$ (by shifting coefficients).
    Decode the already recovered polynomial
    $S_2=Q_{4k+1}(x,H_2)$ first.  By
    \Cref{lem:Q4k+1-from-H2}, this both extracts the parameter block of $S_2$
    and reconstructs its internal quartic $H_4$.  With $(H_2,H_4)$ now
    available, decodability of $S_3$ extracts its parameter block (for $k=1$,
    $S_3=\alpha_1$ is already read directly).  Thus neither auxiliary decoder
    is invoked before the powers it requires have actually been reconstructed;
    \Cref{lem:discharge-side-information} then substitutes those reconstructed
    powers into the two conditional decoders.

    Together with $a$ and $\alpha_0$, these are all parameters, proving
    decodability.

    It remains to verify the causal cutoffs.  For a monic degree-$e$
    polynomial $S$, the square-gadget recurrence first reads $[x^q]S$ in
    row $e+q+1$ and reads its scalar shift in row $e$.  Consequently
    $[x^j]S^2$ is recoverable from output rows at least $j+1$, while
    $[x^j](S+\delta)^2$ is recoverable from rows at least $j$.  This applies
    to $S_2$ on the top window $\{4k+1,\ldots,8k+2\}$.

    For the middle block, recovering $[x^r]\Psi$ from $P'$ uses only the
    output row $r$ and the fixed monic boundary correction at $r=2k$.
    The cutoff conclusion of \Cref{lem:compatible-power} therefore gives
    \[
      \begin{aligned}
      [x^j](S^{(1)}_1)^2
        &\polyfrom([x^r]P_{8k+3}:r\in G_{8k+3},\ r\ge j+1),\\
      [x^j](S^{(2)}_1)^2
        &\polyfrom([x^r]P_{8k+3}:r\in G_{8k+3},\ r\ge j).
      \end{aligned}
    \]
    Finally, after those squares have been subtracted,
    $xS_3+\alpha_0$ is read coefficient by coefficient on
    $\rng{2k+1}$: $[x^j]S_3$ is read in row $j+1$, and $\alpha_0$ in row
    zero.  Substitution in
    \[
      T^{(1)}_{8k+3}=S_2^2-(S^{(1)}_1)^2+S_3,
      \qquad
      T^{(2)}_{8k+3}=(S_2+a)^2-(S^{(2)}_1)^2+\alpha_0
    \]
    gives the required first-component cutoff $j+1$ and second-component
    cutoff $j$.  No internal known power occurs in these formulas, so the
    output compatibility is unconditional.

    For the joint cost, run the smaller circuit once.  The selected gadgets
    of degrees $4k+1$ and $2k-1$ cost $2k$ and $k-1$ products by
    \Cref{lem:odd-gadgets-count} (the second cost is zero when $k=1$), and
    the two displayed difference-of-squares outputs cost one product each.
    The total is $m+2k+(k-1)+2=m+3k+1$.  The first auxiliary call forms and
    records $H_4$, while the smaller circuit's $H_2$ wire is retained.
\end{proof}

\subsection{Finite bases and cost repairs}
\label{sec:special-cases}

These are not extra algebraic existence bases.  Degree~$15$ is the optimized,
fused $k=1$ endpoint of the $8k+7$ branch: it creates the quartic missing
from the one-product degree-$3$ pair without exceeding seven products.
Degrees~$27$ and~$31$ are the two bridges around the unavailable
three-product degree-$7$ joint realization.  Their common top-down decoder is
displayed after the circuits.

\subsubsection{Degree $3$}
\begin{lemma}[A one-product compatible realization for $3$]
    \label{lem:base-three-compatible}
    There exists a splittable pair for $3$ that is compatible on $\rng 3$,
    has a one-realization, and records its monic quadratic $H_2$.
\end{lemma}
\begin{proof}
    Define $H_2[\alpha_1,\alpha_2](x)=(x+\alpha_2)x+\alpha_1$ and set
    \[
      T^{(1)}_3[\alpha_0,\alpha_1,\alpha_2](x)\coloneqq H_2,\qquad
      T^{(2)}_3[\alpha_0,\alpha_1,\alpha_2](x)\coloneqq H_2+\alpha_0.
    \]
    Then $T^{(1)}_3$ and $T^{(2)}_3$ are monic of degree $2$ and
    \[
      P_3=xT^{(1)}_3+T^{(2)}_3
        =x^3+(\alpha_2+1)x^2+(\alpha_1+\alpha_2)x+(\alpha_0+\alpha_1).
    \]
    Hence $\alpha_2=[x^2]P_3-1$, $\alpha_1=[x^1]P_3-\alpha_2$, and
    $\alpha_0=[x^0]P_3-\alpha_1$, so $P_3$ is decodable.  We also verify
    the causal window needed by the induction.  For
    $\Phi_3=xT^{(1)}_3+T^{(2)}_3=P_3$, the relevant coefficients are

    \[
      \begin{array}{lll}
        [x^2]T^{(1)}_3=[x^2]T^{(2)}_3=1,&
        [x^1]T^{(1)}_3=[x^1]T^{(2)}_3=\Phi_{3,2}-1,\\[2pt]
        [x^0]T^{(1)}_3=\Phi_{3,1}-\Phi_{3,2}+1,&
        [x^0]T^{(2)}_3=\Phi_{3,0},
      \end{array}
    \]
    where $\Phi_{3,j}=[x^j]\Phi_3$.  The first-component formulas use
    only rows $\ge j+1$, and the second-component formulas only rows
    $\ge j$, exactly as required in \Cref{def:compatible-pair}.  Thus
    $(T^{(1)}_3,T^{(2)}_3)$ is compatible on $\rng 3$, and in particular it
    is splittable.  The single displayed product computes $H_2$; both
    components are then obtained by additions, so this is a one-realization
    and $H_2$ is a recorded byproduct.
\end{proof}

\subsubsection{Degree $7$: direct septic base}
The canonical construction in \Cref{rem:canonical-split} makes the septic
algebraically splittable.  What the optimal recursion would need, but we do
not have, is a compatible joint realization of its two components using
three multiplications and recording the required power byproducts.  A direct
four-multiplication evaluation scheme is enough for the final polynomial,
and the recursion below is arranged never to call degree~$7$.

\begin{lemma}[A decodable septic in characteristic different from $2$]
\label{lem:septic-base}
Assume $\operatorname{char}(\mathbb F)\ne2$.  Define
\begin{align}
  y&=x(x+\alpha_6),\\
  z&=(\alpha_5+x+y)(\alpha_4+x),\\
  w&=(\alpha_3+z)x,\\
  v&=(\alpha_2+x+z)(\alpha_1+w),\\
  P_7&=\alpha_0+y+w+v.
\end{align}
Then $P_7$ is monic of degree~$7$, uses four multiplications, and is
decodable.
\end{lemma}

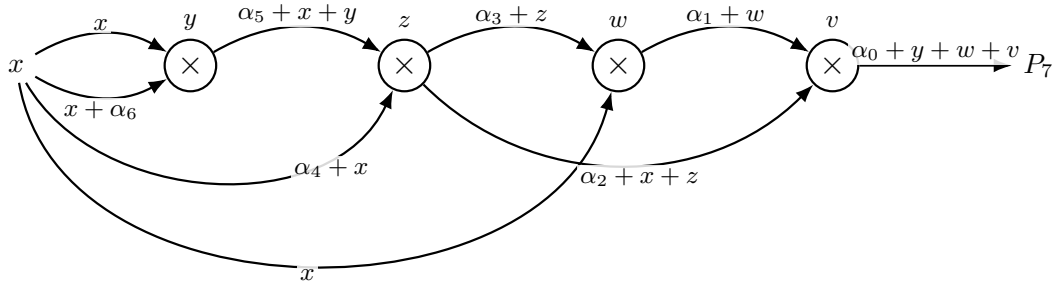
\begin{figure}[H]
  \centering
  \resizebox{0.95\textwidth}{!}{
\begin{tikzpicture}[
    >=Latex, thick,
    mul/.style={circle, draw, inner sep=1.5pt, minimum size=6.5mm, font=\large, fill=white},
    inp/.style={font=\small},
    lbl/.style={font=\footnotesize, inner sep=1pt, fill=white, fill opacity=0.85, text opacity=1}
  ]
  \node[inp] (x) at (0,0) {$x$};
  \node[mul] (my) at (2.2,0) {$\times$};
  \node[mul] (mz) at (4.9,0) {$\times$};
  \node[mul] (mw) at (7.6,0) {$\times$};
  \node[mul] (mv) at (10.3,0) {$\times$};
  \node[inp] (out) at (12.9,0) {$P_7$};
  \node[font=\footnotesize] at (2.2,0.55) {$y$};
  \node[font=\footnotesize] at (4.9,0.55) {$z$};
  \node[font=\footnotesize] at (7.6,0.55) {$w$};
  \node[font=\footnotesize] at (10.3,0.55) {$v$};

  \draw[->] (x) to[bend left=30] node[lbl, above] {$x$} (my);
  \draw[->] (x) to[bend right=30] node[lbl, below] {$x+\alpha_6$} (my);
  \draw[->] (my) to[bend left=30] node[lbl, above] {$\alpha_5+x+y$} (mz);
  \draw[->] (x) to[out=-60, in=-115, looseness=1.0] node[lbl, pos=0.78, below] {$\alpha_4+x$} (mz);
  \draw[->] (mz) to[bend left=30] node[lbl, above] {$\alpha_3+z$} (mw);
  \draw[->] (x) to[out=-80, in=-105, looseness=1.05] node[lbl, pos=0.5, below] {$x$} (mw);
  \draw[->] (mw) to[bend left=30] node[lbl, above] {$\alpha_1+w$} (mv);
  \draw[->] (mz) to[out=-45, in=-135, looseness=1.0] node[lbl, pos=0.55, below] {$\alpha_2+x+z$} (mv);
  \draw[->] (mv) -- node[lbl, above] {$\alpha_0+y+w+v$} (out);
\end{tikzpicture}}
  \caption{The four-multiplication circuit for the septic base $P_7$ of \Cref{lem:septic-base}.
  Every multiplication gate takes two affine combinations of $x$ and previously computed values; the seven parameters $\alpha_0,\ldots,\alpha_6$ enter only as constant terms of these affine forms.}
  \label{fig:septic-circuit}
\end{figure}

\begin{proof}
\Cref{fig:septic-circuit} draws the circuit.
Write
\[
  P_7=x^7+c_6x^6+c_5x^5+c_4x^4+c_3x^3+c_2x^2+c_1x+c_0
\]
and write $z=x^3+z_2x^2+z_1x+z_0$.  The following triangular procedure
recovers the parameters from $(c_0,\ldots,c_6)$:
\begin{align}
 z_2&=c_6/2,\\
 z_1&=(c_5-z_2^2-1)/2,\\
 R&=c_4-1-2z_2z_1-z_2,\\
 \alpha_1&=c_3-z_2-z_2R-z_1^2-z_1,\\
 W&=c_2-z_1-1-z_2\alpha_1-z_1R,\\
 \alpha_6&=c_1-(z_1+1)\alpha_1-W(R+1-W),\\
 \alpha_4&=z_2-1-\alpha_6,\\
 \alpha_5&=z_1-\alpha_4(1+\alpha_6),\\
 z_0&=\alpha_4\alpha_5,\\
 \alpha_3&=W-z_0,\\
 \alpha_2&=R-2z_0-\alpha_3,\\
 \alpha_0&=c_0-(z_0+\alpha_2)\alpha_1.
\end{align}
Indeed, direct expansion gives
\[
 z_2=\alpha_4+1+\alpha_6,\qquad
 z_1=\alpha_4(1+\alpha_6)+\alpha_5,\qquad
 z_0=\alpha_4\alpha_5,
\]
while the coefficient equations for $P_7$ give, successively,
$R=2z_0+\alpha_2+\alpha_3$ and $W=z_0+\alpha_3$ and then the displayed
formulas for $\alpha_1$ and $\alpha_6$.
Substitution recovers every parameter.  The only division is by~$2$, which
is valid under the characteristic assumption.
\end{proof}

\subsubsection{Degree $15$}
\begin{algorithm}[H]
    \caption{
        Seven-product compatible joint realization in degree $15$.
    }

    \begin{algorithmic}
        \State $H_2[\alpha_6, \alpha_7](x) = (x + \alpha_7)x + \alpha_6$
        \State $H_4[\alpha_4, \alpha_5, \alpha_6, \alpha_7](x) = (H_2 + (x + \alpha_5))(H_2 - (x + \alpha_5)) + \alpha_4$
        \\
        \State $S^{(1)}_1 = Q_7[\alpha_8, \ldots, \alpha_{14}](x, H_2, H_4)$
        \State $S^{(1)}_2 = H_2 + \alpha_{3}$
        \State $S^{(1)}_3 = \alpha_1$
        \State $T^{(1)}_{15}[\alpha_0, \ldots, \alpha_{14}](x)
            = (S^{(1)}_1 + S^{(1)}_2)(S^{(1)}_1 - S^{(1)}_2) + S^{(1)}_3
            = (S^{(1)}_1)^2 - (S^{(1)}_2)^2 + S^{(1)}_3$
        \\
        \State $S^{(2)}_1 = H_4$
        \State $S^{(2)}_2 = H_2 + \alpha_2$
        \State $S^{(2)}_3 = \alpha_0$
        \State $T^{(2)}_{15}[\alpha_0, \ldots, \alpha_{14}](x)
            = (S^{(2)}_1 + S^{(2)}_2)(S^{(2)}_1 - S^{(2)}_2) + S^{(2)}_3 + T^{(1)}_{15}
            = (S^{(2)}_1)^2 - (S^{(2)}_2)^2 + S^{(2)}_3 + T^{(1)}_{15}$

        \State \begin{align}
            P_{15}[\alpha_0, \ldots, \alpha_{14}](x)
            &= x T^{(1)}_{15} + T^{(2)}_{15}
            \\&= (x+1)T^{(1)}_{15} + H_4^2 - (H_2 + \alpha_2)^2 + \alpha_0
        \end{align}
    \end{algorithmic}
\end{algorithm}

\subsubsection{Degree $27$}
\begin{algorithm}[H]
    \caption{
        Thirteen-product compatible joint realization in degree $27$.
    }

    \begin{algorithmic}
        \State $H_2[\alpha_2, \alpha_3](x) = (x + \alpha_3)x + \alpha_2$
        \\
        \State $S^{(1)}_1 = Q_{13}[\alpha_{14}, \ldots, \alpha_{26}](x, H_2)$ \Comment{This call records the quartic $H_4$.}
        \State $S^{(1)}_2 = Q_3[\alpha_4, \ldots, \alpha_6](x, H_2)$
        \State $S^{(1)}_3 = \alpha_1$
        \State $T^{(1)}_{27}[\alpha_0, \ldots, \alpha_{26}](x)
            = (S^{(1)}_1 + S^{(1)}_2)(S^{(1)}_1 - S^{(1)}_2) + S^{(1)}_3
            = (S^{(1)}_1)^2 - (S^{(1)}_2)^2 + S^{(1)}_3$
        \\
        \State $S^{(2)}_1 = Q_7[\alpha_7, \ldots, \alpha_{13}](x, H_2, H_4)$
        \State $S^{(2)}_2 = H_2$
        \State $S^{(2)}_3 = \alpha_0$
        \State $T^{(2)}_{27}[\alpha_0, \ldots, \alpha_{26}](x)
            = (S^{(2)}_1 + S^{(2)}_2)(S^{(2)}_1 - S^{(2)}_2) + S^{(2)}_3 + T^{(1)}_{27}
            = (S^{(2)}_1)^2 - (S^{(2)}_2)^2 + S^{(2)}_3 + T^{(1)}_{27}$

        \State \begin{align}
            P_{27}[\alpha_0, \ldots, \alpha_{26}](x)
            &= x T^{(1)}_{27} + T^{(2)}_{27}
            \\&= (x+1)T^{(1)}_{27} + (S^{(2)}_1)^2 - H_2^2 + \alpha_0
        \end{align}
    \end{algorithmic}
\end{algorithm}

\subsubsection{Degree $31$}
\begin{algorithm}[H]
    \caption{
        Fifteen-product compatible joint realization in degree $31$.
    }

    \begin{algorithmic}
        \State $H_2[\alpha_6, \alpha_7](x) = (x + \alpha_7)x + \alpha_6$
        \State $H_4[\alpha_4, \alpha_5, \alpha_6, \alpha_7](x) = (H_2 + (x + \alpha_5))(H_2 - (x + \alpha_5)) + \alpha_4$
        \\
        \State $S^{(1)}_1 = \bar Q_{15}[\alpha_{16}, \ldots, \alpha_{30}](x, H_2, H_4)$
        \State $S^{(1)}_2 = Q_7[\alpha_8, \ldots, \alpha_{14}](x, H_2, H_4)$
        \State $S^{(1)}_3 = Q_3[\alpha_1, \alpha_2, \alpha_3](x, H_2)$
        \State $T^{(1)}_{31}[\alpha_0, \ldots, \alpha_{30}](x)
            = (S^{(1)}_1 + S^{(1)}_2)(S^{(1)}_1 - S^{(1)}_2) + S^{(1)}_3
            = (S^{(1)}_1)^2 - (S^{(1)}_2)^2 + S^{(1)}_3$
        \\
        \State $S^{(2)}_1 = S^{(1)}_1 + \alpha_{15}$
        \State $S^{(2)}_2 = H_4$
        \State $S^{(2)}_3 = \alpha_0$
        \State $T^{(2)}_{31}[\alpha_0, \ldots, \alpha_{30}](x)
            = (S^{(2)}_1 + S^{(2)}_2)(S^{(2)}_1 - S^{(2)}_2) + S^{(2)}_3
            = (S^{(2)}_1)^2 - (S^{(2)}_2)^2 + S^{(2)}_3$

        \State \begin{align}
            P_{31}[\alpha_0, \ldots, \alpha_{30}](x)
            &= x T^{(1)}_{31} + T^{(2)}_{31}
            \\&= x(S^{(1)}_1)^2 + (S^{(1)}_1 + \alpha_{15})^2 - x(S^{(1)}_2)^2 - H_4^2 + x S^{(1)}_3 + \alpha_0
        \end{align}
    \end{algorithmic}
\end{algorithm}

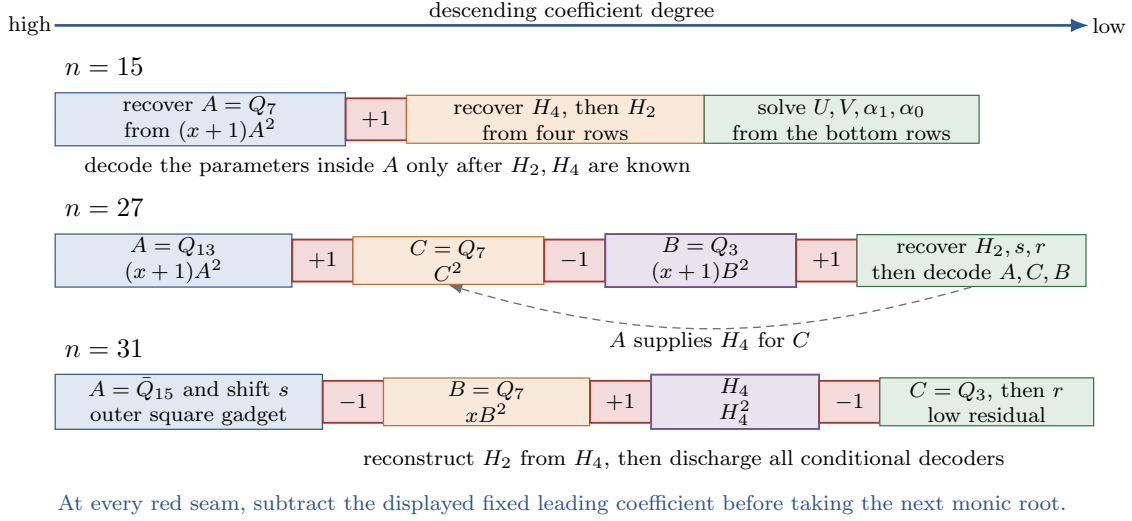
\begin{figure}[H]
  \centering
  \resizebox{\textwidth}{!}{
\begin{tikzpicture}[x=1cm,y=1cm,>=Latex]
  \draw[fp decode] (0,6.25) -- node[fp label, above] {descending coefficient degree}
    (13.55,6.25);
  \node[fp label, anchor=east] at (0,6.25) {high};
  \node[fp label, anchor=west] at (13.55,6.25) {low};

  \node[fp panel title] at (0,5.72) {$n=15$};
  \node[fp active, minimum width=38mm, anchor=west] (a15) at (0,5.02)
    {recover $A=Q_7$\\from $(x+1)A^2$};
  \node[fp seam, minimum width=8mm, anchor=west] at (3.8,5.02) {$+1$};
  \node[fp second, minimum width=39mm, anchor=west] (h15) at (4.6,5.02)
    {recover $H_4$, then $H_2$\\from four rows};
  \node[fp recursive, minimum width=36mm, anchor=west] (l15) at (8.5,5.02)
    {solve $U,V,\alpha_1,\alpha_0$\\from the bottom rows};
  \node[fp label, anchor=west] at (0.35,4.4)
    {decode the parameters inside $A$ only after $H_2,H_4$ are known};

  \node[fp panel title] at (0,3.87) {$n=27$};
  \node[fp active, minimum width=31mm, anchor=west] (a27) at (0,3.17)
    {$A=Q_{13}$\\$(x+1)A^2$};
  \node[fp seam, minimum width=8mm, anchor=west] at (3.1,3.17) {$+1$};
  \node[fp second, minimum width=25mm, anchor=west] (c27) at (3.9,3.17)
    {$C=Q_7$\\$C^2$};
  \node[fp seam, minimum width=8mm, anchor=west] at (6.4,3.17) {$-1$};
  \node[fp pivot, minimum width=25mm, anchor=west] (b27) at (7.2,3.17)
    {$B=Q_3$\\$(x+1)B^2$};
  \node[fp seam, minimum width=8mm, anchor=west] at (9.7,3.17) {$+1$};
  \node[fp recursive, minimum width=30mm, anchor=west] (h27) at (10.5,3.17)
    {recover $H_2,s,r$\\then decode $A,C,B$};
  \draw[fp dependence] (h27.south) to[bend left=16]
    node[fp label, below] {$A$ supplies $H_4$ for $C$} (c27.south);

  \node[fp panel title] at (0,2.02) {$n=31$};
  \node[fp active, minimum width=35mm, anchor=west] (a31) at (0,1.32)
    {$A=\bar Q_{15}$ and shift $s$\\outer square gadget};
  \node[fp seam, minimum width=8mm, anchor=west] at (3.5,1.32) {$-1$};
  \node[fp second, minimum width=27mm, anchor=west] (b31) at (4.3,1.32)
    {$B=Q_7$\\$xB^2$};
  \node[fp seam, minimum width=8mm, anchor=west] at (7.0,1.32) {$+1$};
  \node[fp pivot, minimum width=22mm, anchor=west] (h31) at (7.8,1.32)
    {$H_4$\\$H_4^2$};
  \node[fp seam, minimum width=8mm, anchor=west] at (10.0,1.32) {$-1$};
  \node[fp recursive, minimum width=28mm, anchor=west] (c31) at (10.8,1.32)
    {$C=Q_3$, then $r$\\low residual};
  \node[fp label, anchor=west] at (4.0,0.56)
    {reconstruct $H_2$ from $H_4$, then discharge all conditional decoders};
  \node[fp label, anchor=west, text=fpBlue!80!black] at (0,-0.03)
    {At every red seam, subtract the displayed fixed leading coefficient before taking the next monic root.};
\end{tikzpicture}}
  \caption{The three finite decoders in one visual template.  A coloured
  block denotes a monic polynomial recovered from the indicated square
  window; each red seam is a fixed boundary coefficient contributed by the
  next block.  Recovering a polynomial and decoding its internal parameters
  are deliberately separated: the lower annotations show when the required
  powers have finally become available and the corresponding conditional
  decoder may be discharged.}
  \label{fig:special-case-decoders}
\end{figure}

\begin{lemma}[Finite cost bases]\label{lem:special-cases-splittable}
    For each $n\in\{15,27,31\}$, over an $n$-admissible field the
    special-case construction for $n$ produces a splittable pair.
    Moreover, the direct descending decoders below give the required
    compatibility windows for these three finite constructions; no general
    composition result is needed.  The two components have a joint
    $(n-1)/2$-realization that records their monic quadratic and quartic
    byproducts.
    In particular, $P_{15}$, $P_{27}$, and $P_{31}$ are decodable.
\end{lemma}
\begin{proof}
    We give the finite decoders explicitly.  Write
    $\Phi_n=xT^{(1)}_n+T^{(2)}_n=P_n$ and $G_n=\rng n$.
    For any polynomial $W$ write $W_j=\coeff{W}{j}$, so
    $p_j=\coeff{\Phi_n}{j}$.  Every square root below is the descending
    monic-square recursion (so it uses only divisions by $2$), and all
    quantities described as \text{known} are coefficients of the displayed
    auxiliary powers or quantities recovered in an earlier row block.
    To make explicit why only the top half of a square is needed, if
    $W=x^e+\sum_{r<e}w_rx^r$, then
    \[
      \coeff{W^2}{2e-s}
        =2w_{e-s}+\sum_{r=1}^{s-1}w_{e-r}w_{e-(s-r)}
        \qquad(1\le s\le e).
    \]
    Thus the coefficients in degrees $2e,\ldots,e$ recover the monic
    polynomial $W$ successively.

    \paragraph{The degree-$15$ construction.}
    Put
    \[
      A=S^{(1)}_1,\quad H_2=x^2+bx+c,\quad
      U=H_2+\alpha _3=x^2+bx+d_3,\quad
      V=H_2+\alpha _2=x^2+bx+d_2,
    \]
    and write $H_4=x^4+h_3x^3+h_2x^2+h_1x+h_0$.  The identity
    \[
      P_{15}=(x+1)A^2+H_4^2-(x+1)U^2-V^2+(x+1)\alpha _1+\alpha _0
    \]
    is a relative square shell with $M=x+1$, $\deg A=7$, and error
    \[
      E_A=H_4^2-(x+1)U^2-V^2+(x+1)\alpha_1+\alpha_0.
    \]
    Here $\deg E_A\le8$ and $[x^8]E_A=1$.  Hence
    \Cref{lem:relative-square-shell} recovers $A$ from the top block, with
    the displayed $+1$ seam correction.  Set
    $R=\Phi_{15}-(x+1)A^2$.  Its next four
    coefficients give
    \[
      h_3=R_7/2,\qquad b=h_3/2,\qquad
      h_2=(R_6-h_3^2)/2,
    \]
    \[
      h_1=(R_5+1-2h_3h_2)/2,\qquad
      h_0=(R_4+2b+2-h_2^2-2h_3h_1)/2.
    \]
    The defining relation $H_4=H_2^2-(x+\alpha _5)^2+\alpha _4$
    then yields
    \[
      c=(h_2-b^2+1)/2,\quad
      \alpha _5=bc-h_1/2,\quad \alpha _4=h_0-c^2+\alpha _5^2.
    \]
    Finally put $R'=R-H_4^2$.  The four bottom equations are
    \[
      \begin{aligned}
      d_3&=-(R'_3+b^2+4b)/2,\\
      d_2&=-(R'_2+2b^2+2d_3+2bd_3)/2,\\
      \alpha _1&=R'_1+2bd_3+d_3^2+2bd_2,\\
      \alpha _0&=R'_0+d_3^2+d_2^2-\alpha _1.
      \end{aligned}
    \]
    Hence $\alpha _3=d_3-c$ and $\alpha _2=d_2-c$.  Once $H_2,H_4$
    are known, the unitriangular decoder for $Q_7$ (\Cref{lem:Q-unitriangular})
    recovers the parameters inside $A$; the side information is discharged
    by \Cref{lem:discharge-side-information} because $A,H_2,H_4$ have all
    been recovered from $P_{15}$.

    \paragraph{The degree-$27$ construction.}
    Write
    \[
      A=S^{(1)}_1,\quad C=S^{(2)}_1,\quad B=S^{(1)}_2,
      \quad H_2=x^2+ux+v,
    \]
    where $\deg A=13$, $\deg C=7$, and $\deg B=3$, and put
    $s=\alpha _1$ and $r=\alpha _0$.  The identity
    \[
      P_{27}=(x+1)A^2-(x+1)B^2+(x+1)s+C^2-H_2^2+r
    \]
    gives three consecutive relative square shells.  First set
    \[
      E_A=-(x+1)B^2+(x+1)s+C^2-H_2^2+r.
    \]
    Since $\deg E_A\le14$ and $[x^{14}]E_A=1$,
    \Cref{lem:relative-square-shell} with $M=x+1$ recovers $A$.  Set
    $R=\Phi_{27}-(x+1)A^2$.  Then
    \[
      R=C^2+E_C,\qquad
      E_C=-(x+1)B^2+(x+1)s-H_2^2+r,
    \]
    where $\deg E_C\le7$ and $[x^7]E_C=-1$.  The same lemma, now with
    $M=1$, recovers $C$.  Set $R'=R-C^2=E_C$.  After changing sign,
    \[
      -R'=(x+1)B^2+E_B,\qquad
      E_B=-(x+1)s+H_2^2-r,
    \]
    with $\deg E_B\le4$ and $[x^4]E_B=1$.  A third application with
    $M=x+1$ recovers $B$.  These are exactly the seams $+1,-1,+1$ in
    \Cref{fig:special-case-decoders}.  Finally, with
    $L=R'+(x+1)B^2=(x+1)s-H_2^2+r$, we have
    \[
      u=-L_3/2,\qquad v=-(L_2+u^2)/2,\qquad
      s=L_1+2uv,\qquad r=L_0+v^2-s.
    \]
    Decode $A=Q_{13}$ first using the recovered $H_2$; by
    \Cref{lem:Q4k+1-from-H2} this also reconstructs the quartic $H_4$ used
    by $C$.  Then decode $C=Q_7$ from $(H_2,H_4)$ and $B=Q_3$ from $H_2$.
    These three decoders recover their respective parameter blocks without
    assuming an internal power before it has been reconstructed; this is
    exactly the substitution in \Cref{lem:discharge-side-information}.

    \paragraph{The degree-$31$ construction.}
    Put
    \[
      A=S^{(1)}_1,\quad B=S^{(1)}_2,\quad C=S^{(1)}_3,
      \quad H=H_4,
    \]
    with degrees $15,7,3,4$, respectively, and set
    $s=\alpha _{15}$ and $r=\alpha _0$.  Since
    \[
      P_{31}=xA^2+(A+s)^2+E,
      \qquad E=-xB^2-H^2+xC+r,
    \]
    we have $\deg E\le15=\deg A$ and $[x^{15}]E=-1$.  Therefore
    \Cref{lem:square-gadget-boundary} recovers $A$ and $s$ directly, with
    the displayed fixed boundary correction.  Set
    \[
      R'=\Phi_{31}-xA^2-(A+s)^2.
    \]
    The remaining two monic blocks are again relative square shells:
    \[
      -R'=xB^2+E_B,\qquad E_B=H^2-xC-r.
    \]
    Here $\deg E_B\le8$ and $[x^8]E_B=1$, so
    \Cref{lem:relative-square-shell} with $M=x$ recovers $B$.  Put
    $R''=R'+xB^2$.  Then
    \[
      -R''=H^2+E_H,\qquad E_H=-xC-r,
    \]
    with $\deg E_H\le4$ and $[x^4]E_H=-1$.  The same lemma with $M=1$
    recovers $H$.  The remaining polynomial
    $R'''=R''+H^2=xC+r$ gives
    $r=R'''_0$ and $C=(R'''-r)/x$.  If
    $H=x^4+h_3x^3+h_2x^2+h_1x+h_0$, its parameter block is
    \[
      \alpha _7=h_3/2,\quad
      \alpha _6=(h_2-\alpha _7^2+1)/2,\quad
      \alpha _5=\alpha _7\alpha _6-h_1/2,\quad
      \alpha _4=h_0-\alpha _6^2+\alpha _5^2.
    \]
    The previously proved decoders for $Q_3,Q_7,\bar Q_{15}$ (the latter
    given $(H_2,H_4)$) recover all remaining parameters.  Formally,
    \Cref{lem:discharge-side-information} substitutes the recovered
    $H_2,H_4$ into these conditional decoders.

    It remains to record causality.  The preceding formulas have the following
    row shifts; $t$ denotes a coefficient degree inside the recovered
    polynomial.
    \[
    \begin{array}{c|c|c}
      n&\text{recovered quantity}&\text{row where its coefficient is read}\\ \hline
      15&[x^t]A,\ [x^t]H_4,\
          (\alpha_3,\alpha_2,\alpha_1,\alpha_0)
        &p_{t+8},\ p_{t+4},\ (p_3,p_2,p_1,p_0)\\
      27&[x^t]A,\ [x^t]C,\ [x^t]B,\ (H_2,s,r)
        &p_{t+14},\ p_{t+7},\ p_{t+4},\ (p_3,\ldots,p_0)\\
      31&[x^t]A,\ s,\ [x^t]B,\ [x^t]H_4,\ [x^t]C,\ r
        &p_{t+16},\ p_{15},\ p_{t+8},\ p_{t+4},\ p_{t+1},\ p_0
    \end{array}
    \]
    At a shared boundary the only contribution from the next block is the
    displayed monic coefficient $1$ or $-1$, which is subtracted before that
    block is entered (in the $31$ case this includes the row-$4$ correction
    for $xC$).

    Maintain the descending invariant that a quantity first solved in row
    $r$ is a polynomial in the observed rows $p_s$ with $s\ge r$ and in
    quantities solved above it.  Every displayed square or Cauchy recurrence
    has this form, so the invariant holds across all boundary rows.  The row
    shifts in the table then show that $[x^j]T^{(1)}$ uses only $p_i$ with
    $i\ge j+1$, whereas $[x^j]T^{(2)}$ uses only $p_i$ with $i\ge j$.
    These are exactly the cutoffs in \Cref{def:compatible-pair}.  Since the
    leading coefficient $p_n=1$ is fixed, all three pairs are compatible on
    $\rng n$, hence splittable, and their combined polynomials are decodable.

    Finally, the displayed algorithms are joint circuits for the two
    components.  Charging each shared intermediate only once gives
    \[
    \begin{array}{c|c|c}
      n&\text{products in the joint realization}&\text{total}\\ \hline
      15&H_2:1,\ H_4:1,\ Q_7:3,\ \text{two outer products}:2&7\\
      27&H_2:1,\ Q_{13}:6,\ Q_3:1,\ Q_7:3,\ \text{two outer products}:2&13\\
      31&H_2,H_4:2,\ \bar Q_{15}:7,\ Q_7:3,\ Q_3:1,\
          \text{two outer products}:2&15.
    \end{array}
    \]
    The $Q_{13}$ call in degree~$27$ records the required $H_4$; in the
    other two cases $H_2,H_4$ are displayed intermediates.  Hence all three
    realizations record both byproducts and have exactly $(n-1)/2$ products.
\end{proof}

\subsection{Final construction and decoder}
\label{sec:final}

\begin{figure}[H]
  \centering
  \resizebox{\textwidth}{!}{
\begin{tikzpicture}[x=1cm,y=1cm,>=Latex]
  \node[fp active, rounded corners, minimum width=30mm, minimum height=8mm]
    (n) at (6.75,5.2) {odd target degree $n\ge3$};

  \node[fp known, rounded corners, minimum width=19mm] (seven) at (0.9,3.75)
    {$n=7$\\direct septic};
  \node[fp second, rounded corners, minimum width=30mm] (bases) at (3.45,3.75)
    {$3$ base; $15$ fused $k=1$\\$27,31$ septic bridges};
  \node[fp active, rounded corners, minimum width=23mm] (one) at (6.35,3.75)
    {$n\equiv1\pmod4$\\$T_{2k,2}$ crown};
  \node[fp active, rounded corners, minimum width=23mm] (three) at (9.05,3.75)
    {$n\equiv3\pmod8$\\$8k+3$ shell};
  \node[fp active, rounded corners, minimum width=23mm] (sevenmod) at (11.85,3.75)
    {$n\equiv7\pmod8$\\$8k+7$ shells};

  \coordinate (route) at (6.75,4.55);
  \draw[line width=.65pt, draw=black!75] (n.south) -- (route)
    -- (0.9,4.55) -- (11.85,4.55);
  \draw[fp flow] (0.9,4.55) -- (seven.north);
  \draw[fp flow] (3.45,4.55) -- (bases.north);
  \draw[fp flow] (6.35,4.55) -- (one.north);
  \draw[fp flow] (9.05,4.55) -- (three.north);
  \draw[fp flow] (11.85,4.55) -- (sevenmod.north);

  \node[fp recursive, rounded corners, minimum width=28mm] (small3) at (8.75,2.1)
    {smaller pair at\\$m=(n+1)/4$};
  \node[fp recursive, rounded corners, minimum width=28mm] (small7) at (11.95,2.1)
    {smaller pair at\\$m=(n-3)/4$};
  \draw[fp flow] (small3.north) -- node[fp label, right] {insert} (three.south);
  \draw[fp flow] (small7.north) -- node[fp label, right] {insert} (sevenmod.south);
  \node[fp label, anchor=west] at (8.0,1.4)
    {$m<n$, and the explicit exceptions ensure $m\ne7$};

  \node[fp recursive, minimum width=73mm, minimum height=7mm] (pair) at (4.2,0.55)
    {shared construction of $(T_n^{(1)},T_n^{(2)})$: $(n-1)/2$ products};
  \node[fp gate] (mul) at (8.65,0.55) {$\times$};
  \node[fp active, minimum width=34mm, minimum height=7mm] (out) at (11.3,0.55)
    {$P_n=xT_n^{(1)}+T_n^{(2)}$\\$(n+1)/2$ products total};
  \node[fp label, above=1mm of mul] {multiply by $x$};
  \draw[fp flow] (pair.east) -- (mul.west);
  \draw[fp flow] (mul.east) -- (out.west);
\end{tikzpicture}}
  \caption{Routing of the final strong induction.  The $4k+1$ branch closes
  directly through the $T$ recursion.  The other two congruence classes wrap
  a strictly smaller compatible pair in one or two square shells.  The
  exceptional list removes exactly the cost-optimal recursive calls that
  would otherwise land at the missing three-product degree-$7$ realization
  (and the degree-$15$ byproduct mismatch).
  The bottom row separates the cost of constructing the shared pair from the
  final multiplication by~$x$.  All known-powers instances inside these
  branches use \Cref{alg:constr-known-2n-1}
  (\Cref{thm:construction-height}), giving height
  $2\lceil\log_2 n\rceil+4$ for the odd degrees charted here (the even
  lift of \Cref{thm:construction-count} adds one).}
  \label{fig:final-recursion}
\end{figure}
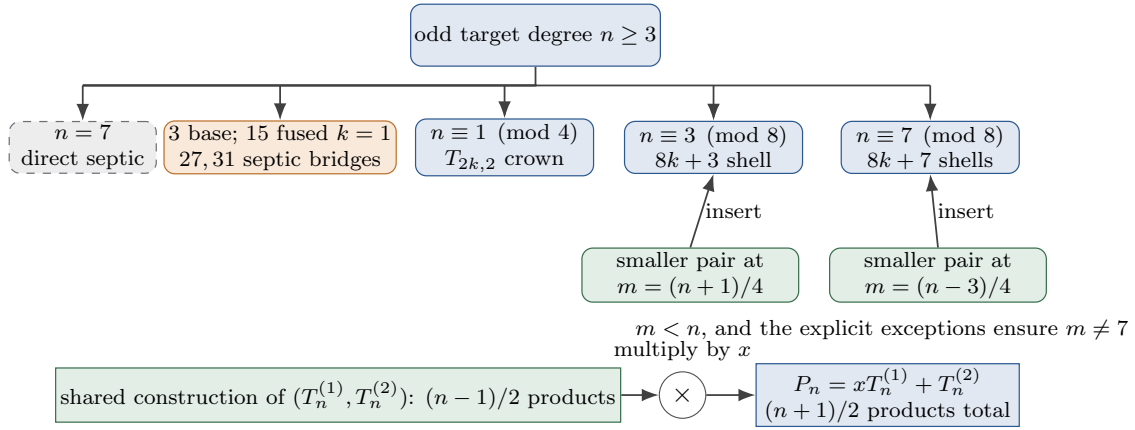

All ingredients are now in place.  The final recursion uses the known-powers
gadgets $Q_{2^t-1}$, the fill gadget $A_{2^l}$, the $T_{k,2^l}$ remainder
recursion, and the three causal closure operations.  Each recursive branch
comes with its own cutoff-respecting decoder: the closure lemmas handle local
combinations, and the stage tables handle $T$.  The decoder summaries used
below are
\Cref{alg:decode-Q-2kminus1} for $Q_{2^t-1}$,
\Cref{alg:decode-fill} for $A_{2^l}$,
and \Cref{alg:decode-Rk2l} for the combined remainder polynomial used in the $T_{k,2^l}$ recursion;
they assemble into \Cref{alg:final-decoder}.
The paper and Lean use the same binary known-powers construction;
\Cref{thm:construction-height} records its complete logarithmic-height bound.
\Cref{fig:final-recursion} charts the routing of the strong induction.

\begin{algorithm}[H]
  \caption{Final odd-degree construction}\label{alg:final-construction}
  \begin{algorithmic}
    \Require odd integer $n\ge 3$
    \Ensure a decodable monic polynomial $P_n$; if $n\ne7$, also a
    jointly $(n-1)/2$-realized splittable pair
    $(T^{(1)}_n,T^{(2)}_n)$ that is compatible with no auxiliary data and
    records the required power byproducts

    \If{$n=7$}
      \State Use the direct septic construction in \Cref{lem:septic-base}.
    \ElsIf{$n\in\{3,15,27,31\}$}
      \State Use the explicit special-case constructions in \Cref{sec:special-cases}.
    \ElsIf{$n\equiv 1\pmod 4$}
      \State Write $n=4k+1$ and construct $(T^{(1)}_{4k+1},T^{(2)}_{4k+1})$ via the $T_{2k,2}$ call; see \Cref{lem:4k+1-splittable}.
    \ElsIf{$n\equiv 3\pmod 8$}
      \State Write $n=8k+3$ and extend the smaller compatible joint realization for $2k+1$ by \Cref{alg:constr-8k+3}.
    \Else
      \State Write $n=8k+7$ with $k\ge 2$ and extend the smaller compatible joint realization for $2k+1$ by \Cref{alg:constr-8k+7} (using $\bar Q_{15}$ or $\bar Q_{8k+7}$ as needed).
    \EndIf
  \end{algorithmic}
\end{algorithm}

\begin{algorithm}[H]
  \caption{Final decoder for $P_n$ (structural recursion)}\label{alg:final-decoder}
  \begin{algorithmic}
    \Require odd $n\ge 3$ and $P_n$ produced by \Cref{alg:final-construction}.
    \Ensure the full parameter list $(\alpha_0,\ldots,\alpha_{n-1})$

    \If{$n=7$}
      \State Apply the triangular decoder in \Cref{lem:septic-base}.
    \ElsIf{$n\in\{3,15,27,31\}$}
      \State Decode by the explicit special-case proofs in \Cref{sec:special-cases}.
    \ElsIf{$n\equiv 1\pmod 4$}
      \State Write $n=4k+1$.
      \State Use the five top-coefficient pivots in \Cref{lem:4k+1-splittable} to derive $H_2,H_4,\tilde H_4=H_4+\rho$ and the five outer parameters.
      \State Derive the pair $(U,V)=T_{k,4}(x,H_2,H_4,\tilde H_4)$ from $P_{4k+1}=xU+V$ using the cutoff decoder in \Cref{lem:4k+1-splittable}.
      \State Form $x(U-H_4^k)+(V-\tilde H_4^k)$ and apply \Cref{alg:decode-Rk2l} at $(k,l)=(k,2)$ to extract $\alpha_0,\ldots,\alpha_{4k-5}$.
    \ElsIf{$n\equiv 3\pmod 8$}
      \State Write $n=8k+3$ and apply the square-gadget extraction in \Cref{lem:8k+3-splittable} to recover the auxiliary polynomials and the squared smaller pair.
      \State Recover $P_{2k+1}$ by monic square roots and recurse on it; this reconstructs $H_2$.
      \State Decode $S^{(1)}_2=Q_{4k+1}$ first via \Cref{lem:Q4k+1-from-H2}, which reconstructs $H_4$; only then decode $S^{(1)}_3=\mathcal Q_{2k-1}$ by the dispatch of \Cref{lem:odd-gadgets-H2H4}, with \Cref{alg:decode-Q-2kminus1} for nested Mersenne blocks.
    \Else
      \State Write $n=8k+7$ with $k\ge 2$ and apply the two square-gadget extractions in \Cref{lem:8k+7-splittable}; this recovers both auxiliary polynomials and isolates $P_{2k+1}$.
      \State Recurse on $P_{2k+1}$, thereby reconstructing the internal powers $H_2,H_4$.
      \State Only now decode the recovered auxiliary gadgets by the $\mathcal Q_d$ dispatch of \Cref{lem:odd-gadgets-H2H4} (i.e.\ \Cref{lem:Q4k+1-from-H2}, \Cref{lem:Q-odd-degree-with-powers}, or \Cref{lem:barQ15,lem:barQ8k+7}), with \Cref{alg:decode-Q-2kminus1} for nested Mersenne blocks.
    \EndIf
  \end{algorithmic}
\end{algorithm}

\begin{theorem}[Cost-optimal compatible pairs in odd degree]
\label{thm:odd-realizable-pairs}
    Let $n\ge3$ be odd with $n\ne7$, and assume that $\mathbb F$ is
    $n$-admissible.  Then there exists a splittable pair
    $(T^{(1)}_n,T^{(2)}_n)$ that is compatible on an explicit window with no
    auxiliary data and has a joint $(n-1)/2$-realization
    (\Cref{def:joint-realization}) of multiplicative height at most
    $2\lceil\log_2n\rceil+3$.  The realization
    records a monic quadratic $H_2$ and, when $n\ge5$, a monic quartic $H_4$.
    In particular, $P_n=xT^{(1)}_n+T^{(2)}_n$ is decodable.
\end{theorem}
\begin{proof}
    We argue by strong induction on odd $n$, proving simultaneously the
    asserted compatible-pair conclusion, the exact joint multiplication
    count, and the byproduct statement.  For the height conjunct, the ledger
    of \Cref{thm:construction-height} threaded through the same
    induction gives $2\lceil\log_2n\rceil+3$ along the pair branches (the
    bound $B(L)=2L+3$ there); the same constant is the height conjunct of
    the machine-checked Lean statement (\texttt{odd\_realizable\_pairs}) for
    the very circuit exhibited here (see the closing remark of
    \Cref{sec:peeled-Q}).
    Since $n$-admissibility implies $m$-admissibility for every $m<n$, all
    recursive pairs and auxiliary gadgets below satisfy their stated
    characteristic hypotheses over the same field.
    The strengthened induction invariant is that every constructed pair records
    its monic quadratic $H_2$ as a polynomial byproduct, and every constructed
    pair of degree at least five also records a monic quartic $H_4$.  These
    polynomials use no extra gates: they are intermediate values already
    displayed in the construction.  The degree-$3$ base records its $H_2$;
    the $4k+1$ and special constructions display both powers; the $8k+3$ step
    inherits $H_2$ and obtains $H_4$ as the byproduct of its
    $Q_{4k+1}(x,H_2)$ call; and the $8k+7$ step inherits both powers from its
    smaller pair.  Thus the byproduct invariant is preserved by every branch.

    Unconditional compatibility is stronger than decodability and is precisely what permits a
    smaller pair to be used before any powers internal to its construction have
    been reconstructed.  The direct proofs in
    \Cref{lem:4k+1-splittable,lem:8k+3-splittable,lem:8k+7-splittable}
    verify the cutoff formulas for each of the three recursive branches; no
    general composition assertion is used.

    The realizability assertion is carried by those same branches, not by a
    separate numerical recurrence.  In either odd induction step, execute the
    smaller joint circuit once, reuse its recorded $H_2,H_4$ wires as inputs
    to the auxiliary gadgets, and then form the two displayed
    difference-of-squares outputs.  Each difference of squares uses the one
    product $(S_3+S_2)(S_3-S_2)$ already shown in the construction algorithm.
    Thus the two output components share the smaller circuit and the power
    byproducts literally; the sums below count the gates of this one joint
    circuit.  The $4k+1$ and finite branches are already displayed as joint
    circuits in the same way.

    The finite cost bases have two distinct causes.  For $n=27$ and
    $n=31$, the relevant recursion would call degree~$7$.  Although the
    septic is algebraically splittable by \Cref{rem:canonical-split}, we do
    not have the three-product compatible joint realization, with recorded
    powers, required by either cost-optimal induction step.  For $n=15$, the smaller degree-$3$
    pair is compatible, but records only $H_2$, whereas the degree-$7$
    auxiliary gadget in the generic $8k+7$ step also requires $H_4$.
    Constructing that quartic separately would exceed the target count; the
    special degree-$15$ circuit instead replaces the generic $Q_3$ block by a
    scalar shift of $H_2$ and spends the saved product on $H_4$.  Thus these
    cases are not artifacts of the decoder proof.

    For $n\equiv 1\pmod 4$ and $n\ge 5$, write $n=4k+1$ and use
    \Cref{lem:4k+1-splittable}.  Its joint-realization clause gives exactly
    $2k=(n-1)/2$ products and records both powers.

    For $n\equiv 3\pmod 8$ we can write $n=8k+3$ with $k\ge 0$.
    If $k=0$ then $n=3$ is supplied by the compatible base
    \Cref{lem:base-three-compatible}.  Hence assume $k\ge1$.
    If $k=3$ then $n=27$ is covered, with its exact cost and recorded powers,
    by \Cref{lem:special-cases-splittable}.
    Otherwise $k\ne 3$, so $2k+1<n$ is odd and $2k+1\ne 7$; by the induction
    hypothesis we have a compatible joint realization for $2k+1$ costing
    $k$ products.  The full conclusion of \Cref{lem:8k+3-splittable} gives
    the required decoder, compatibility, recorded powers, and joint cost
    \[
       k+3k+1=4k+1=\frac{(8k+3)-1}{2}.
    \]

    For $n\equiv 7\pmod 8$ we can write $n=8k+7$ with $k\ge 0$.
    The case $k=0$ is exactly $n=7$ and is excluded.
    If $k=1$ then $n=15$, and if $k=3$ then $n=31$; both are covered, with
    their exact costs and recorded powers, by
    \Cref{lem:special-cases-splittable}.  Otherwise $k\ne1,3$, so $k\ge2$
    and $2k+1<n$ is odd with $2k+1\ne7$.  The induction hypothesis gives a
    compatible joint realization for $2k+1$ costing $k$ products, and
    the full conclusion of \Cref{lem:8k+7-splittable} gives the required
    decoder, compatibility, recorded powers, and joint cost
    \[
       k+3k+3=4k+3=\frac{(8k+7)-1}{2}.
    \]
    The byproduct discussion above verifies in each branch that these are
    realizations of the full strengthened invariant, completing the induction.
\end{proof}

\begin{theorem}[Exact multiplication count]\label{thm:construction-count}
    For every odd $n\ge3$ with $n\ne7$, the construction computes the
    splittable pair $(T^{(1)}_n,T^{(2)}_n)$ using exactly $(n-1)/2$
    multiplications, and therefore computes
    $P_n=xT^{(1)}_n+T^{(2)}_n$ using $(n+1)/2$ multiplications.  The direct
    septic construction uses four multiplications as well.  Consequently, over
    every $n$-admissible field, each monic degree-$n$ polynomial with $n\ge3$
    is covered by a decodable circuit using
    $\lfloor n/2\rfloor+1$ multiplications.
\end{theorem}
\begin{proof}
    For odd $n\ne7$, \Cref{thm:odd-realizable-pairs} supplies the joint
    $(n-1)/2$-realization and its decoder.  Multiplying $T^{(1)}_n$ by $x$
    once gives $P_n$ and raises the cost to $(n+1)/2$.  For $n=7$ the direct
    circuit in \Cref{lem:septic-base} already has the same cost, namely four.

    Finally, if $n\ge4$ is even, first use the odd construction for a monic
    degree-$(n-1)$ polynomial $Q$, at cost $n/2$, and output
    $c_0+xQ$, using one additional product.  This gives $n/2+1$ products.
    Its decoder first reads $c_0$ from the constant coefficient and then
    shifts the remaining coefficients down by one before invoking the decoder
    for $Q$, so this even lift is decodable.  Degrees $1$ and $2$ use
    $x+\alpha_0$ and $(x+\alpha_1)x+\alpha_0$, respectively, and are
    decodable with zero and one product.  Finally,
    \Cref{lem:polynomial-left-inverse-automorphism} turns each of these
    polynomial decoders into a two-sided polynomial preprocessing map.
    Hence the families cover every monic coefficient vector, as asserted.
\end{proof}

\begin{remark}[Exact characteristic restriction]
\label{rem:bad-primes}
    The hypothesis of $n$-admissibility in
    \Cref{thm:odd-realizable-pairs,thm:construction-count} is used only to
    invert the integers listed in \eqref{eq:bad-primes} along the routing
    of \Cref{alg:final-construction}.  Consequently both theorems hold
    verbatim, with the same circuits, the same decoder
    \Cref{alg:final-decoder}, the same multiplication count
    $\lfloor n/2\rfloor+1$ and the same height, over every field whose
    characteristic is not in $\mathrm{Bad}(n)$; equivalently, over every
    field in which each element of $\mathrm{Bad}(n)$ is a unit.  The set
    $\mathrm{Bad}(n)$ is computed by \texttt{tools/bad\_primes.py}, always
    contains~$2$ for $n\ge5$, and its odd elements are at most $(n-1)/4$,
    so characteristic zero or $p>n$ remains a uniform sufficient condition.
    For example $\mathrm{Bad}(n)=\{2\}$ for $5\le n\le12$ and for
    $n=15,17,19,23,31,33,35,39,47$, while $\mathrm{Bad}(13)=\{2,3\}$,
    $\mathrm{Bad}(21)=\{2,5\}$ and $\mathrm{Bad}(29)=\{2,3,7\}$: the
    degree-$47$ circuit is decodable over $\mathrm{GF}(3)$, whereas the
    degree-$13$ crown divides by $2k=6$.  The Lean statement
    \texttt{odd\_realizable\_pairs} certifies the uniform hypothesis that
    $1,\dots,n$ are units (\Cref{sec:formalization-map:units}); the sharper
    set $\mathrm{Bad}(n)$ is read off the displayed pivots and is not
    machine-checked.
\end{remark}

\section{Addition Accounting}
\label{sec:addition-count}

The preceding theorem optimizes the expensive gates, but the displayed
circuits also give a useful addition bound.  We record it separately because
the answer is recursive rather than a single expression such as
$\lfloor n/2\rfloor+1$.

Here is the conclusion before the bookkeeping: the same circuits use
\[
  A_n\le
  \min\left\{2n,\ \frac54n+6\lceil\log_2 n\rceil^2+1\right\}
\]
additions or subtractions.  Thus the uniform constant is~$2$, while the
asymptotic leading constant is~$5/4$.  The first part of this section records
the exact, share-aware recurrences and then proves these two bounds
(the closing subsection treats multiplicative height);
a reader interested only in the headline estimate may skip directly to
\Cref{thm:construction-addition-count,thm:construction-addition-asymptotic}.

Let $\mathcal A$ count binary additions and subtractions of field values.
Negation, copying a wire, and multiplication by a fixed field constant are
free, as in the straight-line-program model of \Cref{sec:model}; every
data-dependent shift such as $U+\alpha$ costs one addition.  A let-bound
subexpression used more than once is charged only once.  Thus this is an
ordinary DAG count, not a count of plus signs after fully expanding a
formula.  If fixed integer multiples must themselves be synthesized from
additions, one may add the length of a chosen addition--subtraction chain at
each displayed multiple $(k-1)U$; this hardware-dependent variant is not the
cost model used here.

Write
\[
 q_1=1,\qquad q_s=5\,2^{s-2}-2\quad(s\ge2)
\]
for the addition count of $Q_{2^s-1}$, and
\[
 f_1=6,\qquad
 f_s=5(2^{s-1}+2^{s-2})-2\quad(s\ge2)
\]
for that of the full fill $A_{2^s}$.  The $s\ge2$ formulas are
\Cref{lem:fill-Q-count}; the two omitted direct bases are
$Q_1=x+\alpha_0$ and
\[
 A_2=(x+\beta_0)\bigl((H_2+\beta_1)S^{(1)}+\alpha_1\bigr)
       +(H_2+\beta_2)S^{(2)}+\alpha_0.
\]

There are two small sharing improvements in the shared $T$ bases.  Every
costed level-one call is supplied as
$\widetilde H_2=H_2+\rho$, so the circuit retains $\rho$ instead of
materializing $\widetilde H_2$ and later subtracting $H_2$ again.  It forms
$\widetilde H_4=H_4+\rho$ directly.  At the following shared odd base, put
\[
 F_1=H_4-(k-1)U_1.
\]
Since $U_2=U_1-\rho$ and $\widetilde H_4=H_4+\rho$, the second factor is
not formed independently:
\begin{equation}
 \widetilde H_4-(k-1)U_2=F_1+k\rho.
 \label{eq:shared-factor}
\end{equation}
Both rewrites preserve the construction identities and all decoder pivots;
they only remove redundant affine gates.  The proof-only quantity
$W_2=W_1+z$ is likewise not materialized, since the circuit directly forms
$\widetilde H_8=H_8+z$.

Let $\tau(k,l)$ be the additions used by this shared schedule for
$(T^{(1)}_{k,2^l},T^{(2)}_{k,2^l})$, with the power inputs already
available (and with the scalar shift retained at the two shared bases).
The algorithms give the exact recurrence
\begin{align}
 \tau(1,l)&=0, \label{eq:T-add-one}\\
 \tau(2m,1)&=\tau(m,2)+5, \label{eq:T-add-even-base}\\
 \tau(2m,l)&=\tau(m,l+1)+2q_{l-1}+8
     &&(l\ge2), \label{eq:T-add-even}\\
 \tau(2m+1,2)&=\tau(m,3)+15
     &&(m\ge1), \label{eq:T-add-odd-base}\\
 \tau(2m+1,l)&=\tau(m,l+1)+q_{l-1}+2q_{l-2}+q_l+16
     &&(m\ge1,\ l\ge3). \label{eq:T-add-odd}
\end{align}
For example, the $5$ in \eqref{eq:T-add-even-base} consists of four
additions for
$(H_2+(x+a))(H_2-(x+a))+e$ and one for $H_4+\rho$.
In \eqref{eq:T-add-odd-base}, the $Q_3$ call costs three, the two affine
inputs $U_1,V_1$ cost three, the common octic costs four, the shifted octic
costs one, and the two factors and two tails cost four by
\eqref{eq:shared-factor}, for a total of fifteen.
The constants $8$ and $16$ in the ordinary branches are obtained by the
same literal count of their two difference-of-squares products and output
assemblies.  Hence the recurrence charges every displayed gate exactly once.

For reference, let $g_d$ denote the addition count of the selected
degree-$d$ auxiliary gadget $\mathcal Q_d$.  The three gadget constructions
give
\begin{align}
 g_1&=1,&g_3&=3,&g_7&=8,\label{eq:g-add-bases}\\
 g_{4k+1}&=\tau(2k,1)+3 &&(k\ge1),\label{eq:g-add-4k1}\\
 g_{8k+3}&=\tau(2k,2)+9 &&(k\ge1),\label{eq:g-add-8k3}\\
 g_{8k+7}&=\tau(k,3)+19 &&(k\ge1).\label{eq:g-add-8k7}
\end{align}
Indeed, \eqref{eq:g-add-4k1} charges one addition for
$\widehat H_2=H_2+\gamma$ and two for
$(x+\beta)S^{(1)}+S^{(2)}$; the scalar difference is passed directly to
the shared base.  The known-powers construction in full generality costs
\[
 q_{l-1}+2+\tau(2k,l)+f_{l-1}
 \tag{known-powers-additions}
\]
for degree $2^{l+1}k+(2^l-1)$; setting $l=2$ gives
\eqref{eq:g-add-8k3}.  Finally, the barred gadget uses five additions for
$H_8$, one for $\widetilde H_8$, thirteen for $A_4$, and
$\tau(k,3)$ internally, proving \eqref{eq:g-add-8k7}.
The low slot in the $8k+3$ construction has one deliberate exception:
when $k=1$ the algorithm uses the scalar $\alpha_1$, not the polynomial
$Q_1=x+\alpha_1$.  Accordingly, put
\[
  g_d^{\mathrm{lo}}=g_d\quad(d\ge3),
  \qquad g_1^{\mathrm{lo}}=0.
\]

Let $a_n$ be the additions used to jointly produce the compatible pair in
odd degree $n\ne7$.  The finite ledgers and the three induction branches are
then
\begin{equation}
 \begin{gathered}
 a_3=3,\qquad a_{15}=23,\\
 a_{27}=43,\qquad a_{31}=43.
 \end{gathered}
 \label{eq:pair-add-bases}
\end{equation}
\begin{align}
 a_{4k+1}&=\tau(2k,1)+2,
 \label{eq:pair-add-4k1}\\
 a_{8k+3}&=a_{2k+1}+g_{4k+1}+g^{\mathrm{lo}}_{2k-1}+7,
 \label{eq:pair-add-8k3}\\
 a_{8k+7}&=a_{2k+1}+g_{2k+1}+g_{4k+3}+6.
 \label{eq:pair-add-8k7}
\end{align}
The first two recursive lines apply for $k\ge1$ (with $k\ne3$ in the
$8k+3$ line), and the last for $k\ge2$, $k\ne3$.
Here $a_{4k+1}$ includes the two additions forming $H_2$; the scalar
$\rho$ is passed to the shared $T$ base without first forming
$\widetilde H_2$.  The constants $7$ and $6$ are the literal costs of the
two outer difference-of-squares assemblies in the $8k+3$ and $8k+7$
algorithms.  In the latter, the two common forms
$S^{(1)}_3\mathbin{\pm}S^{(1)}_2$ are reused and the two scalar coordinates
are $s=a+b,d=b-a$.  This fixed linear reparameterization is invertible under
the admissibility hypothesis, so it does not change decodability.  For the
finite cases, the corresponding ledgers are
\[
\begin{array}{c|l|c}
 n&\text{addition costs, with shared producers charged once}&a_n\\ \hline
 3&H_2:2,\ T^{(2)}=H_2+\alpha_0:1&3\\
 15&H_2:2,\ H_4:4,\ Q_7:8,\ \text{outer assemblies}:9&23\\
 27&H_2:2,\ Q_{13}:23,\ Q_3:3,\ Q_7:8,\
       \text{outer assemblies}:7&43\\
 31&H_2,H_4:6,\ \bar Q_{15}:19,\ Q_7:8,\ Q_3:3,\
       \text{outer assemblies}:7&43.
\end{array}
\]

Finally let $A_n$ count additions in the complete degree-$n$ polynomial.
At top level in degree three we use the direct $Q_3$ circuit with
$H_2=x^2$; it costs three additions, one fewer than first producing the
degree-three pair and then combining it.  (The pair is still used inside
the recursion.)  The displayed septic circuit has eleven additions, but an
invertible affine change of its first three coordinates saves one.  Namely,
with fresh coordinates $\beta_i$, replace its last line by
\[
 P^{\rm add}_7=\beta_0+y+
   (\beta_2+x+z)(\beta_1+w).
\]
This is exactly the septic of \Cref{lem:septic-base} after the substitution
\[
 \alpha_0=\beta_0+\beta_1,\qquad
 \alpha_1=\beta_1,\qquad
 \alpha_2=\beta_2-1,
 \qquad \alpha_i=\beta_i\ (3\le i\le6).
\]
Thus its explicit decoder, followed by
$\beta_0=\alpha_0-\alpha_1$, $\beta_1=\alpha_1$,
$\beta_2=\alpha_2+1$, proves decodability.  The new last line costs two
additions instead of three, so the optimized septic costs ten.  In every
other odd pair case the final combination costs one addition, as does the
even lift, so
\begin{equation}
 \begin{aligned}
 A_1&=1,\qquad A_2=2,\qquad A_3=3,\qquad A_7=10,\\
 A_n&=a_n+1 &&(n\text{ odd},\ n\notin\{3,7\}),\\
 A_n&=A_{n-1}+1 &&(n\ge4\text{ even}).
 \end{aligned}
 \label{eq:full-add-recurrence}
\end{equation}
Equations
\eqref{eq:T-add-one}--\eqref{eq:full-add-recurrence} are an exact,
share-aware addition count for every construction in the theorem.

\begin{theorem}[Uniform addition bound]
\label{thm:construction-addition-count}
The optimized schedules above compute every complete degree-$n$
construction with at most $2n$ additions or subtractions.  More precisely,
$a_n\le2n-1$ for every costed odd compatible pair, and $A_n\le2n$ for
every $n\ge1$.
\end{theorem}
\begin{proof}
We include the short numerical induction so that the bound does not rest on
experimental enumeration.  Substituting the formulas for $q_s$ into
\eqref{eq:T-add-even}--\eqref{eq:T-add-odd} and inducting strongly on $k$
gives
\begin{align}
 \tau(k,2)&\le8(k-1)+2,\label{eq:T-add-bound-2}\\
 \tau(k,l)&\le2(k-1)2^l-\bigl(3\,2^{l-2}-4\bigr)
       &&(k\ge2,\ l\ge3),\label{eq:T-add-bound-high}\\
 \tau(2k,1)&\le8k-1.\label{eq:T-add-bound-1}
\end{align}
For clarity, the fresh nonrecursive costs used in that induction are
$5,10,5\,2^{l-2}+4$ in the three even cases
$l=1,l=2,l\ge3$, and $15,29,5\,2^{l-1}+8$ in the odd cases
$l=2,l=3,l\ge4$.  These are respectively no larger than the difference
between the right-hand side at $(k,l)$ and the right-hand side at the
smaller recursive argument.  This proves the three displayed inequalities;
at $l\ge3$ the worst boundary case is $k=2$, where
\eqref{eq:T-add-bound-high} is an equality.

Splitting $k$ by parity, and once more by its residue modulo four when
needed to select \eqref{eq:g-add-8k3} or \eqref{eq:g-add-8k7}, the same
three bounds give
\begin{align}
 g_{4k+1}+g^{\mathrm{lo}}_{2k-1}&\le12k-3 &&(k\ge1),
 \label{eq:g-pair-bound-3}\\
 g_{2k+1}+g_{4k+3}&\le12k+3 &&(k\ge2).
 \label{eq:g-pair-bound-7}
\end{align}
Here are the only estimates used in that residue check.  If $k\ge3$ is odd,
$g_{4k+1}\le8k-1$ and
$g^{\mathrm{lo}}_{2k-1}\le4k-2$; if $k$ is even,
$g_{4k+1}\le8k+2$ and
$g^{\mathrm{lo}}_{2k-1}\le4k-5$.  The case $k=1$ is checked directly.
For the second line, writing $k=2r$ gives the exact upper ledger
\[
 g_{2k+1}+g_{4k+3}
 \le \tau(r,2)+\tau(r,3)+27\le24r+3=12k+3;
\]
when $k=2r+1$, the estimates
$g_{4r+3}\le8r+3$ and
$g_{8r+7}\le16r+1$ for $r\ge2$ give a stronger bound, and the case $r=1$
is direct.  Thus \eqref{eq:g-pair-bound-3}--\eqref{eq:g-pair-bound-7}
follow without a finite search.

We now use strong induction on the odd target degree.  The four values in
\eqref{eq:pair-add-bases} satisfy $a_n\le2n-1$.  For $n=4k+1$,
\eqref{eq:T-add-bound-1} and \eqref{eq:pair-add-4k1} give
$a_n\le8k+1=2n-1$.  In the $8k+3$ branch, the induction hypothesis and
\eqref{eq:g-pair-bound-3} give
\[
 a_{8k+3}\le(4k+1)+(12k-3)+7=16k+5=2(8k+3)-1.
\]
In the $8k+7$ branch, \eqref{eq:g-pair-bound-7} gives
\[
 a_{8k+7}\le(4k+1)+(12k+3)+6
     =16k+10<2(8k+7)-1.
\]
This proves the pair bound.  Adding the final combination gives
$A_n\le2n$ for odd $n\notin\{3,7\}$; the direct cubic and septic have
$A_3=3$ and $A_7=10$.  The other two small bases are immediate, and an
even lift adds only one gate, so
$A_n=A_{n-1}+1\le2(n-1)+1<2n$.  The theorem follows.
\end{proof}

The uniform factor-two statement is convenient for small degrees, but it
does not describe the asymptotic behaviour of the same schedules.  Their
leading addition constant is in fact $5/4$, not~$2$.

\begin{theorem}[Sharper asymptotic addition bound]
\label{thm:construction-addition-asymptotic}
Put
\[
  L(n)=\lceil\log_2 n\rceil,
  \qquad L(1)=0.
\]
The optimized schedules above satisfy
\[
  A_n\le \frac54n+6L(n)^2+1.
\]
Consequently
\[
  A_n\le
  \min\left\{2n,\ \frac54n+6\lceil\log_2n\rceil^2+1\right\},
\]
so they use at most $\frac54n+O((\log n)^2)$ additions.  In particular,
$A_n\le\frac32n$ for every $n\ge4096$.
\end{theorem}
\begin{proof}
We first sharpen the coarse estimates on the $T$ recursion.  The following
slightly stronger high-level estimate is convenient:
\begin{equation}
 \tau(k,l)\le
 5\,2^{l-2}(k-1)+8L(k)-4
 \qquad(k\ge2,\ l\ge4).
 \label{eq:T-add-sharp-high}
\end{equation}
This follows by strong induction on $k$.  Put $p=2^{l-2}$.  At an even
step the fresh cost is $5p+4$, while at an odd step it is $10p+8$.
The case $k=2$ is an equality.  If $k=2m$ with $m\ge2$, apply the
induction hypothesis to $(m,l+1)$ and use
$L(2m)=L(m)+1$; the resulting bound is four smaller than the right-hand
side of \eqref{eq:T-add-sharp-high}.  If $k=2m+1$, the case $m=1$ is
direct, and for $m\ge2$ the induction hypothesis together with
$L(2m+1)\ge L(m)+1$ gives the claimed bound directly.

The exceptional level $l=3$ has fresh costs $14$ and $29$.  Applying
\eqref{eq:T-add-sharp-high} at level four (and checking $k=1,2,3$
directly) gives
\begin{equation}
 \tau(k,3)\le10(k-1)+8L(k)
 \qquad(k\ge1).
 \label{eq:T-add-sharp-three}
\end{equation}
For $l=2$, use
\[
 \tau(2m,2)=\tau(m,3)+10,
 \qquad
 \tau(2m+1,2)=\tau(m,3)+15.
\]
Together with $L(2m)=L(m)+1$ and
$L(2m+1)\ge L(m)+1$, these give the same bound
\begin{equation}
 \tau(k,2)\le5(k-1)+8L(k).
 \label{eq:T-add-sharp-two}
\end{equation}

Substitution in \eqref{eq:g-add-bases}--\eqref{eq:g-add-8k7} now gives one
uniform estimate for every selected odd gadget:
\[
  g_d\le\frac54d+8L(d).
  \tag{gadget-sharp-bound}
\]
For example,
$g_{4k+1}=\tau(k,2)+8$,
$g_{8k+3}=\tau(2k,2)+9$, and
$g_{8k+7}=\tau(k,3)+19$; the three direct bases are immediate.

We next prove
\[
  a_n\le\frac54n+6L(n)^2
  \tag{pair-sharp-bound}
\]
by strong induction over the costed odd pairs.  The four finite bases satisfy
it directly.  If $n=4k+1$, then
\[
 a_n=\tau(k,2)+7
   \le5(k-1)+8L(k)+7
   \le\frac54n+6L(n)^2.
\]
In either of the remaining branches let $m=2k+1$ be the smaller pair and
let $d_1,d_2$ be the two auxiliary-gadget degrees (with the degree-one low
slot omitted when $k=1$).  In both cases
\[
 d_1+d_2\le n-m-2,
 \qquad L(m)\le L(n)-2.
\]
The outer assembly costs at most seven additions.  Writing $L=L(n)$, the
induction hypothesis and (gadget-sharp-bound) therefore give
\[
 \begin{aligned}
  a_n
  &\le \frac54m+6(L-2)^2
       +\frac54(d_1+d_2)+16L+7\\
  &\le \frac54n+6(L-2)^2+16L+\frac92
   \le \frac54n+6L^2.
 \end{aligned}
\]
The last inequality is $8L\ge57/2$; every recursive target here has
$L\ge4$.  This proves (pair-sharp-bound).

For an odd complete construction, the last combination adds one, except
for the already smaller direct cubic and septic bases.  An even lift also
adds one and satisfies $L(n-1)\le L(n)$.  Hence
$A_n\le\frac54n+6L(n)^2+1$ for every $n$.  Combining this with
\Cref{thm:construction-addition-count} proves the minimum displayed in the
statement.  Finally,
$6L(n)^2+1\le n/4$ for $n\ge4096$ (check the first two dyadic boundary
points; thereafter the right-hand side doubles while the left grows
quadratically in $L$), which gives the last assertion.
\end{proof}

\begin{remark}[What would be needed for an $n+o(n)$ addition bound]
The remaining quarter is structural in the present schedule, not an artefact
of the final estimate: already the Mersenne block has the exact count
\[
 \mathcal A(Q_{2^s-1})=5\,2^{s-2}-2
   =\frac54(2^s-1)-\frac34.
\]
The affine sharing above removes repeated shifts and outer assembly gates,
but it does not change this leading term.  Reaching $n+o(n)$ additions while
retaining the exact $\lfloor n/2\rfloor+1$ multiplication count would
therefore require a genuinely more addition-efficient fill (or a different
known-powers gadget), rather than a sharper analysis of the current DAG.
Rabin and Winograd obtain $n+o(n)$ additions when an additional
$O(\log n)$ multiplications are allowed~\cite{rabin1972number}, and the
constant $\frac54$ is exactly that of Winograd's earlier monic
construction (Knuth's exercise~44, the precursor of theirs;
\Cref{sec:related}): it survived the change of construction while the
logarithmic multiplication term did not.
Eliminating both overheads simultaneously remains open here.
\end{remark}

\subsection*{Multiplicative height}

The construction uses the binary known-powers gadget of
\Cref{alg:constr-known-2n-1} throughout.  Its multiplicative height is
bounded in \Cref{thm:construction-height}: writing
$L=\lceil\log_2 n\rceil$, the complete circuit has height at most $2L+4$
for odd $n\ge3$ and $2L+5$ for even $n\ge4$, with the multiplication
counts recorded above.  This is the maximum number of nonscalar
products on an input-to-output path, not the total gate count.
Knuth's treatment of parallel evaluation~\cite[\S4.6.4]{knuth1997seminumerical}
covers only Estrin's method and the $k$th-order Horner rule, without depth
bounds for preconditioned schemes; \Cref{thm:construction-height} shows that
the minimum multiplication count is compatible with height
$2\lceil\log_2 n\rceil+O(1)$.  We claim no optimality for this height: the
only lower bound is the trivial $\lceil\log_2 n\rceil$, so a factor of two
remains.

\begin{corollary}[All odd degrees have a decodable construction]
\label{cor:all-odd-decodable}
Over every $n$-admissible field, each odd $n\ge3$ has a decodable monic
degree-$n$ construction.  For $n=7$ use
\Cref{lem:septic-base}; for every other odd degree use
\Cref{thm:odd-realizable-pairs}.
\end{corollary}

\section{Formalization Map}
\label{appendix:formalization-map}

This appendix records which statements of the paper are machine-checked in
the Lean~4 development \texttt{FastPoly/\allowbreak } distributed with the source, under
which hypotheses, and where the formalization stops short of the paper.  The
abstract's sentence ``Everything is checked in Lean'' refers to the complete
upper-bound construction stated immediately before it: the decoder,
multiplication count, addition bounds, and logarithmic-height arrangement.
Those claims are joined by the paper-level capstone in
\texttt{PaperMain.lean}.  The sentence does not refer to independent auxiliary
results recorded later in the paper, whose coverage is reported separately
below.
The
Lean statements are made over an abstract commutative ring $R$ and a
commutative $R$-algebra $A$ (Lean: \texttt{[CommRing R] [CommRing A]
[Algebra R A]}); the only characteristic-type assumptions are explicit
\texttt{IsUnit} hypotheses on integers, discussed in
\Cref{sec:formalization-map:units}.  Decoding is expressed as membership in
a generated subalgebra: ``$\theta$ is recovered from $P$'' is
$\theta\in K\sqcup\operatorname{adjoin}_R\{[x^i]P\}$ for the ambient known
algebra $K$, and every recovery proof exhibits the pivot, peel, division, or
block inverse it uses (the repository forbids search and decision
procedures on nontrivial goals).  In the tables, \emph{full} means the
displayed paper statement is proved as stated (up to the deviations noted in
the hypotheses column), \emph{partial} means a strictly weaker or
differently packaged statement is proved, and \emph{not formalized} means
no Lean counterpart exists.  File names are relative to \texttt{FastPoly/\allowbreak }.

\subsection{Main results}
\label{sec:formalization-map:main}

\begingroup
\footnotesize
\setlength{\tabcolsep}{3pt}
\par\noindent\begin{tabular}{p{0.95in}p{2.05in}p{1.75in}p{0.55in}}
\toprule
Paper & Lean (file) & Hypotheses in Lean & Status\\
\midrule
\end{tabular}\par\nopagebreak
\noindent\begin{tabular}{p{0.95in}p{2.05in}p{1.75in}p{0.55in}}
\raggedright \Cref{thm:main} &\raggedright
\texttt{MainArrangementsChecked.of\_\allowbreak charZero},
\texttt{MainArrangementsChecked.of\_\allowbreak charP}
(\texttt{PaperMain.lean});
\texttt{decodedAdditionPolynomial\_\allowbreak exists}
(\texttt{Cost/\allowbreak Additions/\allowbreak DecodedPolynomial.lean});
\texttt{monic\_\allowbreak coefficient\_\allowbreak map\_\allowbreak bijective}
(\texttt{Main.lean});
\texttt{polynomial\_\allowbreak height} (\texttt{HeightFinal.lean});
\texttt{MvPolynomial.algEquivOfDecodable} (\texttt{Automorphism.lean}) &
\raggedright Characteristic zero or $p>n$.  The paper-level capstone constructs
both advertised arrangements at once on one common monic degree-$n$ polynomial:
a fixed program carrying the decoder, multiplication count, and logarithmic height;
and a fixed addition-optimized program carrying its semantics, multiplication bound,
exact literal addition count, and both addition inequalities.  The two scheduling
objectives may use separate fixed syntax trees, but both are proved to compute the
same polynomial;
the coefficient preprocessing map is the polynomial automorphism of
\Cref{lem:polynomial-left-inverse-automorphism}. &
full\\
\end{tabular}\par
\noindent\begin{tabular}{p{0.95in}p{2.05in}p{1.75in}p{0.55in}}
\raggedright \Cref{thm:odd-realizable-pairs} &\raggedright \texttt{odd\_\allowbreak realizable\_\allowbreak pairs}, \texttt{odd\_\allowbreak realizable\_\allowbreak pairs'}
(\texttt{Main.lean}) &
\raggedright \texttt{[Nontrivial A]}; $n$ odd, $3\le n$, $n\ne7$;
\texttt{$\forall$ i, 1 $\le$ i $\to$ i $\le$ n $\to$ IsUnit ((i : $\mathbb Z$) : R)}; any parameter block
\texttt{$\theta$ : $\mathbb N$ $\to$ A}.  Conclusion: \texttt{CompatiblePair $\bot$ T$_1$ T$_2$ (n-1) G};
$H_2$ (and $H_4$ when $n\ge5$) monic of degree $2$ ($4$); every subalgebra
containing the coefficients of $x T_1+T_2$ contains $\theta_0,\dots,\theta_{n-1}$
and the coefficients of $H_2,H_4$; and a program
\texttt{prog : Cost.\allowbreak JointPairProgram R ((n-1)/\allowbreak 2)} with
\texttt{prog.\allowbreak RealizesAt $\theta$ T$_1$ T$_2$ H$_2$ H$_4$} and
\texttt{prog.\allowbreak HeightBounded (2 * Nat.clog 2 n + 3)}. &
full\\
\end{tabular}\par
\noindent\begin{tabular}{p{0.95in}p{2.05in}p{1.75in}p{0.55in}}
\raggedright \Cref{thm:odd-realizable-pairs}, fixed program before keys &\raggedright \texttt{odd\_\allowbreak realizable\_\allowbreak pairs\_\allowbreak free},
\texttt{odd\_\allowbreak realizable\_\allowbreak pairs\_\allowbreak uniform\_\allowbreak family} (\texttt{Main.lean}) &
\raggedright Instantiation at $A=R'[\alpha_0,\dots,\alpha_{n-1}]$
(\texttt{MvPolynomial (Fin n) R'}); the same program syntax realizes the
specialized pair and powers for every key vector over every $R'$-algebra
(\texttt{RealizesFiniteFamily}). &
full\\
\end{tabular}\par
\noindent\begin{tabular}{p{0.95in}p{2.05in}p{1.75in}p{0.55in}}
\raggedright \Cref{thm:construction-count}, count clause &\raggedright \texttt{Cost.pair\_\allowbreak construction\_\allowbreak count},
\texttt{Cost.construction\_\allowbreak count} (\texttt{Cost/\allowbreak Final.lean});
\texttt{Cost.\allowbreak SepticProgram.program} (\texttt{Cost/\allowbreak SepticProgram.lean});
\texttt{Cost.\allowbreak PolynomialProgram.evenLift},
\texttt{evenLift\_\allowbreak realizesAt} (\texttt{Cost/\allowbreak PolynomialProgram.lean}) &
\raggedright Recurrences \texttt{PairCost}/\texttt{PolynomialCost} over $\mathbb N$ (no ring):
$\exists c$, \texttt{PolynomialCost n c} with $c=\lfloor n/2\rfloor+1$ for
$n\ge3$, $c=0,1$ for $n=1,2$.  The literal program with $(n-1)/2$ products is
the \texttt{JointPairProgram} conjunct of the master; the septic program has
four products; the even lift adds one product and one addition. &
full\\
\end{tabular}\par
\noindent\begin{tabular}{p{0.95in}p{2.05in}p{1.75in}p{0.55in}}
\raggedright \Cref{thm:construction-count}, coverage clause;
\Cref{cor:all-odd-decodable} &\raggedright \texttt{monic\_\allowbreak coefficient\_\allowbreak map\_\allowbreak bijective},
\texttt{even\_\allowbreak lift\_\allowbreak bijective},
\texttt{odd\_\allowbreak coefficient\_\allowbreak map\_\allowbreak bijective} (\texttt{Main.lean}) &
\raggedright \texttt{[IsNoetherianRing R'] [Nontrivial R']}; $1,\dots,n$ units; every
$n\ge1$ (affine, quadratic, septic, odd master, even lift).  Conclusion: a
monic degree-$n$ $P$ over the free coordinate algebra whose coefficient
substitution \texttt{MvPolynomial.aeval (fun i => P.coeff i)} is bijective. &
full\\
\end{tabular}\par
\noindent\begin{tabular}{p{0.95in}p{2.05in}p{1.75in}p{0.55in}}
\raggedright \Cref{thm:construction-height} &\raggedright
\texttt{polynomial\_\allowbreak height},
\texttt{odd\_\allowbreak polynomial\_\allowbreak height},
\texttt{septic\_\allowbreak polynomial\_\allowbreak height},
\texttt{RealizedPolynomial.\allowbreak evenLift},
\texttt{polynomial\_\allowbreak height\_\allowbreak of\_\allowbreak charZero},
\texttt{polynomial\_\allowbreak height\_\allowbreak of\_\allowbreak charP}
(\texttt{HeightFinal.lean}); height conjunct \texttt{HeightBounded} of
\texttt{odd\_\allowbreak realizable\_\allowbreak pairs} (\texttt{Main.lean});
\texttt{Circuit.multDepth} (\texttt{Height/\allowbreak Depth.lean}); component
ledgers in \texttt{Height/\allowbreak PeeledCircuit.lean} and
\texttt{Height/\allowbreak TCircuitDepth.lean} &
\raggedright \texttt{[Nontrivial A]}; $3\le n$; $1,\dots,n$ units of $R$ (for the
\texttt{\_of\_charZero}/\texttt{\_of\_charP} forms: a field of characteristic
zero, resp.\ $p>n$).  Conclusion
\texttt{RealizedPolynomial R $\theta$ n (n/2+1) (2 * Nat.clog 2 n + 4 + (n+1) \% 2)}:
one fixed complete-polynomial program is monic of degree $n$, decodes all $n$
parameters, uses exactly $\lfloor n/2\rfloor+1$ products, and has output
\texttt{multDepth} (inputs at depth $0$) at most $2\lceil\log_2 n\rceil+4$ for
odd $n$ and $2\lceil\log_2 n\rceil+5$ for even $n$.  Assembled from the
master's pair conjunct ($T_1,T_2$ at depth $\le2\lceil\log_2 n\rceil+3$, $H_2$
at $\le1$, $H_4$ at $\le2$) by the final product $xT_1+T_2$, the direct septic
($n=7$, lifted for $n=8$), and the even lift $xQ+c_0$.  The stronger
\texttt{DecodedAdditionPolynomial} capstone pairs this height witness with the
addition-certified arrangement on the same semantic polynomial. &
full (count, decoder, height)\\
\end{tabular}\par
\noindent\begin{tabular}{p{0.95in}p{2.05in}p{1.75in}p{0.55in}}
\raggedright \Cref{thm:construction-addition-count},
\Cref{thm:construction-addition-asymptotic} &\raggedright \texttt{Cost.construction\_\allowbreak additions\_\allowbreak checked}
(\texttt{Cost/\allowbreak Additions/\allowbreak Realization.lean}); ledger
\texttt{Cost.construction\_\allowbreak addition\_\allowbreak count}
(\texttt{Cost/\allowbreak Additions/\allowbreak Final.lean}) &
\raggedright For every $n\ge1$ over an $n$-admissible ring and every parameter
environment: one fixed program \texttt{program : PolynomialProgram R mult}
with \texttt{mult}\,$\le\lfloor n/2\rfloor+1$, \texttt{program.RealizesAt}
its degree-$n$ polynomial, and its literal gate count
\texttt{program.additions} satisfying both $\le2n$ and
$4\cdot{}\le5n+24\lceil\log_2n\rceil^2+4$.  Assembled from the
addition-certified base pairs, the recursive $8k+3$/$8k+7$ pair steps, the
certified odd-gadget dispatch, and the odd/even combinators; the program is
the optimized schedule of \Cref{sec:addition-count}, not the master's. &
full\\
\bottomrule
\end{tabular}\par
\endgroup

\subsection{Decoder calculus\texorpdfstring{ (\Cref{appendix:decoder-calculus})}{}}
\label{sec:formalization-map:calculus}

\begingroup
\footnotesize
\setlength{\tabcolsep}{3pt}
\par\noindent\begin{tabular}{p{0.95in}p{2.05in}p{1.75in}p{0.55in}}
\toprule
Paper & Lean (file) & Hypotheses in Lean & Status\\
\midrule
\end{tabular}\par\nopagebreak
\noindent\begin{tabular}{p{0.95in}p{2.05in}p{1.75in}p{0.55in}}
\raggedright \Cref{def:compatible-pair}; visible algebra &\raggedright \texttt{CompatiblePair}, \texttt{Vis}, \texttt{CausalPair}
(\texttt{Recover/\allowbreak Context.lean}) &
\raggedright \texttt{[CommRing R] [CommRing A] [Algebra R A]}; relative to a known
subalgebra $K$ and a window $G$. &
full\\
\end{tabular}\par
\noindent\begin{tabular}{p{0.95in}p{2.05in}p{1.75in}p{0.55in}}
\raggedright scalar pivots \eqref{eq:filtered-pivot} &\raggedright \texttt{mem\_\allowbreak obsAlg\_\allowbreak of\_\allowbreak scalarCert} (\texttt{Recover/\allowbreak Filtered.lean}) &
\raggedright Each pivot slope \texttt{lam j : R} is a unit; well-founded index order. &
full\\
\end{tabular}\par
\noindent\begin{tabular}{p{0.95in}p{2.05in}p{1.75in}p{0.55in}}
\raggedright block pivots with explicit inverse &\raggedright \texttt{mem\_\allowbreak of\_\allowbreak blockCert} (\texttt{Recover/\allowbreak Filtered.lean});
\texttt{mem\_\allowbreak of\_\allowbreak known\_\allowbreak blockCert\_\allowbreak of\_\allowbreak det} (\texttt{Recover/\allowbreak KnownBlock.lean}) &
\raggedright Constant matrix $M$ over $R$ with a supplied $N$, $NM=1$; or entries in the
known algebra with $\det M$ a unit of $R$ (adjugate inverse). &
full\\
\end{tabular}\par
\noindent\begin{tabular}{p{0.95in}p{2.05in}p{1.75in}p{0.55in}}
\raggedright \Cref{def:coefficient-triangular}, \Cref{lem:triangular-shift},
\Cref{lem:triangular-implies-compatible} &\raggedright \texttt{CoeffTriangular}, \texttt{CoeffTriangular.shift\_\allowbreak pivot},
\texttt{CoeffTriangular.toCompatiblePair} (\texttt{Recover/\allowbreak Triangular.lean}) &
\raggedright Unit pivots \texttt{lam j} for $j<d$; the compatibility half is proved. &
full\\
\end{tabular}\par
\noindent\begin{tabular}{p{0.95in}p{2.05in}p{1.75in}p{0.55in}}
\raggedright \Cref{lem:triangular-block-concatenation} &\raggedright \texttt{adjoin\_\allowbreak Ico\_\allowbreak glue}, \texttt{side\_\allowbreak data\_\allowbreak absorb}
(\texttt{Recover/\allowbreak Triangular.lean}) &
\raggedright Proved as the two glue lemmas used at every concatenation site. &
full\\
\end{tabular}\par
\noindent\begin{tabular}{p{0.95in}p{2.05in}p{1.75in}p{0.55in}}
\raggedright \Cref{lem:monic-cauchy-transport} &\raggedright \texttt{coeff\_\allowbreak mul\_\allowbreak monic} (\texttt{Polynomial/\allowbreak TopWindow.lean}) &
\raggedright \texttt{[CommRing A]}; $q$ monic.  No unit hypothesis. &
full\\
\end{tabular}\par
\noindent\begin{tabular}{p{0.95in}p{2.05in}p{1.75in}p{0.55in}}
\raggedright \Cref{lem:peel-monic-factor} &\raggedright \texttt{coeff\_\allowbreak mem\_\allowbreak of\_\allowbreak mul\_\allowbreak monic\_\allowbreak add} (\texttt{Polynomial/\allowbreak PeelMonic.lean}) &
\raggedright $M$ monic with coefficients in $K$.  No unit hypothesis. &
full\\
\end{tabular}\par
\noindent\begin{tabular}{p{0.95in}p{2.05in}p{1.75in}p{0.55in}}
\raggedright \Cref{lem:monic-from-power} &\raggedright \texttt{coeff\_\allowbreak mem\_\allowbreak of\_\allowbreak pow\_\allowbreak add} (\texttt{Polynomial/\allowbreak MonicFromPower.lean}) &
\raggedright \texttt{IsUnit (m : R)} for the root index $m$. &
full\\
\end{tabular}\par
\noindent\begin{tabular}{p{0.95in}p{2.05in}p{1.75in}p{0.55in}}
\raggedright \Cref{lem:monic-division} &\raggedright \texttt{coeff\_\allowbreak quot\_\allowbreak mem}, \texttt{coeff\_\allowbreak rem\_\allowbreak mem}
(\texttt{Polynomial/\allowbreak MonicDivision.lean}) &
\raggedright Divisor monic with coefficients in $K$.  No unit hypothesis. &
full\\
\end{tabular}\par
\noindent\begin{tabular}{p{0.95in}p{2.05in}p{1.75in}p{0.55in}}
\raggedright \Cref{lem:square-gadget}, \Cref{lem:square-gadget-boundary} &\raggedright \texttt{square\_\allowbreak gadget\_\allowbreak mem} (\texttt{Polynomial/\allowbreak SquareGadget.lean});
\texttt{coeff\_\allowbreak mem\_\allowbreak of\_\allowbreak square\_\allowbreak gadget\_\allowbreak relative}
(\texttt{Polynomial/\allowbreak CausalShell.lean}) &
\raggedright \texttt{IsUnit (2 : R)}. &
full\\
\end{tabular}\par
\noindent\begin{tabular}{p{0.95in}p{2.05in}p{1.75in}p{0.55in}}
\raggedright \Cref{lem:scalar-shift-square} &\raggedright \texttt{scalar\_\allowbreak shift\_\allowbreak mem} (\texttt{Polynomial/\allowbreak ScalarShift.lean}) &
\raggedright \texttt{IsUnit lam} and \texttt{IsUnit (2 : R)}. &
full\\
\end{tabular}\par
\noindent\begin{tabular}{p{0.95in}p{2.05in}p{1.75in}p{0.55in}}
\raggedright \Cref{lem:x-alpha-extraction} &\raggedright \texttt{x\_\allowbreak alpha\_\allowbreak mem} (\texttt{Recover/\allowbreak XAlpha.lean}) &
\raggedright A compatible pair with window $G\subseteq[0,n)$.  No unit hypothesis. &
full\\
\end{tabular}\par
\noindent\begin{tabular}{p{0.95in}p{2.05in}p{1.75in}p{0.55in}}
\raggedright \Cref{lem:discharge-side-information},
\Cref{lem:extractable-via-derivable} &\raggedright \texttt{discharge\_\allowbreak side\_\allowbreak information},
\texttt{extractable\_\allowbreak via\_\allowbreak derivable} (\texttt{Recover/\allowbreak Context.lean}) &
\raggedright As stated. &
full\\
\end{tabular}\par
\noindent\begin{tabular}{p{0.95in}p{2.05in}p{1.75in}p{0.55in}}
\raggedright \Cref{lem:polynomial-left-inverse-automorphism} &\raggedright \texttt{MvPolynomial.algEquivOfDecodable} (\texttt{Automorphism.lean});
\texttt{coefficientAlgEquiv\allowbreak OfMonicDecodable} (\texttt{Instantiation.lean}) &
\raggedright \texttt{[IsNoetherianRing R] [Finite $\sigma$]}; decodability of the variables from
the coefficient family gives a two-sided $R$-algebra automorphism. &
full\\
\end{tabular}\par
\noindent\begin{tabular}{p{0.95in}p{2.05in}p{1.75in}p{0.55in}}
\raggedright \Cref{def:joint-realization} &\raggedright \texttt{Cost.\allowbreak JointPairProgram}, \texttt{JointPairProgram.\allowbreak RealizesAt}
(\texttt{Cost/\allowbreak PolynomialCircuit.lean});
\texttt{Cost.\allowbreak MultiplicationProgram} (\texttt{Cost/\allowbreak MultiplicationProgram.lean}) &
\raggedright A straight-line program over $R$ with a stored multiplication count; four
outputs $(T_1,T_2,H_2,H_4)$. &
full\\
\bottomrule
\end{tabular}\par
\endgroup

\subsection{Construction layers\texorpdfstring{ (\Cref{appendix:constructions})}{}}
\label{sec:formalization-map:construction}

\begingroup
\footnotesize
\setlength{\tabcolsep}{3pt}
\par\noindent\begin{tabular}{p{0.95in}p{2.05in}p{1.75in}p{0.55in}}
\toprule
Paper & Lean (file) & Hypotheses in Lean & Status\\
\midrule
\end{tabular}\par\nopagebreak
\noindent\begin{tabular}{p{0.95in}p{2.05in}p{1.75in}p{0.55in}}
\raggedright \Cref{alg:constr-fill}, \Cref{lem:fill-correctness} &\raggedright \texttt{fill\_\allowbreak correct} (\texttt{Section4/\allowbreak FillRec.lean}) &
\raggedright \texttt{[Nontrivial A]}; per-level certificates \texttt{GoodLevel}; known
monic quadratic $H_1$. &
full\\
\end{tabular}\par
\noindent\begin{tabular}{p{0.95in}p{2.05in}p{1.75in}p{0.55in}}
\raggedright \Cref{alg:constr-known-2n-1}, \Cref{lem:Q-unitriangular} &\raggedright \texttt{peel}, \texttt{peel\_\allowbreak correct} (\texttt{Section4/\allowbreak Peeled.lean});
\texttt{peel\_\allowbreak unitriangular} (\texttt{Section4/\allowbreak PeeledCert.lean}) &
\raggedright \texttt{[Nontrivial A]}; a tower of known monic powers of degrees $2^i$. &
full\\
\end{tabular}\par
\noindent\begin{tabular}{p{0.95in}p{2.05in}p{1.75in}p{0.55in}}
\raggedright \Cref{lem:peeled-Q-decodable}, \Cref{lem:peeled-Q-count} &\raggedright \texttt{peel}, \texttt{peel\_\allowbreak correct} (\texttt{Section4/\allowbreak Peeled.lean});
\texttt{peelC\_\allowbreak multiplications}, \texttt{peelC\_\allowbreak additions},
\texttt{peelC\_\allowbreak multDepth} (\texttt{Height/\allowbreak PeeledCircuit.lean}) &
\raggedright Known monic powers and the binary recursion; exact counts
$2^{k-1}-1$ products, $5\cdot2^{k-2}-2$ additions, depth $k$ for $k\ge2$. &
full\\
\end{tabular}\par
\noindent\begin{tabular}{p{0.95in}p{2.05in}p{1.75in}p{0.55in}}
\raggedright \Cref{alg:constr-Tk2l} (structural layer) &\raggedright \texttt{Tpair}, \texttt{TF\_\allowbreak good}, \texttt{Tpair\_\allowbreak good}
(\texttt{Section5/\allowbreak T.lean}) &
\raggedright Monic/degree $k2^l$ for all four branches; no unit hypothesis. &
full\\
\end{tabular}\par
\noindent\begin{tabular}{p{0.95in}p{2.05in}p{1.75in}p{0.55in}}
\raggedright \Cref{lem:Rk2l}, \Cref{lem:Rk2l-leading-coeff},
\Cref{lem:Rk2l-top-boundary} &\raggedright \texttt{Rk2l\_\allowbreak triangular}, \texttt{Rk2l\_\allowbreak top\_\allowbreak two}
(\texttt{Section5/\allowbreak Rk2lTriMaster.lean}); \texttt{Rk2l\_\allowbreak deg},
\texttt{Rk2l\_\allowbreak lead} (\texttt{Section5/\allowbreak Rk2l.lean}) &
\raggedright \texttt{[Nontrivial A]}; $1,\dots,k$ units of $R$; $l\ge2$; tower of known
monic powers; the scalar-difference clause at $l=2$ for odd $k\ge3$. &
full\\
\end{tabular}\par
\noindent\begin{tabular}{p{0.95in}p{2.05in}p{1.75in}p{0.55in}}
\raggedright \Cref{lem:causal-perturbed-T} &\raggedright \texttt{causal\_\allowbreak perturbed\_\allowbreak T}, \texttt{Tpair\_\allowbreak compatiblePair}
(\texttt{Section5/\allowbreak PerturbedT.lean}) &
\raggedright $1,\dots,M$ units of $R$ ($M$ even); $l\ge2$. &
full\\
\end{tabular}\par
\noindent\begin{tabular}{p{0.95in}p{2.05in}p{1.75in}p{0.55in}}
\raggedright \Cref{lem:4k+1-splittable}, \Cref{lem:Q4k+1-from-H2} &\raggedright \texttt{fourk\_\allowbreak decodable}, \texttt{q4k1\_\allowbreak decodable}
(\texttt{Section6/\allowbreak GadgetDecoders.lean}); crown in
\texttt{Section5/\allowbreak FourKPlusOne.lean} &
\raggedright $1,\dots,2k$ units of $R$ and \texttt{IsUnit (2 : R)}; five explicit
$V$-relative pivots. &
full\\
\end{tabular}\par
\noindent\begin{tabular}{p{0.95in}p{2.05in}p{1.75in}p{0.55in}}
\raggedright \Cref{lem:Q-odd-degree-with-powers} &\raggedright \texttt{q\_\allowbreak odd\_\allowbreak degree\_\allowbreak decodable} (\texttt{Section6/\allowbreak QOddDegree.lean}) &
\raggedright $1,\dots,2k$ units of $R$; $l\ge2$; composes \texttt{fill\_\allowbreak correct},
\texttt{causal\_\allowbreak perturbed\_\allowbreak T}, \texttt{peel\_\allowbreak correct}. &
full\\
\end{tabular}\par
\noindent\begin{tabular}{p{0.95in}p{2.05in}p{1.75in}p{0.55in}}
\raggedright \Cref{lem:barQ15}, \Cref{lem:barQ8k+7} &\raggedright \texttt{barredGadgets\_\allowbreak of\_\allowbreak admissible} (\texttt{Examples/\allowbreak BarredGadgets.lean};
\texttt{BarQ15.lean}, \texttt{BarQGeneral.lean}) &
\raggedright \texttt{[Nontrivial A]}; $1,\dots,\mathrm{cap}$ units of $R$; the
$4\times4$ block with determinant $-k^2$ is inverted by adjugate. &
full\\
\end{tabular}\par
\noindent\begin{tabular}{p{0.95in}p{2.05in}p{1.75in}p{0.55in}}
\raggedright \Cref{lem:odd-gadgets-H2H4} &\raggedright \texttt{odd\_\allowbreak gadget\_\allowbreak dispatch} (\texttt{Section6/\allowbreak Dispatch.lean}) &
\raggedright $1,\dots,d$ units of $R$; hypothesis \texttt{BarredGadgets d}, discharged
by the previous row. &
full\\
\end{tabular}\par
\noindent\begin{tabular}{p{0.95in}p{2.05in}p{1.75in}p{0.55in}}
\raggedright \Cref{lem:8k+3-splittable}, \Cref{lem:8k+7-splittable} &\raggedright \texttt{eightk3\_\allowbreak compatible}, \texttt{eightk3\_\allowbreak decodable},
\texttt{eightk7\_\allowbreak compatible}, \texttt{eightk7\_\allowbreak decodable}
(\texttt{Section6/\allowbreak Induction.lean}) &
\raggedright \texttt{IsUnit (2 : R)} plus the smaller pair's compatibility and decoder
and the two gadget decoders as hypotheses. &
full\\
\end{tabular}\par
\noindent\begin{tabular}{p{0.95in}p{2.05in}p{1.75in}p{0.55in}}
\raggedright \Cref{lem:base-three-compatible} &\raggedright \texttt{base\_\allowbreak three\_\allowbreak compatible} (\texttt{Section6/\allowbreak SpecialCases.lean}) &
\raggedright None. &
full\\
\end{tabular}\par
\noindent\begin{tabular}{p{0.95in}p{2.05in}p{1.75in}p{0.55in}}
\raggedright \Cref{lem:septic-base} &\raggedright \texttt{septic\_\allowbreak good}, \texttt{septic\_\allowbreak decodable}
(\texttt{Examples/\allowbreak Septic.lean}); \texttt{Cost.\allowbreak SepticProgram.program}
(four products, ten additions) &
\raggedright \texttt{IsUnit (2 : R)}.  The unpublished impossibility of a three-product
joint septic realization (the reason $n=7$ is exceptional) has no Lean
counterpart. &
full\\
\end{tabular}\par
\noindent\begin{tabular}{p{0.95in}p{2.05in}p{1.75in}p{0.55in}}
\raggedright \Cref{lem:special-cases-splittable} ($15,27,31$) &\raggedright \texttt{decodable} in \texttt{Examples/\allowbreak P15.lean},
\texttt{Examples/\allowbreak P27Full.lean}, \texttt{Examples/\allowbreak P31Full.lean}
(namespaces \texttt{P15}, \texttt{P27Full}, \texttt{P31Full}) &
\raggedright \texttt{[Nontrivial A]}; unit hypotheses as consumed by the inner
\texttt{barQ15}, $Q_7$, $Q_3$ discharges. &
full\\
\end{tabular}\par
\noindent\begin{tabular}{p{0.95in}p{2.05in}p{1.75in}p{0.55in}}
\raggedright \Cref{lem:fill-Q-count}, \Cref{lem:T-multiplication-count},
\Cref{lem:odd-gadgets-count} &\raggedright \texttt{fill\_\allowbreak count}, \texttt{mers\_\allowbreak multiplication\_\allowbreak count},
\texttt{t\_\allowbreak multiplication\_\allowbreak count} (\texttt{Cost/\allowbreak Counts.lean});
\texttt{fourKPlusOne\_\allowbreak multiplication\_\allowbreak count},
\texttt{knownPowersOdd\_\allowbreak multiplication\_\allowbreak count},
\texttt{barredEightKPlusSeven\_\allowbreak multiplication\_\allowbreak count}
(\texttt{Cost/\allowbreak Gadgets.lean}) &
\raggedright Over $\mathbb N$ (schedule recurrences); the realized branch circuits
carry the same counts by construction. &
full\\
\bottomrule
\end{tabular}\par
\endgroup

\subsection{Lower bounds}
\label{sec:formalization-map:lower}

\begingroup
\footnotesize
\setlength{\tabcolsep}{3pt}
\par\noindent\begin{tabular}{p{0.95in}p{2.05in}p{1.75in}p{0.55in}}
\toprule
Paper & Lean (file) & Hypotheses in Lean & Status\\
\midrule
\end{tabular}\par\nopagebreak
\noindent\begin{tabular}{p{0.95in}p{2.05in}p{1.75in}p{0.55in}}
\raggedright Lemma ($6$ parameters), \Cref{sec:lower} &\raggedright \texttt{no\_\allowbreak rationalInverse\_\allowbreak general},
\texttt{no\_\allowbreak rationalInverse\_\allowbreak general\_\allowbreak of\_\allowbreak ringChar\_\allowbreak ne\_\allowbreak two}
(\texttt{LowerBound/\allowbreak General/\allowbreak Main.lean}); normal form
\texttt{exists\_\allowbreak singular\_\allowbreak jacobian},
\texttt{no\_\allowbreak rationalInverse\_\allowbreak affine}
(\texttt{LowerBound/\allowbreak Main.lean}); reduction steps
\texttt{gout\_\allowbreak eq} (\texttt{General/\allowbreak Gauge.lean}),
\texttt{polyJacobian\_\allowbreak outPolyGeneral} (\texttt{Affine.lean}),
\texttt{det\_\allowbreak eq\_\allowbreak zero\_\allowbreak of\_\allowbreak mulVec\_\allowbreak eq\_\allowbreak zero} (\texttt{DQ.lean}),
\texttt{det\_\allowbreak eq\_\allowbreak zero\_\allowbreak along\_\allowbreak orbit} (\texttt{Orbit.lean}),
\texttt{Qval\_\allowbreak Theta} (\texttt{Transversal.lean}); cross-checks
\texttt{det\_\allowbreak eq\_\allowbreak zero\_\allowbreak of\_\allowbreak Qval\_\allowbreak eq} (\texttt{Midpoint.lean}),
\texttt{RationalInverse.\allowbreak transport} (\texttt{Transport.lean}, outside the umbrella) &
\raggedright \texttt{[Field F] [Infinite F]}; \texttt{(2 : F) $\ne$ 0} (resp.\
\texttt{ringChar F $\ne$ 2}); six distinct points; the general three-gate
model \texttt{GCircuit F} (sixteen fixed constants, both first-gate constants
are slots, $A=B=0$ and $A=0\ne B$ allowed) and any affine map $Mp+h_0$ from
six parameters to seven slots, $M$ a $7\times6$ matrix with no rank condition.
The gauge identity, the chain rule for the quadratic slot map, cases
(i)--(iv) of \Cref{appendix:lower} (assembled by the characteristic-free pivot
route, so that $2\ne0$ enters only through the normal-form theorem) and the
first-gate degeneracies are proved; only an everywhere-defined rational left
inverse is excluded.  \texttt{GCircuit} directly encodes the topologically
ordered gate display of \Cref{appendix:lower}; no translation from an
arbitrary-program syntax is claimed. &
full\\
\end{tabular}\par
\noindent\begin{tabular}{p{0.95in}p{2.05in}p{1.75in}p{0.55in}}
\raggedright \Cref{thm:char2-lower} &\raggedright \texttt{no\_\allowbreak surjective\_\allowbreak eval}, \texttt{no\_\allowbreak construction\_\allowbreak general}
(\texttt{LowerBoundChar2/\allowbreak General.lean}); affine case
\texttt{no\_\allowbreak construction} (\texttt{LowerBoundChar2/\allowbreak Main.lean}); sharpness
\texttt{oneGate\_\allowbreak isConstruction} (\texttt{LowerBoundChar2/\allowbreak Sharpness.lean}) &
\raggedright \texttt{[Field F] [Fintype F] [CharP F 2]}; $1<n$; $2n\le|F|$; an $n$-gate
chain \texttt{Circuit F n}; an arbitrary preprocessing map
\texttt{pre : $\Lambda$ $\to$ Slots F n}.  The fibre-counting proof of
\Cref{sec:lower-char2} is formalized as stated; the affine case is a
separate, earlier development. &
full\\
\end{tabular}\par
\noindent\begin{tabular}{p{0.95in}p{2.05in}p{1.75in}p{0.55in}}
\raggedright \Cref{lem:first-char2-circuit-inverse} &\raggedright \texttt{circuit\_\allowbreak eval\_\allowbreak bijective} (\texttt{Examples/\allowbreak Char2Inverse.lean}) &
\raggedright \texttt{[Field F] [CharP F 2] [PerfectRing F 2]}; seven distinct points. &
full\\
\end{tabular}\par
\noindent\begin{tabular}{p{0.95in}p{2.05in}p{1.75in}p{0.55in}}
\raggedright \Cref{lem:char2-small-staircase-butterfly} (degree $9$) &\raggedright
\texttt{coeffMap9Fin\_\allowbreak bijective},
\texttt{P9\_\allowbreak eval\_\allowbreak bijective}
(\texttt{Examples/\allowbreak Char2SmallInverses.lean}) &
\raggedright \texttt{[Field F] [CharP F 2]}; nine distinct points.  The
literal circuit identity, both directions of the displayed coefficient
decoder, and the Vandermonde evaluation step are checked.  The degree-$11$
half of the paper lemma is not yet formalized. &
partial\\
\bottomrule
\end{tabular}\par
\endgroup

\subsection{Unit hypotheses versus \texorpdfstring{$n$}{n}-admissibility}
\label{sec:formalization-map:units}

The paper's standing hypothesis is that $\F$ is $n$-admissible
(\Cref{appendix:constructions}: characteristic zero or characteristic
$p>n$), which \Cref{rem:char2} describes as sufficient rather than sharp.
The Lean master theorem \texttt{odd\_\allowbreak realizable\_\allowbreak pairs} assumes instead,
over an arbitrary commutative ring $R$ (no field, no characteristic),
\begin{center}
  \texttt{$\forall$ i : $\mathbb N$, 1 $\le$ i $\to$ i $\le$ n $\to$ IsUnit (((i : $\mathbb N$) : $\mathbb Z$) : R)},
\end{center}
that is, exactly the integers $1,2,\dots,n$ are units of $R$; this is the
predicate \texttt{Admissible R n} of \texttt{Admissible.lean} up to the
integer cast (\texttt{Int.cast\_\allowbreak natCast}).  Every composite pivot of the
construction, namely the scalar slopes dividing $2k(k-1)$, the block
determinant $-k^2$ of \Cref{lem:barQ8k+7}, and the factor $2$ of every
square gadget, is inverted in Lean as a product of unit factors that are
each at most $n$ (\texttt{Admissible.isUnit\_\allowbreak mul\_\allowbreak cast}); no hypothesis on
an integer larger than $n$ is ever used.  The sub-lemmas consume the same
predicate restricted to their own degree (for instance $1,\dots,k$ in
\texttt{Rk2l\_\allowbreak triangular}, $1,\dots,2k$ in \texttt{fourk\_\allowbreak decodable},
$1,\dots,d$ in \texttt{odd\_\allowbreak gadget\_\allowbreak dispatch}), and several need only
\texttt{IsUnit (2 : R)} or a single root index, as the tables record.  For a
field $\F$ the Lean hypothesis is equivalent to $n$-admissibility: it holds
in characteristic zero and in characteristic $p>n$
(\texttt{admissible\_\allowbreak of\_\allowbreak charZero}, \texttt{admissible\_\allowbreak of\_\allowbreak charP}), and it
fails in characteristic $p\le n$ because $p$ itself is then not a unit.
Thus the Lean statements do not certify the finer claim of \Cref{rem:char2}
that only the pivots actually displayed need be nonzero; they certify the
uniform sufficient condition.  The coverage endpoints
(\texttt{monic\_\allowbreak coefficient\_\allowbreak map\_\allowbreak bijective} and the automorphism of
\Cref{lem:polynomial-left-inverse-automorphism}) additionally assume the
base ring is Noetherian and nontrivial, which every field satisfies.  The
lower bounds use the field-level hypotheses shown in their table: an
infinite field with $2\ne0$ for degree six, and a finite field of
characteristic two for \Cref{thm:char2-lower}.

\paragraph{Build and axioms.}
\begin{sloppypar}
The construction, cost, height, example, and both lower-bound
developments (including \texttt{HeightFinal.lean},
\texttt{LowerBound/\allowbreak General/\allowbreak Main.lean},
\texttt{LowerBoundChar2/\allowbreak General.lean}, and
\texttt{LowerBoundChar2/\allowbreak Sharpness.lean}) are imported by the umbrella
module \texttt{FastPoly.lean}, the default target; \texttt{lake build FastPoly}
completes it with no errors. No file under \texttt{FastPoly/\allowbreak } declares an \texttt{axiom} or
contains a \texttt{sorry}.  \texttt{\#print axioms} on
\texttt{polynomial\_\allowbreak height}, \texttt{polynomial\_\allowbreak height\_\allowbreak of\_\allowbreak charZero},
\texttt{polynomial\_\allowbreak height\_\allowbreak of\_\allowbreak charP} (\texttt{HeightFinal.lean}),
\texttt{decodedAdditionPolynomial\_\allowbreak exists}
(\texttt{Cost/\allowbreak Additions/\allowbreak DecodedPolynomial.lean}), and
\texttt{MainArrangementsChecked.of\_\allowbreak charZero},
\texttt{MainArrangementsChecked.of\_\allowbreak charP} (\texttt{PaperMain.lean}), as well as on
\texttt{no\_\allowbreak surjective\_\allowbreak eval}, \texttt{no\_\allowbreak construction\_\allowbreak general},
\texttt{no\_\allowbreak construction'} (\texttt{LowerBoundChar2/\allowbreak General.lean}),
\texttt{no\_\allowbreak rationalInverse\_\allowbreak general}
(\texttt{LowerBound/\allowbreak General/\allowbreak Main.lean}), and
\texttt{oneGate\_\allowbreak isConstruction}
(\texttt{LowerBoundChar2/\allowbreak Sharpness.lean})
reports only \texttt{propext}, \texttt{Classical.choice}, and
\texttt{Quot.sound}; the same is recorded for
\texttt{construction\_\allowbreak additions\_\allowbreak checked}
(\texttt{Cost/\allowbreak Additions/\allowbreak Realization.lean};
\texttt{FastPoly/\allowbreak ROADMAP.md}) and for the affine characteristic-two
development (module docstring of \texttt{LowerBoundChar2/\allowbreak Main.lean}).
No separate \texttt{\#print axioms} transcript is recorded for the remaining
theorems.
\end{sloppypar}

\section{Proof of the Degree-Six Lower Bound}
\label{appendix:lower}

This appendix proves the $6$-parameter lemma of \Cref{sec:lower}
(\Cref{lem:six-params}): over an infinite field of characteristic $\neq2$, no
straight-line program with at most three nonscalar multiplications and six parameters
has an everywhere-defined rational left inverse from its values at six fixed points.
The notation is that of \Cref{sec:lower}.

\begin{proof}[Proof of \Cref{lem:six-params}]
\noindent\textbf{Proof method (Jacobian obstruction).}
Let $J_F(p)$ be the Jacobian matrix of $F$, i.e., $(J_F)_{k,i}(p)=\frac{\partial P_p(x_k)}{\partial p_i}$.
We will show that for every such three-multiplication program there exists a choice of parameters $p$ such that $\det J_F(p)=0$.
On the other hand, suppose $F$ had an everywhere-defined rational left inverse,
with coordinates $g_i/h_i$ where no $h_i$ has a zero in $\mathbb{F}^6$.  Clearing
denominators, $g_i(F(p))=p_i\,h_i(F(p))$ for every $p\in\mathbb{F}^6$, hence
identically as polynomials because $\mathbb{F}$ is infinite.  Let $p_0$ be a point
with $\det J_F(p_0)=0$ and pick $v\neq0$ with $J_F(p_0)v=0$.  Differentiating the
identity at $p_0$ in the direction $v$, every term on the left and the second term
on the right carry a factor $J_F(p_0)v=0$, leaving
\[
   0=v_i\,h_i(F(p_0))\qquad(1\le i\le 6).
\]
Since $h_i$ vanishes nowhere, $v=0$, a contradiction.  (Note that no
Nullstellensatz step is needed, so $\mathbb{F}$ need only be infinite; ruling out a
left inverse is stronger than ruling out a two-sided one, and by
\Cref{lem:right-left} it also rules out an everywhere-defined rational right
inverse, which is the form in which preprocessing enters the model of
\Cref{sec:model}.)

\medskip
\noindent\textbf{Reduction to a normal form.}
Consider an arbitrary straight-line program in the model of \Cref{sec:lower} that uses
three nonscalar multiplications.
Topologically order the multiplication gates and denote their outputs by $u_1,u_2,u_3$.
Every value available as an input to gate $i$ is obtained from $1$, $x$, the parameters
$p_1,\dots,p_6$ and the earlier $u_j$ by additions and multiplications by fixed field
elements, so each multiplicand at gate $i$ is a fixed linear combination of $x$ and the
earlier $u_j$ plus an affine function of the parameters; the same holds for the output.
(A multiplicand may involve neither $x$ nor any $u_j$: it is then a parameter-affine
constant, and the gate is a multiplication by a parameter, which the model counts.)
Thus the program reads
\begin{align}
   u_1 &= (Ax+a)(Bx+b),\\
   u_2 &= (L_{2,1}u_1+L_{2,0}x+a_2)(R_{2,1}u_1+R_{2,0}x+b_2),\\
   u_3 &= (L_{3,2}u_2+L_{3,1}u_1+L_{3,0}x+a_3)(R_{3,2}u_2+R_{3,1}u_1+R_{3,0}x+b_3),\\
   P &= s_3u_3+s_2u_2+s_1u_1+s_0x+b_1,
\end{align}
where $A,B,L_{i,j},R_{i,j},s_i$ are fixed circuit constants and the seven
\emph{constant slots}
\[
   z=(a,b,a_2,b_2,a_3,b_3,b_1)=H(p)=Mp+h_0
\]
are an affine image of the parameter vector, with $M$ a $7\times6$ matrix over
$\mathbb{F}$ and $h_0\in\mathbb{F}^7$.  (A program with fewer nonscalar multiplications
is included: an unused gate is one whose output has coefficient zero in all later factors
and in the output.)  We write $\lambda=(L_{2,1},R_{2,1},L_{3,1},R_{3,1},s_1)\in\mathbb{F}^5$
for the coefficients with which $u_1$ enters the later factors and the output, and
$\sigma=(a_2,b_2,a_3,b_3,b_1)$ for the last five slots, so that $z=(a,b,\sigma)$.

\smallskip
\noindent\textit{The scalar first gate.}
If $A=B=0$ then $u_1=ab$ does not involve $x$, and substituting it into the later gates
and the output shows that $P$ depends on $z$ only through the five quantities
$\sigma+\lambda\,ab$.  Hence $F$ factors through a polynomial map
$\mathbb{F}^6\to\mathbb{F}^5$, so by the chain rule $J_F(p)$ is the product of a
$6\times5$ and a $5\times6$ matrix and is singular for every $p$.  Otherwise
$(A,B)\neq(0,0)$, and swapping the two factors of the first gate if necessary (which
swaps the slots $a,b$, i.e.\ the first two rows of $M$ and of $h_0$) we may assume
$A\neq0$.

\smallskip
\noindent\textit{The gauge identity and the slot map.}
For $A\neq0$ we use the identity
\begin{equation}
\label{eq:lower-gauge}
   (Ax+a)(Bx+b)=A\cdot x\Bigl(Bx+b+\tfrac{B}{A}a\Bigr)+ab ,
\end{equation}
whose correction $ab$ is free of $x$ (both sides equal $ABx^2+(Ab+Ba)x+ab$).
Writing $u_1'=x(Bx+a_1)$ with $a_1=b+\frac BAa$, we have $u_1=Au_1'+ab$, and
substituting this into the later factors and into the output gives
\[
   L_{i,1}u_1+\dots+a_i=(AL_{i,1})\,u_1'+\dots+(a_i+L_{i,1}ab),\qquad
   s_1u_1+b_1=(As_1)\,u_1'+(b_1+s_1ab),
\]
and likewise for the right factors.  Thus $P$ is computed by the normal form displayed
below, with the fixed constants
\[
   R_{1,0}=B,\qquad L_{i,1}'=AL_{i,1},\qquad R_{i,1}'=AR_{i,1},\qquad s_1'=As_1
\]
(all other constants unchanged) and with the six normal-form slots
\begin{equation}
\label{eq:lower-nu}
   \nu(z)=(a_1,a_2',b_2',a_3',b_3',b_1')
   =\Bigl(b+\tfrac BAa,\ \ \sigma+\lambda\,ab\Bigr),
\end{equation}
i.e.\ $a_2'=a_2+L_{2,1}ab$, $b_2'=b_2+R_{2,1}ab$, $a_3'=a_3+L_{3,1}ab$,
$b_3'=b_3+R_{3,1}ab$, $b_1'=b_1+s_1ab$.  (The rewriting
$(Ax+a)(Bx+b)=A\,x(Bx+b)+a(Bx+b)$ would not do: its correction $aBx+ab$ changes the
$x$-coefficient of every later factor that uses $u_1$ by $L_{i,1}aB$, which depends on the
parameter $a$, whereas the normal form requires those coefficients to be fixed.)
Let $E'\colon\mathbb{F}^6\to\mathbb{F}^6$, $E'(q)=(P'_q(x_0),\dots,P'_q(x_5))$, be the
evaluation map of the normal form with these constants, and let $J(q)$ be its Jacobian.
Then $P_p=P'_{\nu(H(p))}$ identically in $x$, i.e.\ $F=E'\circ Q$ with
\[
\begin{aligned}
   Q&=\nu\circ H\colon\mathbb{F}^6\to\mathbb{F}^6,\\
   Q_1(p)&=H_2(p)+\tfrac BAH_1(p),\qquad
   Q_{1+j}(p)=H_{2+j}(p)+\lambda_j\,H_1(p)H_2(p)\quad(j=1,\dots,5),
\end{aligned}
\]
where $H_i$ is the $i$-th coordinate of $H$; $Q$ is a polynomial map whose first
component is affine and whose other five components have degree two in $p$.  By the
chain rule,
\begin{equation}
\label{eq:lower-chain}
   J_F(p)=J\bigl(Q(p)\bigr)\,DQ(p),\qquad DQ(p)=D\nu\bigl(H(p)\bigr)\,M ,
\end{equation}
so $\det J_F(p)=\det J(Q(p))\cdot\det DQ(p)$: $J_F$ is singular at $p$ as soon as
$DQ(p)$ is singular or $J$ is singular at $Q(p)$.  The $6\times7$ matrix
\[
   D\nu(z)=\begin{bmatrix} \tfrac BA & 1 & 0\\[2pt] \lambda b & \lambda a & I_5\end{bmatrix}
\]
(with $\lambda b$ and $\lambda a$ columns of length five) has rank six, and its kernel is
spanned by the \emph{gauge direction}
\[
   \xi(z)=\Bigl(1,\ -\tfrac BA,\ -\lambda\bigl(b-\tfrac BAa\bigr)\Bigr)\in\mathbb{F}^7 .
\]
The fibres of $\nu$ are the curves (\emph{gauge orbits})
\[
   O_q(u)=\Bigl(u,\ \ a_1-\tfrac BAu,\ \ \sigma^*-\lambda\,u\bigl(a_1-\tfrac BAu\bigr)\Bigr),
   \qquad q=(a_1,\sigma^*)\in\mathbb{F}^6,\ u\in\mathbb{F}:
\]
indeed $\nu(O_q(u))=q$ for all $u$, and every $z=(a,b,\sigma)$ equals $O_{\nu(z)}(a)$.
In particular $\nu$ is surjective, and $Q(p)=Q(p')$ exactly when $H(p)$ and $H(p')$
lie on one gauge orbit.  We now distinguish four cases; in each of (i)--(iii) we exhibit
a point where $DQ$ is singular, and in (iv) $Q$ is invertible and the singular points of
$J_F$ are the preimages of those of $J$.

\smallskip
\noindent\textit{Case (i): $M$ is not injective.}
Pick $v\neq0$ with $Mv=0$.  Then $DQ(p)v=D\nu(H(p))Mv=0$ for every $p$, so
$\det J_F(p)=0$ for every $p\in\mathbb{F}^6$ by \eqref{eq:lower-chain}.

In the remaining cases $M$ is injective.  Let $\omega\in\mathbb{F}^7$ be its cofactor
vector, $\omega_i=(-1)^i\det(M\text{ with row $i$ deleted})$.  Then
$\omega^{\mathsf T}M=0$ (Laplace expansion of $\det[M\mid Me_j]$ along its last column,
a determinant with two equal columns), $\omega\neq0$ because $M$ has a nonzero
$6\times6$ minor, and hence $\operatorname{im}M=\ker\omega^{\mathsf T}$, both having
dimension six.  Therefore
\[
   \operatorname{im}H=\mathcal{H}:=\{z\in\mathbb{F}^7:\ \omega\cdot z=\eta\},\qquad
   \eta:=\omega\cdot h_0 .
\]
Choosing six rows $S$ of $M$ with $\det M_S\neq0$ and putting $M^+=M_S^{-1}\Pi_S$
($\Pi_S$ selects the rows $S$) gives a left inverse, $M^+M=I_6$, with
$H(M^+(z-h_0))=z$ for every $z\in\mathcal{H}$ (write $z-h_0=Mv$).  Write
$\omega=(\omega_a,\omega_b,\omega_\sigma)$ with $\omega_\sigma\in\mathbb{F}^5$, and put
\[
   \kappa=\omega_\sigma\cdot\lambda .
\]
For $q=(a_1,\sigma^*)$ the orbit $O_q$ meets $\mathcal H$ exactly at the roots of the
\emph{orbit polynomial}
\begin{equation}
\label{eq:lower-orbitpoly}
   \omega\cdot O_q(u)-\eta
   =\tfrac BA\kappa\,u^2+\bigl(\omega_a-\tfrac BA\omega_b-\kappa a_1\bigr)u
    +\bigl(\omega_ba_1+\omega_\sigma\cdot\sigma^*-\eta\bigr),
\end{equation}
and lies inside $\mathcal H$ exactly when all three coefficients vanish.

\smallskip
\noindent\textit{Midpoint lemma.}
Let $G\colon\mathbb{F}^n\to\mathbb{F}^m$ be a polynomial map all of whose components
have total degree at most two, and assume $\mathrm{char}(\mathbb{F})\neq2$.  Then for
all $p,p'\in\mathbb{F}^n$
\begin{equation}
\label{eq:lower-midpoint}
   G(p)-G(p')=DG\Bigl(\frac{p+p'}2\Bigr)(p-p') .
\end{equation}
Indeed, a component of $G$ has the form $y\mapsto y^{\mathsf T}Ny+\beta^{\mathsf T}y+\gamma$
with a square matrix $N$, and its derivative at $m$ is
$m^{\mathsf T}(N+N^{\mathsf T})+\beta^{\mathsf T}$.  Expanding
$\frac12(p+p')^{\mathsf T}(N+N^{\mathsf T})(p-p')$, the four mixed terms cancel in pairs
(a $1\times1$ matrix equals its transpose: $p^{\mathsf T}Np'=p'^{\mathsf T}N^{\mathsf T}p$),
and what remains is $p^{\mathsf T}Np-p'^{\mathsf T}Np'$; the affine part contributes
$\beta^{\mathsf T}(p-p')$.  The only division is by $2$.  Consequently, if $G(p)=G(p')$
with $p\neq p'$, then $DG$ has the nonzero kernel vector $p-p'$ at the
$\mathbb{F}$-rational point $(p+p')/2$.

\smallskip
\noindent\textit{Case (ii): two parameter points on one gauge orbit.}
Suppose $p\neq p'$ and $Q(p)=Q(p')$, i.e.\ $H(p)$ and $H(p')$ lie on one gauge orbit.
Since $Q$ has degree two, \eqref{eq:lower-midpoint} gives $DQ(m)(p-p')=0$ at the
rational midpoint $m=(p+p')/2$, so $\det DQ(m)=0$ and $\det J_F(m)=0$ by
\eqref{eq:lower-chain}.  This case arises whenever $\kappa\neq0$ and $B\neq0$: with
\[
   a_1^\dagger=\frac{\frac BA\kappa+\omega_a-\frac BA\omega_b}{\kappa},\qquad
   \sigma^\dagger=\frac{\eta-\omega_ba_1^\dagger}{\kappa}\,\lambda,\qquad
   q^\dagger=(a_1^\dagger,\sigma^\dagger),
\]
the orbit polynomial \eqref{eq:lower-orbitpoly} of $q^\dagger$ is
$\frac BA\kappa\,u(u-1)$, so $O_{q^\dagger}(0)$ and $O_{q^\dagger}(1)$ lie in
$\mathcal H$, and $p=M^+(O_{q^\dagger}(0)-h_0)$, $p'=M^+(O_{q^\dagger}(1)-h_0)$ are two
distinct parameter points (their images $H(p)=O_{q^\dagger}(0)$ and
$H(p')=O_{q^\dagger}(1)$ differ in the first coordinate) with $Q(p)=q^\dagger=Q(p')$.

\smallskip
\noindent\textit{Case (iii): a gauge orbit inside $\operatorname{im}H$.}
Suppose $q$ is such that all three coefficients in \eqref{eq:lower-orbitpoly} vanish,
so that $O_q\subseteq\mathcal H$.  Then every coefficient vector of the polynomial
$u\mapsto O_q(u)-h_0$ lies in $\ker\omega^{\mathsf T}=\operatorname{im}M$ and is fixed
by $MM^+$, so the curve $p(u)=M^+(O_q(u)-h_0)$ satisfies $H(p(u))=O_q(u)$ and
$Q(p(u))=q$ identically in $u$.  Differentiating in $u$ gives $DQ(p(u))\,p'(u)=0$, and
$p'(u)\neq0$ because the first coordinate of $Mp'(u)=\frac{d}{du}O_q(u)$ is $1$.  Hence
$J_F$ is singular along the whole curve: $\det J_F(p(u))=0$ for every $u\in\mathbb{F}$.
(Alternatively, $p(0)\neq p(1)$ and $Q(p(0))=Q(p(1))$, so this is also an instance of
case (ii).)  This case arises when $\kappa\neq0$ and $B=0$, with $q^\dagger$ as in
case (ii), and when $\kappa=0$ and $\omega_a=\frac BA\omega_b$, with $q=\nu(h_0)$: in
the former the orbit polynomial of $q^\dagger$ is $\frac BA\kappa\,u(u-1)=0$; in the
latter it is the constant $\omega_ba_1+\omega_\sigma\cdot\sigma^*-\eta$, which vanishes
for $q=\nu(h_0)$ because $O_{\nu(h_0)}$ passes through $h_0\in\mathcal H$.

\smallskip
\noindent\textit{Case (iv): the transversal case.}
Suppose finally that $\kappa=0$ and $\omega^\circ:=\omega_a-\frac BA\omega_b\neq0$.
Then \eqref{eq:lower-orbitpoly} reads $\omega^\circ u+(\omega_ba_1+\omega_\sigma\cdot\sigma^*-\eta)$,
so every gauge orbit meets $\mathcal H$ in exactly one point, at
\[
   u(q)=-\frac{\omega_ba_1+\omega_\sigma\cdot\sigma^*-\eta}{\omega^\circ}\qquad(q=(a_1,\sigma^*)),
\]
an affine function of $q$.  Define
\[
   \Theta(q)=M^+\bigl(O_q(u(q))-h_0\bigr)
   =M^+\Bigl(\bigl(u(q),\ a_1-\tfrac BAu(q),\ \sigma^*-\lambda\,u(q)\bigl(a_1-\tfrac BAu(q)\bigr)\bigr)-h_0\Bigr),
\]
a polynomial map of degree at most two.  Then $Q\circ\Theta=\mathrm{id}$: since
$O_q(u(q))\in\mathcal H$ we have $H(\Theta(q))=O_q(u(q))$, hence
$Q(\Theta(q))=\nu(O_q(u(q)))=q$.  And $\Theta\circ Q=\mathrm{id}$: for $z=H(p)\in\mathcal H$
and $q=\nu(z)$ we have $z=O_q(a)$ with $a$ the first coordinate of $z$, and
$\omega\cdot z=\eta$ forces $a=u(q)$, so $O_q(u(q))=z=Mp+h_0$ and $\Theta(q)=M^+Mp=p$.
Thus $Q$ is a polynomial automorphism of $\mathbb{F}^6$ with inverse $\Theta$; in
particular $DQ(p)$ is invertible for every $p$, and $\nu$ restricted to
$\mathcal H=\operatorname{im}H$ is a bijection onto the six normal-form slots.
(Conversely, in cases (ii) and (iii) $Q$ is not injective, so $\kappa=0$ and
$\omega_a\neq\frac BA\omega_b$ is exactly the condition for this.)  In this case the
singular points of $J_F$ are the preimages of those of $J$: if $\det J(q_0)=0$ then
$p_0=\Theta(q_0)$ satisfies $Q(p_0)=q_0$ and
$\det J_F(p_0)=\det J(q_0)\cdot\det DQ(p_0)=0$.

In case (iv) the conclusion of the lemma can also be transferred without the Jacobian,
because an everywhere-defined rational left inverse of $F=E'\circ Q$ yields one of $E'$
as soon as $Q$ is surjective (as it is here).  Indeed, let $g_i/h_i$ be such a left
inverse of $F$, with $h_i$ vanishing nowhere and $g_i(F(p))=p_ih_i(F(p))$ for all $p$;
let $d=2$ bound the degrees of the components $Q_j(y)=\sum_{|e|\le d}c_{j,e}\,y^e$ of
$Q$, and put
\[
   \Delta=\prod_{i=1}^6h_i^{\,d},\qquad
   N_j=\sum_{|e|\le d}c_{j,e}\prod_{i=1}^6g_i^{\,e_i}h_i^{\,d-e_i},
\]
so that $N_j/\Delta=Q_j(g_1/h_1,\dots,g_6/h_6)$.  Then $\Delta$ vanishes nowhere, and for
every $q$, choosing $p$ with $Q(p)=q$ and putting $y=E'(q)=F(p)$, we get
$N_j(y)=\sum_ec_{j,e}\,p^e\prod_ih_i(y)^d=Q_j(p)\,\Delta(y)=q_j\,\Delta(y)$; so
$N_j/\Delta$ is an everywhere-defined rational left inverse of $E'$, and a statement
excluding such inverses for every normal form excludes them for $F$.

\smallskip
The four cases are exhaustive: either $M$ is not injective (i), or $\kappa\neq0$
(ii if $B\neq0$, iii if $B=0$), or $\kappa=0$ and $\omega_a=\frac BA\omega_b$ (iii), or
$\kappa=0$ and $\omega_a\neq\frac BA\omega_b$ (iv).  In cases (i)--(iii) we have
exhibited $p_0$ with $\det DQ(p_0)=0$, hence $\det J_F(p_0)=0$, whatever the normal
form; in case (iv) a singular point of $J_F$ is the pull-back $\Theta(q_0)$ of a
singular point $q_0$ of $J$.  It therefore suffices to prove that the normal-form
Jacobian $J$ is singular somewhere, for every choice of the fixed constants; this is
what the rest of the proof does.  From now on $J$ denotes this Jacobian, the six
normal-form slots $(a_1,a_2,b_2,a_3,b_3,b_1)$ are the free parameters, and we drop the
primes.  (The midpoint lemma is the only place where the reduction uses
$\mathrm{char}(\mathbb{F})\neq2$, and it can be avoided: by the Cauchy--Binet formula
$\det DQ(p)=\sum_i\det(D\nu\text{ without column }i)\det(M\text{ without row }i)
=\omega\cdot\xi(H(p))$, because the cofactor vector of $D\nu(z)$ spans $\ker D\nu(z)$ and
has first entry $1$, so it equals $\xi(z)$; thus
$\det DQ(p)=\omega^\circ-\kappa\,\tau(p)$ with $\tau(p)=b-\frac BAa$ evaluated at $H(p)$,
an affine function of $p$ which is non-constant when $\kappa\neq0$ (if it were constant,
$e_2-\frac BAe_1$ would lie in the left kernel $\langle\omega\rangle$ of $M$, forcing
$\omega_\sigma=0$), and one linear pivot then produces $p_0$ with $\det DQ(p_0)=0$.  The
characteristic hypothesis is still used in Case $D=0$ below.)

Thus it remains to treat the program in the following form (all other scalars are fixed circuit constants):
\begin{align}
   u_0 &= x
   \\
   u_1 &= x(R_{1,0} x + a_1)
   \\
   u_2 &= \ell_2\,r_2
   =(L_{2,1} u_1 + L_{2,0} x + a_2)
       (R_{2,1} u_1 + R_{2,0} x + b_2)
      \\
   u_3 &= \ell_3\,r_3
   =(L_{3,2} u_2 + L_{3,1} u_1 + L_{3,0} x + a_3)
       (R_{3,2} u_2 + R_{3,1} u_1 + R_{3,0} x + b_3)
   \\
   P &= s_3 u_3 + s_2 u_2 + s_1 u_1 + s_0 x + b_1.
\end{align}
Note that multiplications by the fixed circuit constants $L_{i,j}$, $R_{i,j}$ and
$s_i$ are scalar multiplications and are not counted.

Define $\bar u_i = \frac{\partial P}{\partial u_i}$.
By the chain rule we have
$\bar u_i = s_i + \sum_{j>i} \bar u_j (L_{j,i} r_j + R_{j,i} \ell_j)$
and so in particular
\begin{align}
   \bar u_3 &= s_3, \\
   \bar u_2 &= s_2 + \bar u_3 (L_{3,2} r_3 + R_{3,2} \ell_3), \\
   \bar u_1 &= s_1 + \bar u_2 (L_{2,1} r_2 + R_{2,1} \ell_2)
      + \bar u_3 (L_{3,1} r_3 + R_{3,1} \ell_3).
\end{align}
and the sensitivities of $P$ to the program parameters are
\[
   \frac{\partial P}{\partial b_1}=1,\qquad
   \frac{\partial P}{\partial b_3}=s_3 \ell_3,\qquad
   \frac{\partial P}{\partial a_3}=s_3 r_3,
\]
\[
   \frac{\partial P}{\partial b_2}=\bar u_2 \ell_2,\qquad
   \frac{\partial P}{\partial a_2}=\bar u_2 r_2,\qquad
   \frac{\partial P}{\partial a_1}=\bar u_1 x.
\]

We split into two cases, depending on
\[
   D = \left|
   \begin{matrix}
      L_{3,2} & R_{3,2} \\
      L_{3,1} & R_{3,1} \\
   \end{matrix}
   \right|.
\]

We already fixed six distinct evaluation points $x_0,\dots,x_5$.
To show the $6\times 6$ Jacobian is singular, it suffices to make two of its rows equal, or to make three of its rows linearly dependent.
First, if $D\neq 0$, we will show that two rows can be set equal.
Second, if $D=0$, we will show that three rows are linearly dependent.

\medskip
\noindent\textbf{Case $D\neq 0$.}
Pick two distinct evaluation points $x_0\neq x_1$ from the fixed set.
We will choose parameters so that the Jacobian rows at $x_0$ and $x_1$ are equal.

\smallskip
\noindent\textbf{Step 1: Make $\ell_3$ and $r_3$ agree at $x_0,x_1$.}
The conditions $\ell_3(x_0)=\ell_3(x_1)$ and $r_3(x_0)=r_3(x_1)$ are equivalent to
\begin{align}
   L_{3,2} (u_2(x_0)-u_2(x_1)) + L_{3,1} (u_1(x_0)-u_1(x_1)) + L_{3,0}(x_0-x_1) &= 0, \\
   R_{3,2} (u_2(x_0)-u_2(x_1)) + R_{3,1} (u_1(x_0)-u_1(x_1)) + R_{3,0}(x_0-x_1) &= 0.
\end{align}
Since $D\neq 0$, the matrix
\(
\begin{bmatrix}
   L_{3,2} & L_{3,1} \\
   R_{3,2} & R_{3,1} \\
\end{bmatrix}
\)
is invertible, so this uniquely determines the required differences
$(u_2(x_0)-u_2(x_1),\,u_1(x_0)-u_1(x_1))$.
Now $u_1(x)=x(R_{1,0}x+a_1)$, so
\[
   u_1(x_0)-u_1(x_1) = (x_0-x_1)(R_{1,0}(x_0+x_1)+a_1),
\]
and we can choose $a_1$ to match the required value.

To match the required value of $u_2(x_0)-u_2(x_1)$, write
\[
   A_k = R_{2,1}u_1(x_k)+R_{2,0}x_k,\qquad
   B_k = L_{2,1}u_1(x_k)+L_{2,0}x_k,
\]
so $u_2(x_k)=(B_k+a_2)(A_k+b_2)$ and hence
\[
   u_2(x_0)-u_2(x_1)
   = (B_0A_0-B_1A_1) + a_2(A_0-A_1) + b_2(B_0-B_1).
\]
If
\(
   \left|
   \begin{matrix}
      L_{2,0} & L_{2,1} \\
      R_{2,0} & R_{2,1} \\
   \end{matrix}
   \right|
   = 0
\)
then the two rows $(L_{2,0},L_{2,1})$ and $(R_{2,0},R_{2,1})$ are proportional.
If $(L_{2,0},L_{2,1})=(0,0)$ then $\ell_2=a_2$ is constant, and choosing $a_2=0$ makes $\frac{\partial P}{\partial b_2}=\bar u_2\ell_2$ vanish identically, so the Jacobian is singular; the case $(R_{2,0},R_{2,1})=(0,0)$ is symmetric with $b_2=0$.
Otherwise $(R_{2,0},R_{2,1})=\alpha(L_{2,0},L_{2,1})$ for some $\alpha$, and choosing $b_2=\alpha a_2$ makes $r_2=\alpha \ell_2$.
In that case $\frac{\partial P}{\partial a_2}=\alpha \frac{\partial P}{\partial b_2}$, so the Jacobian is singular and we are done.
Otherwise this determinant is nonzero, and then $(A_0-A_1,B_0-B_1)\neq (0,0)$ (since $(A_0-A_1,B_0-B_1)$ is an invertible linear transform of $(u_1(x_0)-u_1(x_1),x_0-x_1)$).
Thus we can solve for $(a_2,b_2)$.

\smallskip
\noindent\textbf{Step 2: Make $\bar u_2$ and $\bar u_1$ vanish at $x_0$.}
With $\ell_3(x_0)=\ell_3(x_1)$ and $r_3(x_0)=r_3(x_1)$ we have $\bar u_2(x_0)=\bar u_2(x_1)$.
We now choose $(a_3,b_3)$ so that $\bar u_2(x_0)=0$ and $\bar u_1(x_0)=0$.
If $s_3=0$, then $\frac{\partial P}{\partial a_3}=\frac{\partial P}{\partial b_3}=0$ and the Jacobian is singular, so assume $s_3\neq 0$.
At the fixed point $x_0$, both $\bar u_2(x_0)$ and $\bar u_1(x_0)$ are affine in $(a_3,b_3)$:
\[
	   \bar u_2(x_0) = C_2 + s_3(R_{3,2} a_3 + L_{3,2} b_3),\qquad
	   \bar u_1(x_0) = C_1 + s_3\bigl((R_{3,1}+R_{3,2}T)a_3 + (L_{3,1}+L_{3,2}T)b_3\bigr),
\]
where $C_1,C_2$ do not depend on $(a_3,b_3)$ and
\[
   T = (L_{2,1} r_2 + R_{2,1} \ell_2)(x_0).
\]
Since
\(
\det\!\begin{bmatrix}
	   R_{3,2} & L_{3,2} \\
	   R_{3,1}+R_{3,2}T & L_{3,1}+L_{3,2}T
\end{bmatrix}
= -D\neq 0,
\)
there is a unique solution $(a_3,b_3)$ to $\bar u_2(x_0)=\bar u_1(x_0)=0$.
Then also $\bar u_2(x_1)=\bar u_1(x_1)=0$.

\smallskip
\noindent\textbf{Conclusion.}
With these choices, the sensitivities agree at $x_0$ and $x_1$:
\[
   \frac{\partial P}{\partial b_1}=1,\qquad
   \frac{\partial P}{\partial a_1}=\bar u_1 x=0,\qquad
   \frac{\partial P}{\partial a_2}=\bar u_2 r_2=0,
\]
\[
   \frac{\partial P}{\partial b_2}=\bar u_2 \ell_2=0,\qquad
   \frac{\partial P}{\partial a_3}=s_3 r_3,\qquad
   \frac{\partial P}{\partial b_3}=s_3 \ell_3,
\]
and $\ell_3(x_0)=\ell_3(x_1)$, $r_3(x_0)=r_3(x_1)$.
Thus the Jacobian rows at $x_0$ and $x_1$ are identical, so the Jacobian is singular.

\medskip
\noindent\textbf{Case $D=0$.}

In this case we will show that three rows of the Jacobian are linearly dependent.
Pick three distinct evaluation points $x_0,x_1,x_2$ from the fixed set.
We use the functional
\[
   \langle f \rangle = (x_1 - x_2) f(x_0) + (x_2 - x_0) f(x_1) + (x_0 - x_1) f(x_2),
\]
such that $\langle 1 \rangle = 0$ and $\langle x \rangle = 0$.
(For $x_k=0,1,2$ the weights are $\lambda=(-1,2,-1)$, a second difference.)
Note that $\langle x^2\rangle = (x_0 - x_1)(x_0 - x_2)(x_1 - x_2) \neq 0$.

\smallskip
\noindent\textbf{Subcase $L_{3,2}R_{3,2}=0$.}
Assume $L_{3,2}R_{3,2}=0$.
If $L_{3,2}=R_{3,2}=0$, then the top multiplication does not use $u_2$, so $\ell_3,r_3$ have degree at most $2$ and one checks that all six sensitivity polynomials have degree at most $3$.
Otherwise, exactly one of $L_{3,2},R_{3,2}$ is nonzero; without loss of generality $R_{3,2}\neq 0$ and $L_{3,2}=0$.
Since we are in Case $D=0$, the condition
\(
D=L_{3,2}R_{3,1}-R_{3,2}L_{3,1}=0
\)
forces $L_{3,1}=0$, so $\ell_3$ is affine in $x$.
(The case $L_{3,2}\neq 0$ and $R_{3,2}=0$ is symmetric.)
In either situation, one checks that each of the six sensitivity polynomials has degree at most $4$ in $x$, hence they all lie in $\mathrm{span}\{1,x,x^2,x^3,x^4\}$.
Evaluated at any six distinct points, the corresponding Jacobian columns (evaluation vectors) lie in a subspace of dimension at most $5$, so the $6\times 6$ Jacobian is singular.
Thus we may assume $L_{3,2}R_{3,2}\neq 0$ in the remainder.

We will choose program parameters such that $\bar u_1$ is constant for $k\in\{0,1,2\}$, $\bar u_2(x_k) = 0$ for $k\in\{0,1,2\}$, and $\langle \ell_3 \rangle = \langle r_3 \rangle = 0$.
This is nearly the same strategy as in the previous case, just using three evaluation points.
The only difference is $\bar u_1$ where we use $\langle \bar u_1 x \rangle = \bar u_1 \langle x \rangle = 0$ when $\bar u_1$ is constant.
Another difference is that we will take $a_1=0$, that is, we won't use this parameter to adjust anything.

Note that we can assume
\(
\left|
\begin{matrix}
   L_{2,0} & L_{2,1} \\
   R_{2,0} & R_{2,1} \\
\end{matrix}
\right|
\neq 0
\),
as otherwise the columns $\frac{\partial P}{\partial a_2}$ and $\frac{\partial P}{\partial b_2}$ would be linearly dependent.
Assume also $s_3\neq 0$; otherwise the $a_3$ and $b_3$ columns of the Jacobian are identically $0$.
Since $D=0$ and $L_{3,2}R_{3,2}\neq 0$, we can write
$(R_{3,2}, R_{3,1}) = \alpha (L_{3,2}, L_{3,1})$ for some $\alpha \in \mathbb{F}$.
Under the functional $\langle\cdot\rangle$, constants and the $x$-term vanish, so $\langle r_3 \rangle = \alpha \langle \ell_3 \rangle$.
Therefore, whenever $\langle \bar u_2 \rangle = 0$ we get
\[
   0
   = \langle \bar u_2 \rangle
   = s_3(L_{3,2} \langle r_3 \rangle + R_{3,2} \langle \ell_3 \rangle)
   = s_3(\alpha L_{3,2} + R_{3,2}) \langle \ell_3 \rangle
   = 2s_3 R_{3,2} \langle \ell_3 \rangle,
\]
and since $\mathrm{char}(\mathbb{F})\neq 2$ and $R_{3,2}\neq 0$, it follows that
$\langle \ell_3 \rangle = 0$ and hence also $\langle r_3 \rangle = 0$.

So we focus on showing $\bar u_2 = 0$ point-wise on all three $x_k$.
Let
\[
   f(x) = L_{3,2} r_3(x) + R_{3,2} \ell_3(x),
\]
so that $\bar u_2(x)=s_2+s_3 f(x)$.
\emph{First}, we choose $(a_2,b_2)$ so that $f(x_0)=f(x_1)=f(x_2)$.

Consider the differences
\begin{align}
   f(x_0) - f(x_t)
   &= L_{3,2} (r_3(x_0) - r_3(x_t)) + R_{3,2} (\ell_3(x_0) - \ell_3(x_t))
 \\&=
   2 L_{3,2} R_{3,2} (u_2(x_0) - u_2(x_t))
 \\&\quad
   + (L_{3,2} R_{3,1} + R_{3,2} L_{3,1}) (u_1(x_0) - u_1(x_t))
 \\&\quad
   + (L_{3,2} R_{3,0} + R_{3,2} L_{3,0}) (x_0 - x_t)
   .
\end{align}
In particular, the constant term $L_{3,2}b_3+R_{3,2}a_3$ cancels in $f(x_0)-f(x_t)$, so the constraints $f(x_0)=f(x_t)$ do not depend on $(a_3,b_3)$.
Since neither the $u_1(x_t)$ differences nor the $x_t$ differences depend on $(a_2,b_2)$, we focus on the $u_2(x_t)$ differences.
Let $A_k = R_{2,1} u_1(x_k) + R_{2,0} x_k$ and
$B_k = L_{2,1} u_1(x_k) + L_{2,0} x_k$ such that $u_2(x_k) = (B_k + a_2)(A_k + b_2)$.
Then
\[
   u_2(x_0) - u_2(x_t)
   = (B_0 A_0 - B_t A_t) + a_2 (A_0 - A_t) + b_2 (B_0 - B_t)
   .
\]
To fix $f(x_0) = f(x_t)$ for $t=1,2$, we need to solve the linear system with determinant
\[
   \left|
   \begin{matrix}
      A_0 - A_1 & B_0 - B_1 \\
      A_0 - A_2 & B_0 - B_2 \\
   \end{matrix}
   \right|
   = R_{1,0}
   \left|
   \begin{matrix}
      L_{2,0} & L_{2,1} \\
      R_{2,0} & R_{2,1} \\
   \end{matrix}
   \right|
   (x_0 - x_1)(x_0 - x_2)(x_1 - x_2)
   .
\]
If $R_{1,0}=0$ then $u_1$ is affine in $x$, and one checks that each of the six sensitivity polynomials has degree at most $4$ in $x$; evaluated at any six distinct points, the corresponding $6\times 6$ Jacobian has rank at most $5$ and hence is singular.
So assume $R_{1,0}\neq 0$; then the displayed determinant is nonzero, and we can solve for $(a_2,b_2)$.
Thus $f$ is constant on $\{x_0,x_1,x_2\}$.

\emph{Second}, having fixed $(a_2,b_2)$, we choose $(a_3,b_3)$ to set $\bar u_2(x_0)=0$.
At the fixed point $x_0$ we have
\[
   \bar u_2(x_0) = s_2 + s_3\bigl(C(x_0) + L_{3,2} b_3 + R_{3,2} a_3\bigr),
\]
where $C(x_0)$ does not depend on $(a_3,b_3)$.
Since $L_{3,2}R_{3,2}\neq 0$ we can choose $(a_3,b_3)$ to set $\bar u_2(x_0)=0$.
This choice shifts $f$ by the same constant at all three points, so it preserves $f(x_0)=f(x_1)=f(x_2)$.
Therefore $\bar u_2(x_0)=0$ implies $\bar u_2(x_k)=0$ for all $k\in\{0,1,2\}$.

Finally, we need to make $\bar u_1$ constant.
With $\bar u_2(x_k)=0$ for all $k$, and $D=0$ we can write $(L_{3,1},R_{3,1})=\beta(L_{3,2},R_{3,2})$ for some $\beta\in\mathbb{F}$, and so
\begin{align}
   \bar u_1(x_k)
   &= s_1 + s_3 (L_{3,1} r_3(x_k) + R_{3,1} \ell_3(x_k))
 \\&= s_1 + s_3 \beta (L_{3,2} r_3(x_k) + R_{3,2} \ell_3(x_k))
 \\&= s_1 + s_3 \beta f(x_k)
   ,
\end{align}
but since $f(x_k)$ is constant, so is $\bar u_1(x_k)$.

This means we have
\(
   \langle \frac{\partial P}{\partial a_1} \rangle
   = \langle \bar u_1 x \rangle
   = \bar u_1 \langle x \rangle
   = 0,
\)
which is what we wanted to show.

\smallskip
\noindent\textbf{Conclusion.}
With these choices we have
\[
   \Big\langle \frac{\partial P}{\partial b_1}\Big\rangle=0,\quad
   \Big\langle \frac{\partial P}{\partial a_1}\Big\rangle=0,\quad
   \Big\langle \frac{\partial P}{\partial a_2}\Big\rangle=0,\quad
   \Big\langle \frac{\partial P}{\partial b_2}\Big\rangle=0,\quad
   \Big\langle \frac{\partial P}{\partial a_3}\Big\rangle=0,\quad
   \Big\langle \frac{\partial P}{\partial b_3}\Big\rangle=0,
\]
so the three Jacobian rows at $x_0,x_1,x_2$ are linearly dependent. Hence the Jacobian is singular.
\end{proof}

\paragraph{Formalization.}
The proof of this appendix is machine-checked in Lean~4 in two layers under
\texttt{FastPoly/\allowbreak LowerBound/}, with no \texttt{sorry}; \texttt{\#print axioms}
reports only \texttt{propext}, \texttt{Classical.choice} and \texttt{Quot.sound} for
every theorem named below.  The normal-form half (\texttt{LowerBound/\allowbreak Main.lean})
is \texttt{exists\_singular\_jacobian}: for every circuit in the displayed normal form
(\texttt{Circuit}), every field with $2\neq0$ and every six distinct evaluation points,
$J$ is singular somewhere (the program semantics, the six sensitivity polynomials, the
chain rule for the $\bar u_i$ and both cases of the split on $D$); hence, over an
infinite field, the evaluation map of the normal form has no everywhere-defined rational
left inverse (\texttt{no\_rationalInverse}), also when its six slots are an arbitrary
affine image $Mp+m_0$ of the parameter vector (\texttt{no\_rationalInverse\_affine}).
The reduction of this appendix is formalized in
\texttt{FastPoly/\allowbreak LowerBound/\allowbreak General/}.  Its main theorem
\texttt{no\_rationalInverse\_general} (\texttt{General/\allowbreak Main.lean}; also
\texttt{no\_rationalInverse\_general\_of\_ringChar\_ne\_two}) is \Cref{lem:six-params} as
stated: for every infinite field with $2\neq0$, every six distinct evaluation points, every
program \texttt{GCircuit F} given by the sixteen fixed constants $A,B,L_{i,j},R_{i,j},s_i$
of the general display---both constants of the first gate are slots, and $A=B=0$ and
$A=0\neq B$ are allowed---and every affine map $z=Mp+h_0$ from the six parameters to the
seven constant slots, with $M$ an arbitrary $7\times6$ matrix (\texttt{outPolyGeneral}),
the evaluation map has no everywhere-defined rational left inverse.  The correspondence
with the text is as follows.  The gauge identity \eqref{eq:lower-gauge} and
$P_p=P'_{\nu(H(p))}$ are \texttt{g1\_eq} and \texttt{gout\_eq} (\texttt{Gauge.lean});
$F=E'\circ Q$ and the chain rule \eqref{eq:lower-chain} are
\texttt{outPolyGeneral\_eq\_outPolyOf} and \texttt{polyJacobian\_outPolyGeneral}
(\texttt{Affine.lean}); the gauge direction $\xi\in\ker D\nu$ and case (i) are
\texttt{dnu\_xi} and \texttt{det\_eq\_zero\_of\_mulVec\_eq\_zero} (\texttt{DQ.lean}); the
midpoint identity \eqref{eq:lower-midpoint} for $Q$ and case (ii) are
\texttt{Qval\_sub\_eq\_mulVec} and \texttt{det\_eq\_zero\_of\_Qval\_eq}
(\texttt{Midpoint.lean}); the orbit polynomial \eqref{eq:lower-orbitpoly} and case (iii)
are \texttt{dot\_orbit\_sub} and \texttt{det\_eq\_zero\_along\_orbit}
(\texttt{Orbit.lean}); the inverse $\Theta$ of case (iv) is \texttt{Theta} with
\texttt{Qval\_Theta} (\texttt{Transversal.lean}); and the transport of a left inverse
along the surjective $Q$ is \texttt{RationalInverse.transport} (\texttt{Transport.lean},
a separate target not imported by \texttt{General/\allowbreak Main.lean}).
Two details differ from the text.  Instead of the cofactor vector $\omega$ the
development uses the left-kernel vector normalized at a row $i_0$ with
$\det M_S\neq0$, $S=\{i\neq i_0\}$, namely $\ell_{i_0}=1$ and $\ell_S=-M_{i_0}M_S^{-1}$
(\texttt{lker}, \texttt{LinAlg.lean}), of which every left-kernel vector, $\omega$
included, is a multiple (\texttt{eq\_smul\_lker\_of\_vecMul\_eq\_zero}).  And the
assembly \texttt{exists\_singular\_of\_gauge} follows the characteristic-free route of
the parenthetical remark above: for $\kappa\neq0$ one pivot on $\tau$
(\texttt{exists\_tau\_eq}) makes $\ell\cdot\xi(H(p_0))=0$, so
$\xi(H(p_0))\in\operatorname{im}M$ and $DQ(p_0)$ is singular
(\texttt{det\_eq\_zero\_of\_dot\_xi\_eq\_zero}); for $\kappa=0=\omega^\circ$ this holds at
every $p$; for $\kappa=0\neq\omega^\circ$ the normal-form singular point pulls back
along $\Theta$.  Thus $2\neq0$ enters the general theorem only through
\texttt{exists\_singular\_jacobian}, and the infinitude of $\mathbb{F}$ only through the
Jacobian obstruction (\texttt{RationalInverse.isEmpty\_of\_det\_eq\_zero}); the midpoint
route of case (ii) and the transport route
(\texttt{no\_rationalInverse\_general\_of\_transversal}) are proved as cross-checks and
are not used by \texttt{General/\allowbreak Main.lean}.  The two first-gate degeneracies
are dispatched in \texttt{exists\_singular\_polyJacobian\_general}: $A=B=0$ is treated
with the gauge scalar $0$ in place of $B/A$, so that $L'_{i,1}=R'_{i,1}=s'_1=0$ in the
normal form, rather than by the factorization through $\mathbb{F}^5$; $A=0\neq B$ by
interchanging the two factors (\texttt{outPolyGeneral\_swap}).  The normal-form statement
is recovered as the instance \texttt{no\_rationalInverse\_affine\_of\_general}.  What
remains modelling rather than a theorem is the first paragraph of the reduction: there is
no Lean datatype of straight-line programs, and the statement that, after topologically
ordering the three nonscalar multiplications, every multiplicand is a fixed linear
combination of $x$ and the earlier $u_j$ plus a parameter-affine constant is the
definition of \texttt{GCircuit} and \texttt{outPolyGeneral}, not a derived fact.

\section{A Lower Bound in Characteristic Two}
\label{sec:lower-char2}

The lower bound of \Cref{sec:lower} uses that $2$ is invertible, and it settles a single
degree.  In characteristic~$2$ a different argument is available, and it is both simpler
and much stronger: it rules out $n$ multiplications for bijective evaluation at $2n$
points at \emph{every} $n>1$, with no assumption on the degree or the leading coefficient
of the computed polynomial, and---unlike \Cref{sec:lower}---with no restriction on the
preprocessing: no map from $2n$ parameters to the slots of the chain, whether affine,
rational or arbitrary, makes the evaluation bijective.  The transcendence-degree argument
behind Knuth's Theorems~M and~A is specific to infinite fields (Knuth, answer to
exercise~27~\cite[\S4.6.4]{knuth1997seminumerical}); the counting argument below gives a
lower bound over finite fields of characteristic two instead.  The mechanism is a symmetry of
slot space rather than a Jacobian: in characteristic~$2$ a chain admits a one-parameter
group of slot changes that leaves the computed polynomial \emph{exactly} fixed, and a
second family that translates the argument.  If the evaluation map were onto, its fibres
would be exactly the orbits of the first group, and a set cut out by the two families
could then be counted in two incompatible ways.  The argument, including the statement for
arbitrary preprocessing maps, is machine-checked in Lean (\Cref{rem:char2-formalization}).

\subsection{Model}

Fix a finite field $\F$ of characteristic~$2$ with $Q=\abs{\F}$, and let $n\ge1$.  A
\emph{chain with $n$ multiplications} is given by fixed constants
$\alpha_i,\beta_i,p_{ij},q_{ij},\gamma,r_j\in\F$; on a slot vector
$\mathbf z=(u_1,v_1,\dots,u_n,v_n,w)\in\F^{2n+1}$ it computes
\begin{align}
   G_i &=
   \Bigl(\alpha_i x+\sum_{j<i}p_{ij}G_j+u_i\Bigr)
   \Bigl(\beta_i x+\sum_{j<i}q_{ij}G_j+v_i\Bigr),
   \qquad 1\le i\le n,
   \label{eq:char2-gate}\\
   f_{\mathbf z}(x) &= \gamma x+\sum_{j=1}^{n}r_jG_j+w .
   \label{eq:char2-output}
\end{align}
As in \Cref{sec:lower}, only the $n$ displayed products are charged; additions and
multiplications by the fixed constants are free, and the parameters enter only through the
$2n+1$ additive slots.  For $2n$ distinct points $X=(x_1,\dots,x_{2n})$ let
\begin{equation}
   E_X\colon\F^{2n+1}\to\F^{2n},
   \qquad
   E_X(\mathbf z)=\bigl(f_{\mathbf z}(x_1),\dots,f_{\mathbf z}(x_{2n})\bigr)
   \label{eq:char2-eval}
\end{equation}
be the evaluation map on slot space.  A \emph{$(2n,n)$ construction} is such a chain
together with a \emph{preprocessing map} $\mathbf z\colon\F^{2n}\to\F^{2n+1}$---affine,
polynomial, rational, or an arbitrary function---such that for every $2n$ distinct points
the composite $\Phi_X=E_X\circ\mathbf z\colon\F^{2n}\to\F^{2n}$ is a bijection.  A
bijective $\Phi_X$ makes $E_X$ onto, and conversely any right inverse of an onto $E_X$ is
a preprocessing map with bijective $\Phi_X$; so a chain admits a $(2n,n)$ construction
exactly when $E_X$ is surjective for every $X$.  Over a finite field this is the natural
formulation: it says the $2n$ parameters carry exactly as much information as $2n$
evaluations, which is what the hashing applications of \Cref{sec:experiments} require.

\begin{theorem}[No $(2n,n)$ construction in characteristic two]
\label{thm:char2-lower}
Let $\F$ be a finite field of characteristic~$2$ with $Q=\abs{\F}\ge 2n$ and $n>1$.  Then
no chain with $n$ multiplications is a $(2n,n)$ construction: for every such chain there
are $2n$ distinct points $X$ for which the evaluation map $E_X$ of \eqref{eq:char2-eval}
is not surjective, so no preprocessing map, affine or otherwise, makes $\Phi_X$ a
bijection.
\end{theorem}

The hypothesis $n>1$ cannot be removed: $f_{a,b}(x)=ax+b$ uses one multiplication and its
evaluation at two distinct points is invertible.

\subsection{Proof}

Throughout, write $\mathbf s=(u_2,v_2,\dots,u_n,v_n,w)\in\F^{2n-1}$ for the slots other
than the first two.

\paragraph{The first gate is not scalar.}
If $\alpha_1=\beta_1=0$ then $G_1=u_1v_1$ is a constant, and for each later factor and for
the output it enters through a fixed coefficient $\lambda_j$, so replacing
$s_j$ by $s_j+\lambda_ju_1v_1$ leaves $f$ unchanged.  The polynomial $f_{\mathbf z}$ then
depends only on the $2n-1$ field elements $\mathbf s+\lambda u_1v_1$, so $E_X$ takes at most
$Q^{2n-1}<Q^{2n}$ values and is not surjective for any $X$.  Hence we may assume
$(\alpha_1,\beta_1)\neq(0,0)$, and we write, with $\sigma=\alpha_1v_1+\beta_1u_1$,
\begin{equation}
   G_1=\alpha_1\beta_1x^2+\sigma x+u_1v_1 .
   \label{eq:char2-firstgate}
\end{equation}

\paragraph{Two families of slot changes.}
Let $\lambda,\varepsilon\in\F^{2n-1}$ collect, for each slot of $\mathbf s$, the fixed
coefficient with which $G_1$, respectively $x$, enters the corresponding later factor or
the output.  For $t\in\F$ put $d_t=\sigma t+\alpha_1\beta_1t^2$ and define
\begin{equation}
   \mathcal G_t(u_1,v_1,\mathbf s)=(u_1+\alpha_1t,\;v_1+\beta_1t,\;\mathbf s+\lambda d_t),
   \qquad
   \mathcal T_c(u_1,v_1,\mathbf s)=(u_1+\alpha_1c,\;v_1+\beta_1c,\;\mathbf s+\varepsilon c).
   \label{eq:char2-gauge}
\end{equation}
The first is a \emph{gauge}: expanding
$(\alpha_1x+u_1+\alpha_1t)(\beta_1x+v_1+\beta_1t)=G_1+d_t$, the two cross terms in $x$
cancel \emph{because the characteristic is two}, so $G_1$ changes by the constant $d_t$,
which the corrections $\lambda d_t$ undo in every later factor and in the output.  Hence
$f_{\mathcal G_t(\mathbf z)}=f_{\mathbf z}$ identically.  The second translates the
argument: induction through the gates gives $G_i^{\mathcal T_c(\mathbf z)}(x)=G_i^{\mathbf
z}(x+c)$ and therefore $f_{\mathcal T_c(\mathbf z)}(x)=f_{\mathbf z}(x+c)$.

\paragraph{The gauges form a free group action that commutes with translation.}
Because the characteristic is two,
$\sigma(\mathcal G_t\mathbf z)=\sigma(\mathbf z)+2\alpha_1\beta_1t=\sigma(\mathbf z)$ and
likewise $\sigma(\mathcal T_c\mathbf z)=\sigma(\mathbf z)$; moreover
$d_t+d_s=d_{t+s}$, since the cross term $2\alpha_1\beta_1ts$ vanishes.  Consequently
$\mathcal G_s\circ\mathcal G_t=\mathcal G_{s+t}$ and
$\mathcal T_c\circ\mathcal G_t=\mathcal G_t\circ\mathcal T_c$: the gauges are an action of
the additive group of $\F$ on slot space, and every translation permutes its orbits.  The
action is free, because $\mathcal G_t\mathbf z=\mathbf z$ forces
$\alpha_1t=\beta_1t=0$ and hence $t=0$; so every orbit has exactly $Q$ elements.

\paragraph{If $E_X$ is onto, its fibres are the orbits.}
Fix $2n$ distinct points $X$ and suppose $E_X$ is surjective.  Since
$f_{\mathcal G_t\mathbf z}=f_{\mathbf z}$, the map $E_X$ is constant on orbits, so each of
its $Q^{2n}$ fibres is a nonempty union of orbits and has at least $Q$ elements.  The
fibres partition the $Q^{2n+1}$ slot vectors, so every fibre has exactly $Q$ elements and
is a single orbit:
\begin{equation}
   E_X(\mathbf z)=E_X(\mathbf z')
   \iff
   \mathbf z'=\mathcal G_t\mathbf z\ \text{ for some } t\in\F .
   \label{eq:char2-fibres}
\end{equation}

\paragraph{Counting one set in two ways.}
Fix $c\neq0$.  Since $Q\ge2n$, translation by $c$ has at least $n$ disjoint two-element
orbits: choose $r_1,\dots,r_n$ with $r_1,r_1+c,\dots,r_n,r_n+c$ distinct, let $X$ be that
tuple, and let $\pi$ swap the two entries of each pair.  By
$f_{\mathcal T_c(\mathbf z)}(x)=f_{\mathbf z}(x+c)$,
\begin{equation}
   E_X\circ\mathcal T_c=\pi\circ E_X .
   \label{eq:char2-conjugacy}
\end{equation}
Suppose $E_X$ is surjective, and consider
\[
   \mathcal A=\bigl\{\mathbf z\in\F^{2n+1}:\
   \mathcal T_c\mathbf z=\mathcal G_t\mathbf z\text{ for some }t\in\F\bigr\}.
\]
On one hand, by \eqref{eq:char2-fibres} and \eqref{eq:char2-conjugacy},
$\mathbf z\in\mathcal A$ exactly when $E_X(\mathcal T_c\mathbf z)=E_X(\mathbf z)$, that is,
when $\pi(E_X(\mathbf z))=E_X(\mathbf z)$, that is, when
$E_X(\mathbf z)\in\operatorname{Fix}(\pi)$.  A vector is fixed by $\pi$ exactly when the
entries of every pair agree, so $\abs{\operatorname{Fix}(\pi)}=Q^n$, and each fibre has
$Q$ elements; hence
\[
   \abs{\mathcal A}=Q^{n+1}.
\]
On the other hand $\mathcal A$ can be read off from \eqref{eq:char2-gauge}.  The first two
slots of $\mathcal T_c\mathbf z$ and $\mathcal G_t\mathbf z$ agree only if
$\alpha_1c=\alpha_1t$ and $\beta_1c=\beta_1t$, which forces $t=c$; the remaining slots
then agree exactly when
$\varepsilon c=\lambda d_c(\mathbf z)=\lambda c\bigl(\sigma(\mathbf z)+\alpha_1\beta_1c\bigr)$,
that is,
\[
   \varepsilon=\lambda\bigl(\sigma(\mathbf z)+\alpha_1\beta_1c\bigr)
   \qquad\text{in }\F^{2n-1}.
\]
If $\lambda=0$ this holds for every $\mathbf z$ (when $\varepsilon=0$) or for none.  If
$\lambda\neq0$ it holds for no $\mathbf z$ unless $\varepsilon=\kappa\lambda$ for a scalar
$\kappa$, and then it says $\sigma(\mathbf z)=\kappa+\alpha_1\beta_1c$, a fibre of the
nonconstant linear form $\sigma$, which has $Q^{2n}$ elements.  Hence
\begin{equation}
   \abs{\mathcal A}\in\{0,\;Q^{2n},\;Q^{2n+1}\}.
   \label{eq:char2-fix}
\end{equation}
For $n>1$ the value $Q^{n+1}$ is none of these.  This contradiction shows that $E_X$ is
not surjective for this $X$, which proves \Cref{thm:char2-lower}.\qed

For $n=1$ the two counts agree, $Q^{n+1}=Q^{2n}$, as they must: $ax+b$ is a construction.

\begin{remark}[Formalization]
\label{rem:char2-formalization}
\Cref{thm:char2-lower} is machine-checked in Lean~4
(\texttt{FastPoly/\allowbreak LowerBoundChar2/}).  The fibre-counting proof above is
formalized verbatim in \texttt{General.lean}: \texttt{no\_surjective\_eval}
states that for every $n$-gate chain over a finite field of characteristic two
with $|\F|\ge2n$ and $n>1$ there is an injective point set $X$ at which the
evaluation map on slot space is not surjective, and
\texttt{no\_construction\_general} derives from it that no preprocessing map
whatsoever (an arbitrary function from any parameter type into slot space)
makes evaluation surjective at every $X$; the development proves the model
\eqref{eq:char2-gate}--\eqref{eq:char2-eval}, the gauge identity with its
characteristic-two cancellation, the freeness of the gauge action, the
translation semantics, the orbit-equals-fibre step, and the two counts of the
coincidence set.  An earlier development in the same directory treats the
case of an affine preprocessing map through a transversality lemma and the
conjugacy step \eqref{eq:char2-conjugacy}.  Both top-level statements depend
only on Lean's standard axioms (\texttt{propext}, \texttt{Classical.choice},
\texttt{Quot.sound}).  A companion file exhibits the
one-multiplication chain $f_{a,b}(x)=ax+b$ as a $(2,1)$ construction inside the same
model, so the hypothesis $n>1$ is necessary and the model is not vacuous.
\end{remark}

\begin{remark}[Scope]
\label{rem:char2-scope}
\Cref{thm:char2-lower} concerns an even number of evaluation points and bijective
evaluation over a finite field, and it constrains the chain alone: no preprocessing of
any kind circumvents it.  Here ``preprocessing'' means the map from the parameters to the
$2n+1$ additive slots of \eqref{eq:char2-gate}--\eqref{eq:char2-output}; the constants of
the chain are fixed, and a chain that multiplies by a parameter, as $\theta x$ in
\Cref{sec:model}, lies outside the model and is not covered.  \Cref{sec:lower} concerns rational decoding over an infinite
field of characteristic $\neq2$.  Neither implies the other, and the fixed-degree
constructions of \Cref{appendix:polynomials} are unaffected: they use
$\lfloor n/2\rfloor+1$ multiplications, one more than \Cref{thm:char2-lower} forbids.
\end{remark}

\section{Numerical stability in floating point}
\label{sec:numerical-stability}

The constructions of this paper are exact: over a finite field, or over $\Q$ in
exact rational arithmetic, the schedules evaluate $P$ without error.  It is
nevertheless natural to ask how they behave in floating-point arithmetic,
where Horner's rule and Estrin's scheme are the standard choices.  We show
that adapted-constant schedules---ours included---do not come with the
coefficientwise backward-stability guarantee of Horner's rule: their
floating-point error bound carries an amplification factor $A$, which we
define and measure below and which is unbounded for our schedules and for
the classical adapted schemes, the floating-point counterpart of the
classical observation that adapted-coefficient methods are prone to lose
significance~\cite[\S4.6.4]{knuth1997seminumerical}.

\subsection{Model and a majorant bound for straight-line programs}

We use the standard model~\cite[Ch.~2--3]{higham2002accuracy}: every
floating-point operation satisfies
$\operatorname{fl}(a\circ b)=(a\circ b)(1+\delta)$ with $\abs{\delta}\le u$,
where $u$ is the unit roundoff, and there is no overflow or underflow.  Set
$\gamma_k=ku/(1-ku)$.  The product lemma states that any product of $k$ factors
$(1+\delta_j)^{\pm1}$ equals $1+\theta_k$ with $\abs{\theta_k}\le\gamma_k$.

\paragraph{What is rounded.}
Preprocessing is exact: the coefficients of $P$ are mapped to the parameters
of the schedule in exact rational arithmetic (or in any precision high enough
to be exact for the purpose).  Evaluation is in floating point: the argument
$x$ arrives as a floating-point number, each parameter is rounded once to the
working precision, $\operatorname{fl}(c_j)=c_j(1+\delta_j)$, and every
arithmetic operation of the schedule rounds, with or without fused
multiply--add.  Every evaluation scheme in this paper is a
\emph{straight-line program} $\mathcal C$: a sequence of wires, each either
the input $x$, a constant $c_j$, or the sum, difference, or product of two
earlier wires.  Integer multiples $k\cdot w$ of a wire count as one product.
Constants that are input data (the coefficients $a_i$ in Horner's rule) are
assumed exactly representable, so that Horner's rule has no preprocessing
rounding at all; the preprocessed constants of our schedules, of
Rabin--Winograd, and of Motzkin--Eve carry the one rounding above.

\begin{definition}[Rounding depth and majorant]
\label{def:rounding-depth}
The \emph{rounding depth} $\rho(w)$ of a wire is defined recursively by
$\rho(x)=0$; $\rho(c_j)=0$ if $c_j$ is exact input data and $1$ otherwise;
$\rho(u\pm v)=\max(\rho(u),\rho(v))+1$; and $\rho(u\cdot v)=\rho(u)+\rho(v)+1$.
The \emph{majorant} $M_{\mathcal C}$ is the program obtained from $\mathcal C$
by replacing every constant $c_j$ by $\abs{c_j}$, every integer multiple
$k\cdot w$ by $\abs{k}\cdot w$, and every subtraction by an addition; it is
evaluated exactly, at $\abs{x}$.
\end{definition}

Note that products \emph{add} rounding depths, so that $\rho(v)$ is exactly the
largest number of rounding factors carried by any tree monomial of $v$; a
max-rule for products would undercount squarings.

\begin{theorem}[Majorant bound]
\label{thm:majorant-bound}
Let $\mathcal C$ compute $P(x)$ exactly in exact arithmetic, with output wire
$w$ satisfying $\rho(w)u<1$, and let $\widehat P(x)$ be the value computed in
floating point.  Then
\[
  \abs{\widehat P(x)-P(x)}\;\le\;\gamma_{\rho(w)}\,M_{\mathcal C}(\abs{x}).
\]
\end{theorem}

\begin{proof}
Expand every wire $v$ as a sum of \emph{tree monomials}: $v=\sum_\pi t_\pi$,
where each $t_\pi$ is a product of leaves ($x$'s and constants) along one
multiplicative path of the expression tree, with the sign carried by the
subtractions.  We claim that the computed value satisfies
$\widehat v=\sum_\pi t_\pi\prod_{j\in J_\pi}(1+\delta_j)$ for multisets
$J_\pi$ of rounding errors with $\abs{J_\pi}\le\rho(v)$ (the same $\delta_j$
may recur when a wire is reused).  For the leaves this is the assumption on
the constants.  For $v=u\pm v'$, the computed
$\widehat v=(\widehat u\pm\widehat v')(1+\delta)$ appends one element to every
$J_\pi$, so the depth increases by one.  For $v=u\cdot v'$, every tree
monomial of $v$ is a product $t_\pi s_\sigma$ of one monomial from each
factor, and $\widehat v=\widehat u\,\widehat v'(1+\delta)$ gives it the
multiset $J_\pi\cup J_\sigma\cup\{\delta\}$ of size at most
$\rho(u)+\rho(v')+1$.  By the product lemma each
$\prod_{j\in J_\pi}(1+\delta_j)=1+\theta_\pi$ with
$\abs{\theta_\pi}\le\gamma_{\rho(w)}$, so
$\abs{\widehat P-P}\le\gamma_{\rho(w)}\sum_\pi\abs{t_\pi}$; and
$\sum_\pi\abs{t_\pi}$ is exactly the value of the majorant program at
$\abs{x}$, because taking absolute values of all leaves and integer multiples
and turning all subtractions into additions removes every cancellation between
tree monomials.  No independence among the $\delta_j$ is used.
\end{proof}

\paragraph{Horner and Estrin.}
For Horner's rule the tree monomials are exactly the terms $a_ix^i$, so
$M(\abs{x})=\sum_i\abs{a_i}\abs{x}^i$ and the bound is the classical
$\gamma_{2n}\sum_i\abs{a_i}\abs{x}^i$ (one fewer factor when $P$ is monic and
the leading multiplication is skipped); the same expansion gives the refined
coefficientwise form $\widehat P=\sum_i a_ix^i(1+\eta_i)$ with
$\abs{\eta_i}\le\gamma_{2i+1}$, and $\gamma_{2n}$ for
$i=n$~\cite[\S5.1]{higham2002accuracy}.  Estrin's scheme on a full tree of
degree $n=2^m-1$ also has $M(\abs{x})=\sum_i\abs{a_i}\abs{x}^i$, and counting
the roundings on the path of $a_ix^i$---$m$ additions, $s_2(i)$ explicit
multiplications, and $i-s_2(i)$ factors inherited from the repeated squarings
$x^2,x^4,\dots$, where $s_2(i)$ is the number of ones in the binary expansion
of $i$ (here the additive rule for products matters: the computed
$x^4=x^4(1+\delta_1)^2(1+\delta_2)$ carries three factors)---gives
$\abs{\eta_i}\le\gamma_{i+m}$, a uniform bound $\gamma_{n+m}$ that is
\emph{smaller} than Horner's $\gamma_{2n}$ (again one fewer when monic).  With
a fused multiply--add at every node the product roundings disappear and both
schemes have the uniform bound $\gamma_n$, Horner retaining slightly sharper
coefficientwise constants $\gamma_{\min(i+1,n)}$ versus
$\gamma_{i+m-s_2(i)}$.  Both schemes are therefore \emph{coefficientwise
backward stable}: the computed value is the exact value of a polynomial with
relatively perturbed coefficients, and the relative forward error is at most
$\gamma_\rho\,\kappa(P,x)$ for the condition number
$\kappa(P,x)=\sum_i\abs{a_i}\abs{x}^i/\abs{P(x)}$.  The comparison is of
practical interest in production libraries: the discussion accompanying the
Estrin implementation in Boost.Math~\cite{boostmath932} conjectures bounds of
this shape, and the display above makes them precise---Estrin's tree does not
cost accuracy; its uniform constant $\gamma_{n+m}$ is in fact the smaller
one.

\subsection{Adapted-coefficient schemes and amplification}

For a schedule whose constants are not the coefficients themselves, the tree
monomials are products of $x$ with preprocessed constants, and they cancel: the
majorant can be much larger than $\sum_i\abs{a_i}\abs{x}^i$.  Define the
\emph{schedule amplification}
\[
  A_{\mathcal C}(P,x)\;=\;\frac{M_{\mathcal C}(\abs{x})}{\sum_{i}\abs{a_i}\abs{x}^i}.
\]
Theorem~\ref{thm:majorant-bound} then gives the relative forward bound
$\abs{\widehat P-P}/\abs{P}\le\gamma_{\rho}\,A_{\mathcal C}(P,x)\,\kappa(P,x)$,
and $A_{\mathcal C}$ is precisely the price paid for the multiplications
saved; for Horner and Estrin $A\equiv1$.  For an adapted scheme it is
unbounded as soon as a low coefficient is obtained by cancellation.  Write
$c=(c_1,\dots,c_m)$ for the constants of $\mathcal C$, regarded as
indeterminates, so that the coefficients $a_j=a_j(c)$ of the computed
polynomial $P[c]$ are polynomials in $c$, and call $c$ \emph{admissible} if
the scheme produces it: for our schedules $c=c(\alpha)$ for a key vector
$\alpha\in K^n$, each constant being a key or a signed integer combination
of a few keys; for Rabin--Winograd and Motzkin--Eve, $c$ ranges over the
parameter vectors of the scheme.

\begin{proposition}[Unit pivots make the amplification unbounded]
\label{prop:unit-pivot-unstable}
Suppose that some coefficient has an additive unit pivot,
$a_d=c_i+f(c_{\mathrm{rest}})$ with $f$ independent of $c_i$ (a pivot of
slope $\lambda\ne0$, $a_d=\lambda c_i+f$, is handled identically with
$\lambda c_i$ in place of $c_i$).
\begin{enumerate}[label=(\roman*)]
\item\label{it:pivot-lower} For every $c$ and every $x\ne0$,
  $M_{\mathcal C}(\abs{x})\ge\abs{c_i}\abs{x}^d$, hence
  \[
    A_{\mathcal C}(P[c],x)\;\ge\;\frac{\abs{c_i}\abs{x}^d}{\sum_j\abs{a_j}\abs{x}^j}.
  \]
\item\label{it:pivot-sup} If some admissible $c$ has
  $a_0(c)=\dots=a_d(c)=0$ but a nonzero tree monomial of degree $d$---for
  instance $c_i\ne0$, i.e.\ $f(c_{\mathrm{rest}})\ne0$---then
  $A_{\mathcal C}(P[c],x)\to\infty$ as $x\to0$, and in particular
  $\sup_{P,x}A_{\mathcal C}(P,x)=\infty$.  This is the case whenever $f$
  does not vanish identically on the admissible $c$ (for $c=c(\alpha)$:
  $f(c(\alpha))\not\equiv0$ as a polynomial in $\alpha$) and either $d=0$
  and $c_i$ can be varied independently of
  the other constants, or the coefficient map is onto the monic polynomials
  of degree $n$ and $f(c_{\mathrm{rest}})$ depends on $c$ only through
  $a_{d+1},\dots,a_{n-1}$, as it does at row $d$ of a descending triangular
  decoder.
\end{enumerate}
\end{proposition}

\begin{proof}
\ref{it:pivot-lower}: Group the tree monomials of $P[c]$ by their degree in
$x$.  Those of degree $d$ sum to $a_d(c)\,x^d$, identically in $c$; the ones
among them that contain no leaf $c_i$ sum to $g(c_{\mathrm{rest}})\,x^d$ for
some polynomial $g$, and the ones that do contain $c_i$ sum to
$(a_d-g)\,x^d$.  Setting $c_i=0$ kills every monomial of the second kind, so
$g=a_d|_{c_i=0}=f$, and the tree monomials of degree $d$ containing $c_i$
sum to $c_i\,x^d$.  The majorant is the sum of the absolute values of all
tree monomials (proof of Theorem~\ref{thm:majorant-bound}), so
$M_{\mathcal C}(\abs{x})\ge\abs{c_i}\abs{x}^d$.

\ref{it:pivot-sup}: Write $M_{\mathcal C}(t)=\sum_e m_e t^e$, where
$m_e\ge0$ is the sum of the absolute values of the tree monomials of degree
$e$; the hypothesis says $m_d>0$ (and $m_d\ge\abs{c_i}$ by
\ref{it:pivot-lower}).  At such a $c$ the polynomial is
$P[c]=x^n+\sum_{d<j<n}a_jx^j$, so for $0<\abs{x}\le1$
\[
  A_{\mathcal C}(P[c],x)\;\ge\;
  \frac{m_d\abs{x}^d}{\abs{x}^n+\sum_{d<j<n}\abs{a_j}\abs{x}^j}
  \;\ge\;\frac{m_d}{\abs{x}\bigl(1+\sum_{d<j<n}\abs{a_j}\bigr)}
  \;\longrightarrow\;\infty\qquad(x\to0).
\]
For the sufficient conditions, pick an admissible $c^*$ with
$f(c^*_{\mathrm{rest}})\ne0$, which exists by hypothesis.  If $d=0$
and $c_i$ is free, replace $c^*_i$ by $-f(c^*_{\mathrm{rest}})$: then
$a_0=0$ and $c_i\ne0$.  If the coefficient map is onto and $f$ is
determined by the top coefficients, let $c$ be admissible with
$P[c]=x^n+\sum_{d<j<n}a_j(c^*)\,x^j$: then $a_0(c)=\dots=a_d(c)=0$,
$f(c_{\mathrm{rest}})=f(c^*_{\mathrm{rest}})\ne0$, and $a_d(c)=0$ forces
$c_i=-f(c_{\mathrm{rest}})\ne0$.
\end{proof}

\begin{remark}[The pivot alone does not suffice]
\label{rem:unit-pivot-counterexample}
The hypothesis of part~\ref{it:pivot-sup} cannot be dropped: a unit pivot
with $f\not\equiv0$ does not by itself force $\sup A=\infty$.  The chain
$w_1=x+c_1$, $w_2=x+c_2$, $P=x\,w_1w_2+c_0$ computes
$x^3+(c_1+c_2)\,x^2+c_1c_2\,x+c_0$ with two multiplications, and
$a_2=c_1+c_2$ is a unit pivot in $c_1$ with $f=c_2\not\equiv0$; yet, with
$t=\abs{x}$,
\[
\begin{aligned}
  M_{\mathcal C}(t)-\sum_j\abs{a_j}t^j
  &=\bigl(\abs{c_1}+\abs{c_2}-\abs{c_1+c_2}\bigr)\,t^2
  \le2\min(\abs{c_1},\abs{c_2})\,t^2\\
  &\le2\sqrt{\abs{c_1c_2}}\,t^2
  \le t^3+\abs{c_1c_2}\,t ,
\end{aligned}
\]
so $A_{\mathcal C}\le2$ for all real or complex constants and all $x$, with
equality at $c_1=-c_2=1$, $c_0=0$, $\abs{x}=1$.  What fails is exactly the
hypothesis of part~\ref{it:pivot-sup}: $a_1=c_1c_2$ vanishes only when a
factor does, so no admissible $c$ has $a_1=a_2=0$ with $c_1\ne0$.  Over
$\R$ the coefficient map is not onto (its image is $a_2^2\ge4a_1$); over
$\C$ it is onto, but the correction $c_2$ is a root of $z^2-a_2z+a_1$ and
depends on $a_1$ as well, so the decoder is not triangular.  The
unrestricted claim thus fails already for a two-multiplication cubic, the
multiplication count of $P_3$.
\end{remark}

For the schedules of this paper the hypothesis of part~\ref{it:pivot-sup}
is met at $d=0$: the constant term is the last row of the decoder,
$a_0=\alpha_i+f(\alpha_{\mathrm{rest}})$ for one key $\alpha_i$, and the
degree-$0$ tree monomials of the program do not all vanish when $a_0$ does
(for $P_7$ one has $a_0=\alpha_0+\alpha_1\alpha_2+\alpha_1\alpha_4\alpha_5$
and the sum of the absolute values of the degree-$0$ tree monomials is
$\abs{\alpha_0}+\abs{\alpha_1\alpha_2}+\abs{\alpha_1\alpha_4\alpha_5}$, which
at $a_0=0$ is positive as soon as $\alpha_1\alpha_2\ne0$).  We checked this in exact
rational arithmetic, on the programs exactly as the measurements below
evaluate them, at three random key vectors with nonzero entries and $a_0$
forced to zero through $\alpha_i$, for every odd $5\le n\le63$; for even
$n\le62$ the program is $P_n=xP_{n-1}+\alpha_0$, which at $\alpha_0=0$ has
exactly the amplification of $P_{n-1}$; and $P_3$, whose only degree-$0$
tree monomial is its constant term, has a nonzero degree-$1$ tree monomial
at $x^3$ instead.  The polynomial $x^n$ itself is such an instance: our
chains reach it only with nonzero constants (the zero-key skeleton is not
$x^n$; that of $P_7$ is $x^7+2x^6+2x^5+2x^4+x^3+x^2$, and
\Cref{prop:centering} below shows that below $n-1$ multiplications the
skeleton cannot be a pure product tree), and for $3\le n\le31$ we verified
that its majorant keeps a nonzero term of degree $0$ (odd $n\ge5$), $1$
($n=3$ and even $n\ge6$) or $2$ ($n=4$).  Hence at these degrees
$A_{\mathcal C}(x^n,x)\to\infty$ as $x\to0$, and
$\sup_{P,x}A_{\mathcal C}=\infty$.  Rabin--Winograd's scheme at the degrees
$n=2^m-1$ measured here (where the balanced recursion is exact and no
normalization occurs) and Motzkin--Eve's scheme, with its last constant $c_K$
regarded as a free parameter, both have a free leaf constant that enters
$a_0$ with unit slope and a nonzero correction $f$---the constant term of the
innermost monic remainder, respectively $c_K$---so part~\ref{it:pivot-sup}
applies to them as well.
What the proposition
shows is that the forward bound of Theorem~\ref{thm:majorant-bound}---the
guarantee available for such a schedule---is unbounded relative to
$\sum_i\abs{a_i}\abs{x}^i$: none of these schemes comes with a
coefficientwise backward-stability guarantee of the Horner type,
independently of the multiplication count.  An unbounded upper bound is not
by itself a proof of instability; that the loss is real is shown by the
measurements below, which exhibit the errors.  The proposition is a
statement about the supremum (polynomials whose low coefficients vanish),
not a quantitative bound on a box $\abs{a_j}\le1$; the measurements below
show how large $A$ is on such boxes.
The rounding depth, by contrast,
differs between the schemes only by constant factors: every scheme has
$\rho\ge n-1$, because the leading tree monomial $x^n$ alone passes through
$n-1$ rounded products.  Horner attains $2n-1$, Estrin $n+m-1$, and our
schedules measure $3.6n$--$3.8n$ for $n\le31$
(Table~\ref{tab:numstab})---roughly twice Horner.  Logarithmic multiplicative
height shortens the critical path; it does not reduce the number of rounding
factors.  The decisive quantity is therefore $A$, and it depends on where the
data enter.

\paragraph{Two regimes.}
In the \emph{prescribed-keys} regime the schedule's keys---the parameter
vector $\theta$ of the evaluation circuit $E(x;\theta)$ of
\Cref{sec:model}, written $\alpha$ in the constructions---are the data, as in
hashing, where they are random field elements; the polynomial is induced.  The
constants of the schedule are then the keys themselves or signed sums of a few
of them, and $A$ stays moderate at the degrees relevant to hashing (medians
$10^{0.4}$, $10^{1.4}$, $10^{3.4}$, and $10^{6.8}$ at $n=7$, $15$, $31$, and
$63$ in Table~\ref{tab:numstab}); it grows with $n$ because the induced
coefficients are polynomials in the keys whose terms cancel while the
majorant's do not, but the observed double-precision error never exceeds four
units of $u\sum_i\abs{a_i}\abs{x}^i$ up to $n=63$.  In the
\emph{prescribed-coefficients} regime an arbitrary polynomial is given and the
decoder (\Cref{alg:final-decoder}) recovers the keys.  The coefficient map
$\alpha\mapsto(a_0,\dots,a_{n-1})$ is a polynomial automorphism
(\Cref{lem:polynomial-left-inverse-automorphism}), so the decoder is itself a
polynomial map; but it has high degree---as a polynomial in the top
coefficient, degree $96$, $154$, and $1456$ at $n=15$, $19$, and $23$---and
rational coefficients whose denominators are products of the integer pivots of
\Cref{rem:char2} (powers of two at the degrees measured here, $n\le23$, where every internal index is $k=2$).  The degree compounds along three mechanisms of the
construction: monic square roots of top windows (linear growth), subtraction
of squares (doubling), and synthetic division by $x+\beta$ inside the
known-power blocks (geometric in the block length).  For single-digit integer
input coefficients the keys returned at $n=23$ already have thousands of
binary digits, and the low-row keys are exactly the corrections $-f$ of
\Cref{prop:unit-pivot-unstable} that cancel the chain's own tree monomials:
$A$ grows
explosively---its median already exceeds $10^{43}$ at $n=15$---and
floating-point evaluation is useless beyond small degrees, even though the
same schedule evaluates $P$ exactly in rational arithmetic.

\subsection{Measurements}

Table~\ref{tab:numstab} reports, for random monic polynomials of degree
$n\in\{7,15,31\}$ (prescribed coefficients) and $n\in\{7,15,31,63\}$
(prescribed keys) and random dyadic points $x\in[-2,2]$, the rounding depth
$\rho$ of each schedule, the median and maximum of the amplification $A$, and
the observed double-precision forward error in units of
$u\sum_i\abs{a_i}\abs{x}^i$.  The reference values are computed in exact
rational arithmetic.  In the prescribed-coefficients regime the coefficients
are integers in $[-5,5]$; in the prescribed-keys regime the keys are dyadic
rationals $k/16$, $\abs{k}\le16$.  The schedules are generated by the
reference implementation accompanying the paper.  Rabin--Winograd here uses
balanced splitting with one absorbed coefficient per product; its constants
are integer polynomial expressions in the coefficients whose size roughly
doubles with each level of the balanced recursion (for $n=2^m-1$ no division
occurs at all), so its amplification grows, but far more slowly than the
decoder's.  Motzkin--Eve (Theorem~E of
Knuth~\cite[\S4.6.4]{knuth1997seminumerical}) has numerically computed
constants (double-double root finding, printed to 13--17 digits), so its
error column also includes preprocessing error, and degrees at which the
preprocessing failed verification are omitted.  Preprocessed constants are
converted to double with a single correctly rounded conversion, as in the
model; evaluating every product-and-add with a fused multiply--add changes no
entry of the table at the displayed precision, as the majorant analysis
predicts (it lowers $\rho$ by a constant factor and leaves $A$ unchanged); the one deviation is that in the prescribed-keys regime the induced
coefficients are themselves rounded before Horner, Estrin, and Rabin--Winograd
read them, which is negligible at the reported scale.

\begin{table}[htbp]
\centering\footnotesize
\begin{tabular}{@{}llrrrrrr@{}}
\toprule
regime & scheme & $n$ & $\rho$ & $A$ med. & $A$ max & err.\ med. & err.\ max \\
\midrule
prescribed coefficients & Horner & 7 & 13 & $1$ & $1$ & $0$ & $0$ \\
 & Estrin & 7 & 9 & $1$ & $1$ & $0$ & $0$ \\
 & Rabin--Winograd & 7 & 12 & $6.4$ & $184$ & $0$ & $0$ \\
 & Motzkin--Eve & 7 & 27 & $37$ & $10^{4}$ & $2.3\times10^{3}$ & $2.0\times10^{6}$ \\
 & \textbf{this paper} & 7 & 25 & $10^{12}$ & $10^{27}$ & $0$ & $3.6\times10^{14}$ \\
\addlinespace[2pt]
 & Horner & 15 & 29 & $1$ & $1$ & $0$ & $1.16$ \\
 & Estrin & 15 & 18 & $1$ & $1$ & $0$ & $0.786$ \\
 & Rabin--Winograd & 15 & 22 & $14$ & $10^{3}$ & $0$ & $1.07$ \\
 & Motzkin--Eve & 15 & 59 & $10^{6}$ & $10^{7}$ & $4.6\times10^{6}$ & $8.6\times10^{8}$ \\
 & \textbf{this paper} & 15 & 54 & $10^{43}$ & $10^{71}$ & $2.6\times10^{29}$ & $2.0\times10^{52}$ \\
\addlinespace[2pt]
 & Horner & 31 & 61 & $1$ & $1$ & $0.125$ & $1.37$ \\
 & Estrin & 31 & 35 & $1$ & $1$ & $0$ & $1.13$ \\
 & Rabin--Winograd & 31 & 40 & $10^{4}$ & $10^{8}$ & $0.377$ & $1.1\times10^{3}$ \\
 & \textbf{this paper} & 31 & 118 & $10^{4849}$ & $10^{5756}$ & overflow & overflow \\
\addlinespace[2pt]
\midrule
prescribed keys & Horner & 7 & 13 & $1$ & $1$ & $0$ & $0$ \\
 & Estrin & 7 & 9 & $1$ & $1$ & $0$ & $0$ \\
 & Rabin--Winograd & 7 & 12 & $5.7$ & $10^{5}$ & $0$ & $0$ \\
 & Motzkin--Eve & 7 & 27 & $35$ & $10^{4}$ & $1.5\times10^{3}$ & $1.8\times10^{5}$ \\
 & \textbf{this paper} & 7 & 25 & $2.3$ & $8.4$ & $0$ & $0$ \\
\addlinespace[2pt]
 & Horner & 15 & 29 & $1$ & $1$ & $0.0236$ & $1.09$ \\
 & Estrin & 15 & 18 & $1$ & $1$ & $0$ & $0.557$ \\
 & Rabin--Winograd & 15 & 22 & $10^{4}$ & $10^{8}$ & $738$ & $3.0\times10^{7}$ \\
 & Motzkin--Eve & 15 & 59 & $10^{6}$ & $10^{8}$ & $4.5\times10^{6}$ & $1.8\times10^{9}$ \\
 & \textbf{this paper} & 15 & 54 & $28$ & $292$ & $0$ & $1.4$ \\
\addlinespace[2pt]
 & Horner & 31 & 61 & $1$ & $1$ & $0.083$ & $2.23$ \\
 & Estrin & 31 & 35 & $1$ & $1$ & $0.188$ & $2.05$ \\
 & Rabin--Winograd & 31 & 40 & $10^{27}$ & $10^{47}$ & $6.4\times10^{25}$ & $1.1\times10^{37}$ \\
 & \textbf{this paper} & 31 & 118 & $10^{3}$ & $10^{4}$ & $2.9\times10^{-3}$ & $3.07$ \\
\addlinespace[2pt]
 & Horner & 63 & 125 & $1$ & $1$ & $0.137$ & $4.63$ \\
 & Estrin & 63 & 68 & $1$ & $1$ & $0.121$ & $4.63$ \\
 & Rabin--Winograd & 63 & 74 & $10^{183}$ & $10^{297}$ & $2.2\times10^{115}$ & $5.4\times10^{205}$ \\
 & \textbf{this paper} & 63 & 266 & $10^{7}$ & $10^{10}$ & $9.3\times10^{-5}$ & $3.47$ \\
\addlinespace[2pt]
\bottomrule
\end{tabular}
\caption{Rounding depth $\rho$, schedule amplification $A$, and observed double-precision forward error of the evaluation schemes on random monic polynomials, in two regimes. Errors (err.) are the observed double-precision forward error in units of $u\sum_i|a_i||x|^i$; a coefficientwise backward-stable scheme reports at most about $\rho$. Medians and maxima over 12 polynomials and 3 points each; Motzkin--Eve rows are omitted where its numeric preprocessing failed verification.}
\label{tab:numstab}
\end{table}

Three features stand out.  First, Horner and Estrin behave as the theory
predicts, with errors of at most a few units.  Second, Rabin--Winograd sits in
between when the coefficients are prescribed, its moderate amplification
matching the growth of its integer constants; but those constants are built
from the coefficients, so in the prescribed-keys regime---where the induced
coefficients have denominators up to $16^n$---it loses significance at
$n=31$ and its errors reach $5.4\times10^{205}$ units at $n=63$, while our schedules, whose constants are the
keys, stay within four units.  Third, in the prescribed-coefficients regime the adapted schemes
lose all significance at moderate degrees---our decoder's exact rational keys
are the most extreme case, with every double-precision evaluation overflowing
by $n=31$---whereas in the prescribed-keys regime the same schedules evaluate
the induced polynomial to a small multiple of $u\sum_i\abs{a_i}\abs{x}^i$ at
the degrees shown.  At $n=7$ the keys of most of our sampled prescribed-coefficient
instances happened to be dyadic rationals of at most 53 bits, so the median
error is zero despite the large latent amplification; the maximum,
$3.6\times10^{14}$ units, comes from instances where they were not, and
exactness is not guaranteed even at that degree.

\subsection{Discussion}

The multiplication count $\floor{n/2}+1$ is achieved by an exact change of
coordinates on the coefficient space, and floating-point evaluation of
arbitrary prescribed coefficients is not among its intended uses; there the
methods of choice remain Horner, Estrin, and their compensated
variants~\cite{higham2002accuracy,yu2025compensated}.  The same phenomenon,
to varying degrees, affects the adapted-coefficient schemes since
Motzkin~\cite{motzkin1955evaluation} and Eve~\cite{eve1964evaluation}, and it
is why correctly rounded libraries treat the evaluation scheme as part of the
polynomial generation~\cite{dinechin2007fast,aanjaneya2023fast}.  Conversely,
whenever the keys are the data, the schedules are accurate in floating point
at hashing-relevant degrees; over finite fields both hashing routes of
\Cref{sec:model} are exact, and the floating-point question arises only over
$\R$, where the natural route is to sample the keys directly.

\paragraph{What helps, and what does not.}
We measured the natural modifications of the construction.  The amplification
is carried by the gadget lanes that square: the $n\equiv3\pmod 4$ pairs and
the $n=15,27,31$ specials, versus the $4k+1$ family
($\log_{10}A$ medians $85$ and $1201$ at $n=19$ and $23$ against $1.5$, $4$,
and $14$ at $n=9$, $13$, and $21$).  Routing $n\equiv3\pmod4$ as
$P=x^2P_{n-2}+a_1x+a_0$ with $P_{n-2}$ in the $4k+1$ family costs one extra
multiplication ($\floor{n/2}+2$, still $\log_2(n+1)-3$ below Rabin--Winograd)
and lowers $\log_{10}A$ from $156$ to $28$ at $n=19$ and from $2131$ to $40$
at $n=23$; double-precision results are then exact at $n=11$ and within
$\sim10^2$ units at $n=15$, but no variant is usable in double precision much
beyond $n\approx15$.  A preprocessing-time guard---the majorant evaluated once
in log scale at the largest $\abs{x}$ of the domain, falling back to Horner
when $\rho uA$ exceeds the tolerance---costs no runtime multiplication and is
the practical safeguard.  Rescaling the argument by a power of two is exact (no rounding is
introduced) and is the most effective single modification, at the cost of two
multiplications by constants, $x\cdot s^{-1}$ and $s^n\cdot(\cdot)$---one if
the latter is merged with the leading-coefficient scaling of a non-monic
input; they cannot be absorbed into the chain, whose wire coefficients must
stay integers: decoding $Q(y)=s^{-n}P(sy)$ and evaluating $s^nQ(x/s)$ with
$s=4$ takes the keys at $n=15$ from $10^{44.7}$ to $10^{0.2}$ and $A$ from
$10^{64.5}$ to $10^{6.5}$.  It cannot be pushed
further, and the reason is structural: an automorphism is nearly linear near
its fixed point, and scaling moves every input toward $x^n$, but our chains
with all keys zero do not compute $x^n$ (the zero-key skeleton of $P_7$ is
$x^7+2x^6+2x^5+2x^4+x^3+x^2$), so the structural tree monomials of degree $j$
are amplified by $s^{\,n-j}$ once the keys are small, and $A$ grows again
like $s^n$.  One might hope to remove the floor by \emph{centering} the
construction so that its zero-key chain is a pure product tree; the following
shows this is impossible at any saving over Horner.

\begin{proposition}[Centering costs all the savings]
\label{prop:centering}
Let a chain with $M$ multiplications have the property that every wire is a
monomial in $x$ when all keys are zero.  Then the Jacobian of its coefficient
map at the zero key has rank at most $M+1$.  In particular a polynomial
automorphism with this property has $M\ge n-1$.
\end{proposition}

\begin{proof}
A key $\alpha$ added to a factor of degree $m$ (all keys zero) contributes to
first order $\alpha\cdot c\,x^{n-m}$, where $c$ counts the paths from that
gate to the output, because along every path the remaining factors are
monomials of total degree $n-m$.  A factor is $x$ or one of the $M-1$
non-output gates, so at most $M$ distinct degrees occur, plus the output key at
degree $0$: the linear part spans at most $M+1$ coefficient rows.  An
automorphism has an invertible Jacobian everywhere, so $n\le M+1$.
\end{proof}

So below $n-1$ multiplications the reference point $x^n$ is necessarily
computed with cancellation, the scaling has an optimum ($s\approx4$ at
$n=15$), and the design quantity that remains is the \emph{degree content} of
the skeleton's cancellations---the lower the degrees of the structural tree
monomials, the earlier the $s^{\,n-j}$ growth bites---together with the size
of the keys of $x^n$ itself (in the $4k+1$ family $x^n$ decodes to keys of size
$10^{3.4}$ at $n=13$, a pure normalization artifact of the monic $T_2$ that a
re-normalized gadget would remove).  Dyadic regauging of the wires, by
contrast, is provably a bit-identical no-op; hybrids that use our gadgets only in
low-degree blocks of a balanced Rabin--Winograd tree cannot save
multiplications, because the paper's advantage over Rabin--Winograd is exactly
the $\log_2(n+1)-2$ shared power-chain products of the global construction;
and compensated (double-double) evaluation helps only if the keys themselves
are stored to the higher precision, at $20$--$200\times$ the cost of Horner.
A random search over $2.6\cdot10^5$ four-multiplication septic circuit
structures (about $10^4$ of them monic of degree seven with a full-rank
coefficient map, $37$ decodable by a causal table) separates the two paradigms
without exception.  Thirty-four of the decodable circuits are polynomial
automorphisms---constant pivots, as in this paper---and every one of them is
unstable, the best with median $A=10^{6.6}$ and maximum $10^{17.5}$ on
coefficients in $[-5,5]$; all $4096$ sign variants of $P_7$ are automorphisms
with $A\approx10^{11}$.  The three remaining circuits are dominant but not
surjective (one rational pivot), and they are exactly the stable ones, with
medians $10^{0.6}$, $10^{1.4}$, and $10^{2.1}$.  The best is
\begin{align}
  g_0&=(a_0-x)\,x, & g_1&=(x-g_0+a_1)(x+g_0+a_2),\\
  g_2&=(g_0+a_3)(x+g_1+a_4), & g_3&=x\,(x+g_0-g_1+g_2+a_5),
\end{align}
$P=g_0-g_2+g_3+a_6$: four constant pivots, then a division by a cubic
$\pi(a_4,a_5,a_6)$ in the top coefficients; polynomials with $\pi=0$ are not
in its image, and none of the $2.2\cdot10^4$ sign and key-placement variants
of this topology---all of them equally stable---is an automorphism.  Such a
scheme evaluates every polynomial only after a shift $x\mapsto x+t$ chosen so
that $\pi\ne0$, which is precisely the generic preconditioning of Motzkin,
Eve, and Knuth; it is not an automorphism of the coefficient space, and it is
the automorphism property that this paper is about.  Within that paradigm the
evidence is uniform: stability was never observed, and
\Cref{prop:unit-pivot-unstable}\ref{it:pivot-sup} shows that a polynomial
automorphism with a descending triangular decoder has unbounded
amplification as soon as one of its rows carries a nonzero correction, so
the majorant bound cannot certify coefficientwise backward stability for
any such decoder.

\paragraph{Open problem.}
Mesztenyi and Witzgall \cite{mesztenyi1967stable} observed in 1967 that
schemes with fewer operations lose significance; Rice
\cite{rice1965conditioning} introduced condition numbers of polynomial forms,
and Miller \cite{miller1975computational} the programme of deriving
complexity lower bounds from stability requirements.  We know of no stability lower bound
for preconditioned polynomial evaluation.  Is there a scheme with
$(1-\epsilon)n$ multiplications and rational preprocessing whose amplification
is $n^{O(1)}$ for all real monic $P$ with $\abs{a_i}\le1$ and $\abs{x}\le1$?
We conjecture not: that $A=n^{O(1)}$ forces $n-O(\log n)$ multiplications,
and that $\floor{n/2}+O(1)$ multiplications force $A\ge\exp(\Omega(n))$.
For polynomial automorphisms the question may be more tractable: we conjecture
that every automorphism with $\floor{n/2}+O(1)$ multiplications has a decoder
of degree exponential in $n$, and already at $n=7$ we know of no stable one.
The measured growth of the Rabin--Winograd amplification up to $n=63$ does
not yet distinguish polynomial from exponential, so even the tradeoff form is
open.

\section{Adversarial inputs for heuristic hashes}
\label{app:adversarial}

This appendix records the inputs behind the heuristic rows of
\Cref{tab:injective:adversarial}: for each hash, the message pair we used, the
measured collision rate, and the number of random secrets it was measured over.
All code is in \texttt{tools/bench/adversarial/} of the accompanying repository
(self-contained C++17; every competitor hash is re-implemented from its
published algorithm and checked against published test vectors, e.g.\
\texttt{XXH3} against the official sanity vectors for lengths
$12, 24, 48, 80, 195$ with seed $0$ and seed \texttt{PRIME64}).  Measurements
were taken on an Apple M2 Pro with Apple clang~17 at \texttt{-O3}.

\paragraph{Model and methodology.}
As in \Cref{sec:injective:experiments}, the secret (seed or secret array) is
drawn uniformly at random per trial and hidden; the attacker chooses a
\emph{fixed} pair of messages; the reported quantity is the fraction of secrets
for which the pair collides.  A random function gives $2^{-64}$.  Attacks that
need a hash's \emph{public} default constants (the shipped \texttt{wyhash} and
\texttt{rapidhash} secret arrays, \texttt{XXH3}'s \texttt{kSecret} with seed
$0$) do not count in this model and are listed separately at the end.  A
message is a byte string whose length is a multiple of $16$, read as
little-endian 64-bit words $w_0,w_1,\ldots$; $M$ denotes the all-ones word, and
``complementing'' a word means XORing it with $M$.  Heuristic hashes were run
at $2^{29}$ trials per cell in the site scan and $2^{31}$ trials per cell in the
length scan; the proven baselines at $2^{27}$ and $2^{29}$ respectively.

\paragraph{The multiply--fold differential.}
All four heuristic hashes share a $64\times64\to128$-bit multiply whose two
halves are folded together, $\mathrm{lo}_{64}(uQ)\oplus\mathrm{hi}_{64}(uQ)$
(\texttt{wyhash}, \texttt{rapidhash}, \texttt{XXH3}) or
$\mathrm{lo}_{64}(uQ)+\mathrm{hi}_{64}(uQ)$ (\texttt{MUM}).  The operands
$u,Q$ are a message word combined with a secret or state word, so an attacker
controls only the difference $(\delta,\varepsilon)$ that the two messages impose
on $(u,Q)$, while $(u,Q)$ themselves are uniform.  The relevant quantity is
therefore $P(\delta,\varepsilon)=\Pr_{u,Q}[\mathrm{fold}(u,Q)=
\mathrm{fold}(u\oplus\delta,Q\oplus\varepsilon)]$, which
\texttt{fold\_search.cpp} evaluates exhaustively for $8$- and $16$-bit operands
and by $2^{34}$ samples at $24$--$64$ bits.  For the XOR fold the best
differences are $(M,M)$ and $(M,0)$: at $64$ bits they collide with probability
$2^{-26.7}$ and $2^{-26.8}$, and the exponent scales as $P\approx2^{-0.42n}$ in
the operand width $n$ rather than the ideal $2^{-n}$.  For the additive fold of
\texttt{MUM} the difference $(M,M)$ collides with probability $0.667$ and the
additive difference $(0x55\cdots5,0)$ with probability $0.185$, both
\emph{independent of the operand width}.  Complementing an operand is the single
most powerful difference we found.

\paragraph{\texttt{wyhash} and \texttt{rapidhash} (random secret).}
Versions: \texttt{wyhash} final v4.3 (with \texttt{WYHASH\_\allowbreak CONDOM=1}) and
\texttt{rapidhash} v1.0, both with a uniformly random secret array in place of
the shipped one.  The first step of both hashes is
$\mathrm{mix}(w_0\oplus\mathrm{secret}, w_1\oplus\mathrm{state})$, so
complementing $w_0$ and $w_1$ imposes the $(M,M)$ difference on the first fold.
Measured over $2^{31}$ random secrets per length, the pair collides with
probability $2^{-26.6}$, $2^{-26.7}$, $2^{-26.2}$, $2^{-26.6}$, $2^{-26.4}$
(\texttt{wyhash}) and $2^{-26.4}$, $2^{-26.5}$, $2^{-26.5}$, $2^{-26.2}$,
$2^{-26.8}$ (\texttt{rapidhash}) at $32$, $48$, $64$, $96$ and $160$ bytes: flat
in the message length and unaffected by the random secret, as the
fold analysis predicts.  Complementing a single interior word instead
propagates through two folds and drops to the noise floor of these trial counts.

\paragraph{\texttt{XXH3} (random seed).}
\texttt{XXH3\_64bits\_withSeed} on the $9$--$240$-byte paths, seed drawn at
random per trial.  Complementing $w_0$ and $w_1$ (the operands of the first
\texttt{mix16B}) collides with probability $2^{-26.3}$, $2^{-25.8}$,
$2^{-26.1}$, $2^{-26.2}$, $2^{-26.2}$ at $32$, $48$, $64$, $96$, $160$ bytes
($2^{31}$ seeds per length).  Single-word complements of $w_1$, $w_2$, $w_3$ in
a $32$-byte message give $2^{-26.4}$, $2^{-27.0}$, $2^{-28.0}$ ($2^{29}$
seeds; these three rest on only $6$, $4$ and $2$ observed collisions, so they
locate the rate only to within a factor of about two).  A re-test against the
newest upstream release (v0.8.3, byte-identical to the packaged header) and the
development branch, with $2^{30}$ seeds per cell, reproduced these rates and
sharpened the picture: for the $64$-bit variant the effect lives entirely in
the multiply--fold, so complementing $w_0$ alone, or both words with an
unrelated $w_1$, collides at comparable rates, and the rate then depends on the
rest of the message; one $128$-byte pair reached $2^{-23.3}$ ($97$ and $109$
collisions in two independent runs of $2^{30}$ seeds), which is the entry in
\Cref{tab:injective:adversarial}.  A uniformly random $192$-byte secret in
place of the seed changes nothing.  We reported the differential upstream as
xxHash issue~\#1127.\footnote{\url{https://github.com/Cyan4973/xxHash/issues/1127}}

\paragraph{\texttt{XXH3-128}.}
The $128$-bit variant does not escape the differential.  On its $17$--$128$-byte
path each of the two accumulators adds the multiply--fold of one $16$-byte
chunk and XORs in the \emph{raw word sum} of the other chunk.  Choosing
$w_1=\lnot w_0$ makes that sum invariant under complementing both words, so
one accumulator is unchanged deterministically and the other collides exactly
when the fold does.  Measured over $2^{29}$ random seeds, the full $128$-bit
output collides with probability $2^{-27.0}$, $2^{-27.0}$, $2^{-26.2}$,
$2^{-26.0}$, $2^{-26.4}$ and $2^{-26.0}$ at $32$, $48$, $64$, $100$, $128$ and
$160$ bytes, every collision being a joint collision of both halves;
complementing $w_0$ alone, or both words with an unrelated $w_1$, gave none.
Doubling the output width therefore buys nothing against this input pair, and
\texttt{XXH3-128} scores $28.5$ bits in \Cref{tab:injective:adversarial}, barely above the $27.3$ of \texttt{XXH3}.

\paragraph{\texttt{komihash} (random seed).}
\texttt{komihash} v5.34 (checked against its $63$ published test vectors)
resisted every differential we tried: the $(M,M)$ and $(M,0)$ complements on
the fold operands of its minimal-diffusion path for $8$--$15$-byte messages,
single- and double-word complements in every $16$-byte block position,
length-boundary and zero-padding pairs, and the structural pairs used against
\texttt{MUM}.  Every surface gave $0$ collisions in $2^{32}$ random seeds,
consistent with the $2^{-64}$ baseline.  We therefore list it as ``$\ge32$'' in
\Cref{tab:injective:adversarial}: the search shows only that no pair we found
exceeds $2^{-32}$, and is not a proof.

\paragraph{\texttt{MUM} v3.}
Upstream \texttt{mum.h} (re-implemented in \texttt{hashes.h}) with its default macros, in both the unroll-8 (x86-64) and
unroll-16 (aarch64) configurations.  The additive fold satisfies
$\_\mathrm{mum}(v,p)+\_\mathrm{mum}(\lnot v,p)\in\{2^{64}-1,\,2^{64}-2\}$ (mod~$2^{64}$)
for every $v$ (the second value occurs exactly when the low halves of $vp$ and
$(\lnot v)p$ carry, i.e.\ for a $p/2^{64}$ fraction of the $v$---between $0.15$
and $0.87$ for \texttt{MUM}'s primes; \texttt{selftest.cpp} checks the analogous
identity for complementing both operands), so complementing an input
word toggles that word's contribution to the XOR accumulator by
approximately~$M$.  Complementing any single interior word of a $32$-byte
message therefore collides with probability $2^{-1}$ (the finalizer maps the
flipped and unflipped states together about half the time), and complementing
\emph{two adjacent} interior words $w_1,w_2$ cancels the two toggles and
collides for \emph{every} seed: $2^{-0.0}$ over $2^{29}$ seeds, i.e.\ a
key-free collision.  The shortest key-free pair we know is a single $8$-byte
word: $w=\texttt{0xb0899b7198a95479}$ and $w'=\texttt{0x08eb9f259b16692d}$
satisfy $\_\mathrm{mum}(w,p_0)=\_\mathrm{mum}(w',p_0)$ (found by a Brent
$\rho$ search on $v\mapsto\_\mathrm{mum}(v,p_0)$ and checked over $2^{20}$
seeds), which is what places \texttt{MUM} at $0$ bits in
\Cref{tab:injective:adversarial}.  Two further seed-independent constructions were found: a
pair of values $w_1$ that collide under $\_\mathrm{mum}(\cdot,p_1)$ (a
$\rho$-collision located after $2^{37}$ evaluations), and, at $160$ bytes, a
rotation clique $w_0=p_0\oplus2^k$, $w_1=\mathrm{rotr}(Q,k)\oplus p_1$ of
$K=64$ messages that are pairwise colliding for all $1020$ seeds tried.

\paragraph{The paper's own fold variant.}
For completeness, the recurrence of \Cref{sec:injective} with the field product
replaced by the MUM fold (the throughput data point of
\Cref{sec:injective:experiments}) is subject to the same differential:
complementing the last $b$-word imposes $(M,0)$ on the final fold and collides
with probability $2^{-27.5}$, $2^{-26.7}$, $2^{-27.1}$, $2^{-27.3}$,
$2^{-27.1}$ at $32$--$160$ bytes ($2^{31}$ keys).  This is the variant the text
explicitly excludes from the Schwartz--Zippel guarantee.

\paragraph{Shipped defaults (excluded from the model).}
With the \emph{published} constants, forced collisions exist: placing
$\mathrm{secret}[1]$ at byte offset $\mathrm{len}-16$ annihilates the final
multiplication of \texttt{wyhash} and \texttt{rapidhash} (rate $1.000$ over
$2^{20}$ seeds, at $32$ and $160$ bytes); $w_0=\mathrm{secret}[1]$
(\texttt{wyhash}) or $w_0=\mathrm{secret}[2]$ (\texttt{rapidhash}) zeroes the
block state so that messages differing only in $w_1$ collide; and for
unseeded \texttt{XXH3}, $w_0=\texttt{kSecret}[0..8)$ zeroes
\texttt{mix16B} of the first chunk.  Each of these yields a set of $K=256$
messages that collide pairwise for all $1020$ seeds tried.  A random secret
removes all of them (rate $0$ in the same experiments), which is why they are
excluded from \Cref{tab:injective:adversarial}; \texttt{MUM}'s constructions
have no secret to randomize and remain.

\paragraph{Proven baselines and controls.}
On every differential site, length, annihilation attempt and blocking set
above, the field version of our recurrence over $\F_{2^{64}}$, its Mersenne
variant, and Dietzfelbinger's vector multiply-shift produced $0$ collisions
($2^{24}$--$2^{29}$ trials per cell, and $1020$ seeds for the blocking sets), consistent with their $d/|\F|$ and
$2^{-64}$ bounds.  On two \emph{uniformly random} messages every hash,
heuristic or proven, produced $0$ collisions at $32$ and $160$ bytes
($2^{30}$ trials for the heuristic hashes, $2^{28}$ for the proven ones): the
elevated rates above are entirely input-specific.

\paragraph{A seeded-differential test for SMHasher3.}
SMHasher3 runs every hash under fixed seeds, so the input-specific rates above
are invisible to it: they are probabilities over the \emph{seed} for a fixed
pair of inputs.  We added a test (\texttt{SeedDifferential}, in the paper's
fork of SMHasher3) that draws $2^{24}$ uniformly random seeds ($2^{30}$ in an
extended tier) and counts full-output collisions of fixed structured pairs
derived from a base message whose second word is the complement of its first:
complementing the first two words, single words, or adjacent interior pairs,
at lengths from $16$ bytes to $1$\,KB.  A row fails when its count exceeds
$\max(3,\,E+6\sqrt{E})$ for the expected count $E=N2^{-w}$.  In the extended
tier \texttt{wyhash}, \texttt{rapidhash}, \texttt{XXH3} and \texttt{XXH3-128}
fail at about $2^{-26}$ on the complemented leading pair and every
\texttt{MUM} variant fails on the adjacent pairs at rate $1$, while
\texttt{komihash}, Polymur, SipHash, \texttt{t1ha2}, FarmHash, CityHash, the
Mersenne polynomial hash and ChainHash show no collision in $2^{30}$ seeds.

\paragraph{Caveat.}
The rates reported here are for the worst inputs \emph{we found}, by a
differential search restricted to word complements and additive differences on
the fold operands, together with one-line annihilation constructions.  They are
lower bounds on the true worst-case collision probability of each hash; inputs
with a higher rate may exist.  Each $2^{-26}$--$2^{-27}$ entry rests on roughly
$20$--$35$ observed collisions in $2^{31}$ trials, so its exponent carries a
statistical uncertainty of about $\pm0.3$; the differences between lengths and
between the three XOR-fold hashes are within that noise.  What the experiments
establish is a lower bound of order $2^{-27}$ (\texttt{wyhash},
\texttt{rapidhash}, \texttt{XXH3}) and $1$ (\texttt{MUM}) on the worst case,
against a proven upper bound of $n/2^{64}$ for the three-key field version of
our hash on $n$ word pairs ($(3n-1)/2^{64}$ with a single key word).

\paragraph{Truncated Mersenne outputs.}
A polynomial hash over $\F_p$ with $p=2^{89}-1$ must not be shortened by
keeping the low $64$ bits of the residue.  Two messages that differ only in
the final additive word, by $2^{64}-1$, have residues that differ by exactly
that constant; whenever the sum wraps past $p$, the wrap subtracts
$p\equiv-1\pmod{2^{64}}$ and the low $64$ bits coincide.  This happens
whenever the residue lies in the top $2^{64}$ values of $[0,p)$, so the pair
so, over a random key, the pair collides with probability about $2^{64}/p=2^{-25}$, whatever the remaining words of the message (we measured
$4$ collisions in $2^{26}$ random keys).  The same mechanism gives
$2^{32}/(2^{61}-1)=2^{-29}$ for the low $32$ bits of a residue modulo
$2^{61}-1$ (measured $1$ in $2^{30}$).  The Mersenne rows of
\Cref{tab:injective:adversarial} therefore output the full residue; a
shorter output needs a universal post-hash, for instance one multiply-shift
step or a degree-$k$ chain as in \Cref{sec:main-theorem}.

\providecommand{\F}{\mathbb{F}}
\providecommand{\clnh}{\mathrm{CLNH}}
\providecommand{\lo}{\mathrm{lo}}
\providecommand{\hi}{\mathrm{hi}}
\providecommand{\iver}[1]{[\![#1]\!]}
\providecommand{\cf}[2]{[#1]\,#2}   

\section{Collision probability of \textsc{ChainHash}}
\label{sec:ph}

This section defines the hash family \textsc{ChainHash} implemented in
\texttt{tools/bench/\allowbreak chainhash/\allowbreak chainhash.h} (fast, NEON \texttt{PMULL}) and
\texttt{tools/bench/\allowbreak chainhash/\allowbreak chainhash\_ref.h} (bit-serial reference) and proves
its collision bound.  The family is a three-level composition: a carry-less
NH (CLNH) block hash \cite{lemire2015clhash} applied to sub-blocks of $B/S$
bytes, the last block being processed at $16$-byte pair granularity; the
injective recurrence of \Cref{sec:injective} in its three-variable form, with
$u,y,z$ three independent uniform keys; and a monic degree-$5$ finalizer
evaluated by a three-multiplication circuit whose coefficient map is a
bijection in characteristic~$2$ (\Cref{lem:ph:chain}), applied to the
integer-add twist of the level-2 value by one further key word
(\Cref{sec:ph:twist}).  Every random key
element is drawn uniformly from the field and used directly; there is no
preprocessing step and no heuristic mixing step.

\subsection{Field, notation, and the hash family}
\label{sec:ph:def}

\paragraph{Field.}
Let $\mathcal R=\F_2[X]$ and, for $d\ge1$, let $\mathcal R_{<d}$ denote the
polynomials of degree $<d$, identified with $d$-bit strings (bit $i$ is the
coefficient of $X^i$).  Addition in $\mathcal R$ is bitwise XOR.  The
\emph{carry-less product} of $a,b\in\mathcal R_{<64}$ is their product
$ab\in\mathcal R_{<127}\subseteq\mathcal R_{<128}$ (the \texttt{PMULL}/\texttt{pclmulqdq}
instruction).  Let
\[
  \Pi(X)=X^{64}+X^4+X^3+X+1 ,
\]
which is irreducible over $\F_2$ \cite[\S4.3]{lemire2015clhash}%
\footnote{Re-verified with Rabin's irreducibility test
(\texttt{tools/bench/chainhash/sanity.py}).  The reduction constant $27=0b11011$ in the
code is $\Pi-X^{64}$.}, and let $\F=\F_2[X]/(\Pi)\cong\mathrm{GF}(2^{64})$,
$|\F|=2^{64}$.  Every class in $\F$ has a unique representative in
$\mathcal R_{<64}$, so $64$-bit words denote both elements of $\mathcal R_{<64}$
and elements of $\F$; addition is XOR in both readings.  Multiplication in
$\F$ is $a\cdot b=(ab)\bmod\Pi$, which is what \texttt{gfmul} computes
(\texttt{gf64\_mult} in \texttt{tools/bench/\allowbreak framework/\allowbreak multiplication\_arm.h}:
$ab \equiv ab_{\lo} + r\,ab_{\hi}$ and $r\,ab_{\hi}\equiv (r\,ab_{\hi})_{\lo} + r\,(r\,ab_{\hi})_{\hi}$
with $r=\Pi-X^{64}$, three carry-less products, fully reduced).
A $128$-bit word $w\in\mathcal R_{<128}$ is split as $w=\lo(w)+X^{64}\hi(w)$
with $\lo(w),\hi(w)\in\mathcal R_{<64}$.

\paragraph{Parameters.}
$W$ words per block (\texttt{BLOCK\_WORDS}), $B=8W$ bytes per block, and a
sub-block split $S\in\{1,2\}$ with $8S\mid W$: each block consists of $S$
sub-blocks of $W_s=W/S$ words ($B_s=B/S$ bytes, $W_s/2$ pairs).  The shipped
configurations are $(W,S)=(32,1)$ ($256$-byte blocks, one pair per block) and
$(W,S)=(128,2)$ ($1$\,KB blocks, two $512$-byte sub-blocks); the $64$-byte row
of \Cref{tab:injective:adversarial} is $(W,S)=(8,1)$.  The finalizer degree is
$5$ throughout.

\paragraph{Key.}
$\kappa=(k_0,\dots,k_{W-1})\in\F^{W}$, $\theta=(u,y,z)\in\F^3$,
$c=(c_0,\dots,c_4)\in\F^5$, and the twist word $\tau\in\F$ (\texttt{t\_in} in
the code), all $W+9$ elements independent and uniform on $\F$ ($41$, $137$
and $17$ words for the three configurations above).  Sub-block
position $i\in\{0,\dots,S-1\}$ within a block uses the key segment
$\kappa^{(i)}=(k_{iW_s},\dots,k_{(i+1)W_s-1})$.

\paragraph{Sub-blocks.}
A message $m$ is a byte string of length $\ell=\ell(m)\ge0$.  Its block count
and sub-block count are
\[
  n(m)=\max\{1,\lceil \ell/B\rceil\},\qquad p(m)=S\,n(m).
\]
A \emph{sub-block} is a tuple $g=(g_0,\dots,g_{2w-1})\in(\mathcal R_{<64})^{2w}$
of $64$-bit words for some $0\le w\le W_s/2$; we call $w=w(g)$ its \emph{pair
count} and $(g_{2s},g_{2s+1})$, $0\le s<w$, its \emph{pairs} ($16$ bytes each).
Sub-block $t$ of $m$, $1\le t\le p(m)$, covers the bytes $[(t-1)B_s,\,tB_s)$
of $m$, of which
\[
  r_t=\min\bigl\{B_s,\ \max\{0,\ \ell-(t-1)B_s\}\bigr\}\in[0,B_s]
\]
exist; it is the tuple $m_t\in(\mathcal R_{<64})^{2w_t}$ of the little-endian
$64$-bit words of those $r_t$ bytes followed by $16w_t-r_t\in[0,16)$ zero
bytes, where
\[
  w_t=\lceil r_t/16\rceil\in\{0,1,\dots,W_s/2\}
\]
is its pair count: only the final, partial pair is padded, and only when
$16\nmid r_t$.  Sub-blocks with $tB_s\le\ell$ are \emph{full} ($w_t=W_s/2$);
sub-blocks beyond the data are \emph{empty} ($w_t=0$, $m_t=()$), which happens
for the empty message and, when $S=2$, for the second half of a last block
that ends in its first half.  No byte beyond position $\ell$ is ever read.
Since $\ell$ determines $n(m)$, $p(m)$ and every $r_t$ and $w_t$, and the
bytes of $m_1,\dots,m_{p(m)}$ concatenated begin with $m$, the tuple
$(m_1,\dots,m_{p(m)},\ell)$ determines $m$; in particular two messages of the
same length have sub-blocks of the same pair counts.  Sub-block $t$ sits at
position $i(t)=(t-1)\bmod S$ of its block and is keyed by $\kappa^{(i(t))}$.

\paragraph{Level 1: CLNH.}
For a key segment $\kappa'=(k'_0,\dots,k'_{W_s-1})\in(\mathcal R_{<64})^{W_s}$ and
a sub-block $g=(g_0,\dots,g_{2w-1})$ with pair count $w\le W_s/2$,
\begin{equation}\label{eq:ph:clnh}
  \clnh_{\kappa'}(g)\;=\;\sum_{s=0}^{w-1}(g_{2s}+k'_{2s})\,(g_{2s+1}+k'_{2s+1})\;\in\mathcal R_{<128},
\end{equation}
the sum being XOR of the $128$-bit carry-less products.  It involves only
the first $2w$ key words $k'_0,\dots,k'_{2w-1}$ and is the empty sum $0$ when
$w=0$.  For a full sub-block ($w=W_s/2$) this is \cite[eq.~(7)]{lemire2015clhash},
computed by \texttt{ph\_block}, whose loop XORs the four products of each
$64$-byte group into two independent $128$-bit accumulators (one three-input
XOR per accumulator and group).  For a partial
sub-block, \texttt{ph\_tail} computes exactly the $w$ products of
\eqref{eq:ph:clnh}: full $64$-byte groups and then whole $16$-byte pairs are
loaded in place, and the partial pair (present iff $16\nmid r_t$, and then
the last $r_t-16(w-1)$ bytes of the message) is loaded in place as well, as
the $16$ bytes ending at byte $\ell$ shifted by a \texttt{TBL} byte
permutation whose out-of-range indices supply the zero padding
(\texttt{load\_partial\_end}; $\ell\ge16$ whenever \texttt{ph\_tail}
runs); a message of fewer than $16$ bytes is its own single padded pair,
assembled from overlapping in-bounds lane loads and one \texttt{TBL} gather
(\texttt{load\_small}).  There is no stack copy and no \texttt{memcpy},
and no byte outside $[0,\ell)$ is ever read; the reference \texttt{chainhash\_ref.h}
assembles the same zero-padded words byte by byte (\texttt{word\_at}).  XOR
being associative and commutative, all of these compute \eqref{eq:ph:clnh}
(test T1 of \texttt{test\_chainhash.cpp} compares the two on every
partial-pair/whole-pair/$64$-byte-group combination of the last block and,
for $S=2$, byte by byte across the sub-block boundary).

\paragraph{Level 1$'$: the stream.}
For $m$ with $p=p(m)$ and sub-blocks $m_1,\dots,m_p$ define the
\emph{stream} $\sigma_\kappa(m)=((a_1,b_1),\dots,(a_p,b_p))\in(\F^2)^p$ by
\begin{equation}\label{eq:ph:stream}
  a_t=\lo\bigl(\clnh_{\kappa^{(i(t))}}(m_t)\bigr)+\iver{t=p}\cdot\ell,\qquad
  b_t=\hi\bigl(\clnh_{\kappa^{(i(t))}}(m_t)\bigr),
\end{equation}
where $\ell$ is read as the $64$-bit word (field element) encoding the byte
length.  So the byte length is XORed into the $a$-component of the last pair
only, and every sub-block, empty or not, contributes one pair.

\paragraph{Level 2: the three-key recurrence.}
With $P_0=z$ and, for $t\ge1$,
\begin{equation}\label{eq:ph:rec}
  P_t=a_t+(b_t+y)(P_{t-1}+u)\qquad\text{in }\F ,
\end{equation}
which is \eqref{eq:injective-recurrence} with the three variables $u,y,z$ of
\Cref{def:injective:recurrence} instantiated by the three key elements
(no specialisation), set $V_\kappa(m)=P_{p(m)}$ (the dependence on $\theta$
is suppressed).  We write $P^{(m)}\in\F[U,Y,Z]$ for the formal polynomial
obtained by running \eqref{eq:ph:rec} with indeterminates $U,Y,Z$ in place of
$u,y,z$; then $V_\kappa(m)=P^{(m)}(u,y,z)$, and the coefficients of $P^{(m)}$
depend on $\kappa$ and $m$ only.

\paragraph{Level 3: finalizer.}
For $c\in\F^5$ let $f_c\in\F[X]$ be the polynomial computed by the circuit
\begin{equation}\label{eq:ph:chain5}
  G_1=X\cdot X,\qquad
  G_2=(G_1+c_0)(X+G_1+c_1),\qquad
  f_c=(X+c_2)(G_2+c_3)+c_4 ,
\end{equation}
evaluated exactly as written at $X=v$: three multiplications,
$\lfloor5/2\rfloor+1$ as in \Cref{thm:main} (\texttt{chain\_v<5>} in
\texttt{chainhash.h}, whose comment names the intermediates \texttt{y},
\texttt{z}, \texttt{t}; the code folds each addend into the reduction of
the product it follows).  It is monic of degree $5$ (\Cref{lem:ph:chain}).
The finalizer is applied not to the level-2 value itself but to its
\emph{integer-add twist} $V_\kappa(m)\boxplus\tau$, where $v\boxplus\tau$
denotes the addition modulo $2^{64}$ of the $64$-bit words $v,\tau$ read as
integers (bit $i$ = coefficient of $X^i$ = binary digit $2^i$), carries and
all.  \Cref{sec:ph:twist} explains why the twist is there; the collision and
independence proofs see it only through \Cref{lem:ph:twist}.

\begin{definition}[\textsc{ChainHash}]\label{def:ph:hash}
$H_{\kappa,\theta,c,\tau}(m)=f_c\bigl(V_\kappa(m)\boxplus\tau\bigr)\in\F$.
\end{definition}

\subsection{Statement}

\begin{theorem}[Collision bound]\label{thm:ph:collision}
Fix $W$, $S$ and $n\ge1$, and put $p=Sn$.  For every two distinct messages
$m\ne m'$ with $n(m),n(m')\le n$ (equivalently: of at most $nB$ bytes each),
\[
  \Pr_{\kappa,\theta,c,\tau}\bigl[H_{\kappa,\theta,c,\tau}(m)=H_{\kappa,\theta,c,\tau}(m')\bigr]\;\le\;\frac{p+2}{2^{64}}=\frac{Sn+2}{2^{64}} .
\]
\end{theorem}

For the $256$-byte configuration this reads $(\lceil\ell/256\rceil+2)/2^{64}$
for messages of at most $\ell$ bytes, and for the $1$\,KB configuration
$(2\lceil\ell/1024\rceil+2)/2^{64}$: $3/2^{64}$ and $4/2^{64}$ for messages of
up to $256$ bytes, $4098/2^{64}\approx2^{-52}$ and $2050/2^{64}\approx2^{-53}$
for messages of up to $1$\,MB.  The proof separates the three levels:
\Cref{lem:ph:stream} (level 1), \Cref{lem:ph:level2} (level 2) and
\Cref{lem:ph:finalizer} (level 3), the twist being absorbed by
\Cref{lem:ph:twist}.

\subsection{Level 1: full-width XOR-universality of CLNH}

Lemire and Kaser prove that the \emph{reduced} family $g\mapsto\clnh_{\kappa'}(g)\bmod\Pi$
is XOR-universal and state that the unreduced $128$-bit family is
$2^{-64}$-almost universal \cite[\S5 and Lemma~5]{lemire2015clhash}.  Since
the stream \eqref{eq:ph:stream} keeps \emph{both} halves of the $128$-bit
value and XORs a constant into one of them, we need the unreduced family to
be XOR-universal on its full width, for every $128$-bit constant.  The proof
is their integral-domain argument carried out in $\mathcal R$ instead of a field.
Because the last block is processed at pair granularity, two last sub-blocks can
have \emph{different} pair counts; part (ii) below covers that case.  There
the key products of the extra pairs do not cancel, and a nonzero constant
$C$ is essential (supplied, in \Cref{lem:ph:stream}, by the length XOR).

\begin{lemma}[CLNH is $2^{-64}$-XOR-universal on $128$ bits]\label{lem:ph:clnh}
Let $\kappa'$ be uniform on $(\mathcal R_{<64})^{W_s}$, let $g\in(\mathcal R_{<64})^{2w}$
and $g'\in(\mathcal R_{<64})^{2w'}$ be sub-blocks with pair counts $0\le w\le w'\le W_s/2$,
and let $C\in\mathcal R_{<128}$.
\begin{enumerate}[label=(\roman*)]
\item If $w=w'$ and $g\ne g'$, then
      $\Pr_{\kappa'}\bigl[\clnh_{\kappa'}(g)+\clnh_{\kappa'}(g')=C\bigr]\le 2^{-64}$.
\item If $w<w'$ and $C\ne0$, then
      $\Pr_{\kappa'}\bigl[\clnh_{\kappa'}(g)+\clnh_{\kappa'}(g')=C\bigr]\le 2^{-64}$.
\end{enumerate}
The hypothesis $C\ne0$ in (ii) cannot be dropped: for $w=0$, $w'=1$ and $C=0$
the event is $\{k'_0=g'_0\}\cup\{k'_1=g'_1\}$, of probability $2^{-63}-2^{-128}$.
\end{lemma}
\begin{proof}
(i) Pick $s<w$ with $(g_{2s},g_{2s+1})\ne(g'_{2s},g'_{2s+1})$.  Expanding
\eqref{eq:ph:clnh} in $\mathcal R$ (both sums contain the same $w$ products
and the characteristic is $2$, so the $k'_{2s'}k'_{2s'+1}$ terms cancel for
every $s'$),
\[
  \clnh_{\kappa'}(g)+\clnh_{\kappa'}(g')
  =(g_{2s}+g'_{2s})\,k'_{2s+1}+(g_{2s+1}+g'_{2s+1})\,k'_{2s}
   +\bigl(g_{2s}g_{2s+1}+g'_{2s}g'_{2s+1}\bigr)+T,
\]
where $T$ collects the summands with index $s'\ne s$ and does not involve
$k'_{2s},k'_{2s+1}$.  Suppose $\delta:=g_{2s}+g'_{2s}\ne0$ (otherwise
$g_{2s+1}+g'_{2s+1}\ne0$ and the same argument applies with the roles of
$k'_{2s}$ and $k'_{2s+1}$ exchanged).  Condition on all key words other than
$k'_{2s+1}$.  The event becomes $\delta\,k'_{2s+1}=C'$ for a $C'\in\mathcal R$
that does not depend on $k'_{2s+1}$.  Since $\mathcal R$ is an integral domain
and $\delta\ne0$, the map $k\mapsto\delta k$ is injective, so at most one of
the $2^{64}$ values of $k'_{2s+1}$ satisfies the equation.  The conditional
probability is at most $2^{-64}$; average over the conditioning.

(ii) For $w\le t\le w'$ let $g'^{(t)}:=(g'_0,\dots,g'_{2t-1})$ be the sub-block
formed by the first $t$ pairs of $g'$ (so $g'^{(w')}=g'$, and $g'^{(w)}$ has the
pair count of $g$), and put
$\Delta_t:=\clnh_{\kappa'}(g)+\clnh_{\kappa'}(g'^{(t)})$ and $p_t:=\Pr_{\kappa'}[\Delta_t=C]$;
by \eqref{eq:ph:clnh}, $\Delta_t$ involves only $k'_0,\dots,k'_{2t-1}$.  We show
$p_t\le2^{-64}$ by induction on $t$; the claim is $p_{w'}\le2^{-64}$.
\emph{Base $t=w$.}  If $g\ne g'^{(w)}$ then $p_w\le2^{-64}$ by (i); if
$g=g'^{(w)}$ (in particular whenever $w=0$) then $\Delta_w=0\ne C$ and $p_w=0$.
\emph{Step $t-1\to t$} ($w<t\le w'$).  By \eqref{eq:ph:clnh},
\[
  \Delta_t=\Delta_{t-1}+(g'_{2t-2}+k'_{2t-2})\,(g'_{2t-1}+k'_{2t-1}),
\]
and $\Delta_{t-1}$ does not involve $k'_{2t-2},k'_{2t-1}$.  Split the event
$\{\Delta_t=C\}$ according to whether $k'_{2t-2}=g'_{2t-2}$.
On $\{k'_{2t-2}\ne g'_{2t-2}\}$ condition on all key words other than $k'_{2t-1}$:
the event reads $\delta\,k'_{2t-1}=C'$ with $\delta=g'_{2t-2}+k'_{2t-2}\ne0$ and
$C'$ not depending on $k'_{2t-1}$, which at most one value of $k'_{2t-1}$
satisfies ($\mathcal R$ is an integral domain); hence
$\Pr[\Delta_t=C\wedge k'_{2t-2}\ne g'_{2t-2}]\le(1-2^{-64})\,2^{-64}$.
On $\{k'_{2t-2}=g'_{2t-2}\}$ the last product vanishes and
$\{\Delta_t=C\}=\{\Delta_{t-1}=C\}$, an event independent of
$\{k'_{2t-2}=g'_{2t-2}\}$ (disjoint key words); hence
$\Pr[\Delta_t=C\wedge k'_{2t-2}=g'_{2t-2}]=2^{-64}\,p_{t-1}\le2^{-128}$.
Adding, $p_t\le2^{-64}-2^{-128}+2^{-128}=2^{-64}$.
\end{proof}

\begin{lemma}[Distinct messages give distinct streams w.h.p.]\label{lem:ph:stream}
For all $m\ne m'$,
$\Pr_\kappa[\sigma_\kappa(m)=\sigma_\kappa(m')]\le2^{-64}$.
Moreover $\sigma_\kappa(m)\ne\sigma_\kappa(m')$ holds for \emph{every}
$\kappa$ when $n(m)\ne n(m')$, and also when $\ell(m)\ne\ell(m')$ but the two
padded last sub-blocks coincide (same pair count and the same words;
e.g.\ $m'=m\,\|\,\texttt{0x00}$ with $16\nmid\ell(m)$, and in particular
whenever all padded sub-blocks of $m$ and $m'$ coincide).
\end{lemma}
\begin{proof}
If $n(m)\ne n(m')$ the streams have different lengths $p(m)\ne p(m')$ and
cannot be equal.  Let $n(m)=n(m')$, so $p(m)=p(m')=:p$, with sub-blocks
$m_1,\dots,m_p$ and $m'_1,\dots,m'_p$, lengths $\ell,\ell'$, and last
sub-block pair counts $w_p=w(m_p)$ and $w'_p=w(m'_p)$; sub-block $t$ of both
messages is keyed by the same segment $\kappa^{(i(t))}$.  By
\eqref{eq:ph:stream}, $\sigma_\kappa(m)=\sigma_\kappa(m')$ iff
\begin{equation}\label{eq:ph:stream-eq}
\begin{aligned}
  &\clnh_{\kappa^{(i(t))}}(m_t)=\clnh_{\kappa^{(i(t))}}(m'_t)\qquad(1\le t<p),\quad\text{and}\\
  &\clnh_{\kappa^{(i(p))}}(m_p)+\clnh_{\kappa^{(i(p))}}(m'_p)=C,\qquad\text{where } C:=(\ell+\ell')+X^{64}\cdot0 ,
\end{aligned}
\end{equation}
i.e.\ $C\in\mathcal R_{<128}$ has low word $\ell+\ell'$ (the XOR of the two
$64$-bit length words) and high word $0$.

\emph{Equal lengths, $\ell=\ell'$.}  Then $C=0$, and $m_t$ and $m'_t$ have
the same pair count for every $t$.  Since $(m_1,\dots,m_p,\ell)$ determines
$m$ and $m\ne m'$, there is a $t$ with $m_t\ne m'_t$; its pair count is at
least $1$, because two sub-blocks of pair count $0$ are both the empty tuple
and could not differ.  Hence
$\{\sigma_\kappa(m)=\sigma_\kappa(m')\}\subseteq\{\clnh_{\kappa^{(i(t))}}(m_t)+\clnh_{\kappa^{(i(t))}}(m'_t)=0\}$,
of probability $\le2^{-64}$ by \Cref{lem:ph:clnh}(i) applied to the uniform
segment $\kappa^{(i(t))}$, i.e.\ by the XOR-universality of CLNH over the
common number of pairs.

\emph{Different lengths, $\ell\ne\ell'$.}  Now $\lo(C)=\ell+\ell'\ne0$, so
$C\ne0$; only the last condition of \eqref{eq:ph:stream-eq} is used.
\begin{enumerate}[label=(\alph*)]
\item $w_p=w'_p$ and $m_p=m'_p$: the padded last sub-blocks coincide (this
      includes the situation where all padded sub-blocks of $m$ and $m'$
      coincide, and the case where both last sub-blocks are empty).  The
      last condition reads $0=C\ne0$, so the streams differ for \emph{every}
      $\kappa$, whatever the earlier sub-blocks are: the length XOR separates
      them deterministically.
\item $w_p=w'_p$ and $m_p\ne m'_p$ (so $w_p\ge1$).  Then
      $\{\sigma_\kappa(m)=\sigma_\kappa(m')\}\subseteq\{\clnh_{\kappa^{(i(p))}}(m_p)+\clnh_{\kappa^{(i(p))}}(m'_p)=C\}$,
      of probability $\le2^{-64}$ by \Cref{lem:ph:clnh}(i).
\item $w_p\ne w'_p$ (possible only for different lengths; one of the two
      counts may be $0$).  Then
      $\{\sigma_\kappa(m)=\sigma_\kappa(m')\}\subseteq\{\clnh_{\kappa^{(i(p))}}(m_p)+\clnh_{\kappa^{(i(p))}}(m'_p)=C\}$,
      of probability $\le2^{-64}$ by \Cref{lem:ph:clnh}(ii), which applies
      because $C\ne0$.
\end{enumerate}
In every case $\Pr_\kappa[\sigma_\kappa(m)=\sigma_\kappa(m')]\le2^{-64}$; the
deterministic claims are the case $n(m)\ne n(m')$ and case (a).
\end{proof}

\subsection{Level 2: injectivity and Schwartz--Zippel}

We use \Cref{lem:injective:coeffmap} in its three-variable form, together with
the total-degree bound recorded at the end of its proof.  We restate both
with an explicit decoder, over an arbitrary field (no assumption on the
characteristic; the decoder divides only by monic linear polynomials), and
add the statement for streams of different lengths.

\begin{lemma}[Injectivity and degree of the recurrence; \Cref{lem:injective:coeffmap}]\label{lem:ph:injective}
Let $\F$ be any field and $n\ge1$.  For $(a_1,b_1,\dots,a_n,b_n)\in\F^{2n}$ let
$P_n\in\F[U,Y,Z]$ be given by \Cref{def:injective:recurrence}
($P_0=Z$, $P_i=a_i+(b_i+Y)(P_{i-1}+U)$).
\begin{enumerate}[label=(\roman*)]
\item $P_n=f_1(Y)+Z\,f_2(Y)+U\,f_3(Y)$ with $f_2,f_3$ monic of degree $n$ and
      $\deg f_1\le n-1$.  Hence $P_n$ has total degree $n+1$, and its
      homogeneous part of degree $n+1$ is $(Z+U)Y^n$, the same for every
      parameter vector.
\item The map $(a_1,b_1,\dots,a_n,b_n)\mapsto P_n$ is injective.
\item If $(a_1,b_1,\dots,a_n,b_n)\ne(a'_1,b'_1,\dots,a'_{n'},b'_{n'})$ as
      finite sequences ($n,n'\ge1$), then $P_n-P'_{n'}$ is a nonzero
      polynomial, of total degree at most $n$ when $n=n'$ and exactly
      $\max(n,n')+1$ when $n\ne n'$.
\end{enumerate}
\end{lemma}
\begin{proof}
\emph{Structure.}  Put $g_i:=\prod_{j=i}^{n}(Y+b_j)$ for $1\le i\le n+1$
(so $g_{n+1}=1$, $g_i$ is monic of degree $n-i+1$).  We claim
\begin{equation}\label{eq:ph:structure}
  P_n=f_1(Y)+Z\,f_2(Y)+U\,f_3(Y),\qquad
  f_1=\sum_{i=1}^{n}a_i\,g_{i+1},\quad f_2=g_1,\quad f_3=\sum_{i=1}^{n}g_i .
\end{equation}
Indeed, if $P_{i-1}=f_1^{(i-1)}+Zf_2^{(i-1)}+Uf_3^{(i-1)}$ then
$P_i=a_i+(Y+b_i)\bigl(f_1^{(i-1)}+Zf_2^{(i-1)}+U(f_3^{(i-1)}+1)\bigr)$, i.e.
\[
  f_1^{(i)}=a_i+(Y+b_i)f_1^{(i-1)},\qquad
  f_2^{(i)}=(Y+b_i)f_2^{(i-1)},\qquad
  f_3^{(i)}=(Y+b_i)\bigl(f_3^{(i-1)}+1\bigr),
\]
with $(f_1^{(0)},f_2^{(0)},f_3^{(0)})=(0,1,0)$; unrolling gives
\eqref{eq:ph:structure}.  Since $P_n$ has degree $\le1$ in each of $Z,U$
and no $ZU$ term, $f_1,f_2,f_3$ are determined by $P_n$.
Here $f_2$ is monic of degree $n$, $f_3$ is monic of degree $n$
(its summand $g_1$ dominates), and $\deg f_1\le n-1$; so the monomials of
total degree $n+1$ in $P_n$ are exactly $ZY^n$ and $UY^n$, each with
coefficient $1$.  This proves (i).

\emph{Decoder for the $b$'s.}  We show how to recover $\beta_1$ from a pair
$(G,S)$ of the shape $G=\prod_{j=1}^{m}(Y+\beta_j)$, $S=\sum_{i=1}^{m}\prod_{j=i}^{m}(Y+\beta_j)$,
$m\ge1$; the pair $(f_2,f_3)$ has this shape with $m=n$ and $\beta_j=b_j$.
Write $h_i=\prod_{j=i}^{m}(Y+\beta_j)$, so $G=h_1$, $S-G=\sum_{i=2}^{m}h_i$,
$\deg h_i=m-i+1$ and every $h_i$ is monic.  Then
\begin{equation}\label{eq:ph:pivot}
  \cf{Y^{m-1}}{G}=\sum_{j=1}^{m}\beta_j,\qquad
  \cf{Y^{m-2}}{(S-G)}=\iver{m\ge2}\sum_{j=2}^{m}\beta_j+\iver{m\ge3},
\end{equation}
because $h_2$ (present iff $m\ge2$) contributes its $Y^{m-2}$ coefficient
$\sum_{j\ge2}\beta_j$, $h_3$ (present iff $m\ge3$) contributes its leading
coefficient $1$, and $h_i$ with $i\ge4$ has degree $<m-2$
(for $m=1$ read $\cf{Y^{-1}}{\cdot}=0$).  Hence the explicit pivot
\begin{equation}\label{eq:ph:beta1}
  \beta_1=\cf{Y^{m-1}}{G}-\cf{Y^{m-2}}{(S-G)}+\iver{m\ge3}.
\end{equation}
Knowing $\beta_1$, synthetic division of $G$ by the monic factor $(Y+\beta_1)$
gives $h_2=G/(Y+\beta_1)$ exactly, and $S':=S-G=\sum_{i=2}^{m}h_i$; the pair
$(h_2,S')$ has the same shape with $m-1$ and $(\beta_2,\dots,\beta_m)$.
Iterating recovers $b_1,\dots,b_n$ and, along the way, all $g_i$.
(For $m\ge3$ the two coefficients in \eqref{eq:ph:pivot} are the ones displayed
in the proof of \Cref{lem:injective:coeffmap}; the indicator $\iver{m\ge3}$
is the parenthetical remark there that the constant $1$ in the $Y^{m-2}$
coefficient is absent when $m=2$.)

\emph{Decoder for the $a$'s.}  In $f_1=\sum_{i=1}^{n}a_ig_{i+1}$ the summand
$a_ig_{i+1}$ has degree $n-i$ with leading coefficient $a_i$, so
$a_1=\cf{Y^{n-1}}{f_1}$; subtract $a_1g_2$ and read $a_2=\cf{Y^{n-2}}{(f_1-a_1g_2)}$,
and so on.  This proves (ii).

(iii) If $n=n'$ the two polynomials are distinct by (ii), and by (i) their
homogeneous parts of degree $n+1$ coincide, so
$P_n-P'_n=(f_1-f'_1)+Z(f_2-f'_2)+U(f_3-f'_3)$ with
$\deg(f_2-f'_2),\deg(f_3-f'_3)\le n-1$ (both monic of degree $n$) and
$\deg(f_1-f'_1)\le n-1$: nonzero of total degree at most $n$.  If $n<n'$,
$P_n$ has total degree $n+1<n'+1$ while the degree-$(n'+1)$ part of $P'_{n'}$
is $(Z+U)Y^{n'}\ne0$, so $P_n-P'_{n'}$ is nonzero of total degree exactly
$n'+1$.
\end{proof}

\begin{lemma}[Schwartz--Zippel in three variables]\label{lem:ph:sz}
Let $D\in\F[U,Y,Z]$ be nonzero of total degree $d$ over a finite field $\F$,
and let $u,y,z$ be independent and uniform on $\F$.  Then
$\Pr[D(u,y,z)=0]\le d/|\F|$.
\end{lemma}
\begin{proof}
By induction on the number of variables; for one variable a nonzero
polynomial of degree $d$ has at most $d$ roots.  Write
$D=\sum_{j}D_j(Y,Z)\,U^j$ and let $j_0$ be the largest $j$ with $D_{j_0}\ne0$,
so $\deg D_{j_0}\le d-j_0$.  By induction $\Pr[D_{j_0}(y,z)=0]\le(d-j_0)/|\F|$;
conditioned on any $(y,z)$ with $D_{j_0}(y,z)\ne0$, $D(U,y,z)$ is a nonzero
univariate polynomial of degree $j_0$, which the independent uniform $u$
hits with probability at most $j_0/|\F|$.  Adding the two bounds gives
$d/|\F|$.
\end{proof}

\begin{lemma}[Level-2 collision]\label{lem:ph:level2}
Let $m\ne m'$ with $n(m),n(m')\le n$, and $p=Sn$.  Then
\[
  \Pr_{\kappa,\theta}\bigl[\sigma_\kappa(m)\ne\sigma_\kappa(m')\ \wedge\ V_\kappa(m)=V_\kappa(m')\bigr]
  \;\le\;\begin{cases}
     p/2^{64} & \text{if } n(m)=n(m'),\\
     (p+1)/2^{64} & \text{if } n(m)\ne n(m').
  \end{cases}
\]
\end{lemma}
\begin{proof}
Condition on any $\kappa$ with $\sigma_\kappa(m)\ne\sigma_\kappa(m')$.  The
streams are the parameter sequences of the formal polynomials $P^{(m)}$ and
$P^{(m')}$, of lengths $p(m)$ and $p(m')$, so by \Cref{lem:ph:injective}(iii)
$D:=P^{(m)}-P^{(m')}$ is a nonzero polynomial of total degree $d\le p(m)\le p$
if $p(m)=p(m')$, i.e.\ $n(m)=n(m')$, and $d=\max(p(m),p(m'))+1\le p+1$
otherwise.  Now $V_\kappa(m)=V_\kappa(m')$ iff $D(u,y,z)=0$, and
$\theta=(u,y,z)$ is uniform on $\F^3$ and independent of $\kappa$, so
\Cref{lem:ph:sz} gives $\Pr_\theta[D(u,y,z)=0\mid\kappa]\le d/2^{64}$.
Average over $\kappa$.
\end{proof}

\subsection{Level 3: the finalizer is a uniformly random monic polynomial}

\Cref{thm:main} does not cover characteristic $2$ (\Cref{rem:char2}), so the
bijectivity of the circuit \eqref{eq:ph:chain5} is proved directly, by an
explicit decoder.  Throughout, $\F$ is a field of characteristic $2$.

\begin{lemma}[The degree-$5$ circuit is a bijective parametrisation]\label{lem:ph:chain}
Let $\F$ have characteristic $2$ and write $f_c(X)=\sum_{i\le5}e_iX^i$ for the
polynomial \eqref{eq:ph:chain5}.  Put
\[
  b=c_0+c_1,\qquad d=c_0c_1 .
\]
Then $e_5=1$ and
\begin{equation}\label{eq:ph:chain5-coeffs}
\begin{aligned}
  e_4&=1+c_2, & e_1&=d+c_3+c_0c_2,\\
  e_3&=b+c_2, & e_0&=c_4+c_2\,(d+c_3),\\
  e_2&=c_0+c_2b.
\end{aligned}
\end{equation}
The coefficient map $\varphi_5:\F^5\to\F^5$, $c\mapsto(e_0,\dots,e_4)$, is a
bijection, for every field $\F$ of characteristic $2$ (no perfect-field
hypothesis is needed; in particular for every finite field
$\mathrm{GF}(2^k)$).  Consequently, if $c$ is uniform on $\mathrm{GF}(2^k)^5$
then $f_c$ is a uniformly random monic polynomial of degree $5$.
\end{lemma}
\begin{proof}
\emph{Expansion.}  Over any commutative ring, $G_1=X^2$ and
\[
  G_2=(X^2+c_0)(X^2+X+c_1)=X^4+X^3+bX^2+c_0X+d ,
\]
whence $G_2+c_3=X^4+X^3+bX^2+c_0X+(d+c_3)$.  Multiplying by $X+c_2$ and
adding $c_4$ gives $e_5=1$ and \eqref{eq:ph:chain5-coeffs}: each $e_i$ for
$i\le4$ is the coefficient of $X^{i-1}$ of that quartic plus $c_2$ times its
coefficient of $X^i$.

\emph{Pivot coordinates.}  From here on characteristic $2$ is used, so subtraction is written as addition.  The change of parameters
\begin{equation}\label{eq:ph:chain5-q}
  q=(q_0,\dots,q_4):=(c_2,\ b,\ c_0,\ c_3,\ c_4)
  =(c_2,\ c_0+c_1,\ c_0,\ c_3,\ c_4)
\end{equation}
is a bijection of $\F^5$, with inverse
\begin{equation}\label{eq:ph:chain5-cq}
  c=(q_2,\ q_1+q_2,\ q_0,\ q_3,\ q_4):
\end{equation}
indeed $c_0=q_2$, then $c_1=b+c_0=q_1+q_2$, and the remaining entries are
copied.  Substituting \eqref{eq:ph:chain5-cq} into
\eqref{eq:ph:chain5-coeffs}, with
$d=c_0c_1=q_2\,(q_1+q_2)=:\delta$ (a polynomial in $q_1,q_2$), the rows
read, from $X^4$ down to $X^0$,
\begin{equation}\label{eq:ph:chain5-rows}
\begin{aligned}
  e_4&=q_0+1,\\
  e_3&=q_1+q_0,\\
  e_2&=q_2+q_0q_1,\\
  e_1&=q_3+\delta+q_0q_2,\\
  e_0&=q_4+q_0\,(\delta+q_3).
\end{aligned}
\end{equation}
Row $i$ (the coefficient $e_{4-i}$) has the form $q_i+K_i(q_0,\dots,q_{i-1})$,
where $K_i$ is a polynomial in the earlier coordinates only
($K_0=1$, $K_1=q_0$, $K_2=q_0q_1$, $K_3=\delta+q_0q_2$,
$K_4=q_0(\delta+q_3)$): the system is unitriangular with every pivot the
identity.  (The degree-$7$ circuit that this replaces had two Frobenius
pivots $q_i\mapsto q_i^2$ and therefore needed $\F$ perfect; here no root is
ever taken.)

\emph{Decoder.}  Given $(e_0,\dots,e_4)$, recover $q$ top-down, each line
using only $e$ and the coordinates already recovered:
\begin{equation}\label{eq:ph:chain5-decoder}
\begin{aligned}
  q_0&=e_4+1,\qquad q_1=e_3+q_0,\qquad q_2=e_2+q_0q_1,\\
  q_3&=e_1+\delta+q_0q_2\qquad\text{with }\delta=q_2\,(q_1+q_2),\\
  q_4&=e_0+q_0\,(\delta+q_3),
\end{aligned}
\end{equation}
then $c$ from \eqref{eq:ph:chain5-cq}.  Line by line,
\eqref{eq:ph:chain5-decoder} composed with \eqref{eq:ph:chain5-rows} is the
identity in both directions (substituting the rows into the decoder gives
$q_0,\dots,q_4$ back; substituting the decoder into the rows gives
$e_4,\dots,e_0$ back), using only the ring operations of $\F$.  Hence
$\varphi_5$ is a bijection over every field of characteristic $2$, and a
bijection of $\F^5$ maps the uniform distribution to itself.
\end{proof}

\begin{remark}[Sanity checks, and the relation to \Cref{thm:main}]\label{rem:ph:thm-main}
The decoder \eqref{eq:ph:chain5-decoder} \emph{is} the proof; the following
checks only guard against transcription errors.  The expansion
\eqref{eq:ph:chain5-coeffs}, the coordinate change
\eqref{eq:ph:chain5-q}--\eqref{eq:ph:chain5-cq}, the row structure
\eqref{eq:ph:chain5-rows} (unit pivots, no later coordinate in any row) and
both compositions decoder$\circ$rows and rows$\circ$decoder were verified
symbolically (the expansion over $\F_2[c_0,\dots,c_4][X]$, the compositions over $\F_2[q_0,\dots,q_4]$ and $\F_2[e_0,\dots,e_4]$; \texttt{tools/bench/chainhash/verify5.py});
test T5 of \texttt{test\_chainhash.cpp} expands the transcribed circuit over
$\mathrm{GF}(2^{64})[X]$, checks that it is monic of degree $5$, and checks
decode$\circ$encode and encode$\circ$decode on random parameter vectors and
random monic polynomials; and an exhaustive run of the same decoder
(\texttt{tools/bench/chainhash/\allowbreak exh5.c}) over $\mathrm{GF}(2)$,
$\mathrm{GF}(4)$, $\mathrm{GF}(8)$, $\mathrm{GF}(16)$ and $\mathrm{GF}(32)$,
i.e.\ over all $2^{5k}$ parameter vectors for $k\le5$ ($32^5=33{,}554{,}432$
for $k=5$), found no decoding failure, no two parameter vectors with the same
coefficient vector, and no point at which the gate-by-gate circuit and
Horner's rule on \eqref{eq:ph:chain5-coeffs} disagree.
\Cref{thm:main} asserts rational preprocessing (a rational right inverse on a dense set of coefficient vectors) for characteristic $0$ or
$p>n$, which does \emph{not} cover $\F=\mathrm{GF}(2^{64})$ (\Cref{rem:char2});
the circuit \eqref{eq:ph:chain5} reaches the same count
$\lfloor5/2\rfloor+1=3$ with a decoder whose pivots are all units, so the
statement holds over every field of characteristic $2$ and not only over the
perfect ones.  The point of \Cref{sec:main-theorem:kwise} is that sampling
the circuit parameters $c$ uniformly is then equivalent to sampling the
standard Wegman--Carter key of \eqref{eq:kwise:monic-poly}, the coefficient vector (written $c$ there and $e=\varphi_5(c)$ here).
\end{remark}

\begin{lemma}[Finalizer]\label{lem:ph:finalizer}
Let $c$ be uniform on $\F^5$, $\F=\mathrm{GF}(2^{64})$.
\begin{enumerate}[label=(\roman*)]
\item For fixed $v\ne v'\in\F$: $\Pr_c[f_c(v)=f_c(v')]=2^{-64}$ exactly.
\item For fixed pairwise distinct $v_1,\dots,v_t\in\F$ with $t\le5$, the vector
      $(f_c(v_1),\dots,f_c(v_t))$ is uniform on $\F^t$.
\end{enumerate}
\end{lemma}
\begin{proof}
By \Cref{lem:ph:chain}, $e=\varphi_5(c)$ is uniform on $\F^5$ and
$f_c(X)=X^5+\sum_{i<5}e_iX^i$.
(i) $f_c(v)-f_c(v')=(v-v')\,e_1+\bigl[(v^5-v'^5)+\sum_{i\ne1}e_i(v^i-v'^i)\bigr]$.
Conditioned on $(e_i)_{i\ne1}$ the bracket is a constant and $e_1\mapsto(v-v')e_1+\text{const}$
is a bijection of $\F$ ($v-v'\ne0$), so exactly one value of the uniform $e_1$
gives $0$: conditional probability $2^{-64}$, hence unconditional $2^{-64}$.
(ii) $(f_c(v_j))_{j\le t}=(v_j^5)_j+Me$ where $M=(v_j^i)_{j\le t,\,i<5}\in\F^{t\times5}$.
The $t\times t$ minor with columns $i=0,\dots,t-1$ is a Vandermonde matrix in
distinct points, hence invertible, so $M$ has rank $t$ and $e\mapsto Me$ is
surjective with all fibres of equal size $|\F|^{5-t}$; the image of the uniform
$e$ is uniform on $\F^t$, and adding the constant vector preserves uniformity.
\end{proof}

\paragraph{The twist.}
\Cref{def:ph:hash} feeds the finalizer with $V_\kappa(m)\boxplus\tau$ rather
than with $V_\kappa(m)$.  The proofs below use the finalizer only through the
following lemma, which holds for every bijection of its input.

\begin{lemma}[Bijections of the finalizer input are free]\label{lem:ph:twist}
Let $\psi:\F\to\F$ be a bijection and let $c$ be uniform on $\F^5$,
$\F=\mathrm{GF}(2^{64})$.
\begin{enumerate}[label=(\roman*)]
\item For fixed $v\ne v'\in\F$: $\Pr_c[f_c(\psi(v))=f_c(\psi(v'))]=2^{-64}$ exactly.
\item For fixed pairwise distinct $v_1,\dots,v_t\in\F$ with $t\le5$, the
      vector $\bigl(f_c(\psi(v_j))\bigr)_{j\le t}$ is uniform on $\F^t$.
\end{enumerate}
The same holds for a keyed family $\psi_\tau$ when the key $\tau$ is
independent of $c$, conditionally on every value of $\tau$.
\end{lemma}
\begin{proof}
Distinct points have distinct images under a bijection, so
$\psi(v)\ne\psi(v')$ and $\psi(v_1),\dots,\psi(v_t)$ are pairwise distinct,
and (i), (ii) are \Cref{lem:ph:finalizer}(i), (ii) at these points.  For a
keyed $\psi_\tau$ condition on $\tau$: $c$ is still uniform on $\F^5$.
\end{proof}

\begin{corollary}\label{cor:ph:twist}
For every $\tau\in\F$ the map $\psi_\tau(v)=v\boxplus\tau$ is a bijection of
$\F$, with inverse $v\mapsto v\boxminus\tau$ (subtraction modulo $2^{64}$),
so \Cref{lem:ph:twist} applies to the finalizer stage
$v\mapsto f_c(v\boxplus\tau)$ of \Cref{def:ph:hash}, $\tau$ being independent
of $c$.
\end{corollary}

\subsection{Proof of \texorpdfstring{\Cref{thm:ph:collision}}{the collision bound}}

Let $m\ne m'$ with $n(m),n(m')\le n$, $p=Sn$, and abbreviate
$\sigma=\sigma_\kappa(m)$, $\sigma'=\sigma_\kappa(m')$, $V=V_\kappa(m)$, $V'=V_\kappa(m')$.
Define the events
\[
\begin{aligned}
  E_1&=\{\sigma=\sigma'\},\qquad
  E_2=\{\sigma\ne\sigma'\ \wedge\ V=V'\},\\
  E_3&=\{V\ne V'\ \wedge\ f_c(V\boxplus\tau)=f_c(V'\boxplus\tau)\}.
\end{aligned}
\]
If $H(m)=H(m')$ then either the streams coincide ($E_1$), or they differ but
the level-2 values coincide ($E_2$), or the level-2 values differ and the
finalizer collides ($E_3$); so $\{H(m)=H(m')\}\subseteq E_1\cup E_2\cup E_3$ and
\[
  \Pr[H(m)=H(m')]\le\Pr[E_1]+\Pr[E_2]+\Pr[E_3].
\]
$E_1$ depends on $\kappa$ only; $E_2$ depends on $(\kappa,\theta)$; for $E_3$,
condition on any $(\kappa,\theta,\tau)$ with $V\ne V'$: $c$ is independent of
$(\kappa,\theta,\tau)$ and $v\mapsto v\boxplus\tau$ is a bijection
(\Cref{cor:ph:twist}), so \Cref{lem:ph:twist}(i) gives conditional
probability exactly $2^{-64}$, whence $\Pr[E_3]\le2^{-64}$.
If $n(m)=n(m')$, then $\Pr[E_1]\le2^{-64}$ by \Cref{lem:ph:stream} and
$\Pr[E_2]\le p/2^{64}$ by \Cref{lem:ph:level2}, for a total of
$(1+p+1)/2^{64}$.  If $n(m)\ne n(m')$, then $E_1$ is impossible
(\Cref{lem:ph:stream}) and $\Pr[E_2]\le(p+1)/2^{64}$ (\Cref{lem:ph:level2}),
for a total of $(0+(p+1)+1)/2^{64}$.  In both cases
$\Pr[H(m)=H(m')]\le(p+2)/2^{64}$.  In the terminology of
\Cref{def:injective:universal}, \textsc{ChainHash} is
$\frac{Sn+2}{2^{64}}$-almost universal on messages of at most $nB$ bytes. \qed

\subsection{\texorpdfstring{$5$}{5}-wise independence}

\begin{theorem}[$5$-wise independence up to level-1/2 collisions]\label{thm:ph:kwise}
Fix $W$, $S$, $n\ge1$, $p=Sn$, and $t\le5$ pairwise distinct messages
$m_1,\dots,m_t$ with $n(m_i)\le n$.  Let $\mathcal D$ be the event
(determined by $(\kappa,\theta)$) that the level-2 values
$V_\kappa(m_1),\dots,V_\kappa(m_t)$ are pairwise distinct, i.e.\ that no pair
suffers an $E_1$ or $E_2$ collision.  Then
\begin{enumerate}[label=(\roman*)]
\item $\Pr[\neg\mathcal D]\le\binom t2\,\frac{p+1}{2^{64}}$;
\item conditioned on any $(\kappa,\theta)\in\mathcal D$ and on any $\tau$,
      the vector $(H(m_1),\dots,H(m_t))$ is uniform on $\F^t$: the outputs
      are independent and uniform, in particular $5$-wise independent;
\item consequently, for every $T\subseteq\F^t$,
      $\bigl|\Pr[(H(m_1),\dots,H(m_t))\in T]-|T|/2^{64t}\bigr|\le\binom t2\frac{p+1}{2^{64}}$.
\end{enumerate}
\end{theorem}
\begin{proof}
(i) For each pair $i<j$, $\{V_\kappa(m_i)=V_\kappa(m_j)\}\subseteq E_1^{ij}\cup E_2^{ij}$
with the events of the previous proof, of probability at most
$2^{-64}+p/2^{64}$ or $0+(p+1)/2^{64}$ by \Cref{lem:ph:stream,lem:ph:level2};
take a union bound over the $\binom t2$ pairs.
(ii) $H(m_i)=f_c(V_\kappa(m_i)\boxplus\tau)$ and $c$ is independent of
$(\kappa,\theta,\tau)$; apply \Cref{lem:ph:twist}(ii) with
$\psi_\tau(v)=v\boxplus\tau$ (\Cref{cor:ph:twist}) to the distinct points
$V_\kappa(m_1),\dots,V_\kappa(m_t)$.
(iii) $\Pr[\cdot\in T]=\Pr[\mathcal D]\,|T|/2^{64t}+\Pr[\neg\mathcal D\wedge\cdot\in T]$,
and both $\Pr[\neg\mathcal D]\,|T|/2^{64t}$ and $\Pr[\neg\mathcal D\wedge\cdot\in T]$ lie in $[0,\Pr[\neg\mathcal D]]$.
\end{proof}

\subsection{The input twist}
\label{sec:ph:twist}

\Cref{def:ph:hash} applies the finalizer not to the level-2 value
$V_\kappa(m)$ but to its \emph{integer-add twist}: with one further key word
$\tau\in\F$, uniform and independent of $(\kappa,\theta,c)$,
\begin{equation}\label{eq:ph:twist-def}
  H_{\kappa,\theta,c,\tau}(m)=f_c\bigl(V_\kappa(m)\boxplus\tau\bigr),
\end{equation}
where $v\boxplus\tau$ is the addition modulo $2^{64}$ of the $64$-bit words
$v,\tau$ read as integers (bit $i$ = coefficient of $X^i$ = binary digit
$2^i$), carries and all; the key thus has $W+9$ words ($41$, $137$ and $17$
for the three configurations).  This subsection explains why the twist is
there, recalls that it costs nothing in
\Cref{thm:ph:collision,thm:ph:kwise}, and states precisely what it changes.

Throughout, $\F=\mathrm{GF}(2^{64})$ is identified with $\F_2^{64}$.  Every
map $F:\F\to\F$ has an \emph{algebraic normal form}: each of its $64$ output
bits is a unique polynomial in the $64$ input bits $v_0,\dots,v_{63}$ of
degree $\le1$ in each variable, and the \emph{$\F_2$-degree} $\deg_{\F_2}F$ is
the largest degree of these $64$ polynomials.  $F$ is $\F_2$-affine iff
$\deg_{\F_2}F\le1$; the field operations $v\mapsto v+a$, $v\mapsto av$ and the
Frobenius $v\mapsto v^2$ are $\F_2$-linear or affine, and a product of two
maps of $\F_2$-degrees $d,d'$ has $\F_2$-degree at most $d+d'$ (each output bit
of a field product is a bilinear form in the bits of the factors).

\paragraph{Why degree $5$.}
The finalizer's degree does not enter the collision bound; it buys the
$k$-wise independence of \Cref{thm:ph:kwise}.  Five is a principled stop:
$5$-wise independence is what linear probing provably needs.  Pagh, Pagh and
Ru\v{z}i\'c \cite{pagh2009linear} show that $5$-wise independent hashing
gives linear probing expected constant time per operation, and
P\u{a}tra\c{s}cu and Thorup \cite{patracscu2016kindependence} show that
$4$-wise independence does not suffice: there are $4$-wise independent
families under which linear probing takes expected $\Omega(\log n)$ time on
some inputs.  The step from degree $5$ to degree $7$ costs one field
multiplication per message, i.e.\ one quarter of the finalizer, with no
change to the theorems of this appendix except \Cref{lem:ph:chain}, whose degree-$7$ form has two Frobenius-root pivots and therefore needs a perfect field.

\paragraph{Why not degree $5$ alone: the Frobenius observation.}
The reason the earlier design used degree $7$ (\Cref{rem:ph:single-key}) is
not $k$-wise independence but a different kind of structure, which the
following observation makes precise.  For $e\ge0$ let $\mathrm{wt}(e)$ be the
number of ones in the binary expansion of $e$.

\begin{lemma}[$\F_2$-degree of polynomials over $\mathrm{GF}(2^k)$]\label{lem:ph:popcount}
Over $\F=\mathrm{GF}(2^k)$ the map $v\mapsto v^e$ has $\F_2$-degree at most
$\mathrm{wt}(e)$.  Hence every $f\in\F[X]$ of degree $\le6$ has $\F_2$-degree at most
$2$, and all its discrete derivatives $D_af(v)=f(v+a)+f(v)$ are $\F_2$-affine
in $v$.
\end{lemma}
\begin{proof}
$v^e=\prod_{i:\,e_i=1}v^{2^i}$ is a product of $\mathrm{wt}(e)$ Frobenius powers,
each $\F_2$-linear in $v$, so each of its output bits is a polynomial of
degree $\le\mathrm{wt}(e)$ in the bits of $v$; multiplication by the coefficients
and addition are $\F_2$-linear, and $\mathrm{wt}(e)\le2$ for every $e\le6$ (the first
exponent of weight $3$ is $7$).  A derivative of a degree-$\le2$ map has
degree $\le1$.
\end{proof}

(The $\F_2$-degree of $v\mapsto v^e$ is in fact exactly $\mathrm{wt}(e)$ for
$0\le e<2^k$; only the upper bound is used.)  So the degree-$3$ and
degree-$5$ circuits, whatever their parameters, are quadratic maps of the
$64$ bits of $v$, and $X^7=X^4X^2X$ is the first odd-degree monomial that is
cubic.

\paragraph{What the tests see.}
This structure is invisible to $k$-wise independence---a random monic
polynomial of degree $5$ is exactly $5$-wise independent, quadratic or
not---but SMHasher3's fixed-seed keysets are built to expose it: Zeroes
(all-zero messages of consecutive lengths), Sparse (messages of up to $1280$ bytes with few
set bits), Permutation (concatenations of up to $23$ words from fixed sets of $2$, $8$ or $15$ words), TwoBytes (all $2$--$20$-byte messages with one or two nonzero
bytes), Bitflip (a message and its single-bit flips) and SeedZeroes (zero
messages under seeds with at most two set bits) all consist of messages
that differ from one another in few bit positions of one block.  For a fixed
key and a fixed length within one block, $V_\kappa(m)=a_1+(b_1+y)(z+u)$ with
$(a_1,b_1)$ the halves of $\clnh_{\kappa}(m_1)+\ell$, and each summand
$(g_{2s}+k_{2s})(g_{2s+1}+k_{2s+1})$ of \eqref{eq:ph:clnh} is $\F_2$-biaffine
in the two message words: $V_\kappa$ is an $\F_2$-quadratic function of the
message bits, and $\F_2$-affine on any keyset that varies only one word of
each pair (one or two bytes inside a word; the length word in Zeroes within
a $16$-byte window of lengths).  Composed with an $\F_2$-quadratic finalizer,
the hash restricted to such a keyset is an $\F_2$-quadratic (in general, a
degree-$\le4$) function of a handful of Boolean variables---all its second
derivatives $h(m)+h(m+a)+h(m+b)+h(m+a+b)$ are constant over the keyset---and
the distribution tests, which histogram $8$--$22$-bit windows of the output
over the keyset, detect this reliably at the keyset sizes used (the worst
bias reported for the untwisted degree-$5$ hash was $195.9\times$ the
expected value on a $16$-bit window of the Permutation keyset).  We offer
this as the explanation of the pattern of failures, not as a theorem about
the tests.

\paragraph{Measurements.}
Full SMHasher3 suite ($200$ tests), $256$-byte configuration, M2~Pro
(\texttt{tools/bench/chainhash/twist\_results.md}); ``cycles'' is the
small-key average over $1$--$31$-byte messages:
\begin{center}
\begin{tabular}{llrl}
finalizer & twist & tests passed & cycles/hash \\\hline
degree $7$ (4 mult.) & none & $200/200$ & $95.7$ \\
degree $5$ (3 mult.) & none & $178/200$ & $82.0$ \\
degree $5$ (3 mult.) & input, $v\boxplus \tau$ & $200/200$ & $81.8$ \\
degree $5$ (3 mult.) & input and output & $200/200$ & $83.2$ \\
degree $3$ (2 mult.) & input, $v\boxplus \tau$ & $183/200$ & $66.5$ \\
degree $3$ (2 mult.) & input and output & $183/200$ & $67.4$ \\
\end{tabular}
\end{center}
The untwisted degree-$5$ hash fails $22$ tests, all in the keysets Zeroes,
Sparse, Permutation, TwoBytes and Bitflip; the twisted degree-$3$ hash fails
$17$, in Zeroes, Permutation and SeedZeroes.  An output twist $h\boxplus\tau_{\mathrm{out}}$ adds
nothing and was dropped.  The twist costs one integer addition, below the
timing noise.  The cycle counts are those of the block level and tail loader
of the time of the experiment; the optimised implementation shipped since
(\Cref{rem:ph:gaps}(A7)) computes the same function bit for bit and measures
$66$ cycles per small key ($72$ for the $1$\,KB configuration) and $20.7$
($17.5$) bytes/cycle in bulk in the same benchmark (SMHasher3's own, on $256$\,KB keys; the harness figures of \Cref{tab:injective:adversarial} use $16$\,KB messages and order the two configurations the other way).  Degree $5$ with the
input twist is therefore $15\%$ faster
than degree $7$ on short messages with the same test record; the bulk
throughput on $256$\,KB messages does not depend on the finalizer (the six
runs report $16.0$--$19.0$ bytes/cycle, varying with machine load, not with
the variant).  The $\F_2$-degree bound below
applies to degree $3$ exactly as to degree $5$ (and \texttt{anf\_check.py}
measures the same degree for both), so the $\F_2$-degree does not separate
the two; what provably separates them is $3$-wise versus $5$-wise
independence.  We do not claim to know which property the failing tests
measure.

\paragraph{The twist is free.}
Instead of over-provisioning the degree, \textsc{ChainHash} inserts a fixed
bijection between the recurrence and the finalizer.  \Cref{lem:ph:twist} and
\Cref{cor:ph:twist} say that this costs nothing: the proofs of
\Cref{thm:ph:collision,thm:ph:kwise} use the finalizer only through
\Cref{lem:ph:twist}, applied to the level-2 values with
$(\kappa,\theta,\tau)$ fixed, and they hold verbatim, with the same bounds,
for every bijection in place of $\psi_\tau$---including the identity, i.e.\
the untwisted hash.  The twist changes neither the collision bound, nor the
$5$-wise independence, nor the key size beyond one word.

\paragraph{What the twist changes.}
The XOR of a constant is $\F_2$-affine and would change nothing; integer
addition is not, because of its carry chain.

\begin{proposition}[Algebraic normal form of $v\boxplus \tau$]\label{prop:ph:twist-bits}
Write $s=v\boxplus \tau$ with bits $s_i$, and let $\gamma_i\in\F_2$ be the carry
into position $i$ ($\gamma_0=0$).  Over $\F_2$,
\begin{equation}\label{eq:ph:carry}
  s_i=v_i+\tau_i+\gamma_i,\qquad
  \gamma_{i+1}=v_i\tau_i+(v_i+\tau_i)\,\gamma_i,\qquad
  \gamma_i=\sum_{j<i}v_j\tau_j\prod_{l=j+1}^{i-1}(v_l+\tau_l),
\end{equation}
so that, for a fixed $\tau$,
\[
  s_0=v_0+\tau_0,\qquad
  s_1=v_1+\tau_1+\tau_0v_0,\qquad
  s_2=v_2+\tau_2+\tau_1v_1+\tau_0\tau_1v_0+\tau_0v_0v_1 .
\]
If $\tau\ne0$ has lowest set bit $j_0$, then $\deg_{\F_2}s_i=\max(1,\,i-j_0)$
as a function of $v$.  In particular $\psi_\tau$ is $\F_2$-affine iff $\tau=0$ or
$j_0\ge62$, and for odd $\tau$ the top bit $s_{63}$ has $\F_2$-degree $63$.
\end{proposition}
\begin{proof}
The recursion is the full adder: the sum bit is the XOR of the three inputs
and the carry out is their majority,
$v_i\tau_i+v_i\gamma_i+\tau_i\gamma_i=v_i\tau_i+(v_i+\tau_i)\gamma_i$.  Unrolling it
gives the sum in \eqref{eq:ph:carry}: $\gamma_i=1$ iff some position $j<i$
generates a carry ($v_j=\tau_j=1$) and every position strictly between $j$ and
$i$ propagates it ($v_l+\tau_l=1$); these events are disjoint for different
$j$ (generation at $j$ means $v_j+\tau_j=0$, so $j$ does not propagate a carry
from below), and a disjoint OR is a sum over $\F_2$.  The three displayed
bits are the cases $i\le2$.  For the degree, fix $\tau$ with lowest set bit
$j_0$: the summands with $j<j_0$ vanish, the summand $j=j_0$ is
$v_{j_0}\prod_{l=j_0+1}^{i-1}(v_l+\tau_l)$, which contains the monomial
$v_{j_0}v_{j_0+1}\cdots v_{i-1}$ of degree $i-j_0$, and no summand with
$j>j_0$ involves $v_{j_0}$, so this monomial survives in $\gamma_i$; every
monomial of $\gamma_i$ has degree $\le i-j_0$.  Hence
$\deg\gamma_i=i-j_0$ for $i>j_0$ and $\gamma_i=0$ for $i\le j_0$, and
$s_i=v_i+\tau_i+\gamma_i$ has degree $\max(1,i-j_0)$.  All bits are affine iff
$63-j_0\le1$.
\end{proof}

\begin{proposition}[The twisted finalizer is not quadratic]\label{prop:ph:twist-degree}
Let $\tau\ne0$ have lowest set bit $j_0\le61$, and let $h_c=f_c\circ\psi_\tau$,
$h_c(v)=f_c(v\boxplus \tau)$, with $f_c$ the degree-$5$ polynomial
\eqref{eq:ph:chain5} and $e=\varphi_5(c)$ its coefficients.
\begin{enumerate}[label=(\roman*)]
\item $\deg_{\F_2}h_c\le63$ for every $c$; indeed $\deg_{\F_2}(f\circ\psi)\le63$
      for every bijection $\psi$ of $\F$ and every $f\in\F[X]$ of degree
      $<2^{64}-1$.
\item For $2\le r\le63-j_0$, the coefficient of the monomial
      $v_{j_0}v_{j_0+1}\cdots v_{j_0+r-1}$ in the algebraic normal form of
      $h_c$ is
      \[
      \begin{aligned}
        \sum_{j<2^r}f_c\bigl(\tau\boxplus 2^{j_0}j\bigr)&=e_1\,w_r+R_r ,\\
        \text{where}\qquad
        w_r&=\bigoplus_{j<2^r}\bigl(\tau\boxplus2^{j_0}j\bigr)=2^{j_0+r}\bigl(\mu\oplus(\mu+1)\bigr)\ne0 ,
      \end{aligned}
      \]
      $\mu=\lfloor \tau/2^{j_0+r}\rfloor$, the sum $\mu+1$ and the XOR are
      taken modulo $2^{64-j_0-r}$, and $R_r$ depends on $(e_2,e_3,e_4,\tau)$
      but not on $e_1$.
\item Consequently, for uniform $c$,
      $\Pr_c\bigl[\deg_{\F_2}h_c<63-j_0\bigr]\le2^{-64}$: all but a $2^{-64}$
      fraction of the keys $c$ give $\F_2$-degree exactly $63$ when $\tau$ is
      odd, and at least $3$ whenever $j_0\le60$.  By contrast
      $\deg_{\F_2}f_c\le2$ and $\deg_{\F_2}\bigl(f_c\circ(v\mapsto v\oplus \tau)\bigr)\le2$
      for every $c$ and $\tau$ (\Cref{lem:ph:popcount}).
\end{enumerate}
\end{proposition}
\begin{proof}
We use M\"obius inversion: for $F:\F_2^{64}\to\F_2^{64}$ and $I\subseteq\{0,\dots,63\}$,
the coefficient of $\prod_{i\in I}v_i$ in the algebraic normal form of $F$ is
$\sum_{v\in\mathrm{span}\{2^i:i\in I\}}F(v)$, the sum of $F$ over the
$2^{|I|}$ vectors supported on $I$ (setting the other variables to $0$ kills
exactly the monomials that involve them).

(i) Take $I=\{0,\dots,63\}$: the coefficient of $v_0\cdots v_{63}$ is
$\sum_{v\in\F}f(\psi(v))=\sum_{x\in\F}f(x)$, and
$\sum_{x\in\F}x^e=0$ for $0\le e<2^{64}-1$ (for $e=0$ it is $2^{64}\cdot1=0$;
for $e\ge1$, with $g$ a generator of $\F^\times$, it is
$\sum_{i<2^{64}-1}g^{ie}=(g^{e(2^{64}-1)}-1)/(g^e-1)=0$, as $g^e\ne1$ for $0<e<2^{64}-1$).

(ii) Take $I=\{j_0,\dots,j_0+r-1\}$; the vectors supported on $I$ are
$v=2^{j_0}j$, $j<2^r$, and $v\boxplus \tau=\tau\boxplus2^{j_0}j$.  Write
$f_c=X^5+e_3X^3+L$ with $L(x)=e_4x^4+e_2x^2+e_1x+e_0$.  Since $x\mapsto x^4$
and $x\mapsto x^2$ are $\F_2$-linear and $e_0$ is added $2^r$ times,
$\sum_jL(x_j)=e_4w_r^4+e_2w_r^2+e_1w_r$ with $w_r=\bigoplus_jx_j$, so
$R_r=\sum_j\bigl(x_j^5+e_3x_j^3\bigr)+e_4w_r^4+e_2w_r^2$.  To compute $w_r$,
write $\tau=2^{j_0}\tau'$ with $\tau'$ odd and $\tau'=2^r\mu+\rho$, $0<\rho<2^r$, $\rho$ odd; then
$\tau\boxplus2^{j_0}j=2^{j_0}\bigl((\tau'+j)\bmod2^{64-j_0}\bigr)$, and the $2^r$
integers $\tau'+j$ are $2^r\mu+\rho,\dots,2^r\mu+2^r-1$ followed by
$2^r(\mu+1),\dots,2^r(\mu+1)+\rho-1$, with $\mu+1$ modulo $2^{64-j_0-r}$.  Their XOR is
$2^r\mu\cdot[\,2^r-\rho\text{ odd}\,]\oplus2^r(\mu+1)\cdot[\,\rho\text{ odd}\,]\oplus\bigoplus_{j<2^r}j
=2^r\bigl(\mu\oplus(\mu+1)\bigr)$, because $\rho$ and $2^r-\rho$ are odd and
$\bigoplus_{j<2^r}j=0$ for $r\ge2$.  Finally $\mu\ne\mu+1$ modulo
$2^{64-j_0-r}\ge2$, so $\mu\oplus(\mu+1)$ is a nonzero integer below
$2^{64-j_0-r}$ and $w_r=2^{j_0+r}(\mu\oplus(\mu+1))\ne0$ in $\F_2^{64}$.

(iii) By \Cref{lem:ph:chain}, $e=\varphi_5(c)$ is uniform on $\F^5$, so
$e_1$ is uniform and independent of $(e_0,e_2,e_3,e_4)$.  For $r=63-j_0$ the
coefficient in (ii) is $e_1w_r+R_r$ with $w_r\ne0$, which vanishes for
exactly one value of $e_1$; otherwise $h_c$ has a nonzero monomial of degree
$63-j_0$.  For odd $\tau$ this is degree $63$, the maximum by (i).  The
contrast statements are \Cref{lem:ph:popcount}, since $v\mapsto v\oplus \tau$
is $\F_2$-affine.
\end{proof}

The bad twist words---$\tau=0$ and those with $j_0\ge62$, for which $\psi_\tau$ is
affine and $h_c$ quadratic for every $c$---are four of the $2^{64}$, and a
uniform $\tau$ has $j_0\le60$ with probability $1-2^{-61}$.  The proposition
is silent on what the tests measure; it says that the specific structure of
\Cref{lem:ph:popcount}, on which the earlier degree-$7$ decision was based,
is removed by the twist at the cost of one addition rather than one field
multiplication.

\paragraph{Numerical check.}
\texttt{tools/bench/chainhash/anf\_check.py}
computes the algebraic normal form of the finalizer restricted to $10$ input
bits (the other $54$ bits fixed at random) by the M\"obius transform, for
$20$ random keys.  On the cube of bits $0$--$9$ with an odd $\tau$ it finds
$\F_2$-degree $2$ for the degree-$5$ circuit alone, $3$ for the former
degree-$7$ circuit, $2$ for $f_c(v\oplus \tau)$, and $10$---the maximum a
$10$-variable function can have---for $f_c(v\boxplus \tau)$, for the degree-$5$
and the degree-$3$ circuit alike; on the cube of bits $j_0,\dots,j_0+9$ for
a random $j_0\le50$ it again finds $10$, and on a cube of bits below $j_0$
(where $\psi_\tau$ is affine) it finds $2$.  The same script checks
\eqref{eq:ph:carry} and the three displayed bits exhaustively over $8$-bit
$v,\tau$, the degree formula $\max(1,i-j_0)$ for all $8$-bit $\tau\ne0$, the value
of $w_r$ on $3{,}094$ random $(\tau,r)$ pairs, and that the top coefficient of
\Cref{prop:ph:twist-degree}(ii) equals the sum $\sum_jf_c(\tau\boxplus2^{j_0}j)$ and
moves by $\Delta e_1\cdot w_r$ when $e_1$ alone is changed, on $40$ random
instances.

To summarise: the twist changes only the algebraic structure of the
finalizer as a function of the bits of its input.  The collision bound and
the $5$-wise independence are those of \Cref{thm:ph:collision,thm:ph:kwise}
for every bijection of the finalizer input (\Cref{lem:ph:twist}), and
\Cref{prop:ph:twist-degree} says what changes.  Whether that change is
adequate for a given test suite is an empirical question; for SMHasher3 the
answer was $200/200$ for degree $5$ and $183/200$ for degree $3$.

\subsection{Remarks}

\begin{remark}[What each level contributes]
The bound is $2^{-64}$ (CLNH) $+\ p/2^{64}$ (recurrence) $+\ 2^{-64}$
(finalizer), with $p=Sn$ the number of recurrence steps, i.e.\ of field
multiplications between the block level and the finalizer.  The finalizer's
degree does not enter the collision bound: it buys the $5$-wise guarantee of
\Cref{thm:ph:kwise}, at the price of three multiplications per message; the
$\F_2$-algebraic structure of the output map is the business of the twist
(\Cref{sec:ph:twist}), not of the degree.  The recurrence term is the $n/|\F|$ of \Cref{sec:injective} with
$n$ now counting sub-blocks of $B_s$ bytes rather than pairs of words, which
is why the total is far below the $(\ell/16+O(1))/2^{64}$ one would get from
the recurrence alone on the raw words.
\end{remark}

\begin{remark}[The single-key variant]\label{rem:ph:single-key}
An earlier version of \textsc{ChainHash} used the single-key specialisation
\eqref{eq:injective-single-key}, $P_0=x$, $P_t=a_t+(b_t+x^3)(P_{t-1}+x^2)$,
with $S=1$ and a finalizer of degree $3$ or $5$.  Its level-2 term was
$(3p-1)/2^{64}$ for equal block counts (the three leading coefficients of the
univariate polynomial are message-independent) and $(3p+2)/2^{64}$ otherwise,
for totals of $(3p+1)/2^{64}$ and $(3p+3)/2^{64}$.  The three resident key
words that held $x,x^2,x^3$ now hold $u,y,z$, and the total-degree bound of
\Cref{lem:ph:injective}(iii) replaces the univariate one, dividing the
level-2 term by about three.  The low-degree finalizers were dropped for the
reason given in \Cref{sec:ph:twist}: over $\mathrm{GF}(2^{64})$ the
$\F_2$-degree of $v\mapsto v^e$ is the number of ones in the binary expansion
of $e$, so a polynomial of degree $3$ or $5$ is quadratic over $\F_2$ and the
corresponding hashes fail SMHasher3's distribution tests
(\Cref{lem:ph:popcount}).  The design then used, for a while, a degree-$7$
finalizer evaluated with four multiplications (the first odd degree whose
leading monomial is cubic), with a decoder whose pivots included two
Frobenius roots; the present degree-$5$ circuit with the input twist reaches
the same test record with three multiplications and unit pivots only.
\end{remark}

\begin{remark}[The length XOR]
Unlike CLHASH, which folds the length in with a separate random word
$k''\star|M|$ \cite[Alg.~4]{lemire2015clhash}, \textsc{ChainHash} XORs the
raw byte length into $a_p$ without any key.  This is sound because, for
$\ell\ne\ell'$ and equal block counts (\Cref{lem:ph:stream}):
(a) when the padded last sub-blocks coincide, the streams differ
\emph{deterministically} (case~(a)) and the recurrence is injective on streams;
(b) when they have the same pair count but differ, the relevant event is the
full-width XOR-universality of \Cref{lem:ph:clnh}(i) with the constant
$C=(\ell+\ell')+X^{64}\cdot0$ (case~(b)); and
(c) when they have different pair counts (case~(c)) the key product
$(g'_{2s}+k'_{2s})(g'_{2s+1}+k'_{2s+1})$ of each extra pair does not cancel and
vanishes with probability about $2^{-63}$, so the sum is \emph{not}
$2^{-64}$-XOR-universal against $C=0$; it is exactly the nonzero constant
$\ell+\ell'$ supplied by the length XOR that makes \Cref{lem:ph:clnh}(ii)
apply.  Without the length XOR, pair granularity would cost about a factor
of two in the level-1 term for such pairs, and $m$ and $m\,\|\,\texttt{0x00}$
(with $16\nmid\ell(m)$) would collide outright.  The XOR must land in a
component of the stream that is fed to the injective recurrence; it does.
\end{remark}

\begin{remark}[The keys range over the whole field]
No lemma needs $u$, $y$ or $z$ to be nonzero; the Schwartz--Zippel count in
\Cref{lem:ph:sz} is over all of $\F^3$ and includes the degenerate keys
(e.g.\ $y=0$, where the recurrence becomes $P_t=a_t+b_t(P_{t-1}+u)$) among the
at most $d\,|\F|^2$ zeros.  Restricting the keys to $\F^\times$ would give the
slightly worse $d/(2^{64}-1)$.
\end{remark}

\begin{remark}[Assumptions, and the gap between the theorem and the code]\label{rem:ph:gaps}
\begin{enumerate}[label=(A\arabic*)]
\item \emph{Ideal key.}  \Cref{thm:ph:collision,thm:ph:kwise} are about the family
      indexed by a uniform $(\kappa,\theta,c,\tau)\in\F^{W+9}$ ($8(W+9)$ random
      bytes; $328$, $1{,}096$ and $136$ bytes for $W=32,128,8$).  The
      implementation derives $(\kappa,\theta,c,\tau)$ from a $64$-bit seed with
      \texttt{splitmix64}; the resulting $2^{64}$-member subfamily is not
      covered by the theorems (no pseudorandomness assumption is made or
      needed for the ideal-key statements; cf.\ the counting argument of
      \cite[\S2.5]{lemire2015clhash}).
\item \emph{Exact field arithmetic.}  \texttt{gfmul} is assumed to be the field
      product in $\F_2[X]/(\Pi)$ (irreducibility of $\Pi$: \cite[\S4.3]{lemire2015clhash},
      re-checked in \texttt{tools/bench/chainhash/sanity.py}); the fast and reference
      multiplies are compared in test T7, and the fast hash against the
      bit-serial reference in tests T1/T4 of \texttt{test\_\allowbreak chainhash.cpp}.
      Equivalence of \texttt{chainhash.h} with \Cref{def:ph:hash} is tested, not proved.
\item \emph{Level 1 citation.}  \cite{lemire2015clhash} states XOR-universality
      for the \emph{reduced} $64$-bit CLNH and $2^{-64}$-almost universality for
      the $128$-bit output; the full-width XOR-universality used here is
      \Cref{lem:ph:clnh}, proved above (same integral-domain argument as their Lemma~5).
\item \emph{Level 2 citation.}  \Cref{lem:injective:coeffmap} of \Cref{sec:injective}
      is used in the restated form \Cref{lem:ph:injective}, whose indicator
      $\iver{m\ge3}$ in \eqref{eq:ph:pivot}--\eqref{eq:ph:beta1} is the
      parenthetical remark in that proof that the constant $1$ in the
      $y^{n-2}$ coefficient of $f_3$ is present only for $n\ge3$.  The
      total-degree bound $n/|\F|$ stated at the end of that proof is
      \Cref{lem:ph:injective}(iii) with \Cref{lem:ph:sz}.
\item \emph{Level 3 citation.}  \Cref{thm:main} does not apply in characteristic $2$;
      \Cref{lem:ph:chain} proves the needed bijectivity for the degree-$5$
      circuit directly, and \Cref{lem:ph:twist} carries it through the twist.  Extending \textsc{ChainHash} to another degree needs
      a characteristic-$2$ bijective circuit of that degree with its own
      decoder (\Cref{rem:char2}).
\item \emph{Message model.}  Messages are byte strings; the length is XORed as
      a $64$-bit value, so $\ell<2^{64}$.  $n$ counts $B$-byte blocks and
      $p=Sn$ counts sub-blocks; a message of $\ell$ bytes costs
      $\lceil\ell/16\rceil$ carry-less products at level 1, $p(m)$ field
      multiplications at level 2 and $3$ in the finalizer (plus one integer
      addition for the twist).  The empty message
      is one block of $S$ empty sub-blocks, each with CLNH sum the empty sum
      $0$, hashed with $\ell=0$ (so every stream pair is $(0,0)$).
\item \emph{Evaluation schedule.}  \texttt{chainhash.h} computes
      \Cref{def:ph:hash} exactly (bit for bit against the reference, tests
      T1/T4) but orders the field operations for latency.  Every field
      element lives in lane~$0$ of a NEON register, and every XOR that follows
      a multiplication (the $a_t$ of a recurrence step, the finalizer
      constants) is folded into that product's reduction.  Blocks
      $1,\dots,n-1$ run without length logic on the shifted state
      $Q_t=P_t+u$, i.e.\ $Q_t=(a_t+u)+(b_t+y)Q_{t-1}$, with the \texttt{EXT}
      that positions $b_t+y$ kept off the loop-carried chain; the last block
      is peeled and its final step takes $(a_p+\ell,\,b_p+y)$, yielding
      $P_p$ itself.  A message of at most one sub-block ($256$ or $512$
      bytes) takes a loop-free leaf path in which the key-only parts of the
      $S$ steps are precomputed constants of the key; for $S=2$ the two steps
      are fused into $P_2=(\ell+yu+y^2(z+u))+ay+by(z+u)$, two independent
      products and one reduction.  For sub-blocks of at most $16$ words the
      recurrence state is carried unreduced as a $128$-bit polynomial and
      reduced once at the end.  In the $256$-byte configuration the block
      key is register-resident and a block costs $56$ SIMD instructions and
      $8$ paired loads.  None of this changes the function.
\end{enumerate}
\end{remark}

\end{document}